\documentclass[journal]{IEEEtran}
\IEEEoverridecommandlockouts
\usepackage{cite}
\usepackage{amsmath,amssymb,amsfonts}
\usepackage{algorithmic}
\usepackage{graphicx}
\usepackage{textcomp}
\usepackage{xcolor}
\usepackage{graphicx}
\usepackage{epstopdf}
\usepackage{float}
\usepackage[tight,TABTOPCAP]{subfigure}
\usepackage{multirow}
\usepackage{diagbox} 
\usepackage{makecell}
\usepackage{amsmath}
\usepackage{array}
\usepackage{tabu}
\usepackage{url}
\usepackage{threeparttable}
\usepackage[T1]{fontenc}
\usepackage{xcolor,cite,etoolbox}
\makeatletter
\pretocmd\@bibitem{\color{black}\csname keycolor#1\endcsname}{}{\fail}
\newcommand\citecolor[1]{\@namedef{keycolor#1}{\color{red}}}
\makeatother

\newcolumntype{I}{!{\vrule width 3pt}}
\newlength\savedwidth

\newlength\savewidth

\def\BibTeX{{\rm B\kern-.05em{\sc i\kern-.025em b}\kern-.08em
    T\kern-.1667em\lower.7ex\hbox{E}\kern-.125emX}}

\begin{document}

\title{Gradient-based Feature Extraction From Raw Bayer Pattern Images}

\author{Wei Zhou,~\IEEEmembership{Student Member,~IEEE}, Ling Zhang, Shengyu Gao and Xin Lou,~\IEEEmembership{Member,~IEEE}
\thanks{This work was supported in part by the Natural Science Foundation of China (No. 61801292) and in part by the Shanghai Sailing Program (No. 18YF1416600). (Corresponding author: Xin Lou.)

The authors are with the School of Information Science and Technology, ShanghaiTech University, Shanghai 201210, China. W. Zhou is also with the Shanghai Institute of Microsystem and Information Technology, Chinese Academy of Sciences and University of Chinese Academy of Sciences.

This paper has supplementary downloadable material available at http://ieeexplore.ieee.org., provided by the author. The material includes profiling results of openISP framework in \cite{RN66} and detailed description of transform matrix $H$ in key point matching experiments. Contact \{zhouwei1, zhangling, gaosy, louxin\}@shanghaitech.edu.cn for further questions about this work.
}}

\maketitle

\begin{abstract}
In this paper, the impact of demosaicing on gradient extraction is studied and a gradient-based feature extraction pipeline based on raw Bayer pattern images is proposed. It is shown both theoretically and experimentally that the Bayer pattern images are applicable to the central difference gradient-based feature extraction algorithms with negligible performance degradation, as long as the arrangement of color filter array (CFA) patterns matches the gradient operators. The color difference constancy assumption, which is widely used in various demosaicing algorithms, is applied in the proposed Bayer pattern image-based gradient extraction pipeline. Experimental results show that the gradients extracted from Bayer pattern images are robust enough to be used in histogram of oriented gradients (HOG)-based pedestrian detection algorithms and shift-invariant feature transform (SIFT)-based matching algorithms. By skipping most of the steps in the image signal processing (ISP) pipeline, the computational complexity and power consumption of a computer vision system can be reduced significantly.
\end{abstract}


\section{Introduction}
\label{section:sec1}
\begin{figure}[t]
\centering
\subfigure[]{
\label{fig:mosiacl} 
\includegraphics[width=1in]{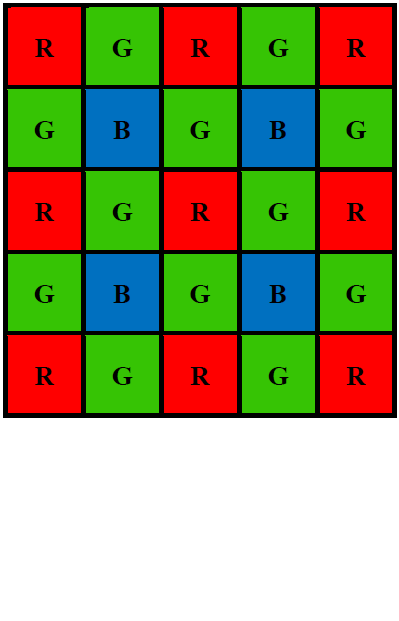}}
\subfigure[]{
\label{fig:ISP} 
\includegraphics[height=2in]{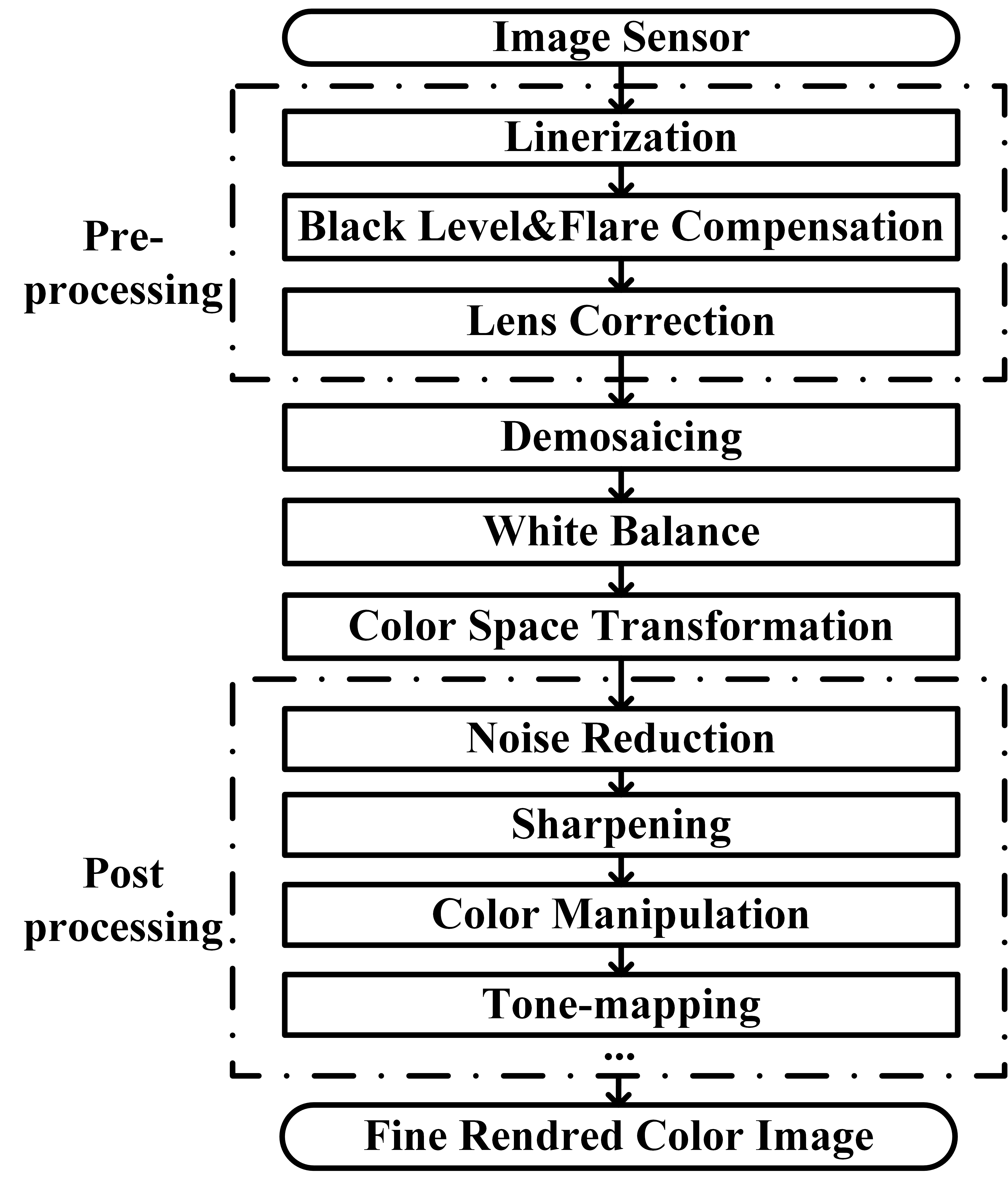}}
\caption{(a) The RGGB Bayer CFA pattern. (b) The conventional ISP pipeline.}
\label{fig:mosaic and ISP}
\vspace{-3mm}
\end{figure}

Computer vision studies how to extract useful information from digital images and videos to obtain high-level understanding. As an indispensable component, image sensors convert the outside world scene to digital images that are consumed by computer vision algorithms. To produce color images, the information from three channels, i.e., red (R), green (G) and blue (B), are needed. There are two primary technology families used in today’s color cameras: the mono-sensor technique and the three-sensor technique. Although three-sensor cameras are able to produce high-quality color images, their popularity is limited by the high manufacturing cost and large size \cite{RN1}. As an alternative, the mono-sensor technique is employed in most of the digital color cameras and smartphones nowadays. In a mono-sensor color camera, images are captured with one sensor covered by a color filter array (CFA), e.g., the Bayer pattern\cite{RN28} shown in Fig. \ref{fig:mosiacl}, such that only one out of three color components is captured by each pixel element. This single channel image is converted to a color image by interpolating the other two missing color components at each pixel. This process is referred to as demosaicing, which is a fundamental step in the traditional image signal processing (ISP) pipeline. Apart from the demosaicing step, other ISP stages are usually determined by the manufacturers according to the application scenarios\cite{RN4}.

Almost all the existing computer vision algorithms take images processed by the ISP pipeline as inputs. However, the existing ISP pipelines are designed for photography with a goal of generating high-quality images for human consumption. Although pleasing scenes can be produced, no additional information is put in by the ISP. In addition, it has been shown that the ISP pipeline may introduce cumulative errors and undermine the original information from image sensors\cite{RN14}. For example, as the demosaicing process smoothes the image, the information entropy of the image decreases\cite{RN17}. Moreover, it has been shown that ISP algorithms are computation intensive and consume a significant portion of processing time and power in a computer vision system \cite{RN55,RN23}. Profiling statistics of major steps in an ISP pipeline was presented in \cite{RN23}, which show that the demosaicing step involves a lot of memory access (which may be a bottleneck) and the denoising steps consumes more computation than others (see supplemental material). If certain ISP steps are not necessary, we can skip them to reduce the computational complexity and power consumption of the system. Therefore, for computer vision applications, the configuration or even the necessity of the complete ISP pipeline needs to be reconsidered.

The optimal configuration of the ISP pipeline for different computer vision applications remains an open problem\cite{RN23,RN56,RN57}. In a recent paper, Buckler et. al. use an empirical approach to study the ISP's impact on different vision applications\cite{RN23}. Extensive experiments based on eight existing vision algorithms are conducted and a minimal ISP pipeline consisting of denoise, demosaicing and gamma compression is proposed. But all the conclusions in \cite{RN23} are drawn based on experimental results without detailed theoretical analysis. There are also some studies that try to bypass the traditional ISP and extract the high-level global features such as edge and local binary pattern (LBP), from Bayer pattern images \cite{RN3,RN10,RN45}. Moreover, it is experimentally shown in\cite{RN20} and\cite{RN22} that the Bayer pattern images can be applied directly in some local feature descriptors such as scale-invariant feature transform (SIFT) and speeded up robust features (SURF) with negligible performance degradation.

It is noted that all the aforementioned works are experiment based, such that the applicability of their results to other vision algorithms is unclear. The basic analysis of extracting gradient-based feature from raw Bayer images is introduced in \cite{RN65}. In this paper, the impact of demosaicing on gradient-based feature extraction is studied. It is shown both theoretically and experimentally that the raw Bayer pattern images are applicable to the central difference gradient-based feature extraction algorithms with negligible performance degradation. Therefore, instead of demosaicing the Bayer pattern images before gradient computation, we propose to extract gradients directly from the Bayer pattern images by taking advantage of the color difference constancy assumption, which is widely used in demosaicing algorithms.

The reminder of the paper is organized as follows. Section \ref{section:sec2} presents the background information, including the ISP pipeline and several gradient-based high-level vision features. Section \ref{section:sec3} presents the derivation of the gradient-based feature extraction from the Bayer pattern images. Experimental results are presented in Section \ref{section:sec4} followed by the discussion in Section \ref{section:sec5} and conclusions in Section \ref{section:sec6}.

\section{Background}
\label{section:sec2}
\subsection{The Conventional ISP Pipeline}
\label{section:sec2.1}

Shown in Fig. \ref{fig:ISP} is an ISP pipeline from Adobe DNG converter \cite{RN36}. Although the specific algorithms and their orders may vary for different manufacturers, the basic steps in Fig. \ref{fig:ISP} are usually covered. The details of the functionality of each step is illustrated in Table \ref{tab:ISP}.

\subsection{Demosaicing}
\label{section:sec2.2}
Demosaicing is a crucial step to convert a single-channel Bayer pattern image to a three-channel color image by interpolating the other two missing color components at each pixel. It has a decisive effect on the final image quality. In order to minimize the color artifacts, sophisticated demosaicing algorithms are always computation hungry.

The problem of demosaicing a Bayer pattern image has been intensively studied in the past decades and a lot of algorithms have been proposed \cite{RN29,RN30,RN7,RN6}. All these algorithms can be grouped into two categories. The first category considers only the spatial correlation of the pixels and interpolates the missing color components separately using the same color channel. Although these single-channel interpolation algorithms may achieve fairly good results in the low frequency (smooth) regions, they always fail in the high-frequency regions, especially in the areas with rich texture information or along the edges\cite{RN1}.

To improve the demosaicing performance, the other category of algorithms takes the nature of the images' high spectral inter-channel correlation into account. Almost all these algorithms are based on either the color ratio constancy assumption \cite{RN7}  or the color difference constancy assumption \cite{RN6}. According to the color image model in \cite{RN7}, which is a result of viewing Lambertian non-flat surface patches, the three color channels can be expressed as 
\begin{equation}
\label{equ:three chaneel}
I^{k}\left (x, y\right )=\rho _{k}\left ( x, y  \right )\left \langle \overrightarrow{N}\left (x, y\right ),\overrightarrow{l} \right \rangle,
\end{equation}
where $ \rho $  is the reflection coefficient, $\overrightarrow{N}(x,y)$ is the surface's normal vector at location $(x,y)$, $\overrightarrow{l}$ is the incident light vector, $I(x,y)$ is the intensity at location $(x,y)$ and $ k\subseteq \left \{ R,G,B \right \}$ indicates one of the three channels. Note that a Lambertian surface is equally bright from all viewing directions and does not absorb any incident light\cite{RN62}.

At a given pixel location, the ratio of any two color components, denoted by $k$ and $k'$, is given by
\begin{equation}
\label{equ:color ratio constancy assumption1}
\frac{I^{k}\left ( x, y\right )}{I^{{k}'}\left ( x, y \right )} = \frac{\rho _{k}\left (x, y  \right )\left \langle \overrightarrow{N}\left (x, y  \right ),\overrightarrow{l} \right \rangle}{\rho _{{k}'}\left ( x, y \right )\left \langle \overrightarrow{N}\left (x, y  \right ),\overrightarrow{l} \right \rangle}= \frac{\rho _{k}\left (x, y  \right )}{\rho _{{k}'}\left (x, y  \right )}.
\end{equation}
Suppose that objects are made up of one single material, i.e., the reflection coefficient $ \rho $  for each channel is a constant, the ratio of $\rho _{k}\left (x, y  \right )/\rho _{{k}'}\left (x, y  \right )$ reduces to a constant, such that \eqref{equ:color ratio constancy assumption1} can be simplified as
\begin{equation}
\label{equ:color ratio constancy assumption2}
\frac{I^{k}\left ( x, y \right )}{I^{{k}'}\left ( x, y \right )} = constant.
\end{equation}
Equation \eqref{equ:color ratio constancy assumption2} is referred to as color ratio constancy. In the same manner, the color difference constancy assumption is given by
\begin{equation}
\label{equ:color difference constancy assumption}
\begin{split}
&\!I^{k}\left ( x, y \right )\!-\!I^{{k}'}\left ( x, y \right )\! \\= &\!\rho _{k}\left (x, y  \right )\left \langle \overrightarrow{N}\left (x, y  \right ),\overrightarrow{l} \right \rangle\! -\!\rho _{{k}'}\left (x, y  \right )\left \langle \overrightarrow{N}\left (x, y  \right ),\overrightarrow{l} \right \rangle \!\\=&\!\left [ \rho _{k}\left ( x, y \right )\!-\!\rho _{{k}'}\left ( x, y \right ) \right ] \left \langle \overrightarrow{N}\left ( x, y \right ),\overrightarrow{l} \right \rangle\! \\=&C(x,y).
\end{split}
\end{equation}
Note that the direction and amplitude of the incident light are assumed to be locally constant, such that the color component difference $C(x,y)$ is also a constant within a neighborhood of $(x,y)$\cite{RN1}.

The color ratio and difference constancy assumptions are widely used in various demosaicing algorithms \cite{RN13}. In practical applications, the color difference constancy assumption always is preferred due to its superior peak signal to noise ratio (PSNR) performance. As will be shown, in this work, the color difference constancy can be utilized to directly extract the gradient information from the Bayer pattern images.

\begin{table}[t]
\caption{The Functionalities of the Steps in ISP Pipeline }
\label{tab:ISP}
\begin{center}
\setlength\arrayrulewidth{0.6pt}
\renewcommand{\arraystretch}{1.2}
\setlength{\tabcolsep}{2.8pt}{
\begin{tabular}{|c|m{6cm}|}
\hline
\textbf{ISP Steps}                 &\textbf{\makecell[c]{Functionality}}                                                                                                \\ \hline
Linerization                    &  To transform the raw data into linear space.                                                                 \\ \hline
\makecell[c]{Black Level \& \\Flare Compensation} &To compensate the noises contributed by black level current and flare.                                      \\ \hline
Lens Correction                 & To compensate lens distortion and uneven  light fall.                                                     \\ \hline
Demosaicing                    & To convert a single-channel Bayer pattern image a three-channel color image.                           \\ \hline
White Balance                   & To remove unrealistic color casts such that white objects are rendered white.                                     \\ \hline
\makecell[c]{Color Space \\Transformation}     & To transform the camera color space to a standard color space. \\ \hline
Noise Reduction                 & To suppress noises introduced in preceding steps.                                                            \\ \hline
Sharpening                      & To enhance the edges for clarity improvement.                                         \\ \hline
Color Manipulation              & To generate different styles of photos.                        \\ \hline
Tone-mapping                    & To compress the dynamic range of images while preserving the visual effect.                                      \\ \hline
\end{tabular}}
\end{center}
\vspace{-3mm}
\end{table}

\begin{figure}[t]
\centering
\subfigure[]{
\label{fig:Gradient operators:center} 
\includegraphics[width=1in]{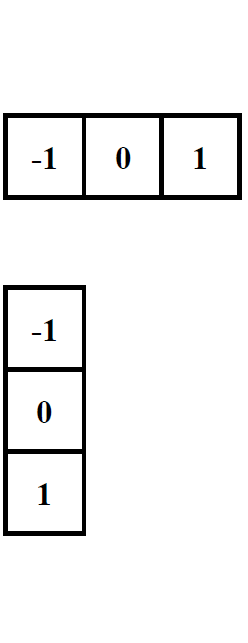}}
\hspace{10mm}
\subfigure[]{
\label{fig:Gradient operators:sobel} 
\includegraphics[width=1in]{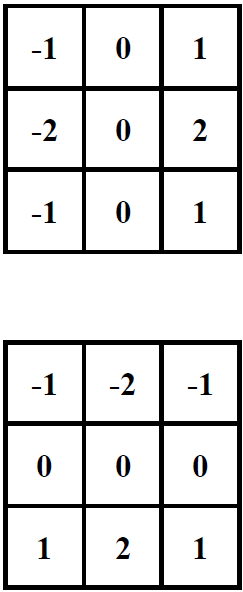}}
\caption{Gradient operators. (a) The central difference operator and (b) the Sobel operator.}
\label{fig:Gradient operators}
\vspace{-3mm}
\end{figure}

\subsection{High-level Features}

In the past decades, many different feature descriptors such as Harr-like features \cite{RN31}, LBP \cite{RN32}, SIFT \cite{RN9} and histograms of oriented gradients (HOG) \cite{RN24} have been proposed for object detection. In this work, we mainly focus on the central difference gradient-based feature descriptors, study their applicability on Bayer pattern images and analyze the corresponding performances. Without loss of generality, HOG and SIFT are taken as examples in the analysis and experiments. The results can be extended to other descriptors, such as SURF \cite{RN33}, Color-SIFT \cite{RN11}, Affine-SIFT \cite{RN34} and F-HOG \cite{RN40}, as long as the central difference is used for gradient computation.

SIFT is a local feature descriptor which detects key points in images. The computation of SIFT can be divided into five steps\cite{RN19} as
\begin{enumerate}
\item Scale space construction. The scale space is approximated by the difference-of-Gaussian (DoG) pyramid, which is computed as
\begin{equation}
\label{equ:DOG}
\begin{split}
D\left ( x,y,\sigma \right )&\!=\left( G\left ( x,y,l^{i} \sigma  \right )\!-\!G\left ( x,y,l^{i-1} \sigma  \right )\right )\!\ast \!I(x,y)\\
                             &\!=\!L\left ( x,y,l^{i} \sigma  \right )\!-\!L\left ( x,y,l^{i-1} \sigma  \right ).
\end{split}
\end{equation}
Here, $G\left ( x,y,\sigma  \right )=\frac{1}{2\pi\sigma^{2}}e^{-\frac{(x^{2}+y^{2})}{2\sigma^{2}}}$ is the Gaussian function, $l=2^{\frac{1}{s}}$ is a constant multiplicative factor whose value is determined by the number of scales $s$, $\ast$ denotes the convolution operator, $i$ indicates the $i$-th layer in DoG pyramid and $L\left ( x,y,l^{i} \sigma\right )$ is the convolution of the original image with the Gaussion function at scale $l^{i}\sigma$.
\item Extremum detection. To detect the local maxima and minima by comparing each pixel with its neighbors in a $3\times3$ neighbourhood among the current scale, scale above and scale below.
\item Key point localization. To perform a refinement of key point candidates identified in the previous step. The unstable key points such as points with low contrast or poorly localized along an edge are rejected.
\item Orientation determination. To assign one or more orientations to each key point. A histogram is created for a region centered on the key point with radius of $3\sigma _{0}$, where $\sigma _{0}$ is 1.5 times that of the scale of the key point. The direction with the highest bar in the histogram is regarded as the dominant direction and directions with heights of larger than 80\% of the highest bar is regarded as the auxiliary directions.
\item Key point description. To construct a descriptor vector for each key point. A gradient histogram with 8 bins is created for each $16\times16$ pixel region around the key point. The key point descriptor is constructed by concatenating the histograms of a set of $4\times4$ regions around the key point.
\end{enumerate}

HOG is a feature descriptor initially proposed for pedestrian detection \cite{RN24}. It counts the number of occurrences of gradient orientation in a detection window. The key steps of HOG feature generation are similar to steps 4 and 5 in the SIFT descriptor. The main difference is that orientation histograms in HOG are usually computed on an $8 \times 8$ cell and summarized as a global feature by a sliding window.

\begin{figure}
\centering
\hspace{-5mm}
\subfigure[]{
\label{fig:feacture exaction conv} 
\includegraphics[width=1.3in]{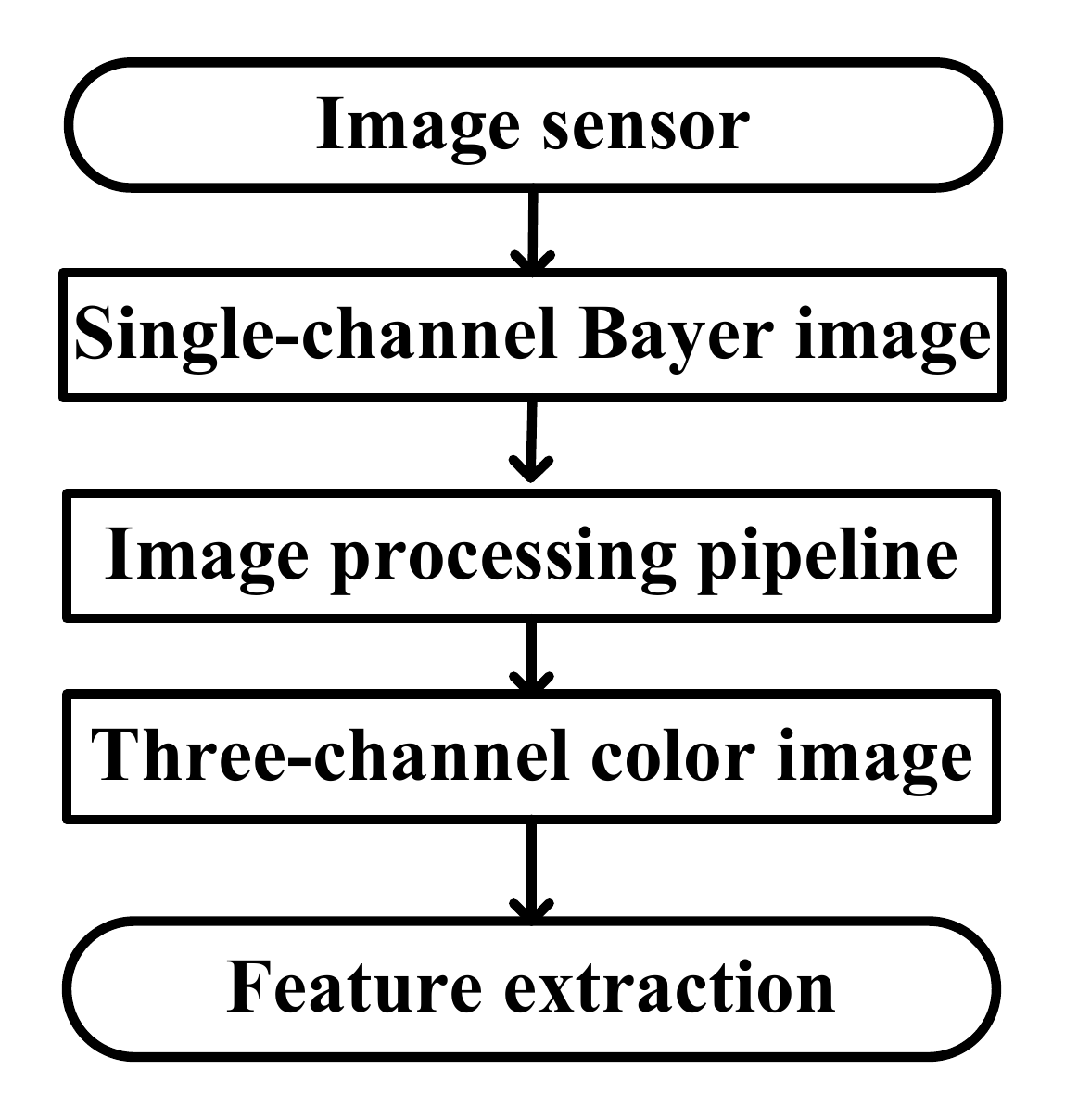}}
\hspace{1mm}
\subfigure[]{
\label{fig:feacture exaction mine} 
\includegraphics[width=1.5in]{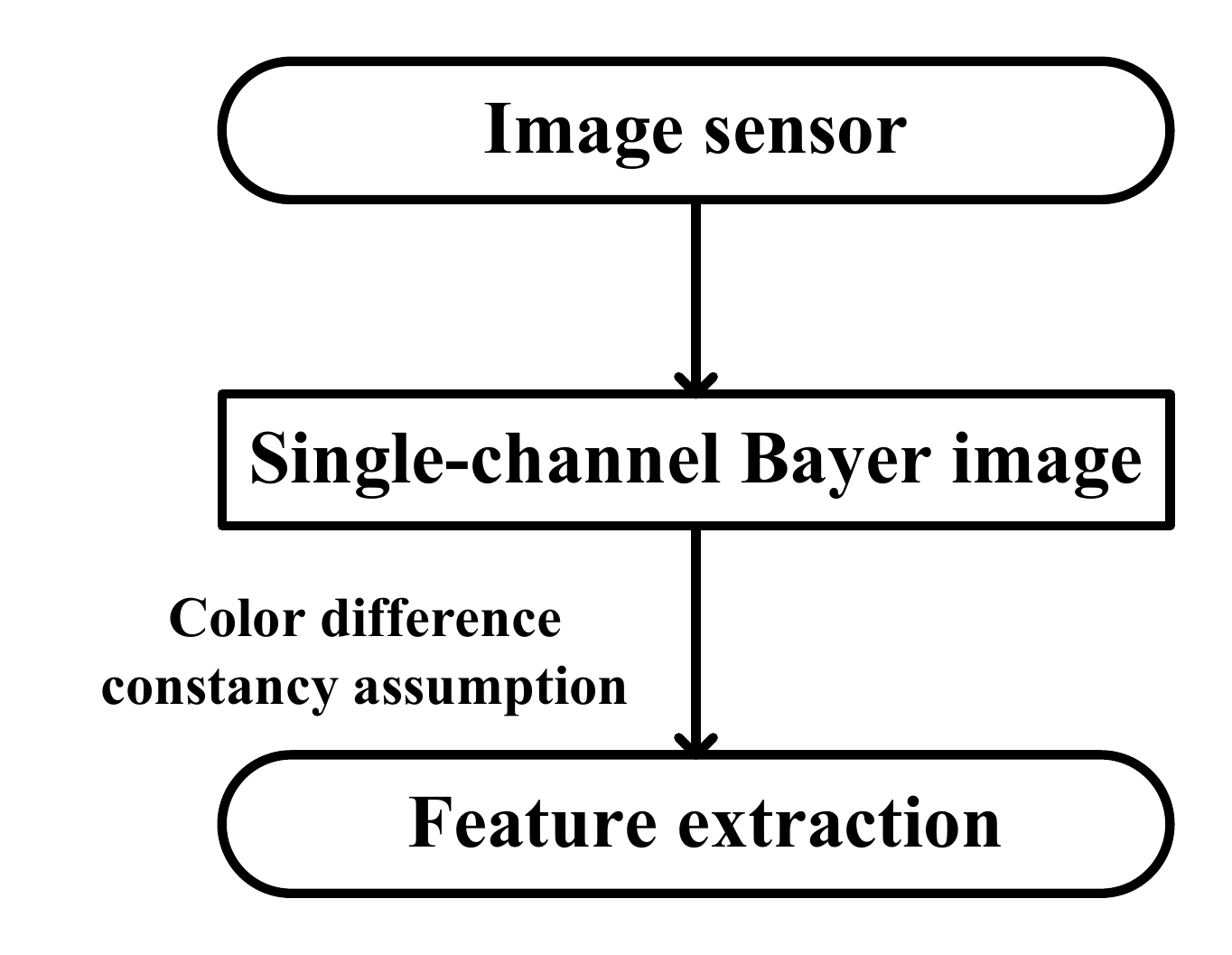}}
  \caption{Feature extraction pipelines. (a) The conventional pipeline. (b) The proposed pipeline.}
  \label{fig:feacture exaction}
  \vspace{-3mm}
\end{figure}

\begin{figure}[t]
\centering
\includegraphics[width=2in]{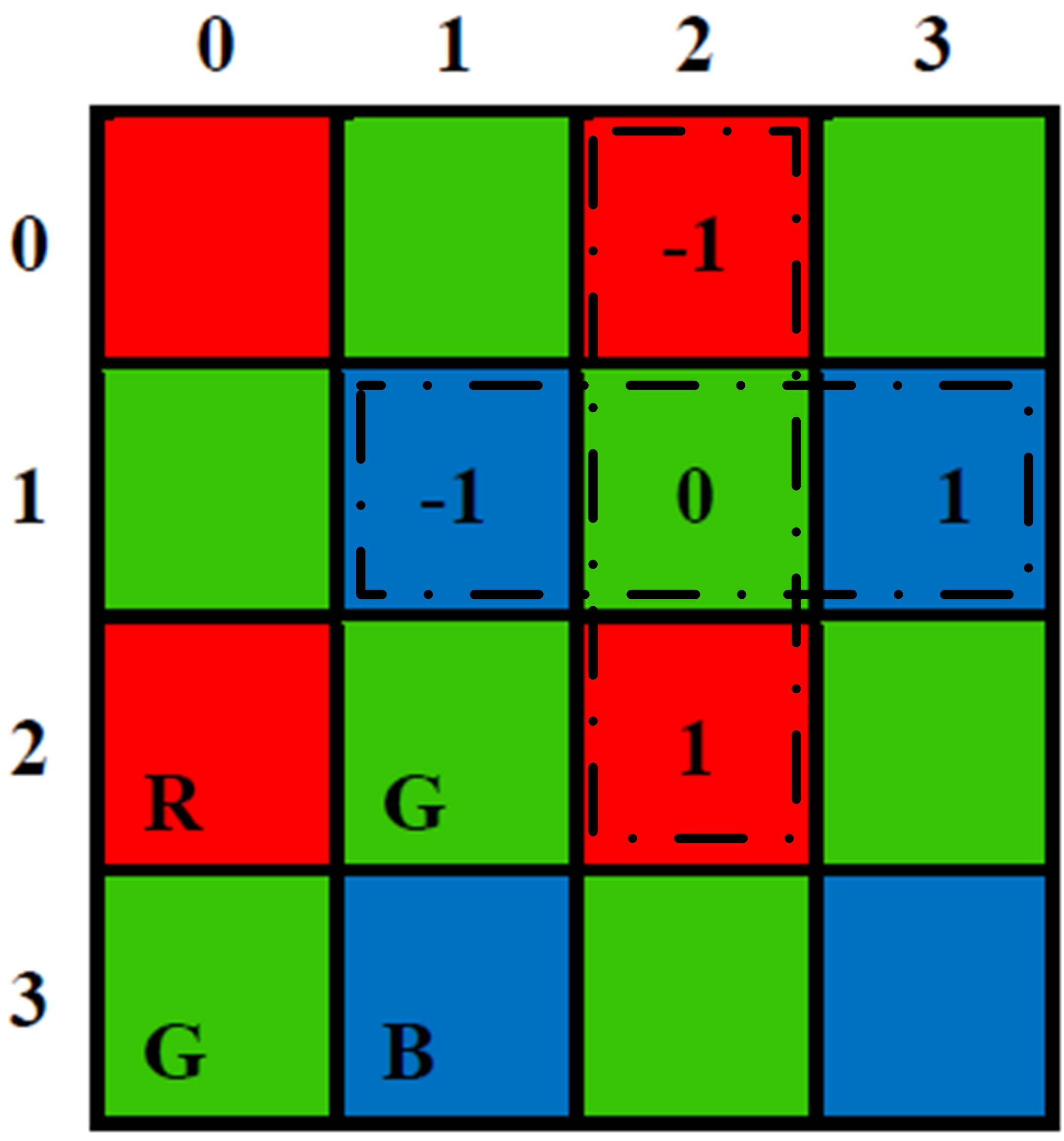}
\caption{Gradient computation based on Bayer pattern image.}
\label{fig:gradientCompute}
\end{figure}

\begin{figure*}[t]
\centering
\subfigure[]{
\label{fig:edgeCase:orig} 
\includegraphics[width=1.4in]{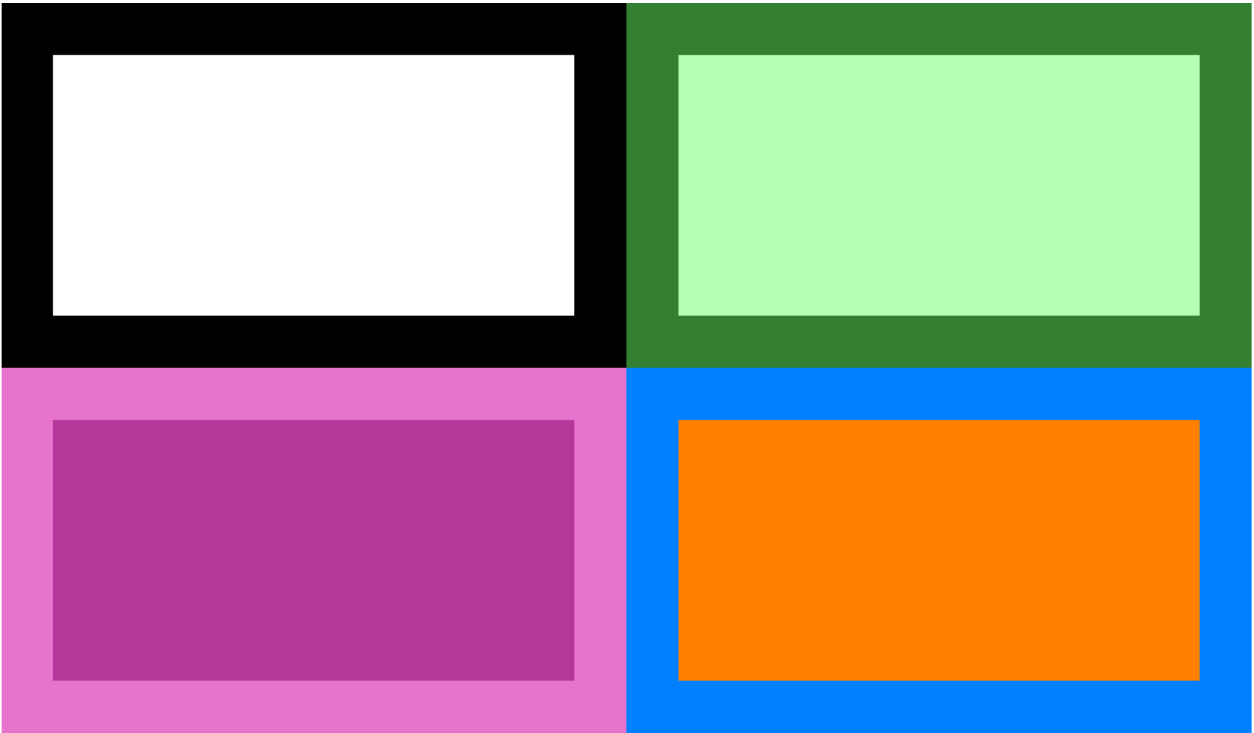}}
\hspace{-3mm}
\subfigure[]{
\label{fig:edgeCase:G_R} 
\includegraphics[width=1.4in]{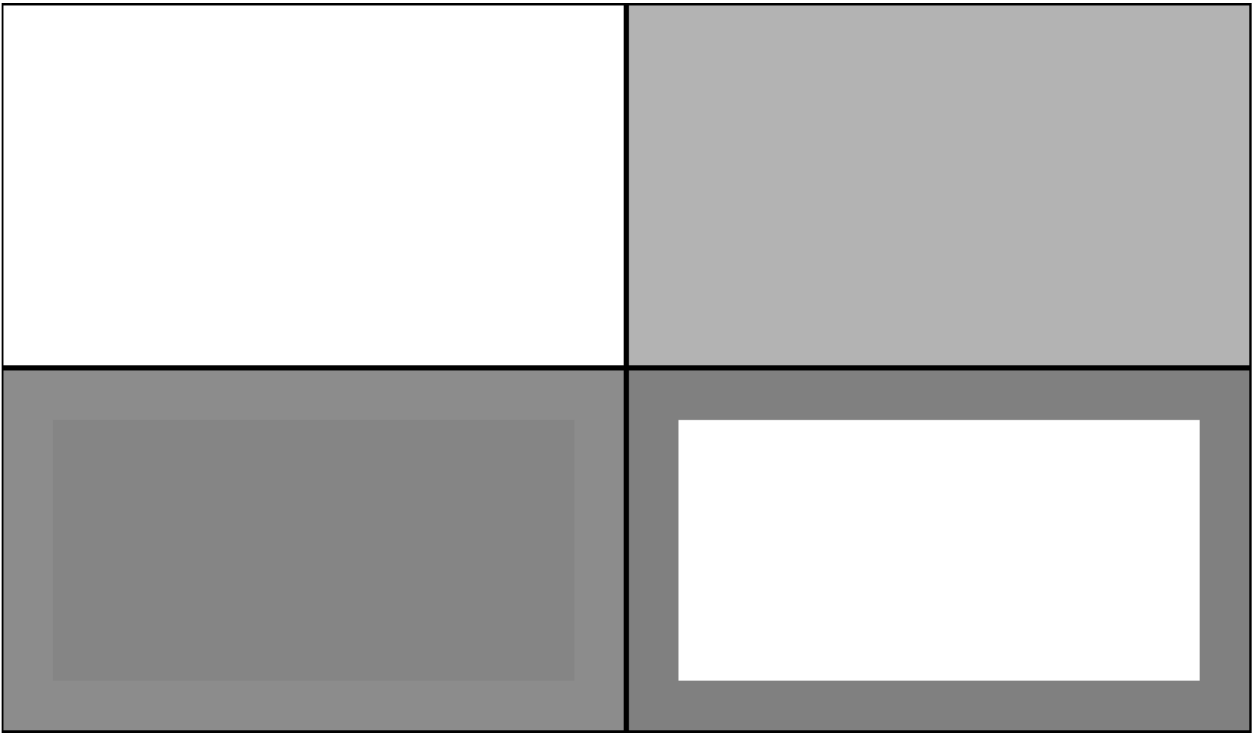}}
\hspace{-3mm}
\subfigure[]{
\label{fig:edgeCase:G_B} 
\includegraphics[width=1.4in]{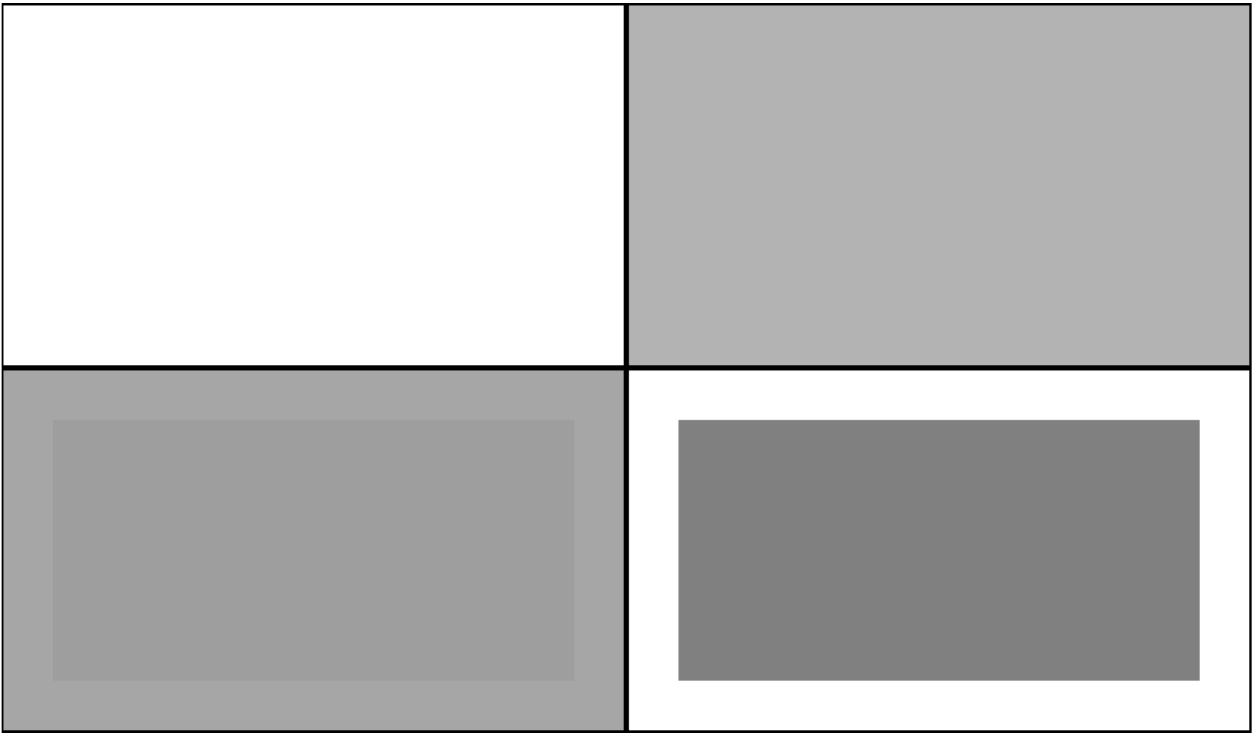}}
\hspace{-3mm}
\subfigure[]{
\label{fig:edgeCase:G_R_m} 
\includegraphics[width=1.4in]{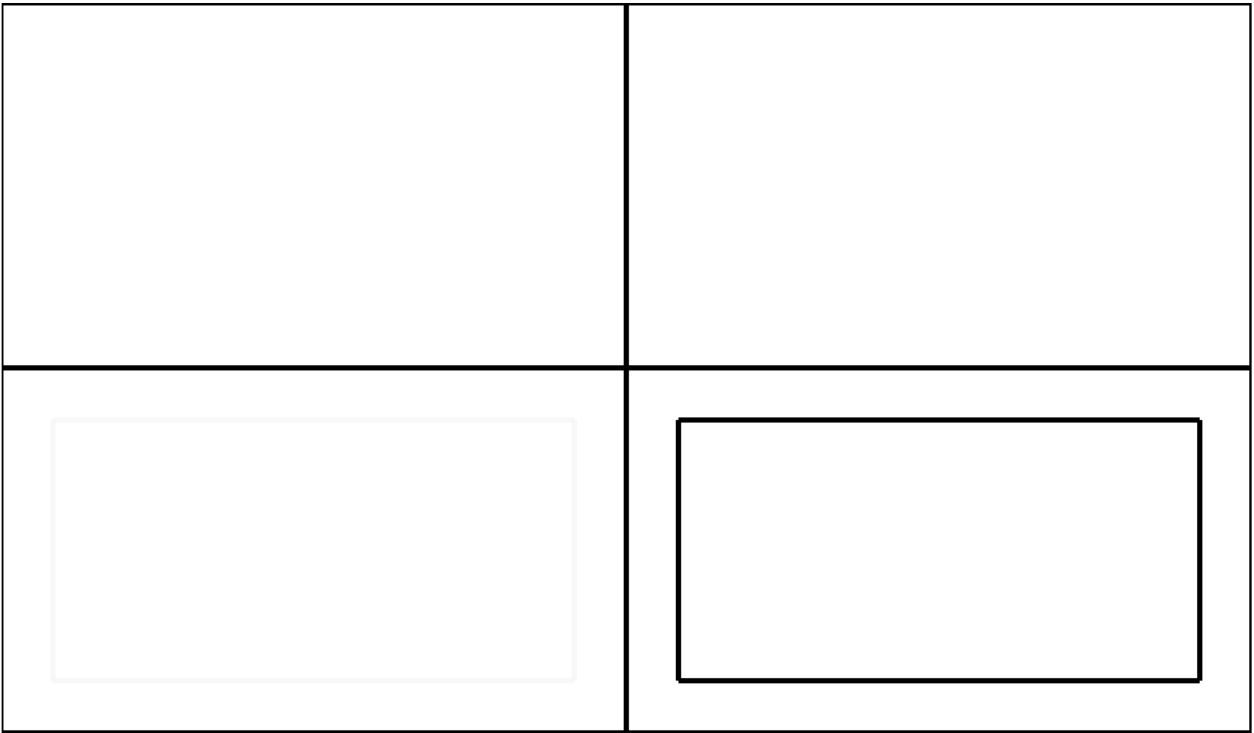}}
\hspace{-3mm}
\subfigure[]{
\label{fig:edgeCase:G_B_m} 
\includegraphics[width=1.4in]{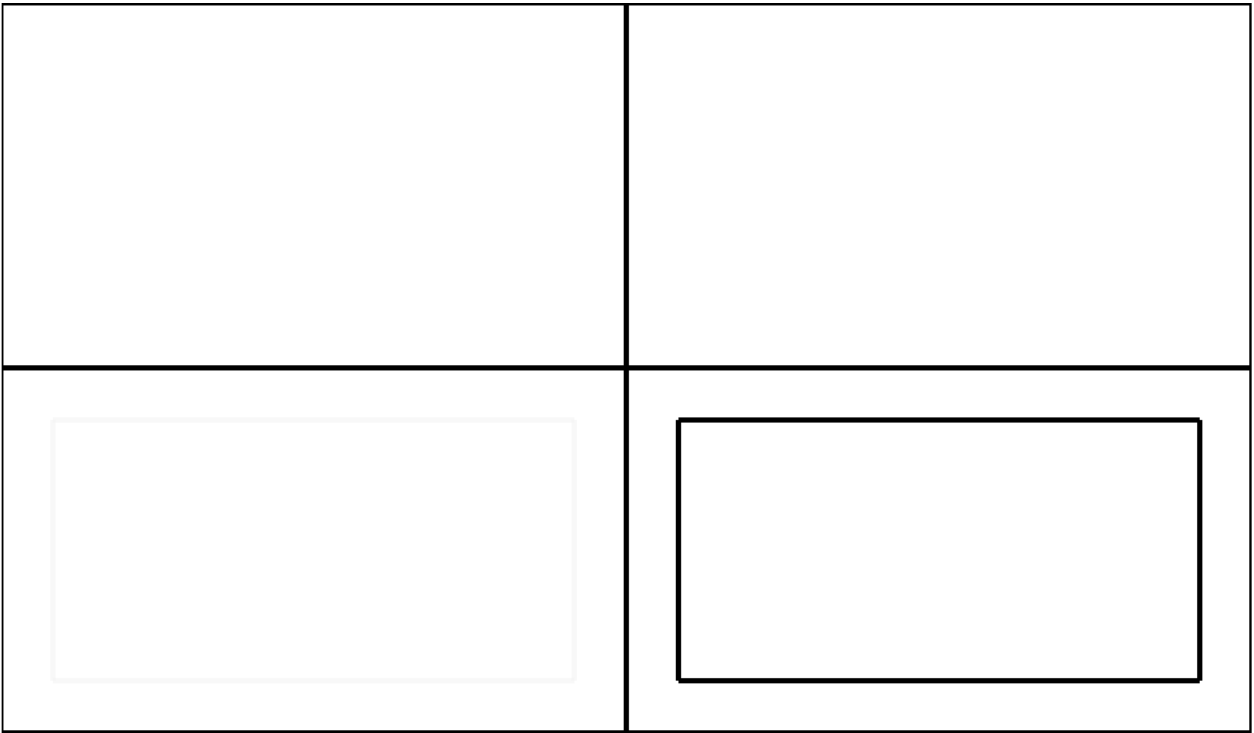}}
\caption{Examples of color transitions that violate (the bottom-right image) and agree with (top and bottom-left images) the model assumption in \eqref{equ:cdc}. (a) Top-left: background with color [0.0, 0.0, 0.0] (black) and foreground color with [1.0, 1.0, 1.0] (white); Top-right: background with color [0.2, 0.5, 0.2] and foreground with color [0.7, 1.0, 0.7]; Bottom-left: background with color [0.9, 0.45, 0.8] and foreground with color [0.7, 0.22, 0.6]. Bottom-right: background with color [0.0, 0.5, 1.0] and foreground with color [1.0, 0.5, 0.0]. (b)-(c) G channel $-$ R channel and G channel $-$ B channel of (a). (d)-(e) The gradient map of (b) and (c).}
\label{fig:edgeCase}
\hspace{-1mm}
\end{figure*}

\section{Gradient and Multiscale Models for Bayer Pattern Images}
\label{section:sec3}
\subsection{Gradient Extraction from Bayer Pattern Images}
\label{section:sec3.1}
Image gradient measures the change of intensity in specific directions. Mathematically, for a two-dimensional function $f(x,y)$, the gradients can be computed by the derivatives with respect to $x$ and $y$. For a digital image where $x$ and $y$ are discrete values, the derivatives can be approximated by finite differences.

There are different ways to define the difference of a digital image, as long as the following three conditions are satisfied: (i) zero in constant intensity area; (ii) non-zero along the ramps and (iii) nonzero at the onset of an intensity step or ramp \cite{RN8}. One of the most commonly used image gradient computation is the central difference based approach as

\begin{equation}
\label{equ:central differencex}
G_{x}(x,y)= I\left ( x+1,y \right )-I\left ( x-1,y \right ),
\end{equation}
\begin{equation}
\label{equ:central differencey}
G_{y}(x,y)= I\left ( x,y+1 \right )-I\left ( x,y-1 \right ).
\end{equation}
Here $I(x,y)$ is the intensity at location $(x,y)$, $G_{x}$ and $G_{y}$ represent the gradients in the horizontal and vertical directions, respectively. The computation of \eqref{equ:central differencex} and \eqref{equ:central differencey} can be implemented by the convolution of the templates in Fig. \ref{fig:Gradient operators:center} with the images.

The fundamental idea of the proposed Bayer pattern image based gradient extraction is illustrated in Fig. \ref{fig:feacture exaction mine}. Instead of demosaicing the Bayer pattern images before difference computation as shown in Fig. \ref{fig:feacture exaction conv}, we propose to take advantage of the color difference constancy assumption directly for gradient extraction based on Bayer pattern images. Note that by convolving the filter templates in Fig. \ref{fig:Gradient operators:center} directly with a Bayer pattern image, all the three conditions for a valid difference definition mentioned are satisfied. To illustrate this, let us consider the example in Fig. \ref{fig:gradientCompute}.

\begin{figure*}[t]
\centering
\subfigure{
\includegraphics[width=0.6in]{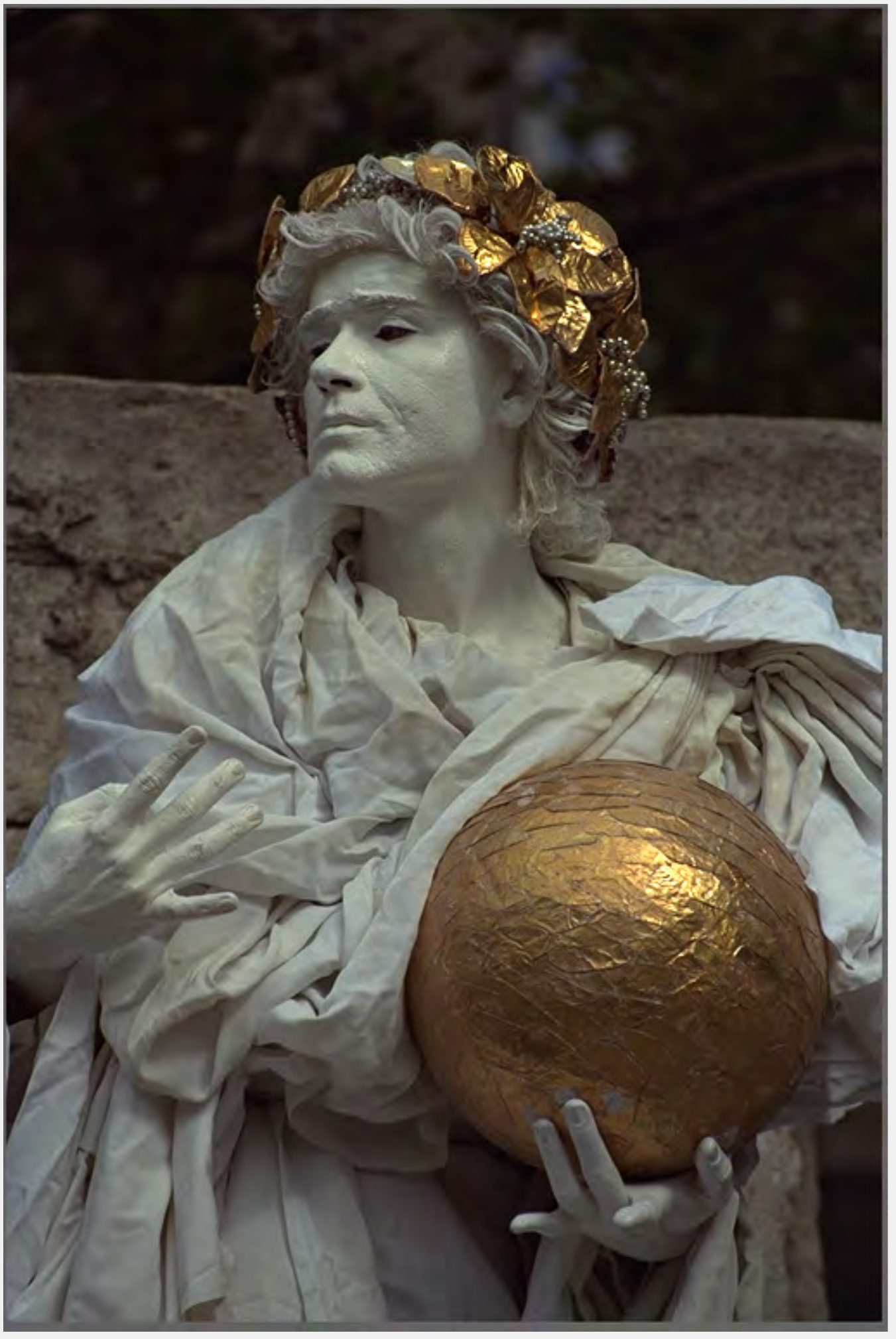}}
\hspace{-3.8mm}
\subfigure{
\includegraphics[width=0.6in]{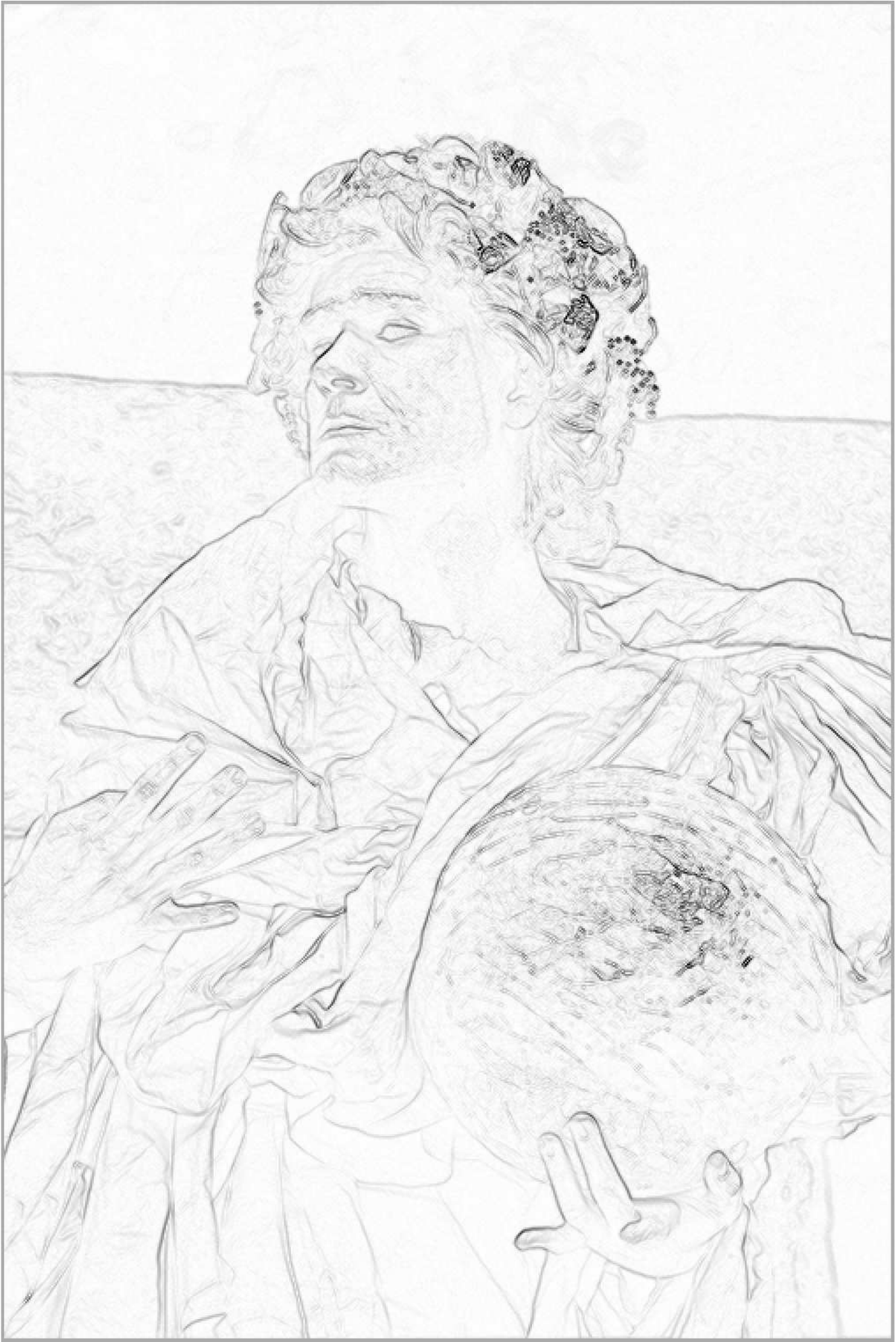}}
\hspace{-3.8mm}
\subfigure{
\includegraphics[width=0.6in]{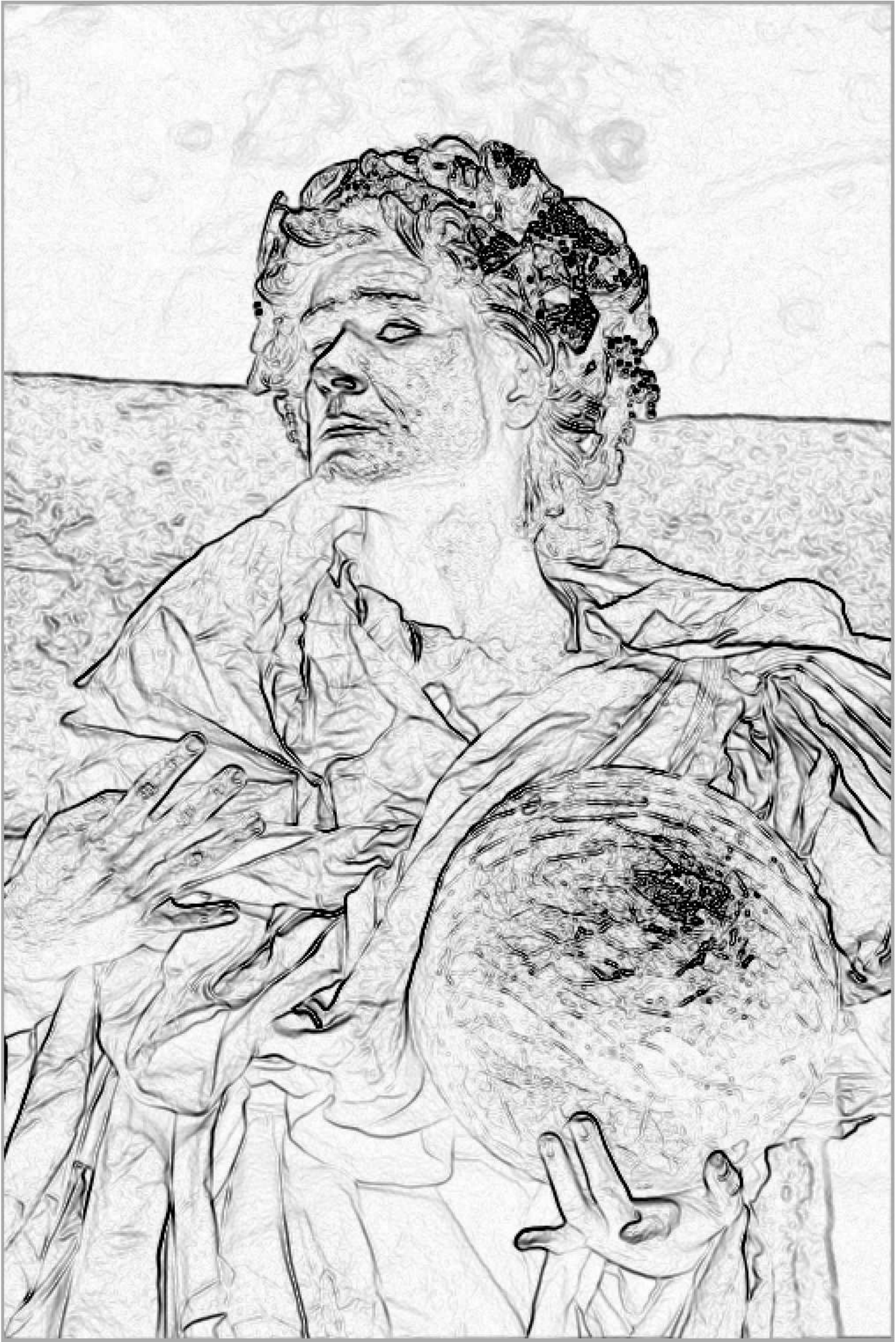}}
\hspace{-3.8mm}
\subfigure{
\includegraphics[width=0.6in]{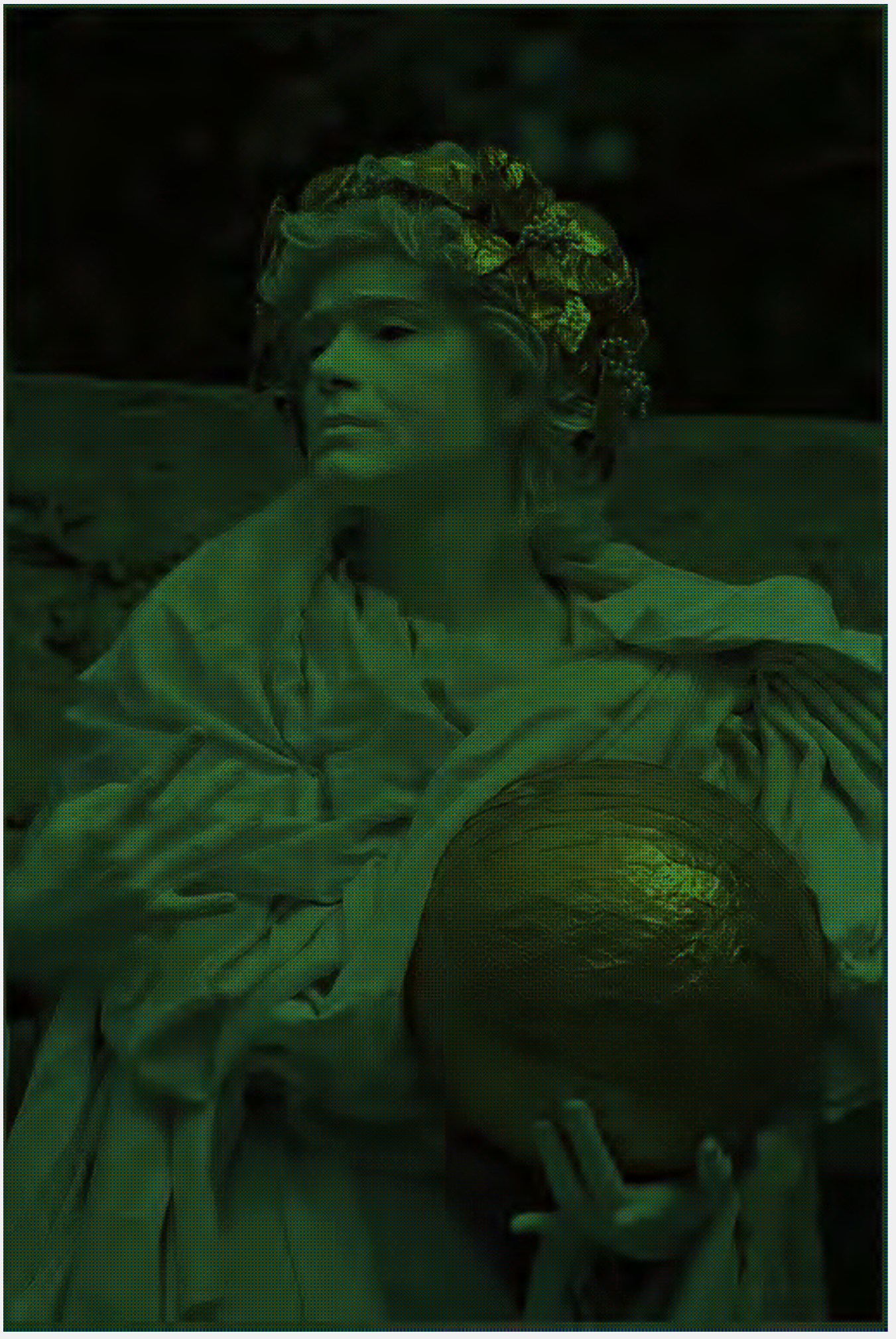}}
\hspace{-3.8mm}
\subfigure{
\includegraphics[width=0.6in]{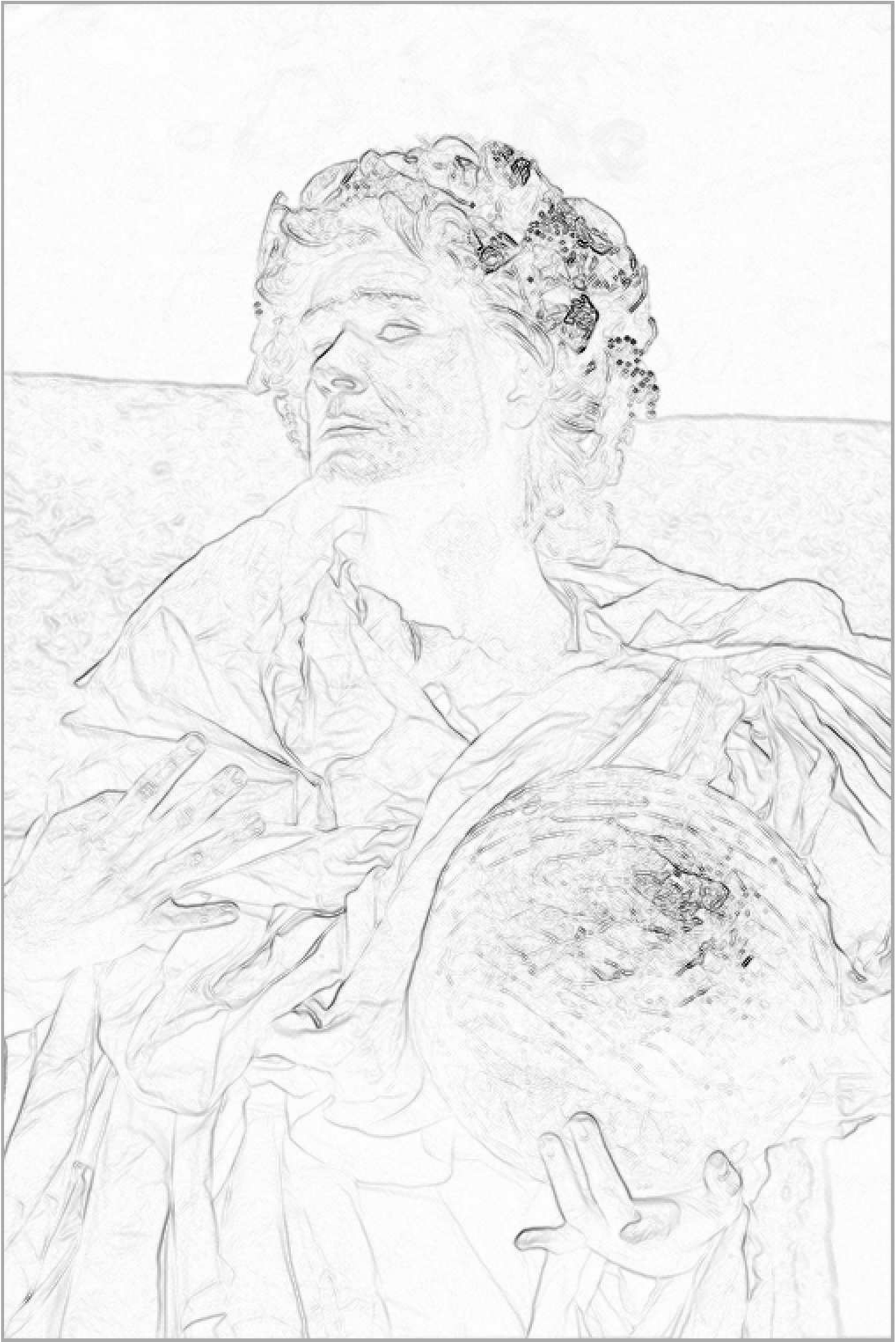}}
\hspace{-3.8mm}
\subfigure{
\includegraphics[width=0.6in]{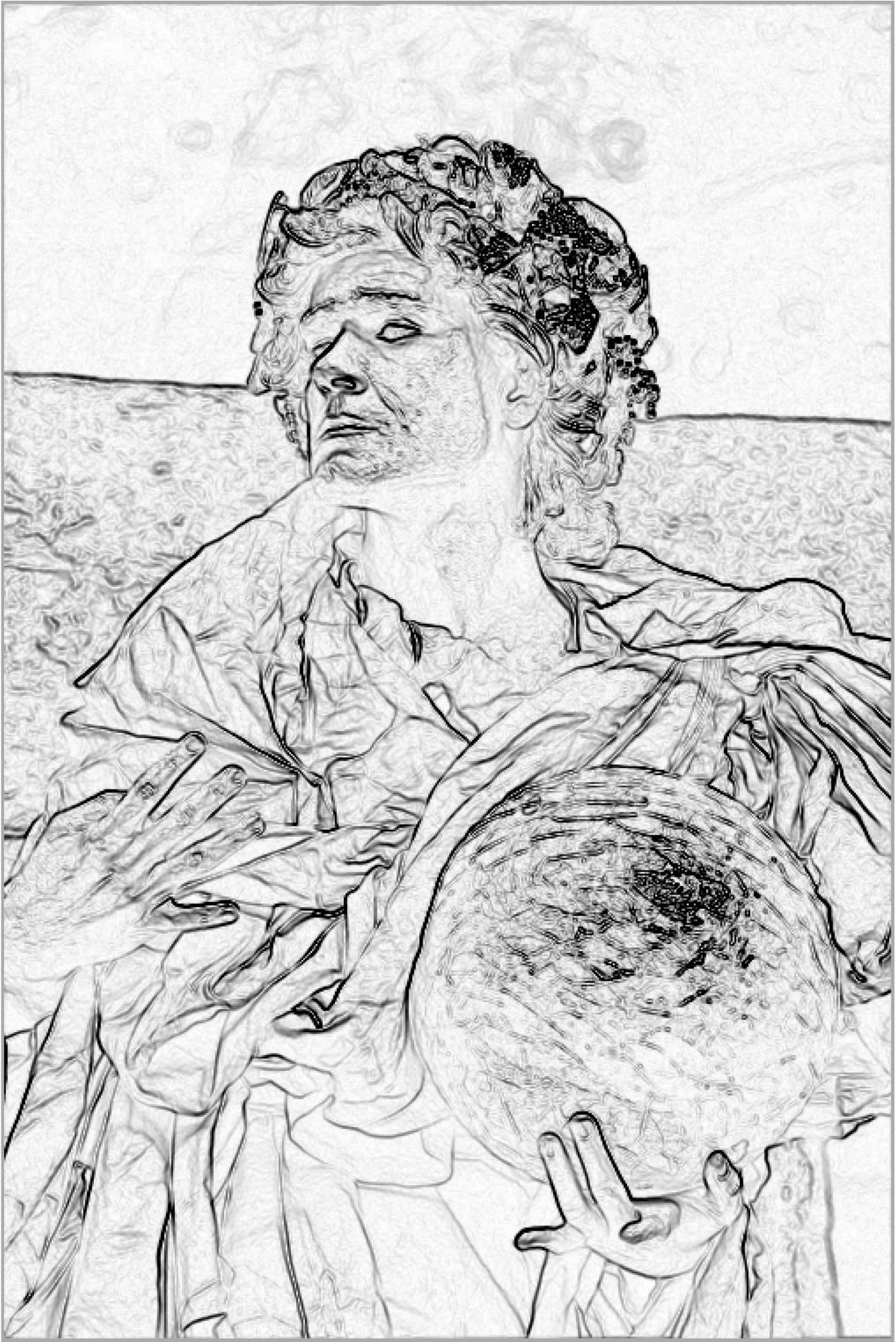}}
\hspace{-3.8mm}
\subfigure{
\includegraphics[width=0.6in]{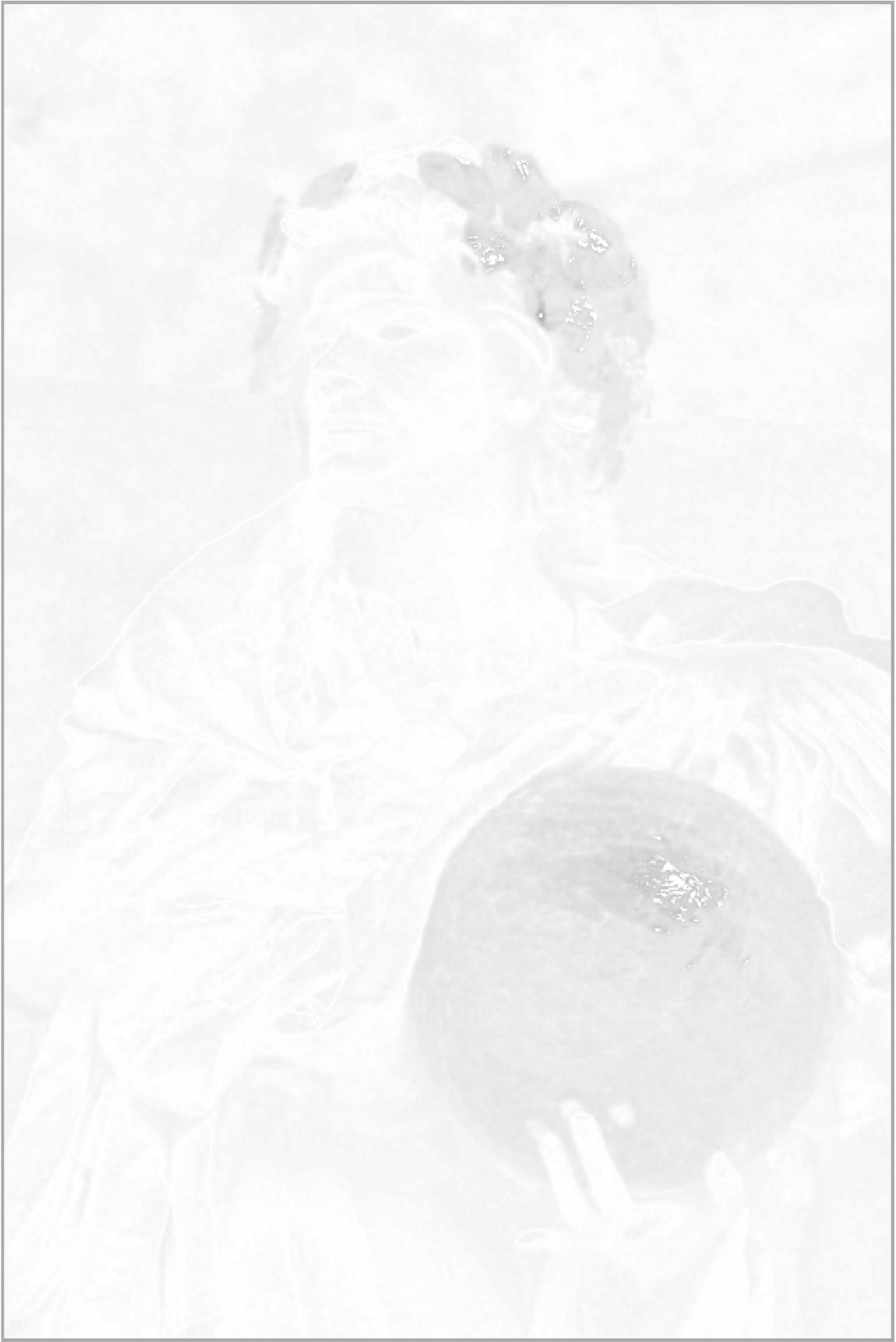}}
\hspace{-3.8mm}
\subfigure{
\includegraphics[width=0.6in]{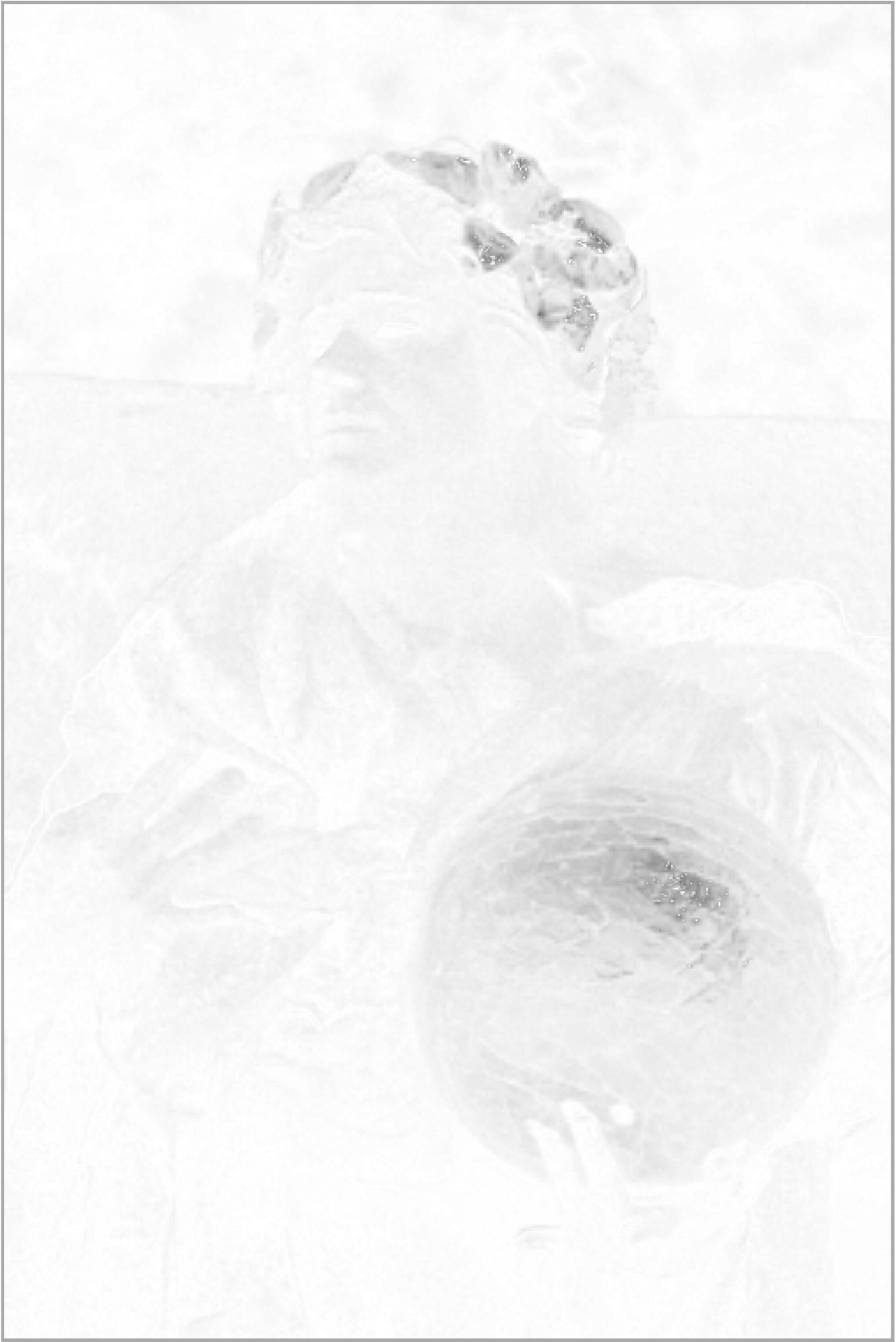}}
\hspace{-3.8mm}
\subfigure{
\includegraphics[width=0.6in]{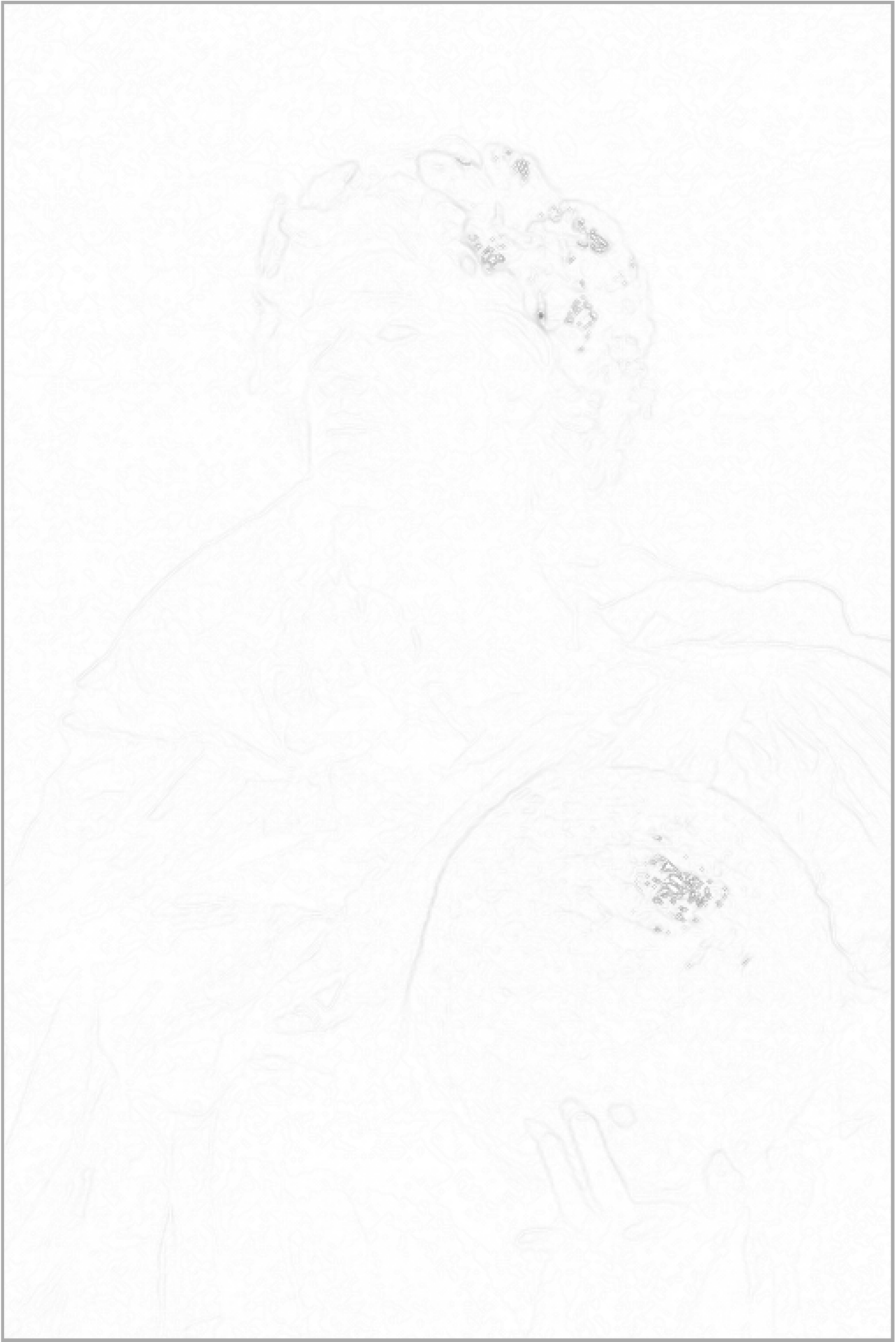}}
\hspace{-3.8mm}
\subfigure{
\includegraphics[width=0.6in]{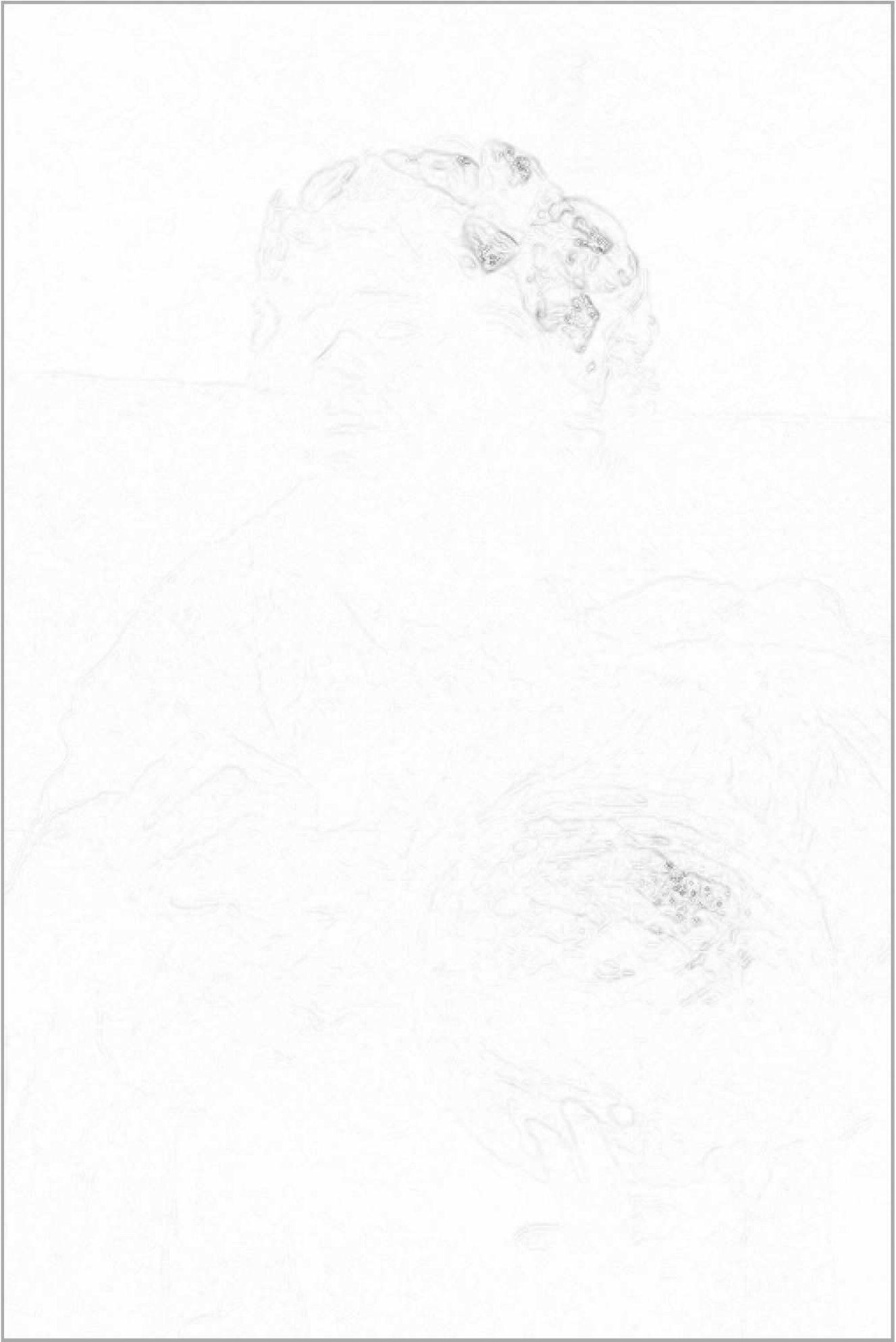}}
\hspace{-3.8mm}
\subfigure{
\includegraphics[width=0.6in]{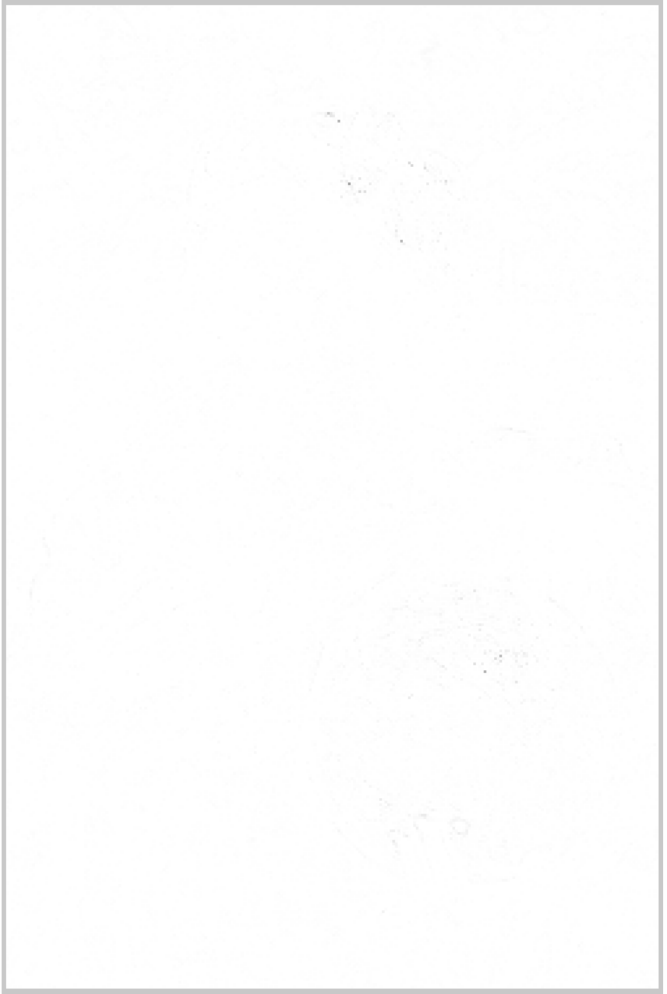}}
\hspace{-3.8mm}
\subfigure{
\includegraphics[width=0.6in]{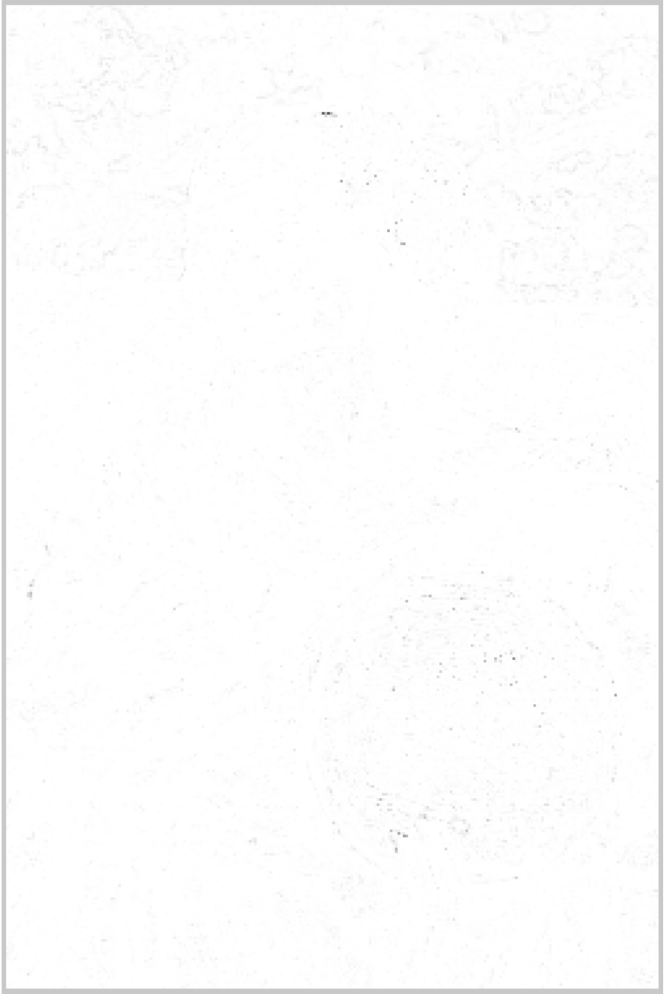}}\\
\setcounter{enumi}{1}
\vspace{-2.5mm}
\addtocounter{subfigure}{-12}
\subfigure[]{
\label{fig:gradient and missmap:k03gray} 
\includegraphics[width=0.6in]{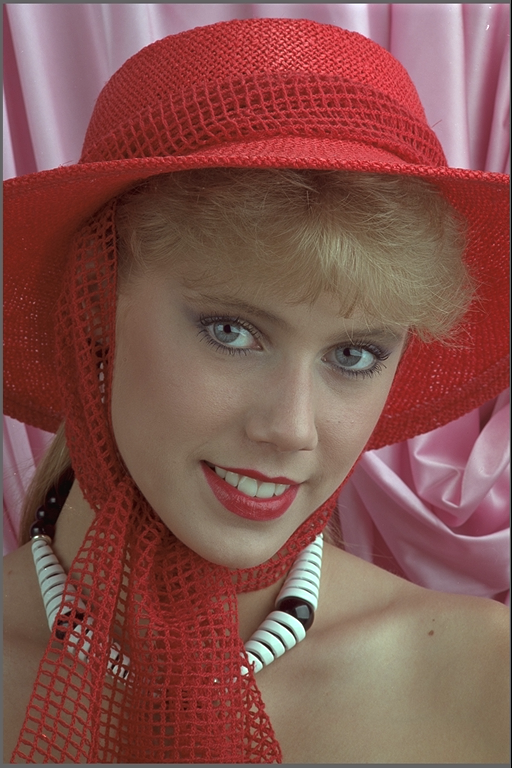}}
\hspace{-3.8mm}
\subfigure[]{
\label{fig:gradient and missmap:grayCenter} 
\includegraphics[width=0.6in]{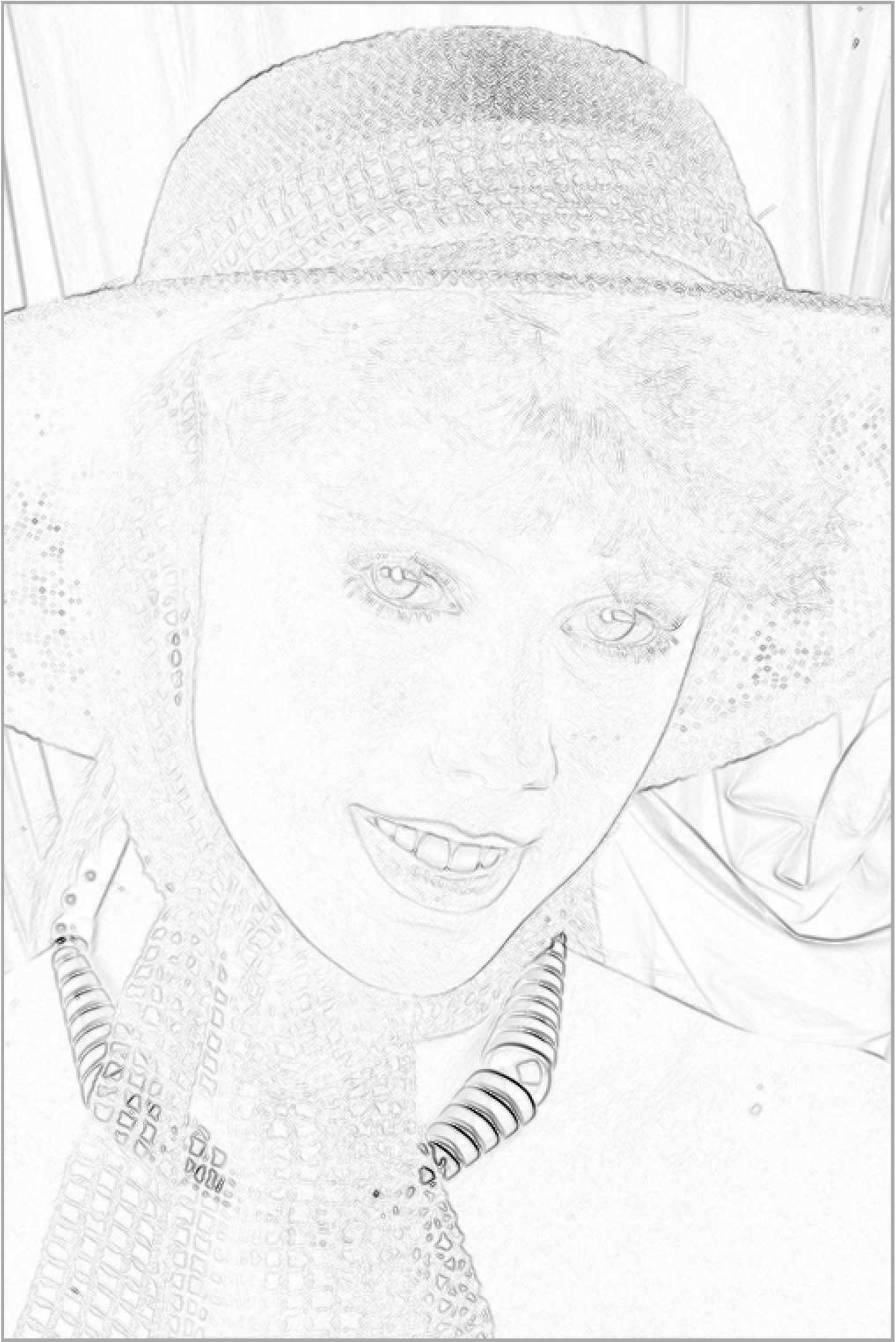}}
\hspace{-3.8mm}
\subfigure[]{
\label{fig:gradient and missmap:graySobel} 
\includegraphics[width=0.6in]{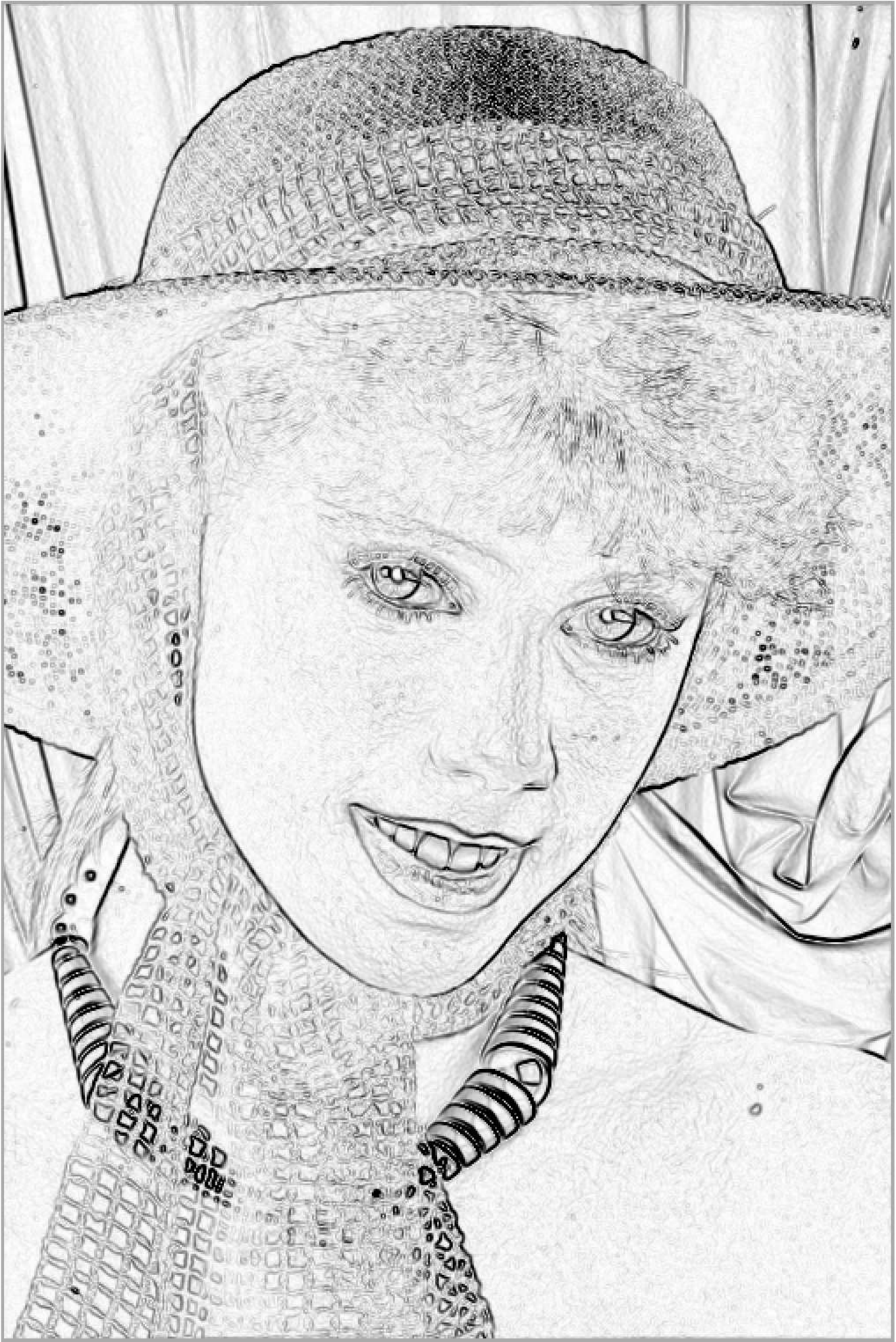}}
\hspace{-3.8mm}
\subfigure[]{
\label{fig:gradient and missmap:k03bayer} 
\includegraphics[width=0.6in]{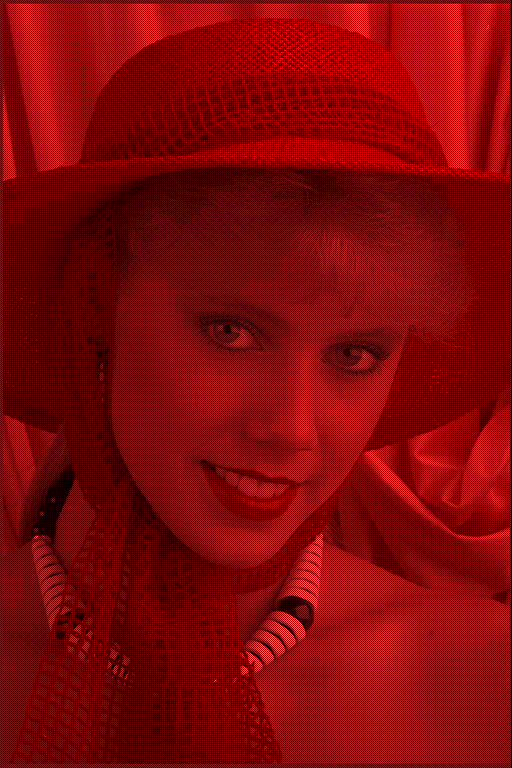}}
\hspace{-3.8mm}
\subfigure[]{
\label{fig:gradient and missmap:BayerCenter} 
\includegraphics[width=0.6in]{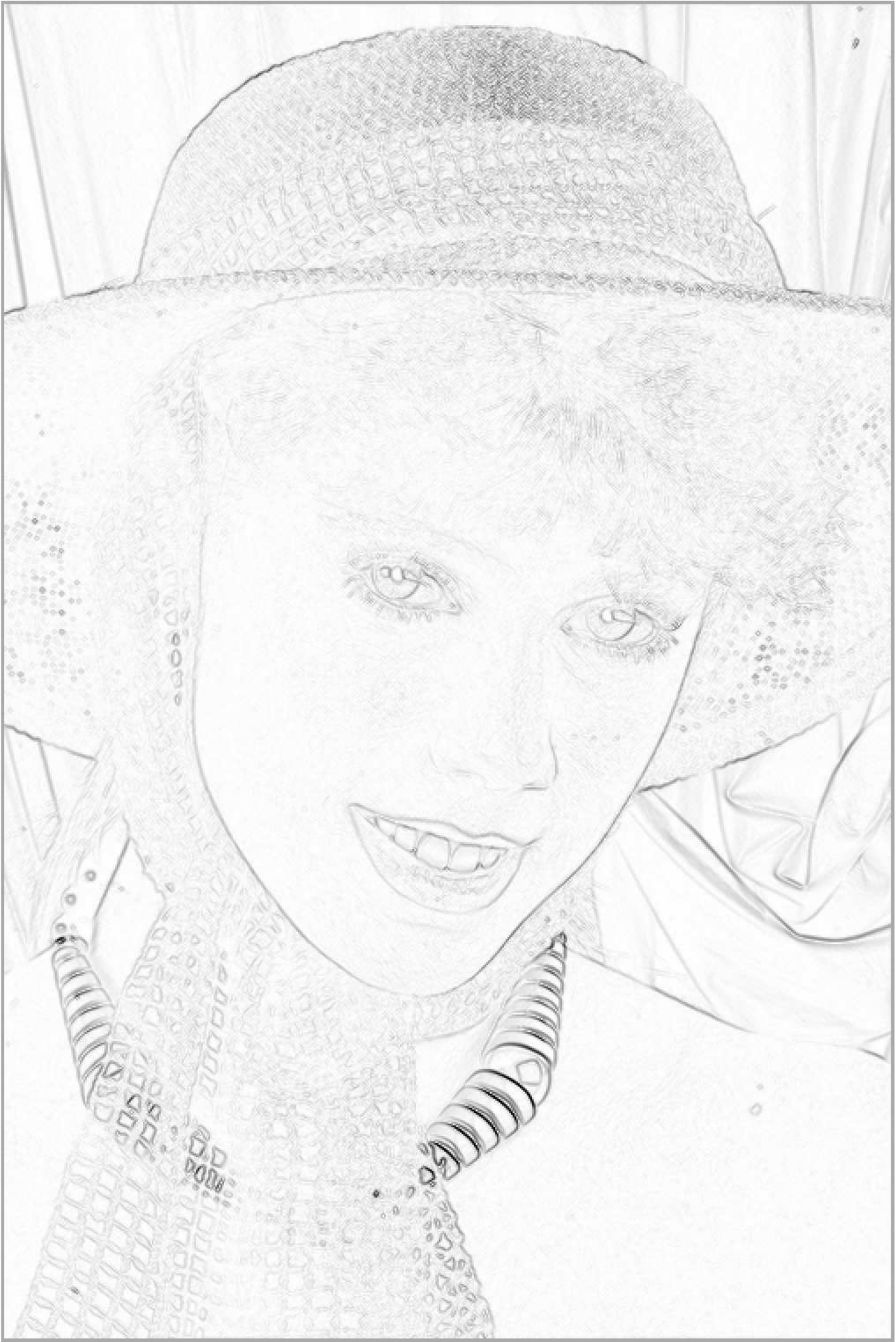}}
\hspace{-3.8mm}
\subfigure[]{
\label{fig:gradient and missmap:BayerSobel} 
\includegraphics[width=0.6in]{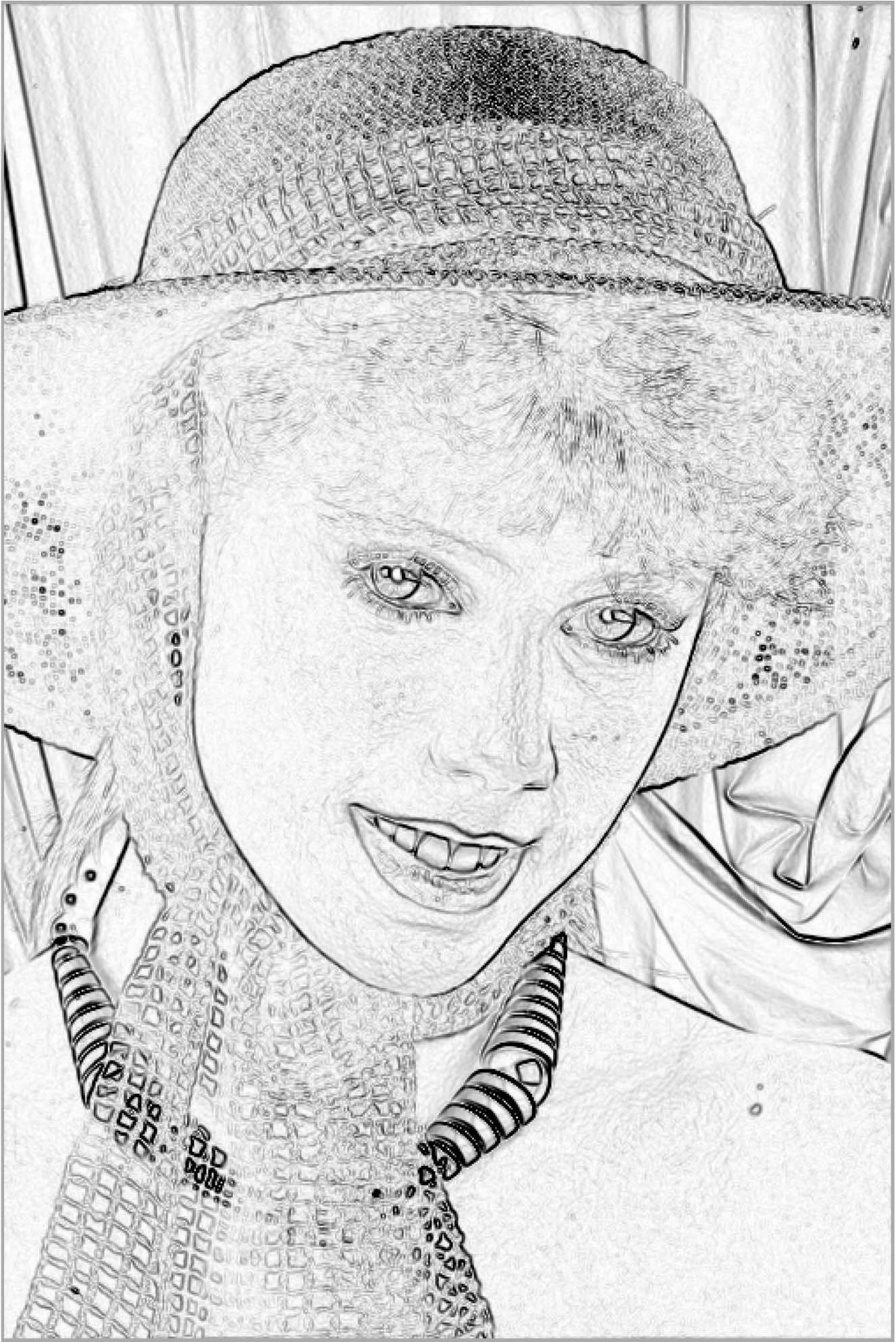}}
\hspace{-3.8mm}
\subfigure[]{
\label{fig:gradient and missmap:G_R} 
\includegraphics[width=0.6in]{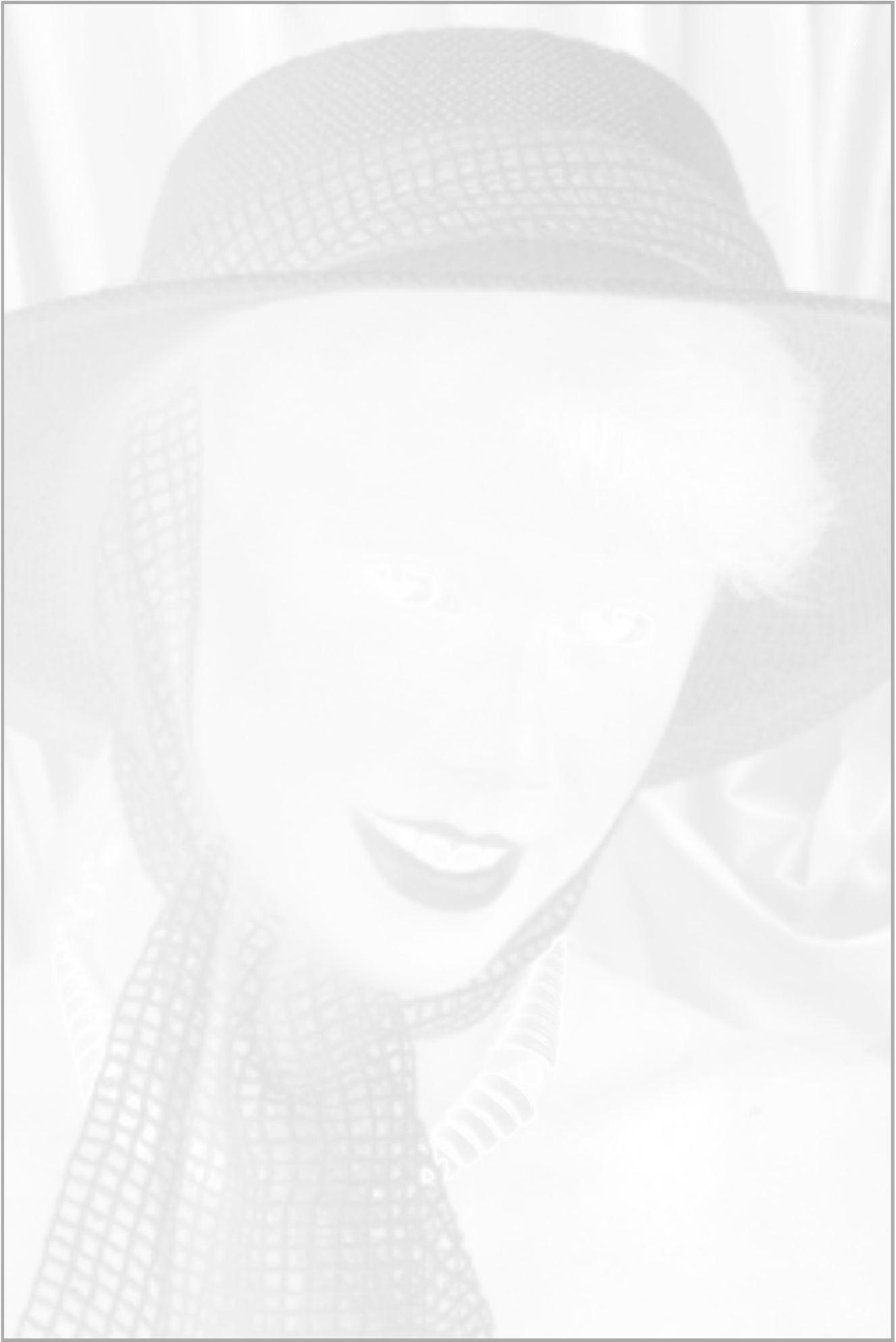}}
\hspace{-3.8mm}
\subfigure[]{
\label{fig:gradient and missmap:G_B} 
\includegraphics[width=0.6in]{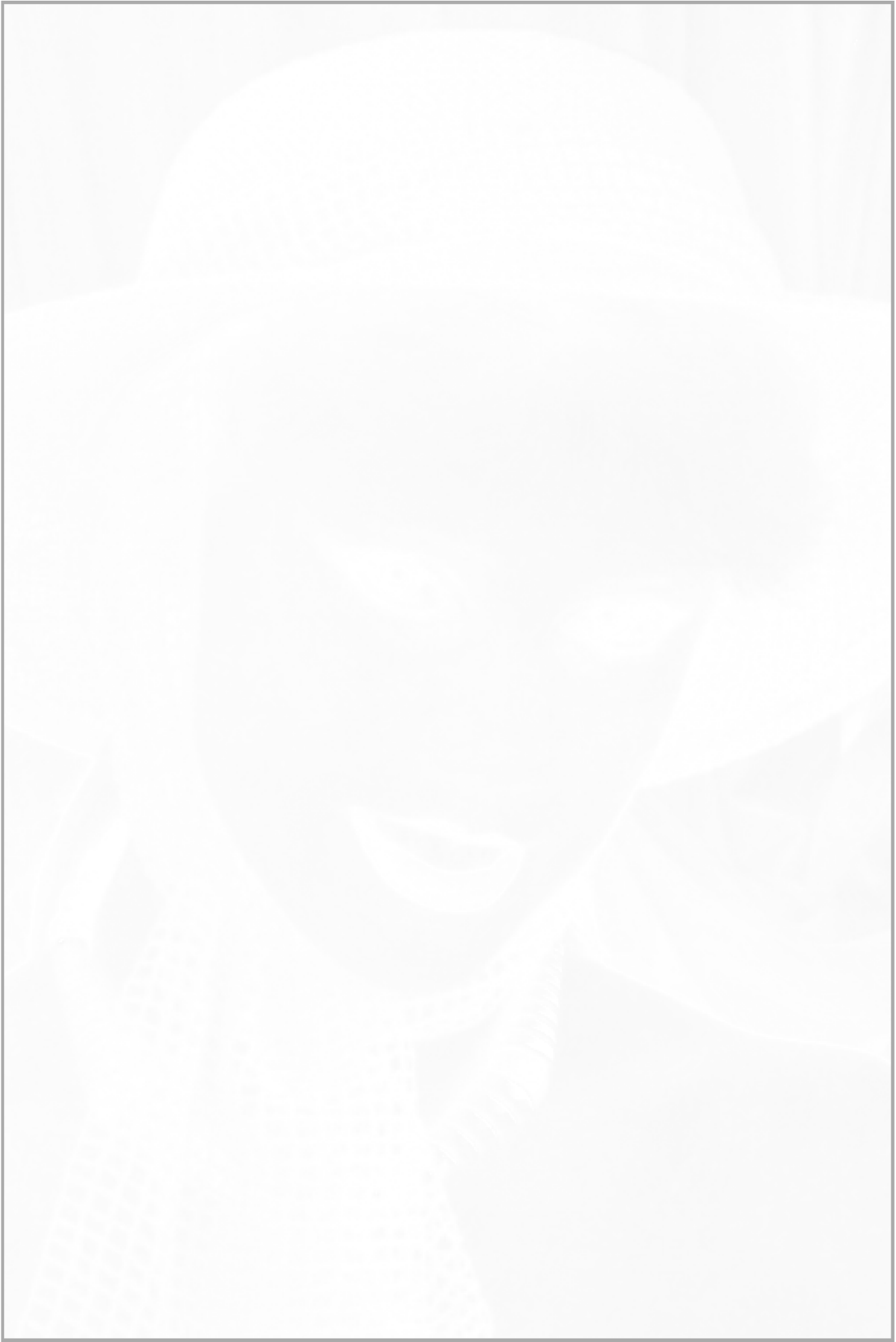}}
\hspace{-3.8mm}
\subfigure[]{
\label{fig:gradient and missmap:G_R_m} 
\includegraphics[width=0.6in]{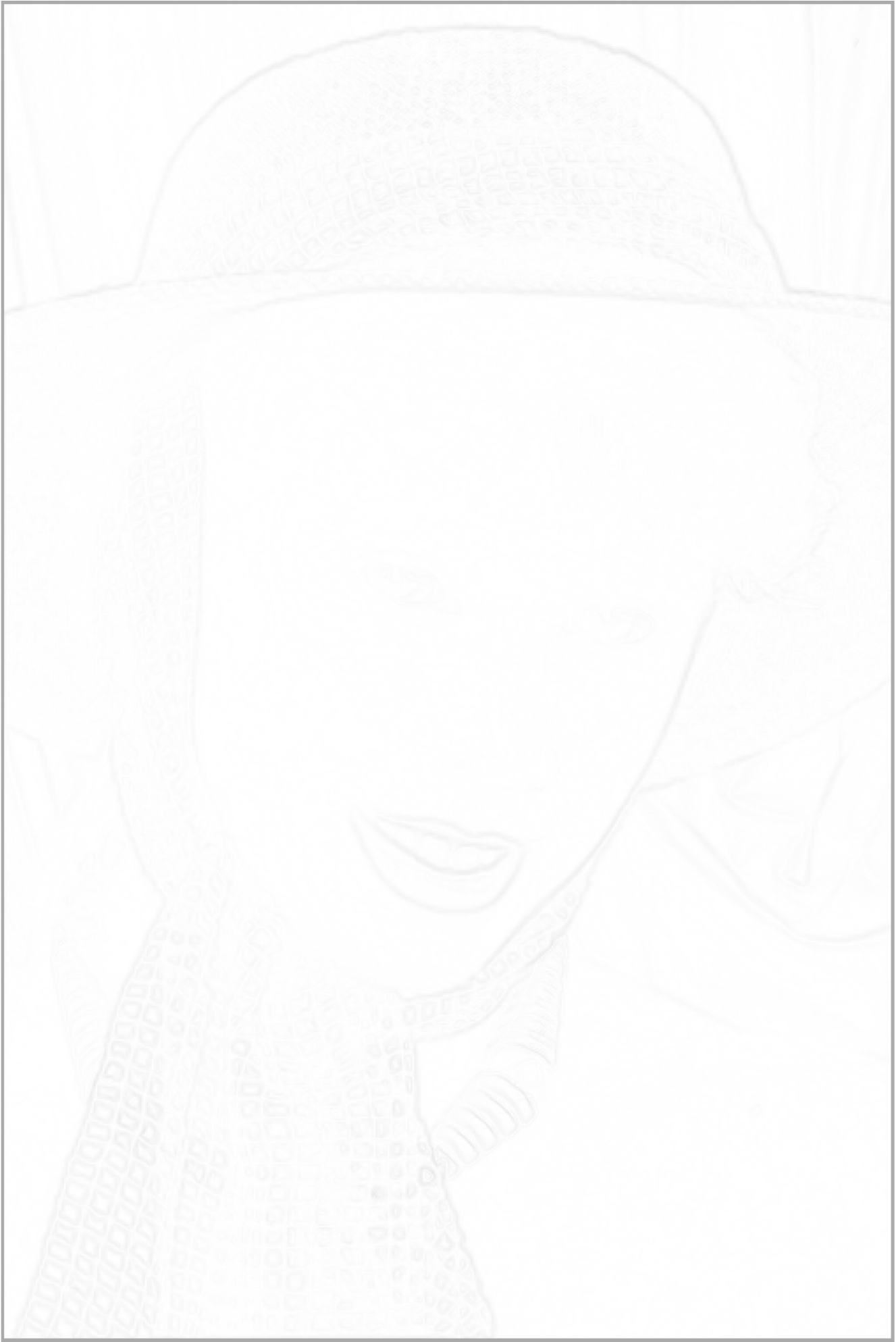}}
\hspace{-3.8mm}
\subfigure[]{
\label{fig:gradient and missmap:G_B_m} 
\includegraphics[width=0.6in]{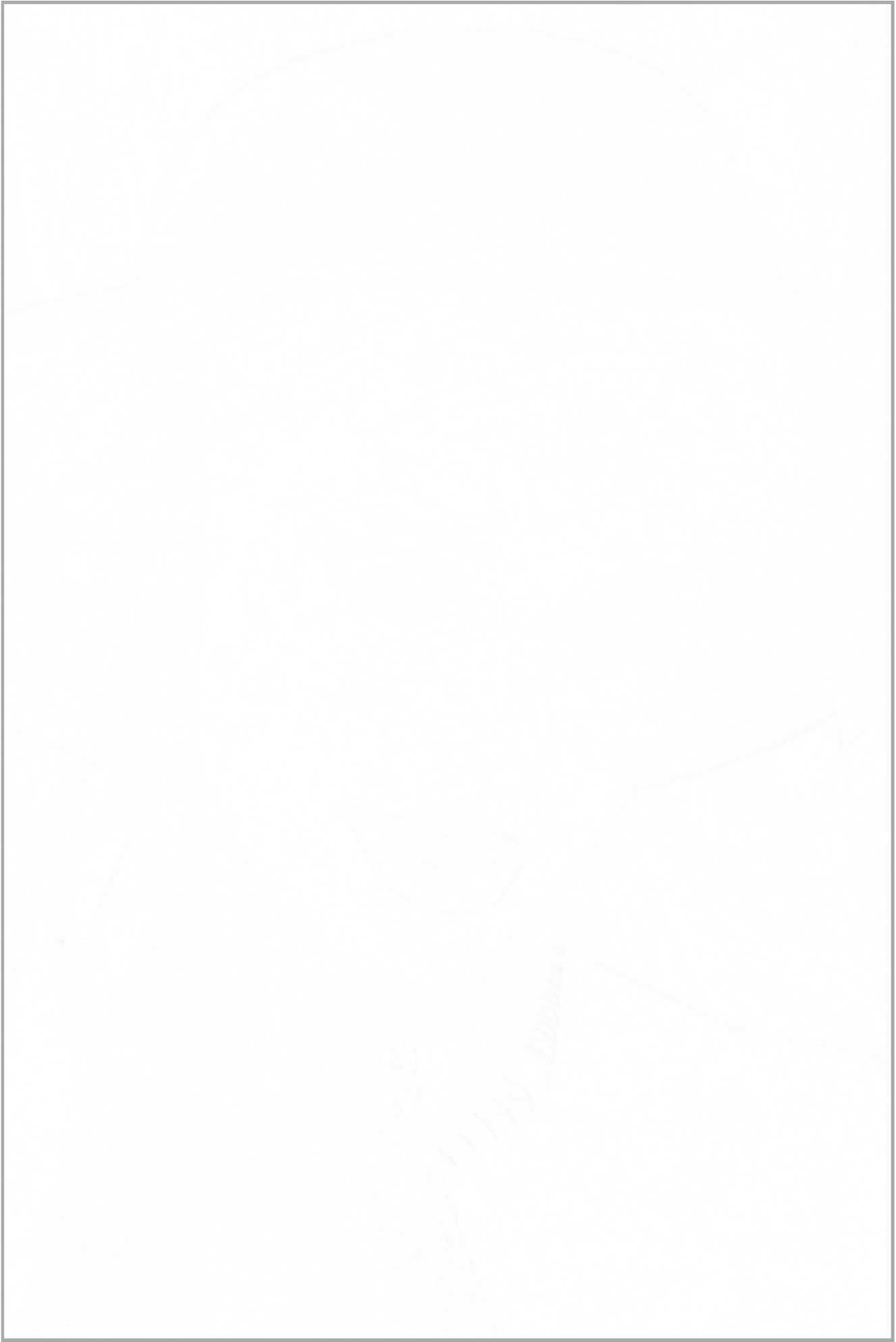}}
\hspace{-3.8mm}
\subfigure[]{
\label{fig:gradient and missmap:GMS_center} 
\includegraphics[width=0.6in]{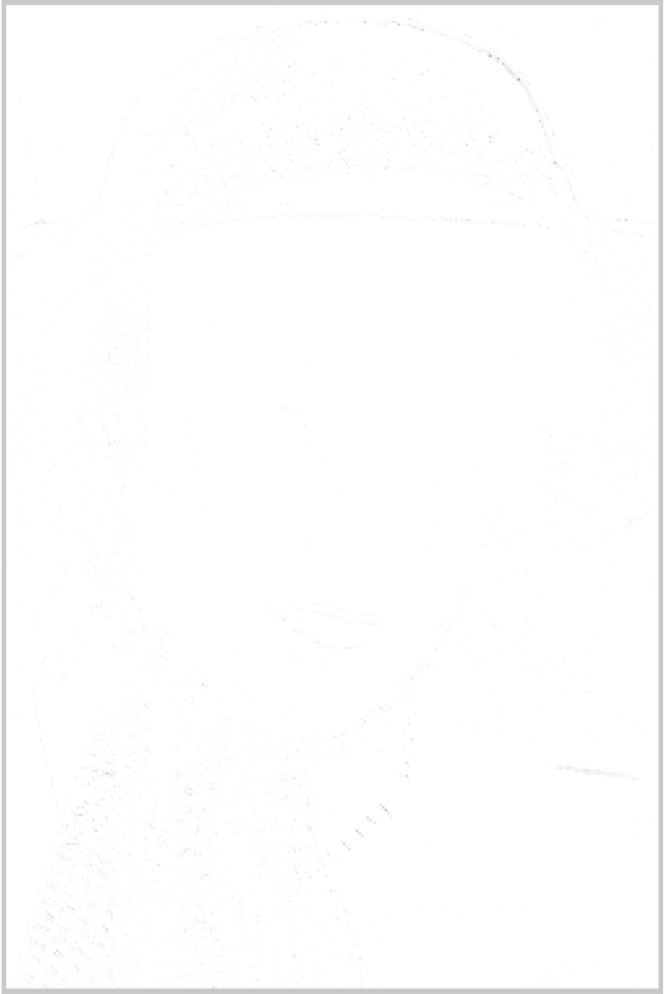}}
\hspace{-3.8mm}
\subfigure[]{
\label{fig:gradient and missmap:GMS_sobel} 
\includegraphics[width=0.6in]{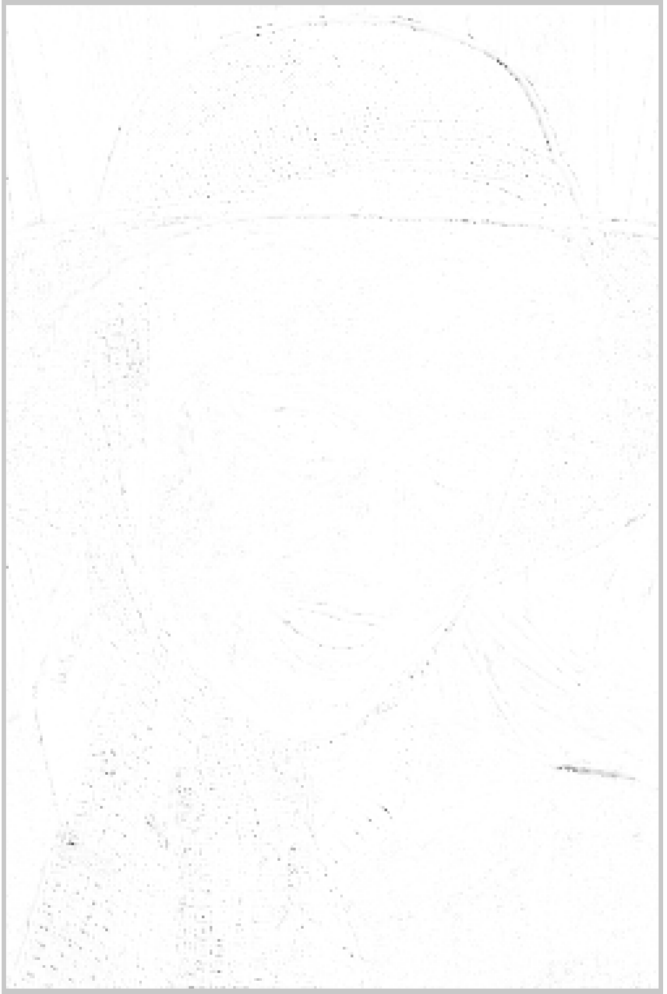}}
\caption{Comparison of gradients extracted from color images and their Bayer version. (a) Image Kodim17 (top) and Image Kodim04 (bottom).
(b)-(c): Gradient magnitude maps generated from (a) using the central difference operator and the Sobel operator in Fig. \ref{fig:Gradient operators}.
(d) The resampled Bayer versions of Kodim17 and Kodim04 (displayed as a three-channel image).
(e)-(f): Gradient magnitude maps generated from (d) using the central difference operator and the Sobel operator in Fig. \ref{fig:Gradient operators}.
(g)-(h): The difference images generate from (a) by (G channel $-$ R channel) and (G channel $-$ B channel).
(i)-(j): Gradient magnitude maps generated from (g) and (h) using operators in Fig. \ref{fig:Gradient operators:center}.
(k) The gradient magnitude similarity (GMS) maps \eqref{equ:GMS} between (b) and (e) with GMSD=0.004 (top) and GMSD=0.007 (bottom).
(l) The GMS maps between (c) and (f) with GMSD=0.011 (top) and GMSD=0.022 (bottom).}
\label{fig:gradient and missmap}
\vspace{-0.3mm}
\end{figure*}

\begin{figure*}[t]
\centering
\subfigure{
\includegraphics[width=1.1in]{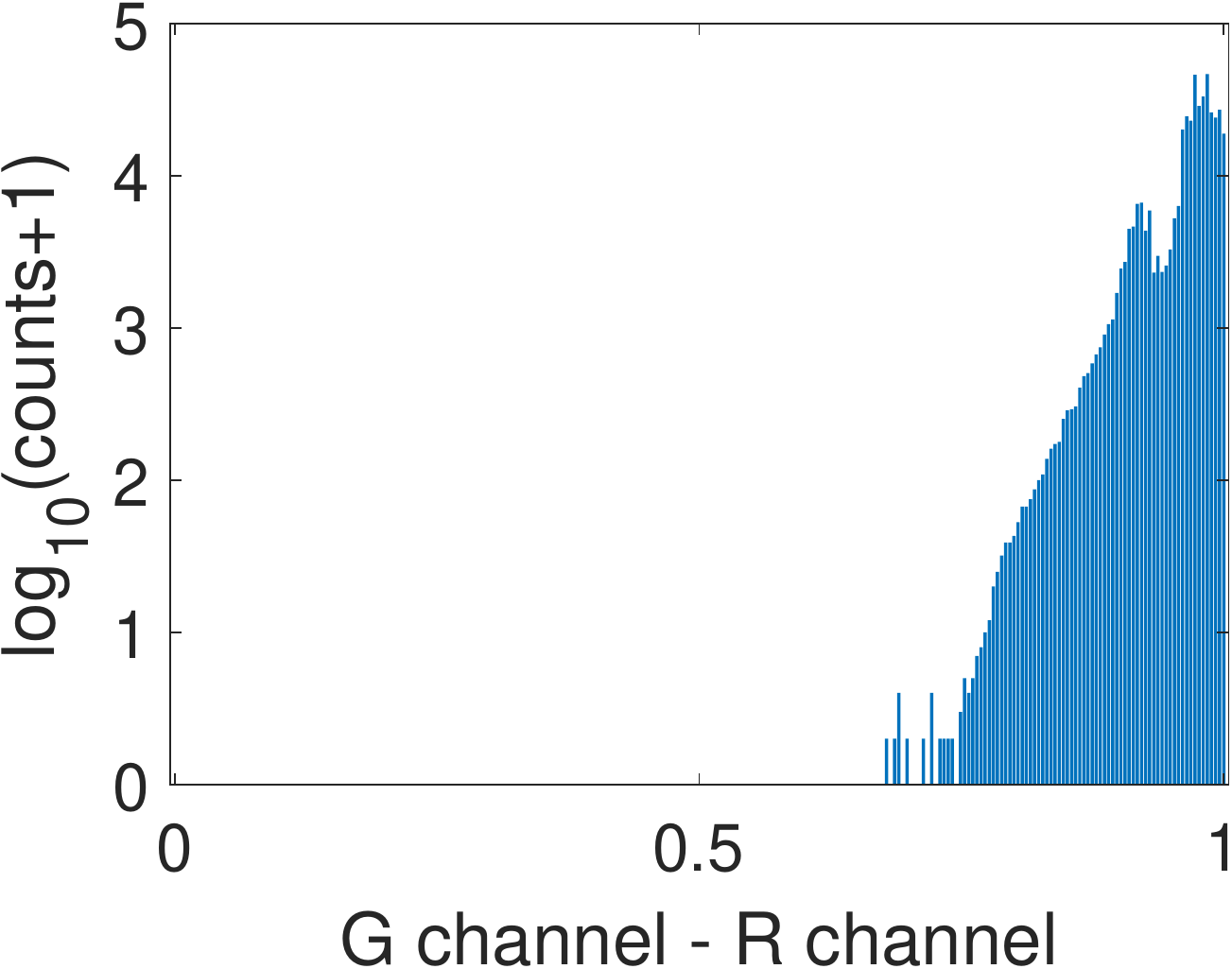}}
\subfigure{
\includegraphics[width=1.1in]{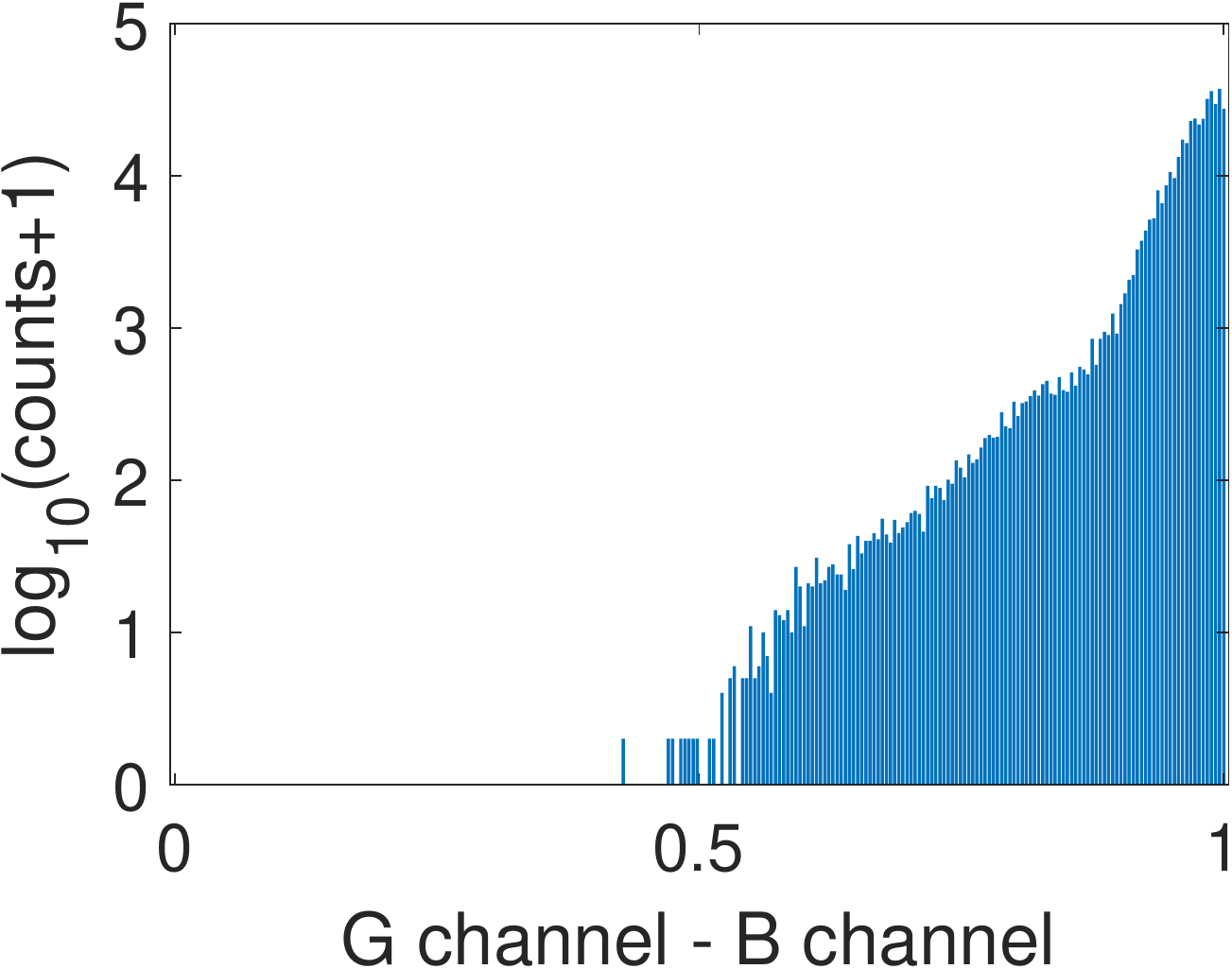}}
\subfigure{
\includegraphics[width=1.1in]{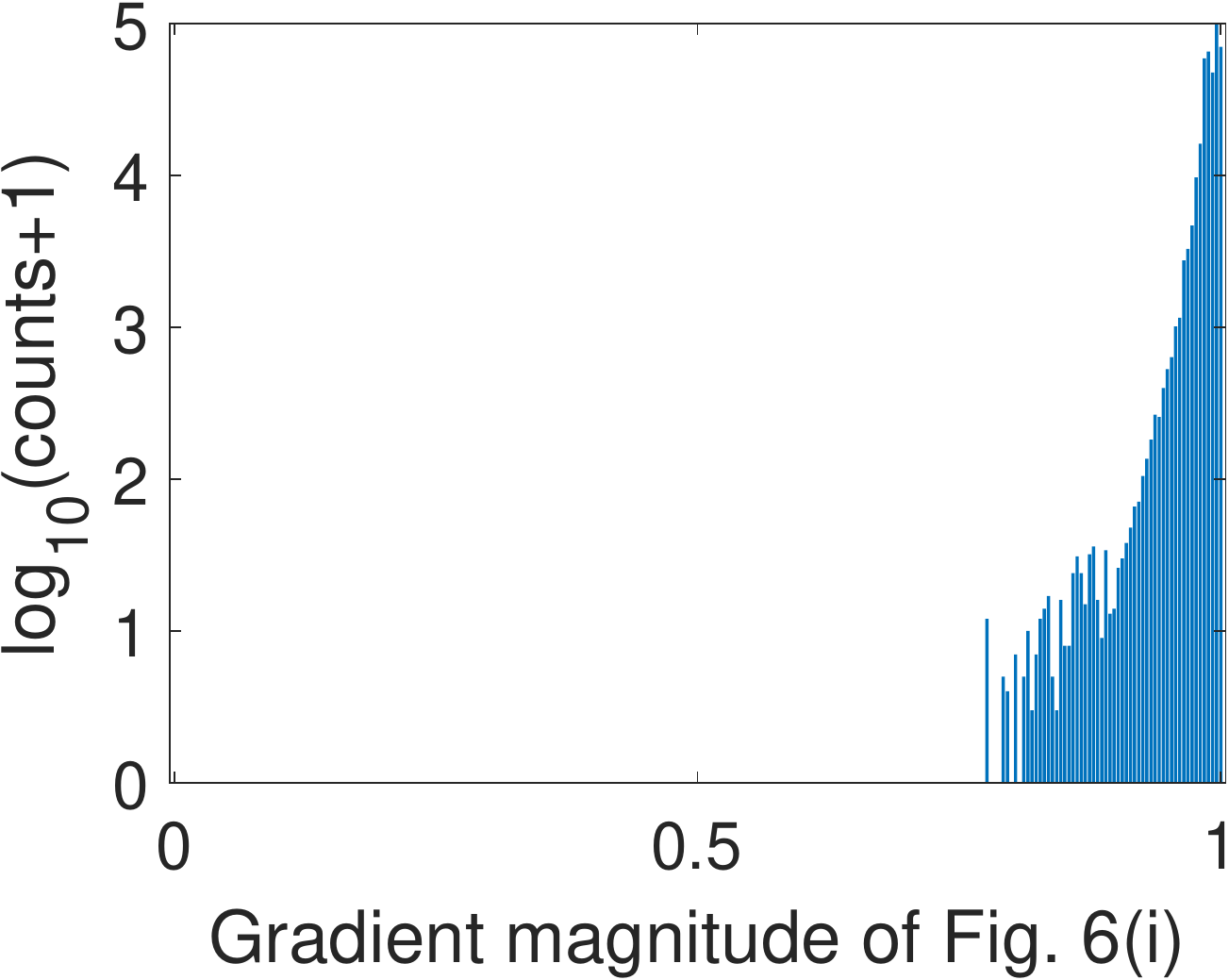}}
\subfigure{
\includegraphics[width=1.1in]{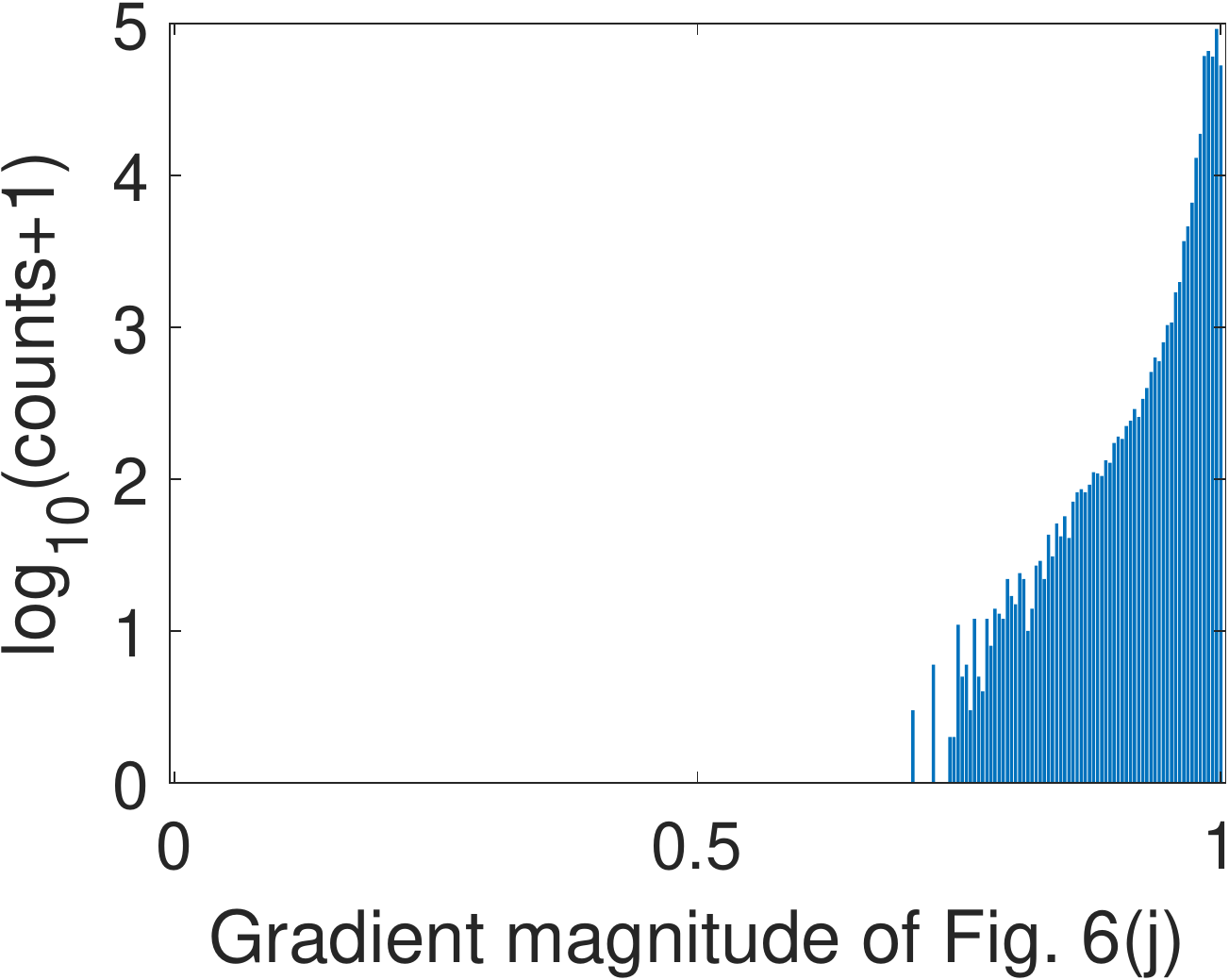}}
\subfigure{
\includegraphics[width=1.1in]{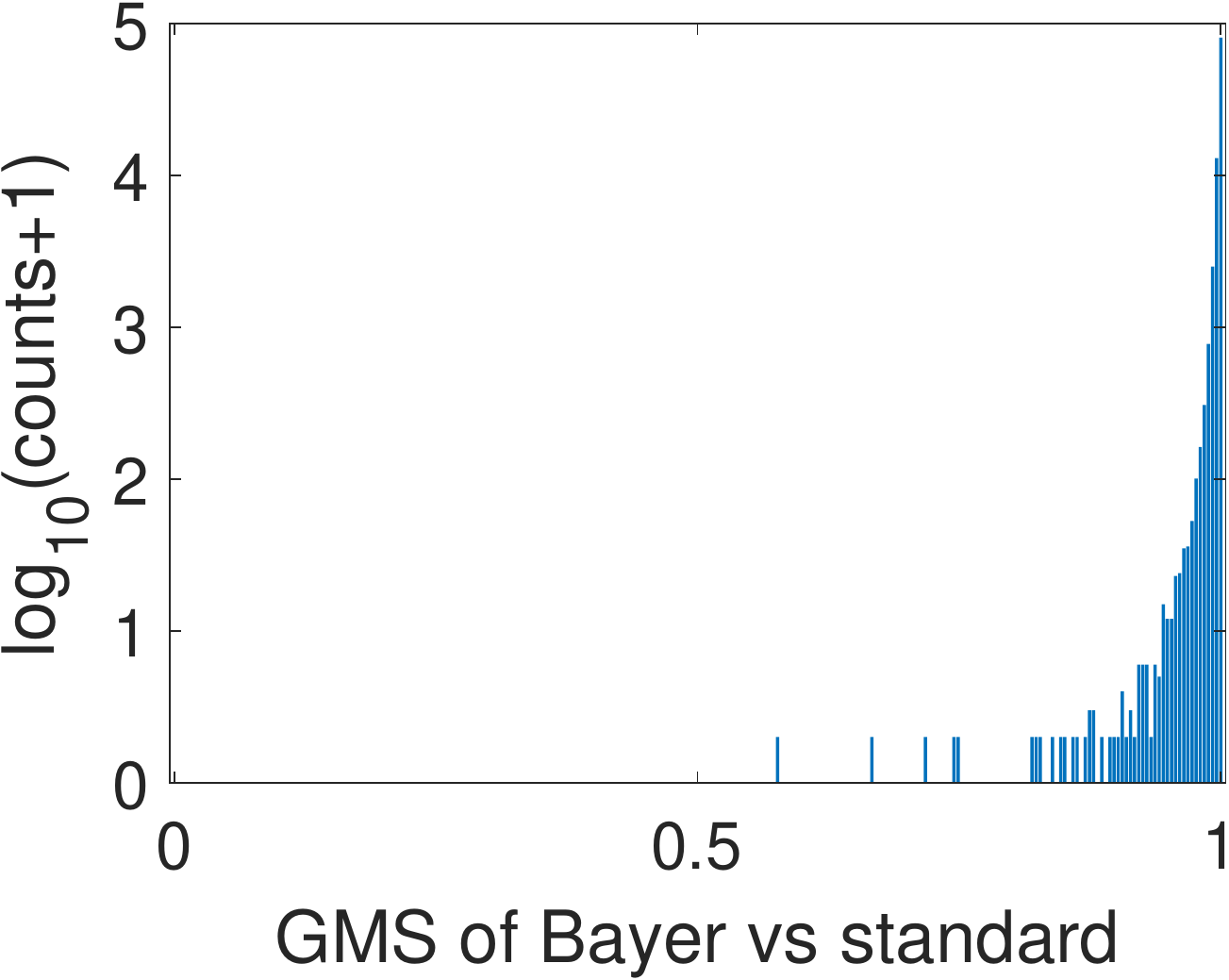}}
\subfigure{
\includegraphics[width=1.1in]{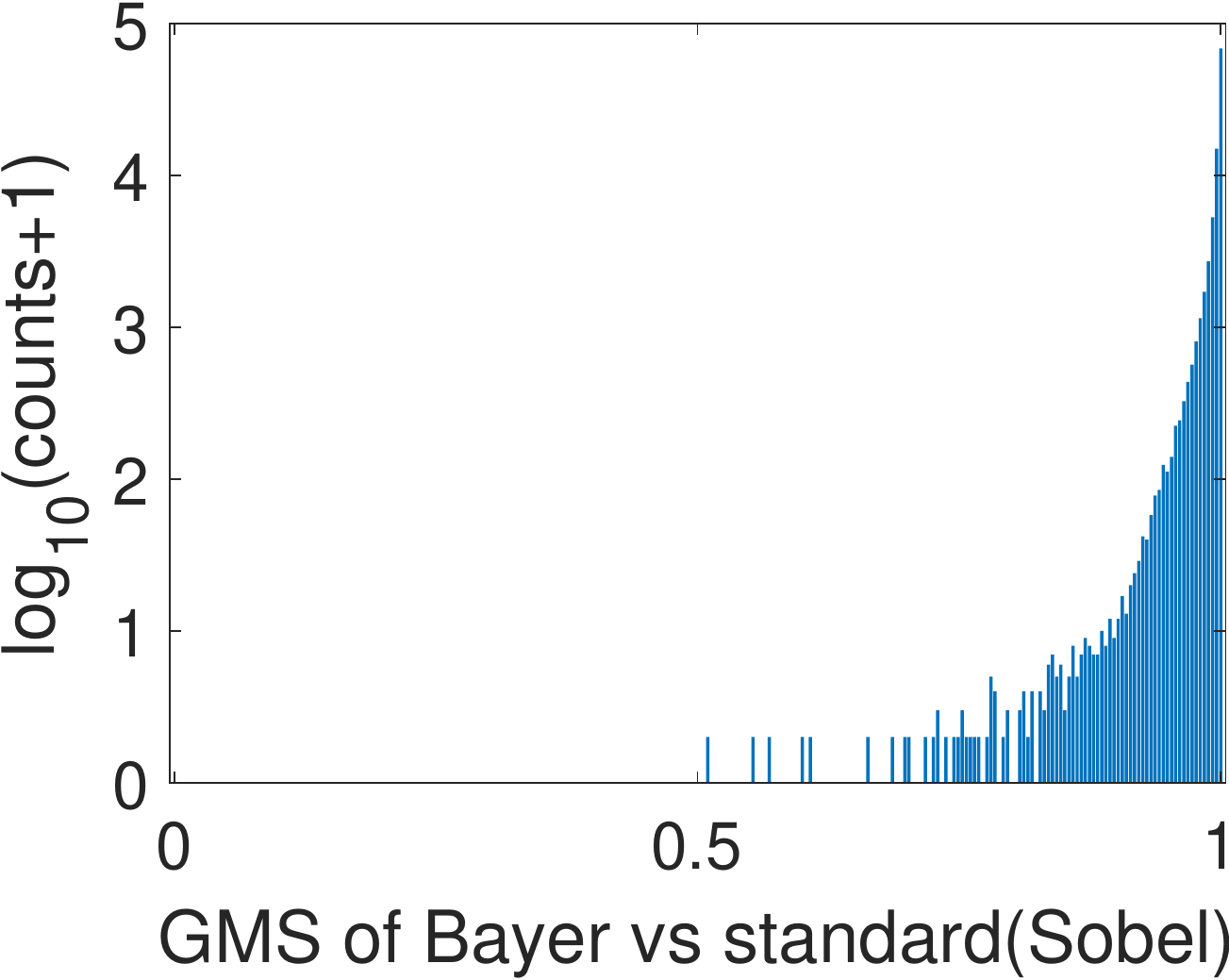}}
\setcounter{enumi}{1}
\addtocounter{subfigure}{-6}
\subfigure[]{
\label{fig:imhist:G_R2} 
\includegraphics[width=1.1in]{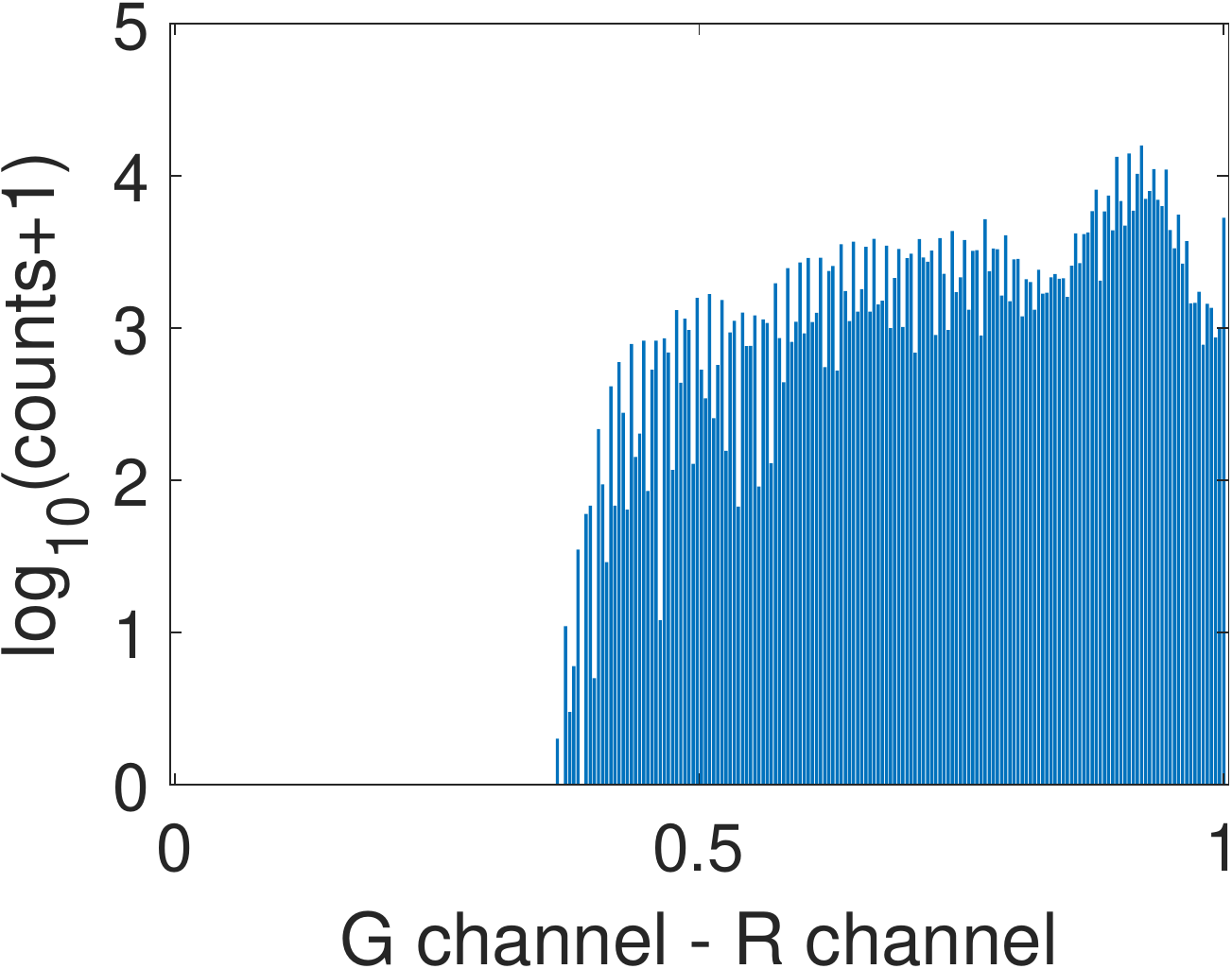}}
\subfigure[]{
\label{fig:imhist:G_B2} 
\includegraphics[width=1.1in]{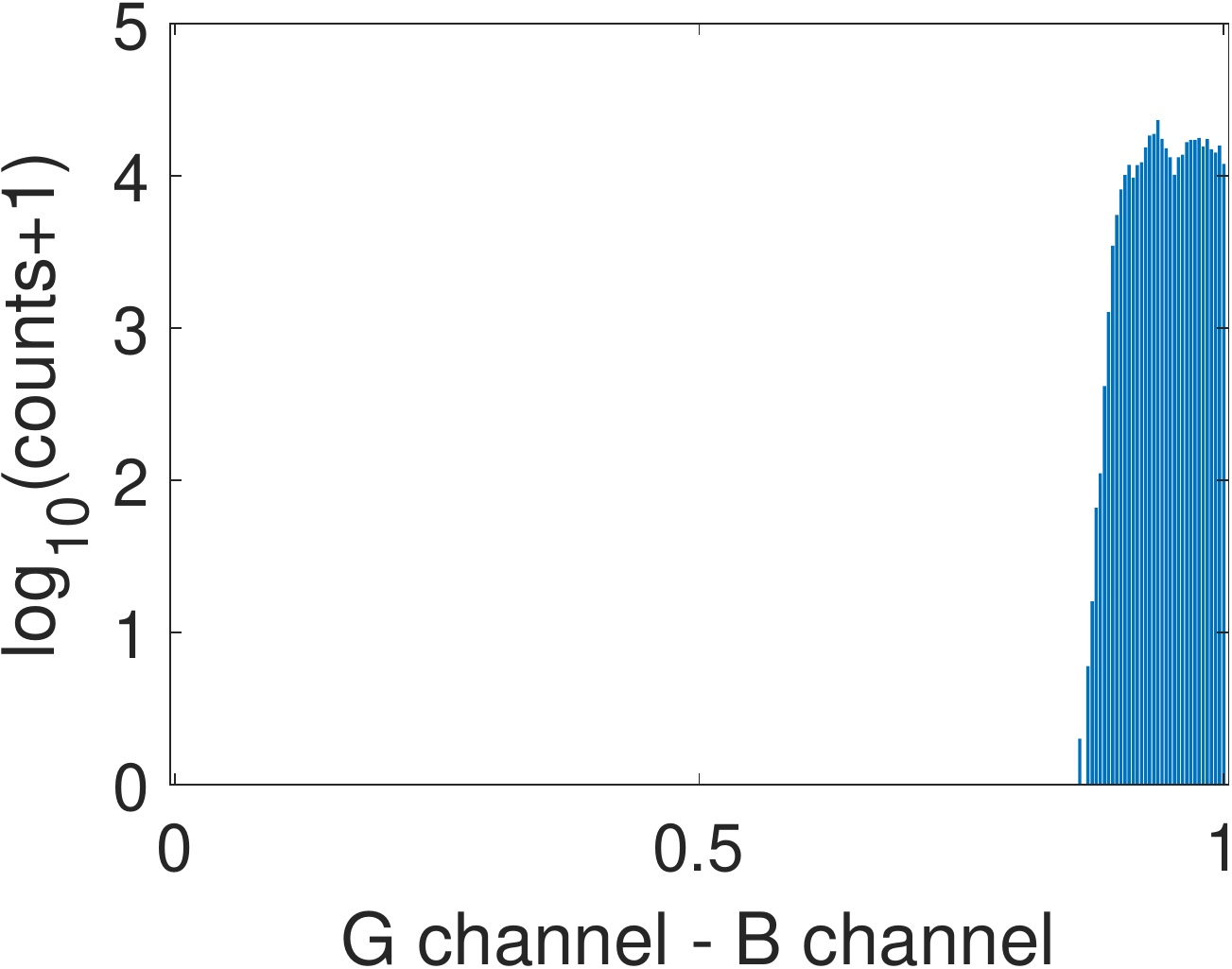}}
\subfigure[]{
\label{fig:imhist:G_R_M2} 
\includegraphics[width=1.1in]{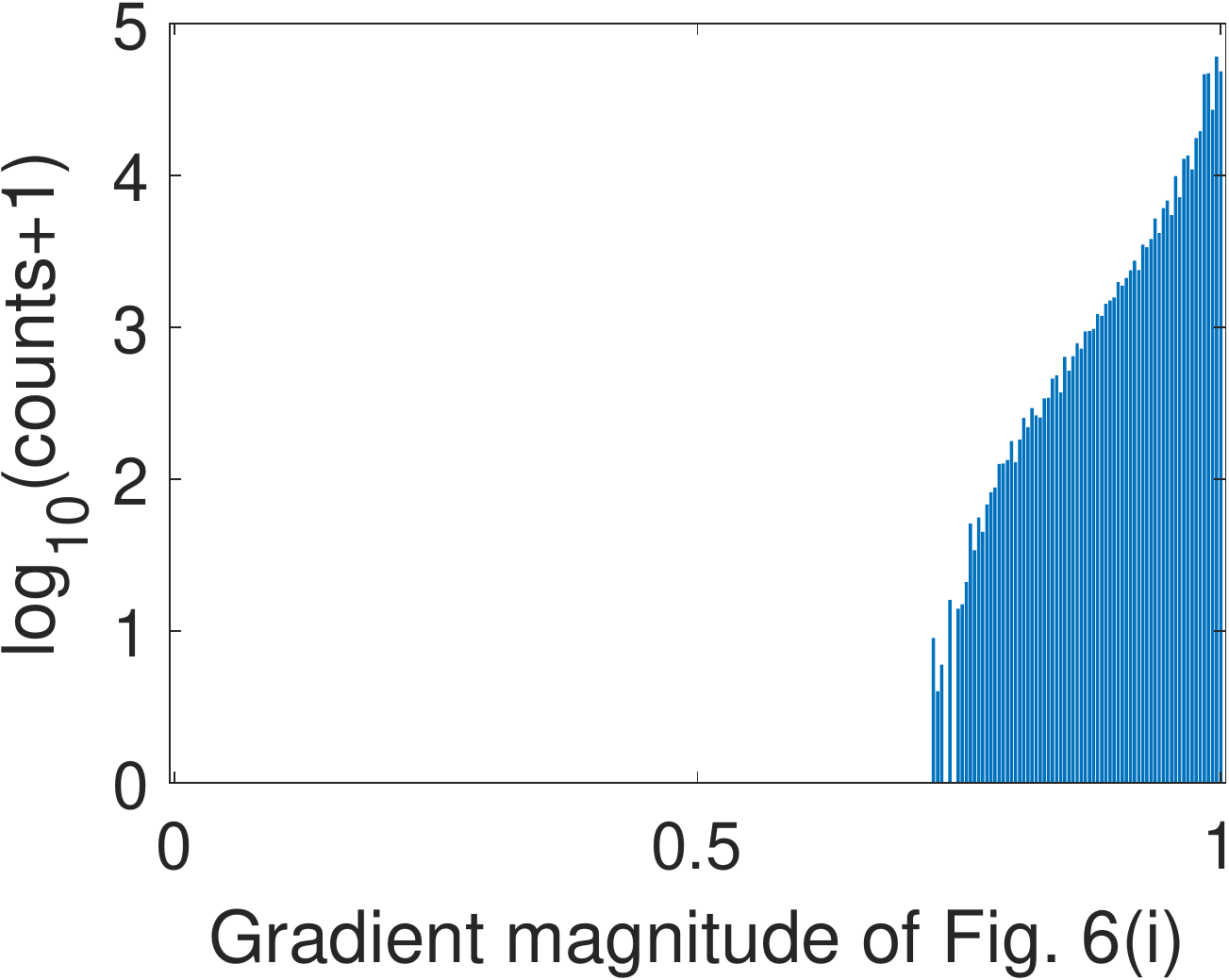}}
\subfigure[]{
\label{fig:imhist:G_B_M2} 
\includegraphics[width=1.1in]{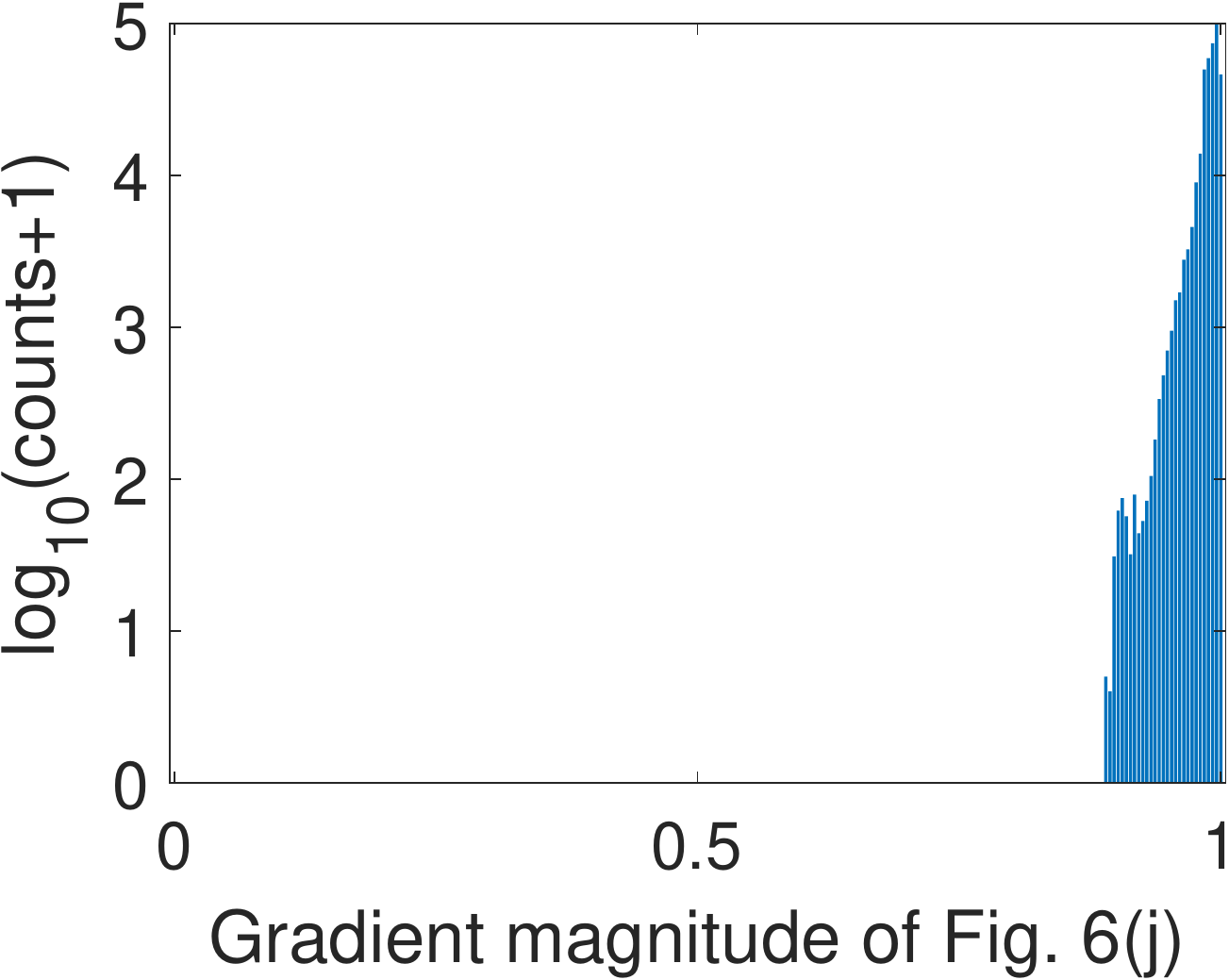}}
\subfigure[]{
\label{fig:imhist:GMSD__center} 
\includegraphics[width=1.1in]{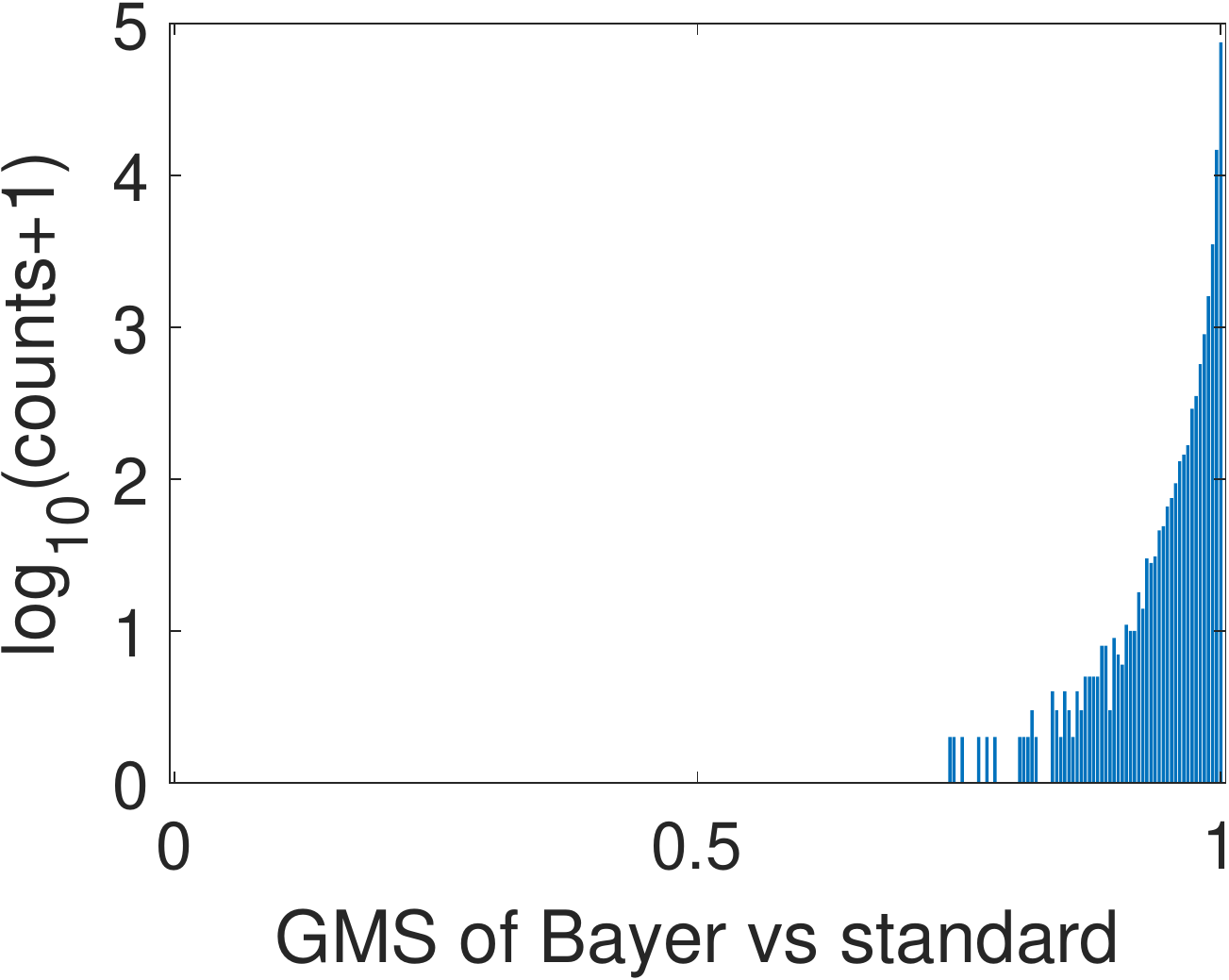}}
\subfigure[]{
\label{fig:imhist:GMSD__sobel} 
\includegraphics[width=1.1in]{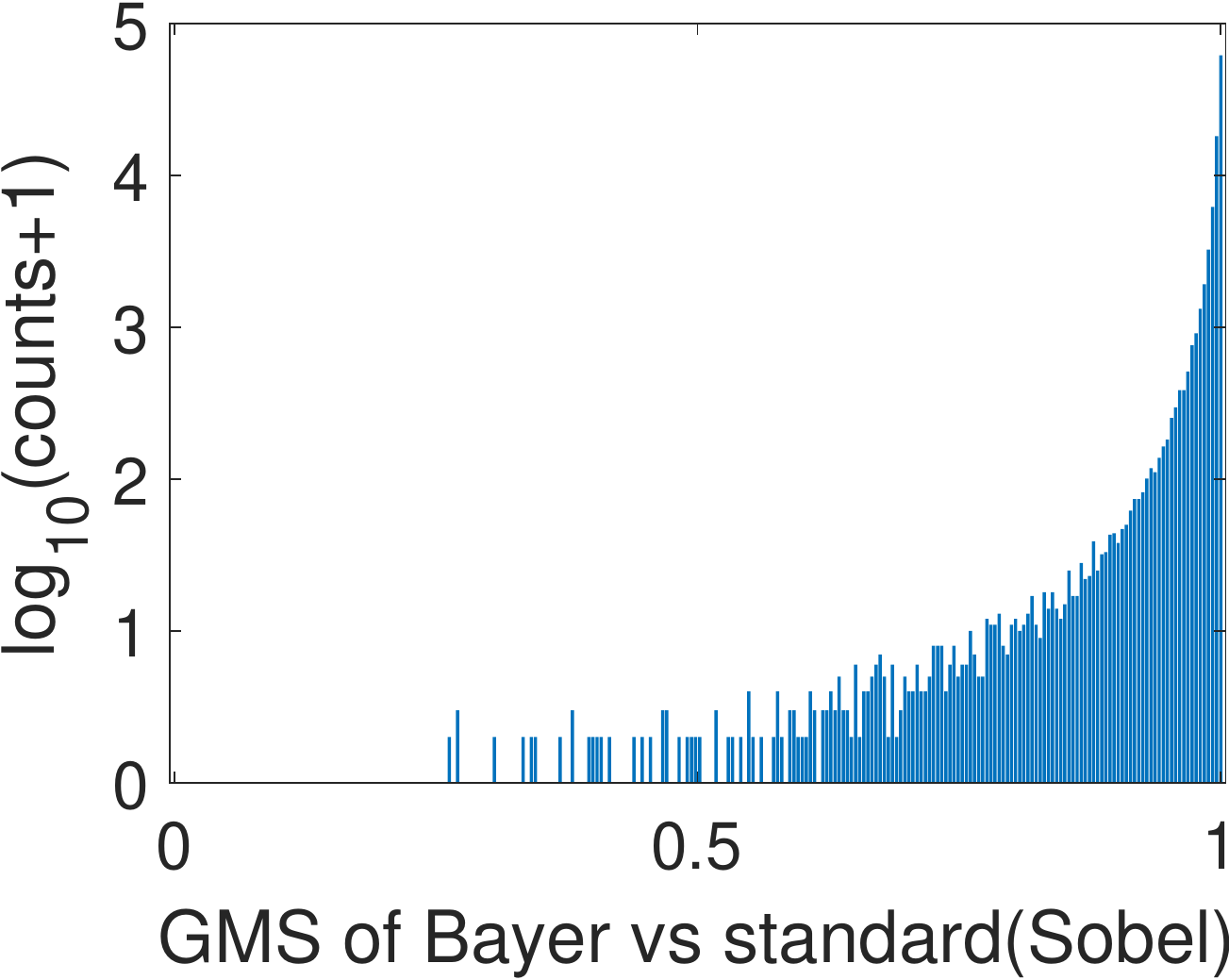}}

\caption{The gray-level histograms of (g) - (l) in Fig. \ref{fig:gradient and missmap}.}
\label{fig:imhist}
\vspace{-3mm}
\end{figure*}

As we can see, the two input pixels for coefficients 1 and -1 in the convolution templates are from the same channel, i.e., differences are always computed on homogeneous pixels. As shown in Fig. \ref{fig:gradientCompute}, applying the convolution templates at locations $(1,2)$ generates
\begin{equation}
\label{equ:c1}
G_{x\left ( 1,2 \right )}^{B}=I^{B}_{\left ( 1,3 \right )}-I^{B}_{\left ( 1,1 \right )},
\end{equation}
\begin{equation}
\label{equ:c2}
G_{y\left ( 1,2 \right )}^{R}=I^{R}_{\left ( 2,2 \right )}-I^{R}_{\left ( 0,2 \right )},
\end{equation}
where $G^{B}$ and $G^{R}$ are the gradients of the blue and red channels, respectively.

In the demosaicing tasks, it is a common practice to interpolate the G channel first followed by the R/B channels. This is because there are twice as many G channel pixels as R/B channel pixels in Bayer pattern images. The color difference constancy assumption in \eqref{equ:color difference constancy assumption} can then be used to estimate the missing pixels of the R and B channels.

\begin{equation}
\label{equ:general_color_difference_constancy1}
I^{G}(x,y)=I^{k}\left ( x,y \right )+C^{k}(x,y).
\end{equation}
Here, $k$ represents either R or B channel, $C^{k}(x,y)$ is the difference between the R/B channel and the G channel at pixel location $(x,y)$, which needs to be estimated in demosaicing tasks\cite{RN46}.

Consider two pixels within a small neighborhood at locations $(x,y)$ and $(x',y')$, according to \eqref{equ:general_color_difference_constancy1}, we have
\begin{equation}
\label{equ:red combine1}
\begin{split}
&I^{G}(x,y)-I^{G}(x',y')\\=&I^{k}(x,y)-I^{k}( x',y')+C^{k}(x,y)-C^{k}(x',y')\\
=&I^{k}(x,y)-I^{k}( x',y')+\delta^{k}(x,y,x',y').
\end{split}
\end{equation}
Where $\delta^{k}(x,y,x',y')=C^{k}(x,y)-C^{k}(x',y')$. The value of $\delta^{k}(x,y,x',y')$ is crucial in our analysis and will be discussed in detail.

Generally, there are flat areas (e.g. background) and texture areas (e.g., corners and edges) in a natural image, these two situations will be discussed separately.

For the flat areas, the difference between two pixels is negligible such that
\begin{equation}
\begin{split}
\label{equ:apply1}
I^{G}(x,y)-I^{G}(x',y')=I^{k}(x,y)-I^{k}(x',y')\approx 0,
\end{split}
\end{equation}
i.e., $\delta^{k}(x,y,x',y')\approx 0$. This means the intensity difference between channels is approximately constant across nearby pixel locations, i.e., the color changes are small in the neighborhood of flat areas.

Importantly, a constant $C^k$ also means that some non-smooth color transitions (i.e., texture) are included as well. For example, for the two synthetic images in the top row of Fig. \ref{fig:edgeCase:orig}, suppose $(x,y)$ is a point in the background and $(x',y')$ is another point in the foreground. These two images correspond to the situation of $C^{k}(x,y)=C^{k}(x',y')$. Fig. \ref{fig:edgeCase:G_R}-\ref{fig:edgeCase:G_B} are the difference images of $G-R$ and $G-B$, respectively, while Fig. \ref{fig:edgeCase:G_R_m}-\ref{fig:edgeCase:G_B_m} are the corresponding gradient maps. Note that all the gradient maps and difference images are displayed as inverse images (1 $-$ original gray value), where 1 means difference or gradient is zero and the the corresponding location is displayed as white (likewise for Fig. \ref{fig:gradient and missmap} and \ref{fig:Channels}). Then,
\begin{enumerate}
\item Fig. \ref{fig:edgeCase:orig} top-left: $I^{R}=I^{G}=I^{B}$ for both background and foreground. This results in $C^{k}(x,y)=0$, which further leads to $\delta^{k}(x,y,x',y')=0$, as shown in Fig. \ref{fig:edgeCase:G_R}-\ref{fig:edgeCase:G_B_m}, top-left images.
\item Fig. \ref{fig:edgeCase:orig} top-right: $I^{R}=I^{B}$ for both background and foreground. This results in a constant $C^{k}(x,y)$ for both background and foreground, which further leads to $\delta^{k}(x,y,x',y')=0$, as shown in Fig. \ref{fig:edgeCase:G_R}-\ref{fig:edgeCase:G_B_m}, top-right images.
\end{enumerate}
For the above two cases, although there are obvious edges in the original images, we still have $\delta^{k}(x,y,x',y')=0$.

For the more extreme texture areas, the analysis is more complex. To analyze the areas with complex textures, \eqref{equ:red combine1} can be further rewritten as
\begin{equation}
\label{equ:red combine2}
\delta^{k}(x,y,x',y')\!=\!(I^{G}(x,y)\!-\!I^{k}(x,y))\!-\!(I^{G}(x',y')\!-\!I^{k}( x',y')).
\end{equation}
Note that image's gradients are always computed among a small neighborhood. Considering the central difference-based horizontal gradient computation at pixel location $(x,y)$, we have
\begin{equation}
\begin{split}
\label{equ:delta_gradient}
&\delta^{k}(x+1,y,x-1,y)\\=&(I^{G}(x\!+\!1,y)\!-\!I^{k}(x\!+\!1,y))\!-\!(I^{G}(x\!-\!1,y)\!-\!I^{k}(x\!-\!1,y))\\
      =&I^{G-k}(x+1,y)-I^{G-k}(x-1,y)\\
      =&G_{x}^{G-k}(x,y).
\end{split}
\end{equation}
Here, $G-k$ represents the difference image of the G channel and the R/B channel. It can be observed from \eqref{equ:delta_gradient} that $\delta^{k}(x+1,y,x-1,y)$ is exactly the gradient of the difference image at location $(x,y)$. It has been shown in  \cite{RN48} that the difference images are slowly-varying over a spatial domain, meaning that the gradient $G_{x}^{G-k}(x,y)$ in \eqref{equ:delta_gradient} is negligible, i.e., $\delta^{k}(x+1,y,x-1,y)$ approximates to zero.
This can be illustrated by the bottom-left image in Fig. \ref{fig:edgeCase:orig}. Suppose $(x,y)$ is a point on the background border such that $(x+1,y)$ and $(x-1,y)$ are two points in the foreground and background, respectively. In this case, the following relationship holds (\ref{fig:edgeCase:orig}, bottom-left).
\begin{equation}
\begin{split}
\label{equ:relation}
\left\{\begin{array}{l}
C^{k}(x+1,y) \neq C^{k}(x-1,y),\\
Sgn(C^{k}(x+1,y))=Sgn(C^{k}(x-1,y)).
\end{array}\right.
\end{split}
\end{equation}
Here, $Sgn(\cdot )$ is the signum function. Let $k=B$, the results of \eqref{equ:delta_gradient} is
 \begin{equation}
\begin{split}
\label{equ:relation2}
&\delta^{B}(x+1,y,x-1,y)\\=&(I^{G}(x\!+\!1,y)\!-\!I^{B}(x\!+\!1,y))\!-\!(I^{G}(x\!-\!1,y)\!-\!I^{B}(x\!-\!1,y))\\
                      =&(0.45-0.8)-(0.22-0.6)\\
                      =&0.03
\end{split}
\end{equation}
The corresponding result is illustrated in Fig. \ref{fig:edgeCase:G_B_m} bottom-left, where the edge is negligible.

Note that there are also failure cases. For example, the bottom-right image in Fig. \ref{fig:edgeCase:orig} satisfies
\begin{equation}
\begin{split}
\label{equ:relation3}
\left\{\begin{array}{l}
C^{k}(x+1,y) \neq C^{k}(x-1,y),\\
Sgn(C^{k}(x+1,y))\neq Sgn(C^{k}(x-1,y)).
\end{array}\right.
\end{split}
\end{equation}
In this case, the result of \eqref{equ:delta_gradient} will be $\delta^{k}(x+1,y,x-1,y)=1$, which is illustrated in Fig. \ref{fig:edgeCase:G_R_m} and \ref{fig:edgeCase:G_B_m}, bottom-right. Details of failure cases will be discussion in Section \ref{section:sec4.3.1}.

Fig. \ref{fig:gradient and missmap} illustrates two situations: image with dull colors (top) and image with high saturation colors (bottom). Fig. \ref{fig:gradient and missmap:G_R} and Fig. \ref{fig:gradient and missmap:G_B} are the difference images of G channel $-$ R channel and G channel $-$ B channel ($C^{k}(x,y)$ in Eq. \ref{equ:color ratio constancy assumption1}), respectively, while Fig. \ref{fig:gradient and missmap:G_R_m} and Fig. \ref{fig:gradient and missmap:G_B_m}  are the corresponding gradient magnitude maps of the difference images $(\delta^{k}(x+1,y,x-1,y)$ in \eqref{equ:delta_gradient}) computed using the central difference operator.
The corresponding gray-level histograms of these images are shown in Fig. \ref{fig:imhist:G_R2}-\ref{fig:imhist:G_B_M2}. It can be found that for images with dull colors, the differences between G and R channels are distributed in a much smaller range than that of high saturation color images, while the differences between G and B channels are distributed in a larger range. For all these images, most of $\delta^{k}(x+1,y,x-1,y)$ are distributed in a small range, as shown in Fig. \ref{fig:imhist:G_R_M2} and \ref{fig:imhist:G_B_M2}. The distribution in Fig. \ref{fig:imhist:G_R_M2} bottom is wider than the other three plots, which are caused by texture areas such as hat and hair edges in Fig. \ref{fig:gradient and missmap:G_R_m}. As we can see, apart from these small exceptions, Fig. \ref{fig:gradient and missmap:G_R_m} and Fig. \ref{fig:gradient and missmap:G_B_m} are almost all white. Therefore, for most cases, $\delta^{k}(x,y,x',y')$ is small and can be ignored if pixel locations $(x,y)$ and $(x', y')$ are within a small neighborhood.

As a result of the above discussion, \eqref{equ:red combine1} can be rewritten as
\begin{equation}
\label{equ:move1}
I^{G}\left ( x,y \right )-I^{G}\left ( {x}',{y}'\right )\approx I^{R}\left ( x,y \right )-I^{R}\left ({x}',{y}' \right ),
\end{equation}
\begin{equation}
\label{equ:move2}
I^{G}\left ( x,y \right )-I^{G}\left ( {x}',{y}'\right )\approx I^{B}\left ( x,y \right )-I^{B}\left ({x}',{y}' \right ).
\end{equation}
Combining the gradient definition of \eqref{equ:central differencex} and \eqref{equ:central differencey} with \eqref{equ:move1} and \eqref{equ:move2}, we have

\begin{equation}
\label{equ:cdc}
G\approx G^{G}\approx G^{R}\approx G^{B},
\end{equation}
meaning that the gradients of natural images can be computed using any one of the three channels as long as the color difference constancy holds. Combining \eqref{equ:cdc} with \eqref{equ:c1} and \eqref{equ:c2}, we have
\begin{equation}
\label{equ:conbinex}
G_{x\left ( 1,2 \right )}=G^{B}_{x\left ( 1,2 \right )}=I^{B}_{\left ( 1,3 \right )}-I^{B}_{\left ( 1,1 \right )},
\end{equation}
\begin{equation}
\label{equ:conbiney}
G_{y\left ( 1,2 \right )}=G^{R}_{x\left ( 1,2 \right )}=I^{R}_{\left ( 2,2 \right )}-I^{R}_{\left ( 0, 2 \right )}.
\end{equation}
Therefore, even though two color components are missing at each pixel, the gradients of location $(1,2)$ can be computed directly from the Bayer pattern image using the blue and red channel. The gradients of any other pixel locations can be computed in the same manner.

Generally, the above conclusion can be extended to other symmetrical first-order differential operators (with alternating zero and nonzero coefficients) on any kind of Bayer pattern. Let us take the Sobel operators in Fig. \ref{fig:Gradient operators:sobel} as an example. Applying the Sobel operators in Fig. \ref{fig:Gradient operators:sobel} to the pixel location $(1,2)$ of the Bayer pattern image results
\begin{equation}
\begin{split}
\label{equ:sobelx}
G_{x\left ( 1,2 \right )}^{'}\!=\!I_{\left ( 0,3 \right )}^{G}\!+\!2\!\times\!I_{\left ( 1,3 \right )}^{B}\!+\!I_{\left ( 2,3 \right )}^{G}\!-\!I_{\left ( 0,1 \right )}^{G}\!-\!2\!\times \!I_{\left ( 1,1 \right )}^{B}\!-\!I_{\left ( 2,1 \right )}^{B}\\
=\!( I_{\left ( 0,3 \right )}^{G}\!-\!I_{\left ( 0,1 \right )}^{G})\!+\!2\! \times\!(I_{\left ( 1,3 \right )}^{B}\!-\!I_{\left ( 1,1 \right )}^{B})\!+\! (I_{ ( 2,3 )}^{G}\!-\!I_{ ( 2,1)}^{G}),
\end{split}
\end{equation}
\begin{equation}
\begin{split}
\label{equ:sobely}
G_{y\left ( 1,2 \right )}^{'}\!=\!I_{\left ( 2,1 \right )}^{G}\!+\!2\!\times\!I_{\left ( 2,2 \right )}^{R}\!+\!I_{\left ( 2,3 \right )}^{G}\!-\!I_{\left ( 0,1 \right )}^{G}\!-\!2\!\times \!I_{\left ( 0,2 \right )}^{R}\!-\!I_{\left ( 0,3 \right )}^{G}\\
=\!( I_{\left ( 2,1 \right )}^{G}\!-\!I_{\left ( 0,1 \right )}^{G} )\!+\!2\! \times\!(I_{\left ( 2,2 \right )}^{R}\!-\!I_{\left ( 0,2 \right )}^{R})\!+\! ( I_{\left ( 2,3 \right )}^{G}\!-\!I_{\left ( 0,3 \right )}^{G} ).
\end{split}
\end{equation}
As for gradient computation using \eqref{equ:central differencex} and \eqref{equ:central differencey}, differences are always computed on homogeneous pixels for Sobel-based differential operations in \eqref{equ:sobelx} and \eqref{equ:sobely}, i.e., pixel values are always subtracted from pixel values of the same channel. Moreover, according to the color difference constancy assumption, \eqref{equ:sobelx} and \eqref{equ:sobely} can be rewritten as
\begin{equation}
\begin{split}
\label{equ:sobelx_extention}
G_{x\left ( 1,2 \right )}^{'}\!
& \approx \!( I_{\left ( 0,3 \right )}^{\widehat{R}}\!-\!I_{\left ( 0,1 \right )}^{\widehat{R}})\!+\!2\! \times\!(I_{\left ( 1,3 \right )}^{\widehat{R}}\!-\!I_{\left ( 1,1 \right )}^{\widehat{R}})\!+\! (I_{ ( 2,3 )}^{\widehat{R}}\!-\!I_{ ( 2,1)}^{\widehat{R}})\\
& \approx \!( I_{\left ( 0,3 \right )}^{G}\!-\!I_{\left ( 0,1 \right )}^{G})\!+\!2\! \times\!(I_{\left ( 1,3 \right )}^{\widehat{G}}\!-\!I_{\left ( 1,1 \right )}^{\widehat{G}})\!+\! (I_{ ( 2,3 )}^{G}\!-\!I_{ ( 2,1)}^{G})\\
& \approx \!( I_{\left ( 0,3 \right )}^{\widehat{B}}\!-\!I_{\left ( 0,1 \right )}^{\widehat{B}})\!+\!2\! \times\!(I_{\left ( 1,3 \right )}^{B}\!-\!I_{\left ( 1,1 \right )}^{B})\!+\! (I_{ ( 2,3 )}^{\widehat{B}}\!-\!I_{ ( 2,1)}^{\widehat{B}}),
\end{split}
\end{equation}
\begin{equation}
\begin{split}
\label{equ:sobely_extention}
G_{y\left ( 1,2 \right )}^{'}\! & \approx
\!( I_{\left ( 2,1 \right )}^{\widehat{R}}\!-\!I_{\left ( 0,1 \right )}^{\widehat{R}} )\!+\!2\! \times\!(I_{\left ( 2,2 \right )}^{R}\!-\!I_{\left ( 0,2 \right )}^{R})\!+\! ( I_{\left ( 2,3 \right )}^{\widehat{R}}\!-\!I_{\left ( 0,3 \right )}^{\widehat{R}} )\\
& \approx \!( I_{\left ( 2,1 \right )}^{G}\!-\!I_{\left ( 0,1 \right )}^{G} )\!+\!2\! \times\!(I_{\left ( 2,2 \right )}^{\widehat{G}}\!-\!I_{\left ( 0,2 \right )}^{\widehat{G}})\!+\! ( I_{\left ( 2,3 \right )}^{G}\!-\!I_{\left ( 0,3 \right )}^{G} )\\
& \approx \!( I_{\left ( 2,1 \right )}^{\widehat{B}}\!-\!I_{\left ( 0,1 \right )}^{\widehat{B}} )\!+\!2\! \times\!(I_{\left ( 2,2 \right )}^{\widehat{B}}\!-\!I_{\left ( 0,2 \right )}^{\widehat{B}})\!+\! ( I_{\left ( 2,3 \right )}^{\widehat{B}}\!-\!I_{\left ( 0,3 \right )}^{\widehat{B}} ),
\end{split}
\end{equation}
where $\widehat{R}$, $\widehat{G}$ and $\widehat{B}$ represent the missing color components at the corresponding locations. Therefore, the Sobel-based gradients can also be extracted directly from the Bayer pattern images as long as the color difference constancy holds.

In terms of different Bayer patterns, they are merely different arrangements of the RGB pixels, while the alternating pattern of $R$, $G$ and $B$ at each row and column are preserved. For example, discarding the first column of the Bayer pattern in Fig. \ref{fig:gradientCompute} generates the {GRBG} Bayer pattern. Therefore, different Bayer patterns do not have any impact on the applicability of the discussed differential operators to Bayer pattern images. Moreover, the discussed gradient extraction method can be directly extended to other special CFA patterns with alternating color filter arrangements, e.g., RYYB, RGB-IR, as long as the arrangement of CFA patterns matches the gradient operators such that gradient operations are performed on the same color channel, i.e., subtract or add operations are performed on the same color channel, and the coefficients of the subtract or add terms in the gradient operator are equal such that the gradients compute from R/B channel can be approximated to G channel.

To validate the proposed Bayer pattern image-based gradient extraction, the differential operators in Fig. \ref{fig:Gradient operators} are applied to true color images Kodim17 and Kodim04 from the Kodak image dataset\cite{franzen} and the corresponding resampled Bayer version.
The generated gradient maps are shown in Fig. \ref{fig:gradient and missmap}. For display purpose, images in Fig. \ref{fig:gradient and missmap:k03gray} are shown as color images while the gradient magnitude maps in Fig. \ref{fig:gradient and missmap:grayCenter} and \ref{fig:gradient and missmap:graySobel} are computed from the corresponding gray-scale images generated using Fig. \ref{fig:gradient and missmap:k03gray}.
The Bayer pattern images in Fig. \ref{fig:gradient and missmap:k03bayer} are presented as three-channel images to illustrate its Bayer ``mosaic'' structure. For the clearness of presentation, all the gradient maps and difference images in Fig. \ref{fig:gradient and missmap} are displayed as inverse images. As illustrated in Fig. \ref{fig:gradient and missmap}, the gradient maps generated from the Bayer pattern images look almost the same as that generated from the true color version.
To compare these gradient maps, two GMS maps are presented in Fig. \ref{fig:gradient and missmap:GMS_center} and \ref{fig:gradient and missmap:GMS_sobel}, and the corresponding distributions of GMS values are presented in Fig.\ref{fig:imhist:GMSD__center} and \ref{fig:imhist:GMSD__sobel}.
As can be seen, the GMS maps are almost pure white, and the histograms are distributed in a small range, meaning that the compared gradient maps are very close to each other.
Overall, gradients of Bayer images yield a good approximation of the image gradients, except for a few pixels around certain color edges.

\begin{figure}[t]
\centering
\subfigure[]{
\label{fig:superpixel and blur:kodim01} 
\includegraphics[width=1.6in]{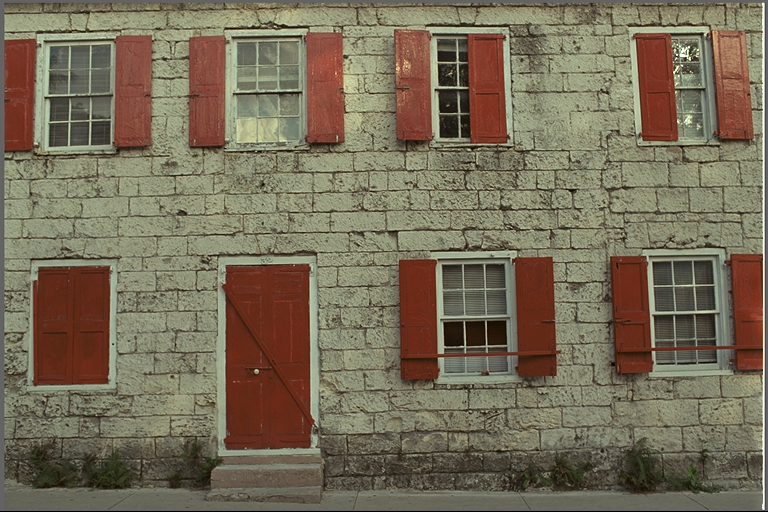}}
\subfigure[]{
\label{fig:superpixel and blur:colorBlur} 
\includegraphics[width=1.6in]{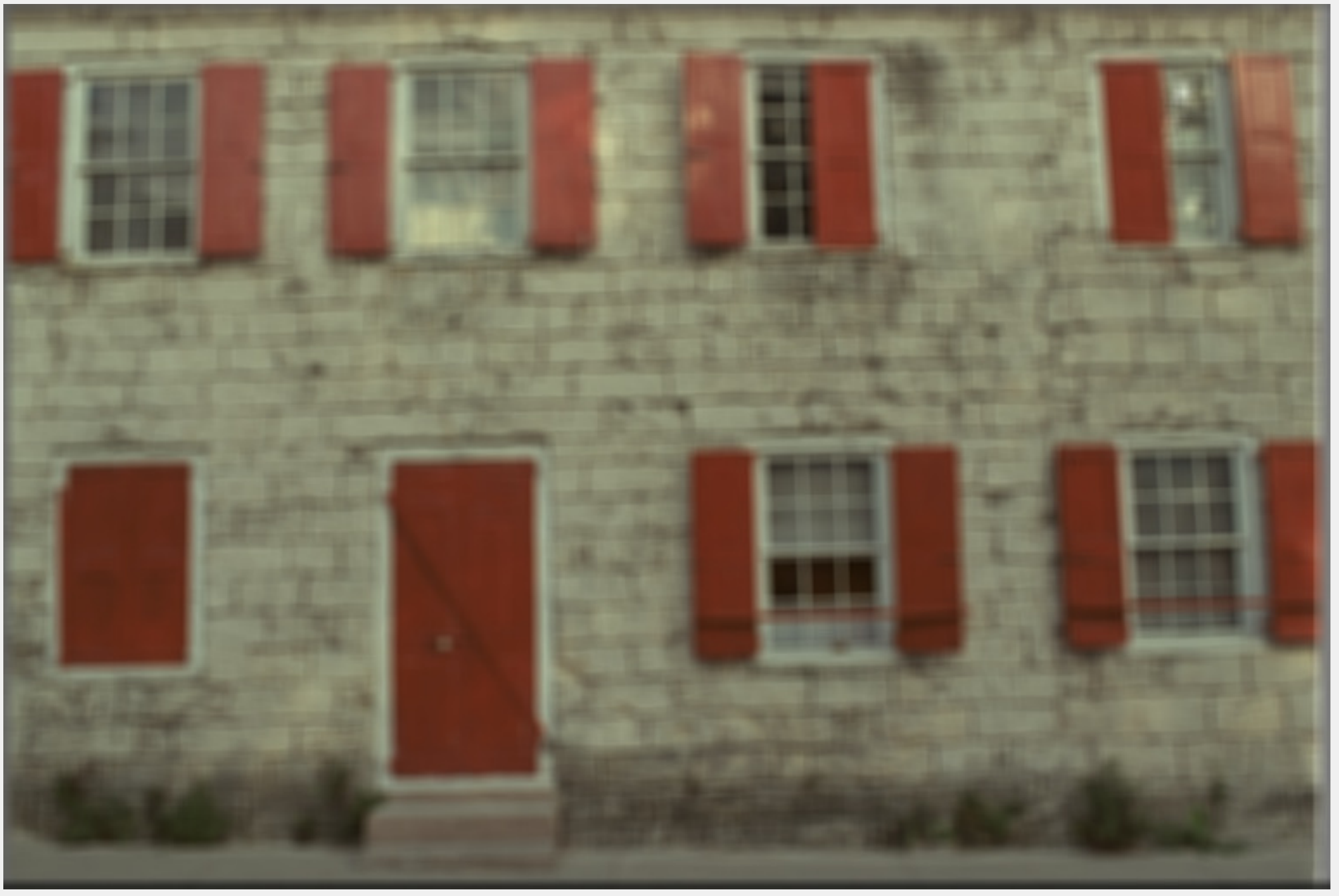}}\\
\subfigure[]{
\label{fig:superpixel and blur:bayer} 
\includegraphics[width=1.6in]{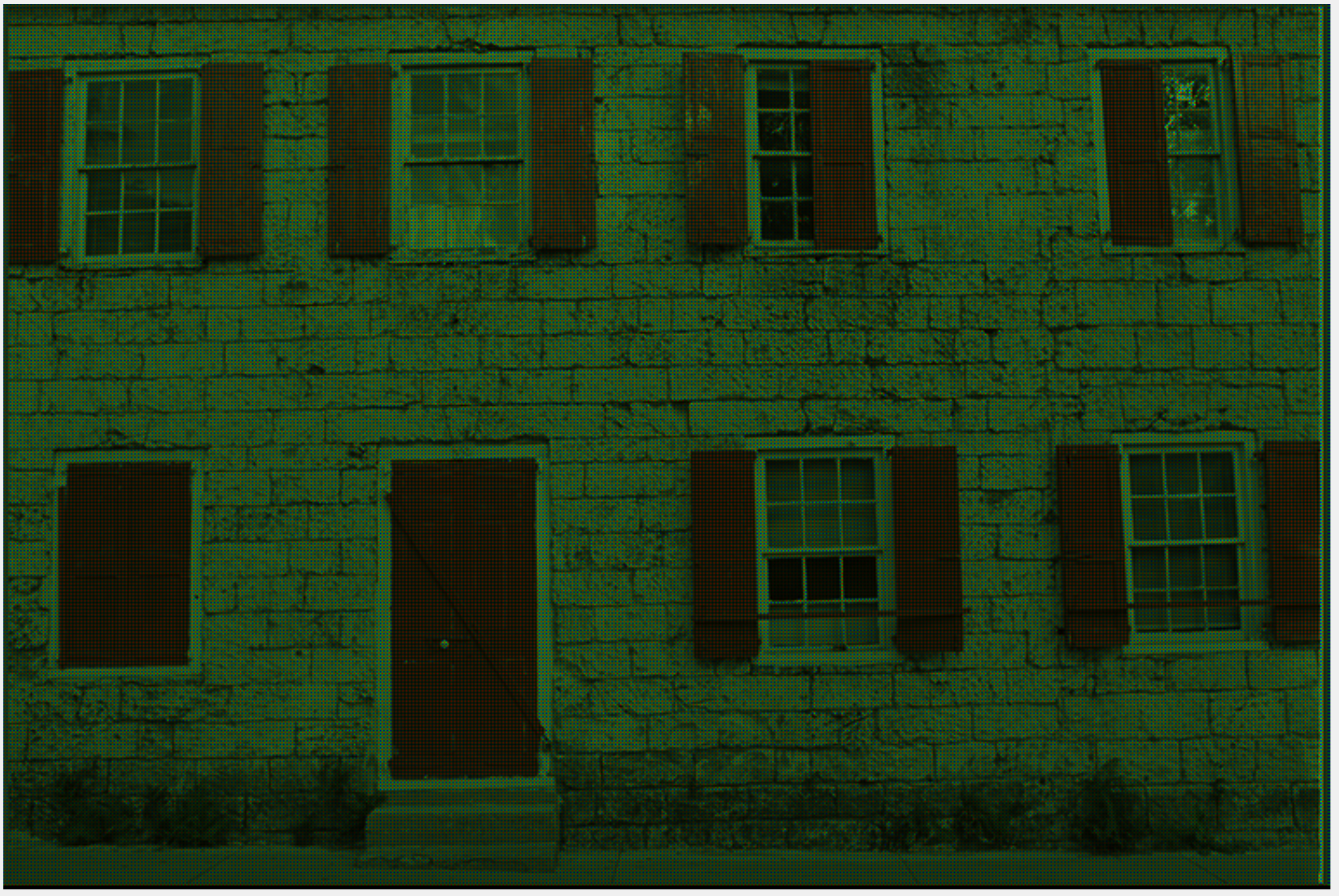}}
\subfigure[]{
\label{fig:superpixel and blur:bayerBlur} 
\includegraphics[width=1.6in]{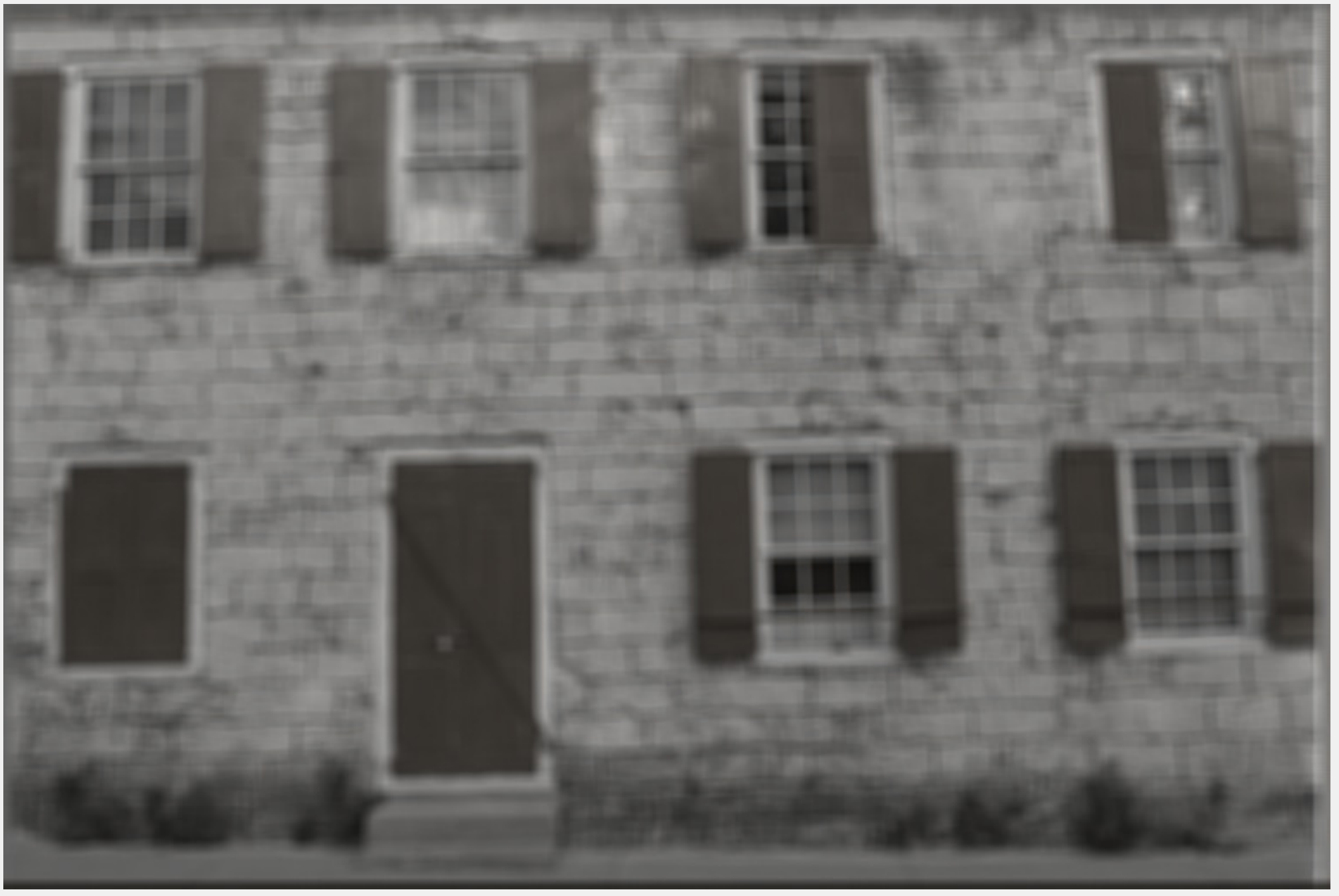}}\\
\centering
\subfigure[]{
\label{fig:superpixel and blur:mosaic2} 
\includegraphics[width=1.6in,height=1.6in]{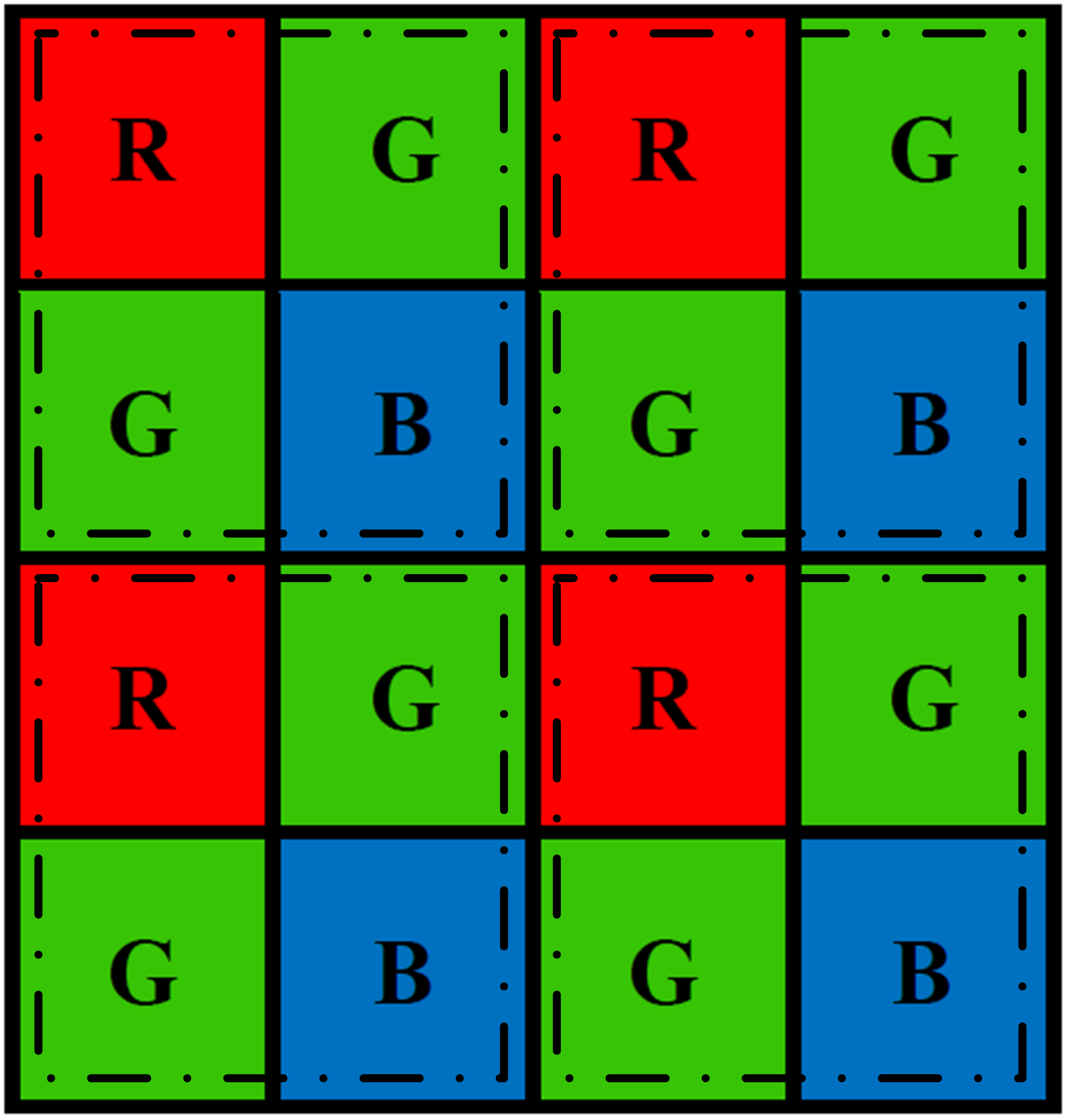}}
\subfigure[]{
\label{fig:superpixel and blur:superpixelBlur} 
\includegraphics[width=1.6in]{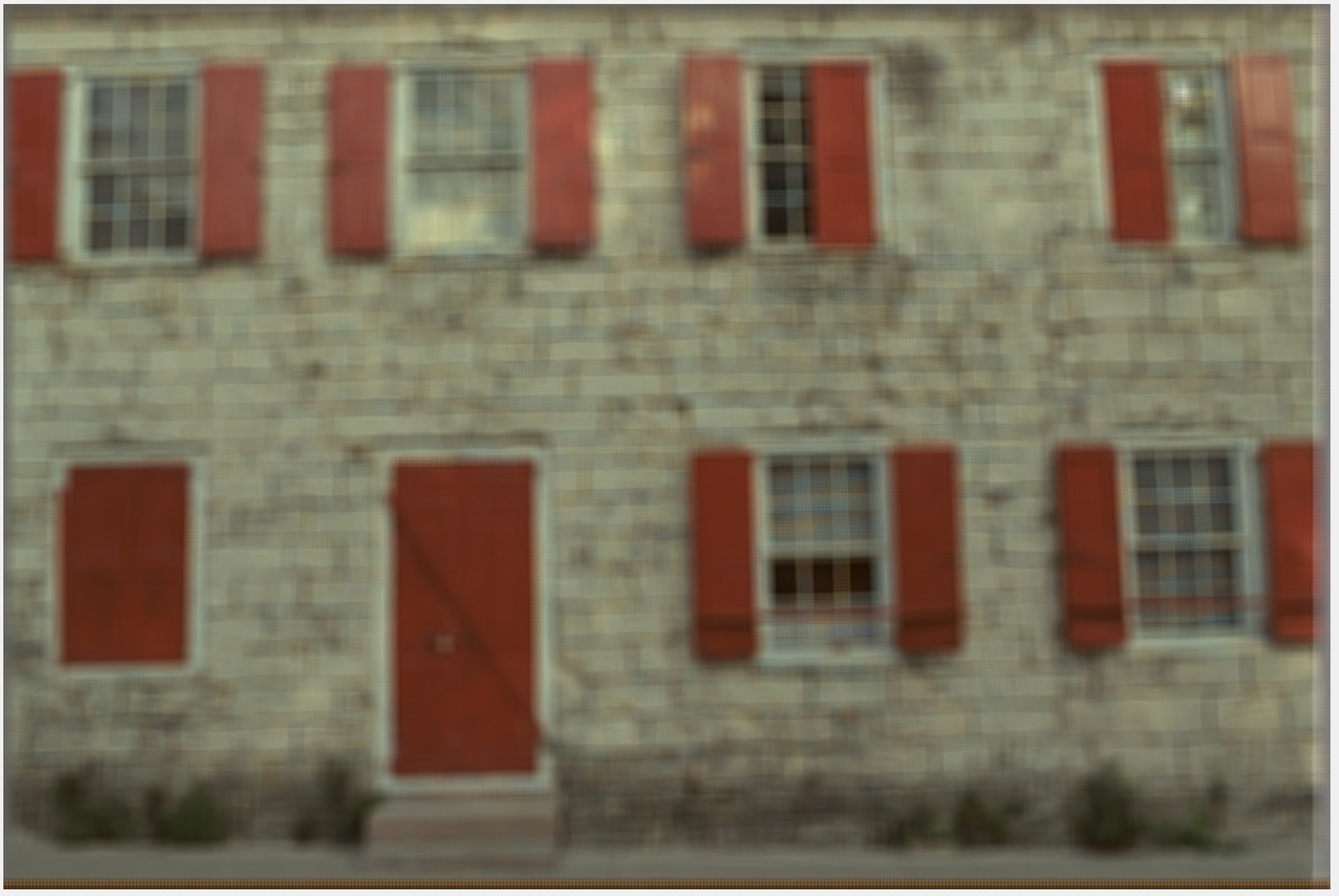}}
\caption{(a) The true color image Image Kodim01. (b) Gaussian blurred version of (a). (c) Bayer version of (a). (d) Direct Gaussian blurred version of (c) then demosaicing. (e) The $2\times2$ super-pixel structure. (f) Super-pixel structure-based Gaussian blurred version of (c) then demosaicing. }
\label{fig:superpixel and blur}
\vspace{-3mm}
\end{figure}

\subsection{The Multiscale Model for Bayer Pattern Images}
\label{section:sec3.2}
In SIFT, the scale-space is approximated by a DoG pyramid. The construction of the DoG pyramid can be divided into two parts: Gaussian blurring at different scales and resizing of the blurred images. Due to the special alternating pixel arrange of Bayer pattern images, directly Gaussian blur the images will destroy the ``mosaic structure''. This phenomenon
is illustrated in Fig. \ref{fig:superpixel and blur}. If the Bayer pattern image is directly Gaussian filtered, the resulting image (after demosaicing) looks like a ``three-channel grayscale image'' as shown in Fig. \ref{fig:superpixel and blur:bayerBlur}, meaning that the Bayer image is treated as a single channel image, ignoring the ``mosaic structure'' from the beginning and, effectively, loosing/destroying the color information. Thus, smoothing on Bayer pattern image directly will blend the channels, which is akin to a RGB-to-gray conversion. Moreover, loss of the color information makes some of the algorithms in the SIFT family such as ``C-SIFT'' and ``RGB-SIFT'' no longer applicable.

To address the above mentioned problem, the super-pixel approach as illustrated in Fig. \ref{fig:superpixel and blur:mosaic2} is used in this work. A super-pixel is a compound pixel consisting of a complete Bayer pattern. The Bayer pattern image can therefore be regarded as a ``continuous'' image filling with super-pixels. Operating on the super-pixel structure preserves the Bayer pattern of the original images. Fig. \ref{fig:superpixel and blur:superpixelBlur} shows the Gaussian blurred image (after demosaicing) based on the super-pixel structure. As can be seen, it is close to that generated by the full color approach. Moreover, the super-pixel structure can also be used for resizing when constructing the scale space. The detailed comparison results will be presented in Section \ref{section:sec4.3}.

\section{Experiments}
\label{section:sec4}
In this section, experimental results are presented to demonstrate the effectiveness of the proposed  Bayer pattern image-based gradient extraction. The datasets used in the experiments are introduced first, followed by the details of the experiments setup and evaluation results.
\subsection{Datasets}
There are five datasets used for differents experiments in this work. Among these five datasets, four are commonly used benchmarks in different image processing and computer vision tasks such as demosaicing, pedestrian detection. A brief description of these datasets is presented in Table \ref{tad:datasets}.

\begin{table*}[t]
\centering
\setlength\arrayrulewidth{0.6pt}
\renewcommand{\arraystretch}{1.2}
\setlength{\tabcolsep}{1pt}{
\caption{Notation of Different Datasets Used in Experiments.}
\label{tad:datasets}
\begin{tabular}{|c|l|c|c|c}
\cline{1-4}
\multirow{2}{*}{\textbf{Datasets}}                                                   & \multicolumn{1}{c|}{\multirow{2}{*}{\textbf{Brief Introduction}}}                                                                                                                 & \multicolumn{2}{c|}{\textbf{Generation of the Corresponding Color/Bayer Versions}}                                                                & \multirow{2}{*}{\textbf{}} \\ \cline{3-4}
                                                                                     & \multicolumn{1}{c|}{}                                                                                                                                               & Color                 & Bayer                                                                                                                     &                            \\ \cline{1-4}
\begin{tabular}[c]{@{}c@{}}The Kodak lossless \\ true color image suite \cite{franzen}\end{tabular} & A popular standard test suite for demosaicing algorithms.                                                                                                           & -                     & \multicolumn{1}{l|}{\begin{tabular}[c]{@{}l@{}}Resampling according to the corresp-\\onding Bayer pattern.\end{tabular}} &                            \\ \cline{1-4}
\begin{tabular}[c]{@{}c@{}}The SHTech \\ pedestrian dataset\end{tabular}             & \begin{tabular}[c]{@{}l@{}}Our own pedestrian dataset shoot by a Huawei Honor 8 mobile \\ phone with the FreeDcam APP\cite{freexperia} to by pass the entire ISP.\end{tabular} & ISP pipeline in \cite{RN36}. & -                                                                                                                         &                            \\ \cline{1-4}
\begin{tabular}[c]{@{}c@{}}The PASCALRAW\\ dataset \cite{RN53}\end{tabular}               & A  recently published raw image dataset for object detection.                                                                                                       & ISP pipeline in \cite{RN36}. & -                                                                                                                         &                            \\ \cline{1-4}
\begin{tabular}[c]{@{}c@{}}The INRIA \\ pedestrian dataset \cite{RN25} \end{tabular}              & A popular dataset for pedestrian detection algorithms.                                                                                                              & -                     & \multicolumn{1}{l|}{\begin{tabular}[c]{@{}l@{}}Reverse ISP pipeline introduced in \cite{RN23}.\end{tabular}}   &                            \\ \cline{1-4}
\begin{tabular}[c]{@{}c@{}}The See-in-the-Dark \\ (SID) dataset\cite{RN52}\end{tabular}   & \begin{tabular}[c]{@{}l@{}}A recently published raw image dataset shoot under low light\\  conditions.\end{tabular}                                                 & -                     & -                                                                                                                         &                            \\ \cline{1-4}
\end{tabular}}
\end{table*}

\subsection{Experiments Setup and Evaluation Criteria}
\label{section:sec4.2}
\subsubsection{Gradient Map and Multiscale Model}
In our experiment, the operators in Fig. \ref{fig:Gradient operators:center} are used to extract the gradients from color images and their corresponding Bayer versions. For color images, gray scale images are generated for gradient extraction. To blur and resize the Bayer pattern images, the super-pixel structure discussed in Section \ref{section:sec3.2} is utilized.

To estimate the differences among gradient maps, blurred images and resized images, some image quality assessment methods are used in these experiments.

The gradient magnitude similarity deviation (GMSD) is proposed in \cite{RN49} to evaluate the similarity of gradient magnitudes. Given two gradient maps, the GMSD is defined by
\begin{equation}
\label{equ:GMSD}
GMSD=\sqrt{\frac{1}{H\times W}\sum_{x=1}^{H}\sum_{y=1}^{W}(GMS(x,y)-GMSM)^{2}},
\end{equation}
where,
\begin{equation}
\label{equ:GMSM}
GMSM=\frac{1}{H\times W}\sum_{x=1}^{H}\sum_{y=1}^{W}GMS(x,y),
\end{equation}
\begin{equation}
\label{equ:GMS}
GMS(x,y)=\frac{2m_{1}(x,y)m_{2}(x,y)+c}{m_{1}^{2}(x,y)+m_{2}^{2}(x,y)+c}.
\end{equation}
Here, $W$ and $H$ are the width and height of the images, $m_{j}(x,y)$ is the gradient magnitude of the $j$-th image at pixel location $(x,y)$, defined by $m(x,y)=\sqrt{G_{x}(x,y)+G_{y}(x,y)}$, and $c$ is a small value set to $0.0026$ to avoid divisions by 0. According to \cite{RN49}, the smaller the GMSD is, the closer the gradient maps are.

Mean squared error (MSE) is the simplest and most commonly used full-reference quality metric. It is an evaluation that is computed by averaging the squared intensity differences of distorted and reference image pixels. For two given images, the MSE is given by
 \begin{equation}
\label{equ:MSE}
MSE=\frac{1}{H\times W}\sum_{x=1}^{H}\sum_{y=1}^{W}\left ( I_{1}\left ( x,y \right ) -I_{2}\left ( x,y \right ) \right ),
\end{equation}
where $W$ and $H$ is the width and height of the image. The MSE can be converted to PSNR by
\begin{equation}
\label{equ:psnr}
PSNR=10\log_{10}\left ( \frac{\left (2^{n}-1\right )^{2}}{MSE} \right ),
\end{equation}
where $n$ represent the pixel bit depth of images. For images with 8-bit pixel depth, the typical values of PSNR for lossy images are between 30 and 50 dB\cite{RN39}.

Structural similarity (SSIM) is also a full-reference quality metric which compares luminance, contrast, structure among two images \cite{RN16}. The SSIM ranges from 0 to 1, where 1 means that the two compared images are identical. Due to the fact that SSIM is a metric for local region comparison, the mean SSIM (MSSIM) is usually used in practice.

\subsubsection{Influence of Noise}
Noise reduction, which has a deterministic impact on the quality of imaging, is a critical step in image processing pipelines. Basically, there are two kinds of noise in an image, i.e., signal-independent noise (e.g., bad-pixels, dark currents) and signal-dependent noise (e.g., photon noise). For modern cameras, the signal-dependent noise, which is affected by lighting conditions and exposure time \cite{RN50,RN51}, is the dominant noise source. In \cite{RN50}, image noise is modeled as additive noise, which is a mixture of Gaussian and Poissonian process that obeys the distribution of
\begin{equation}
\label{equ:gaussian}
\eta _{h}\sim \mathit{N}(0,ay(x)+b).
\end{equation}
 Here, $\eta _{h}$ is the signal noise, $y(x)$ is the noise-free signal and $a,b$ are two parameters. Note that the dataset used in pedestrian detection experiments is all shoot under sufficient illumination and proper exposure. To study the influence of noise on the proposed Bayer pattern image-based gradient feature extraction pipeline, we use the See-in-the-Dark (SID) dataset introduced  in \cite{RN52} and the model in \eqref{equ:gaussian} to obtain a set of different noise parameters under low light conditions (2650 parameter pairs in total) and randomly choose parameter pairs for each image in pedestrian detection datasets to generate the corresponding noisy images.
\subsubsection{HOG Descriptor} To compare the performance of HOG descriptors extracted from color images and Bayer pattern images, the traditional HOG $+$ support vector machine (SVM) framework proposed in \cite{RN24} is used to detect pedestrians from color images and their Bayer versions. The INRIA, SHTech and PASCALRAW dataset are used in the pedestrian detection task, where models are trained and tested on each dataset separately.
Precision-recall curve along with average precision are used to present the detection results\cite{RN59}.

\begin{table}[t]
\caption{Comparison Results of Bayer Image Based and Color Image Based Gradients}
\label{tab:gradient evaluation}
\centering
\setlength\arrayrulewidth{0.6pt}
\renewcommand{\arraystretch}{1.3}
\setlength{\tabcolsep}{16.3pt}{
\begin{tabular}{|c|c|c|c|c}
\cline{1-4}
\multirow{2}{*}{\textbf{Datasets}} & \multicolumn{3}{c|}{\textbf{Average}} &  \\ \cline{2-4}
                          & MSSIM    & PSNR    & GMSD    &  \\ \cline{1-4}
Kodak                     & 0.975    & 38.276  & 0.069   &  \\ \cline{1-4}
SHTech                    & 0.850    & 34.683  & 0.119  &  \\ \cline{1-4}
PASCALRAW                 & 0.9367   & 37.36   & 0.127   &  \\ \cline{1-4}
INRIA                     & 0.817    & 30.288  & 0.148   &  \\ \cline{1-4}
\end{tabular}}
\begin{tablenotes}[flushleft] 
\item For the INRIA dataset, gamma compression with scale factor of 2 and exponent of 0.5 is used.
\end{tablenotes}

\vspace{-3mm}
\end{table}

\begin{table}[t]
\centering
\setlength\arrayrulewidth{0.6pt}
\renewcommand{\arraystretch}{1.3}
\setlength{\tabcolsep}{8pt}{
\caption{Comparison Results of the True Color Images and Images Generated Using Different Demosaicing Algorithms}
\label{tab:demosaic methods}
\begin{tabular}{|c|c|c|c|c|}
\hline
\multirow{2}{*}{\textbf{\makecell[c]{Methods}}} & \multicolumn{3}{c|}{\textbf{Average}} \\ \cline{2-4}
 & MSSIM & PSNR & GMSD  \\ \hline
\makecell[c]{Nearest Neighbor} & 0.8865  & 25.744 &0.082  \\ \hline
\makecell[c]{Linear Interpolation} & 0.945  & 29.255 &0.089  \\ \hline
\makecell[c]{Cubic Interpolation} & 0.952  & 29.354 &0.084 \\ \hline
\makecell[c]{Adaptive Color Plane Interpolation} & 0.976  & 34.452 &0.070 \\ \hline
\makecell[c]{Hybrid Interpolation} & 0.990 & 39.010 &0.065 \\ \hline
\end{tabular}
}
\vspace{-3mm}
\end{table}

\begin{figure}[t]
\centering
\subfigure[]{
\label{fig:HOG:originalhuman} 
\includegraphics[width=0.8in,height=1.9in]{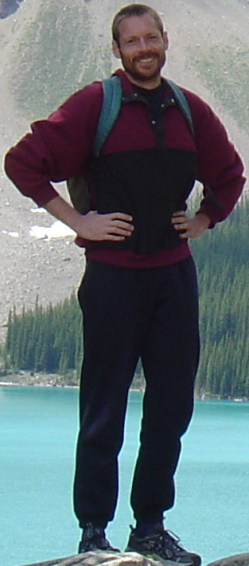}}
\hspace{-1mm}
\subfigure[]{
\label{fig:HOG:singlehuman} 
\includegraphics[width=0.8in,height=1.9in]{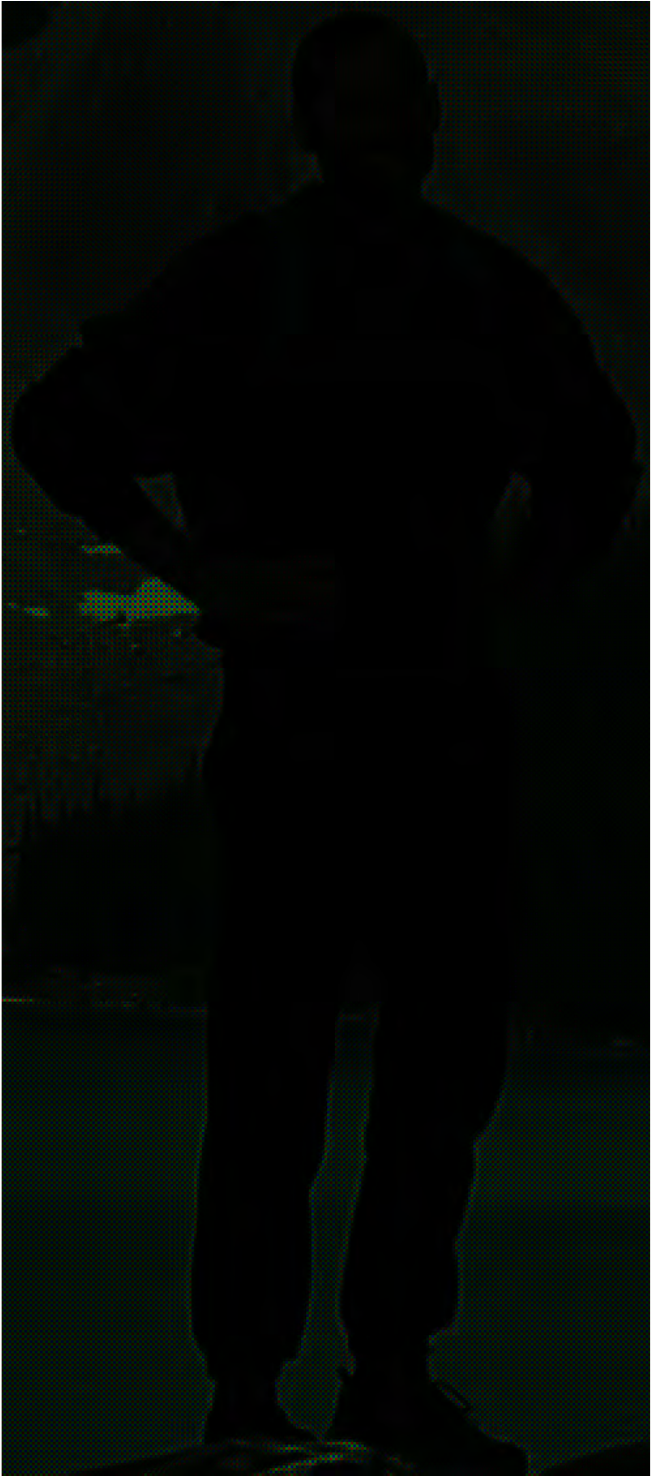}}
\hspace{-1mm}
\subfigure[]{
\label{fig:HOG:singleGamma} 
\includegraphics[width=0.8in,height=1.9in]{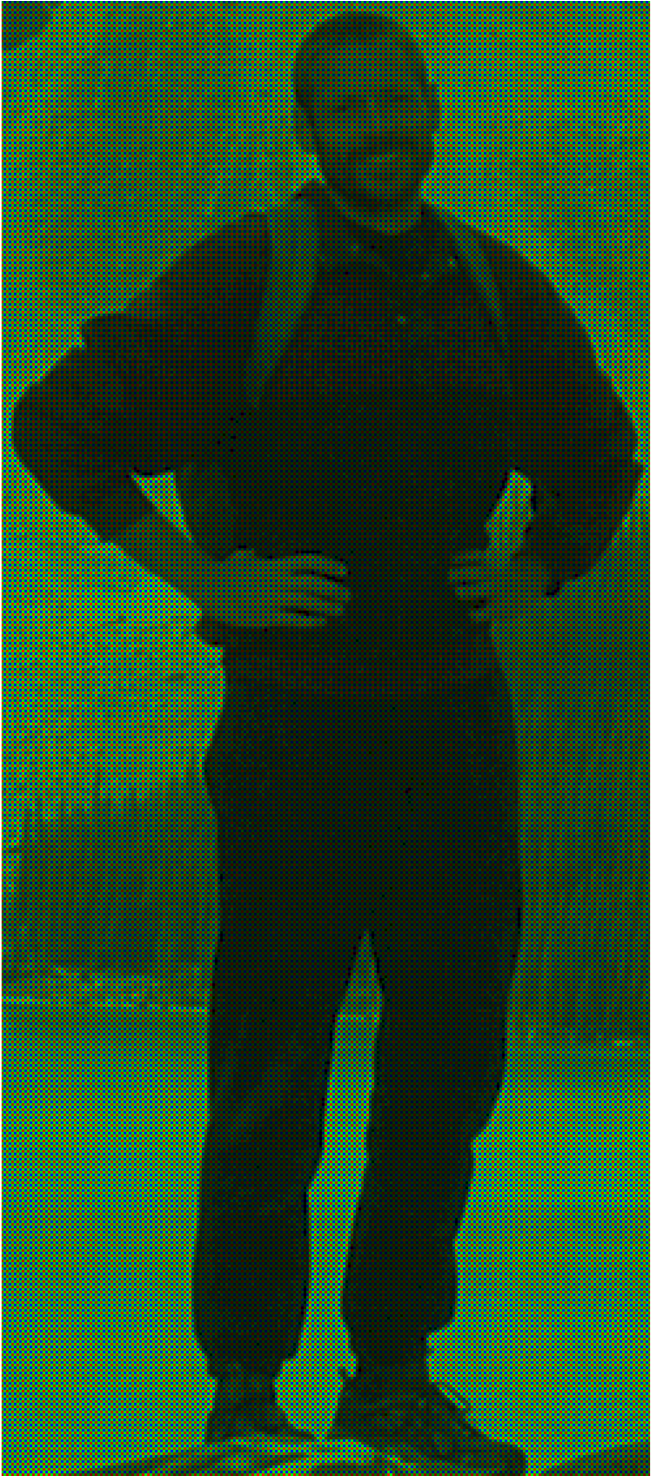}}\\
\hspace{-1mm}
\subfigure[]{
\label{fig:HOG:originalHOG} 
\includegraphics[width=0.8in,height=1.9in]{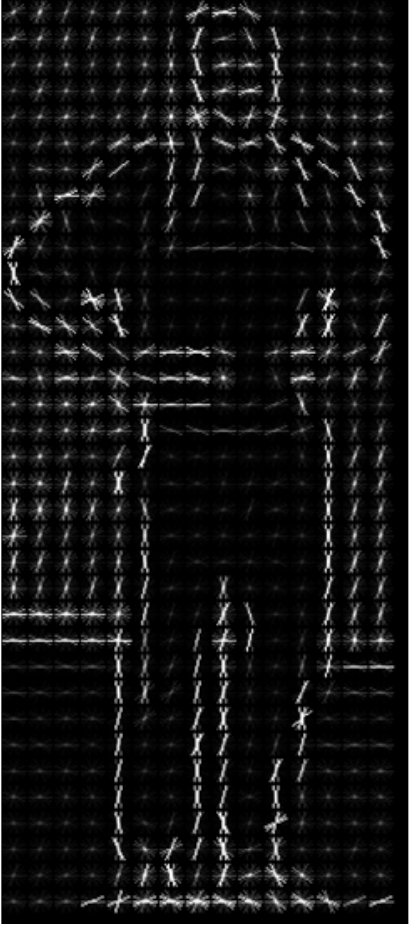}}
\hspace{-1mm}
\centering
\subfigure[]{
\label{fig:HOG:singleHOG} 
\includegraphics[width=0.8in,height=1.9in]{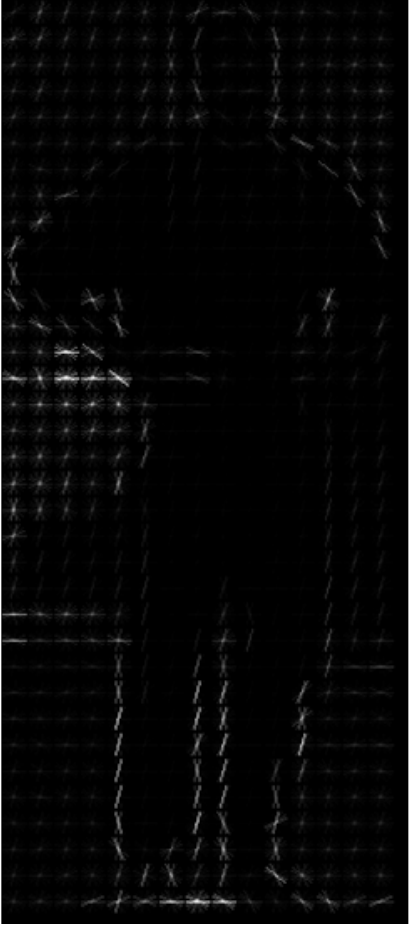}}
\hspace{-1mm}
\subfigure[]{
\label{fig:HOG:singleGammaHOG} 
\includegraphics[width=0.8in,height=1.9in]{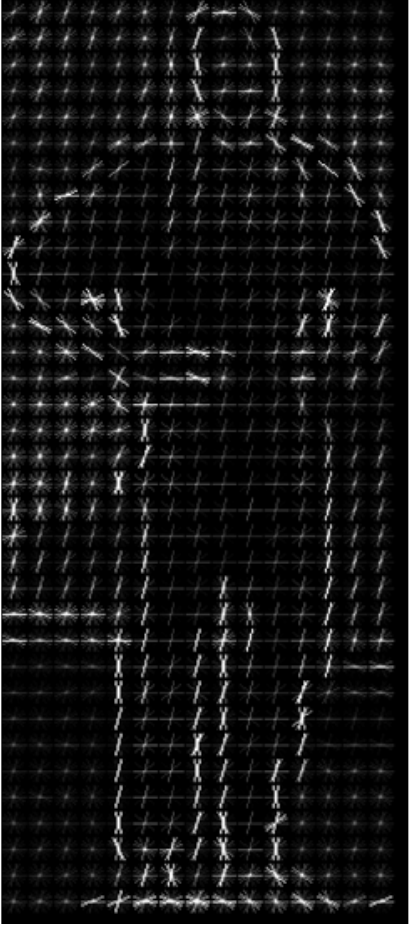}}\\
\caption{ Comparison of HOG features. (a) An image from the INRIA pedestrian dataset. (b) The converted Bayer version of (a) using the reverse pipeline in \cite{RN23}. (c) The Bayer version image after gamma compression. (d)-(f): Visualization of the generated HOG descriptors.}
\label{fig:HOG}
\vspace{-3mm}
\end{figure}

\begin{figure}
  \centering
  \includegraphics[width=3.5in]{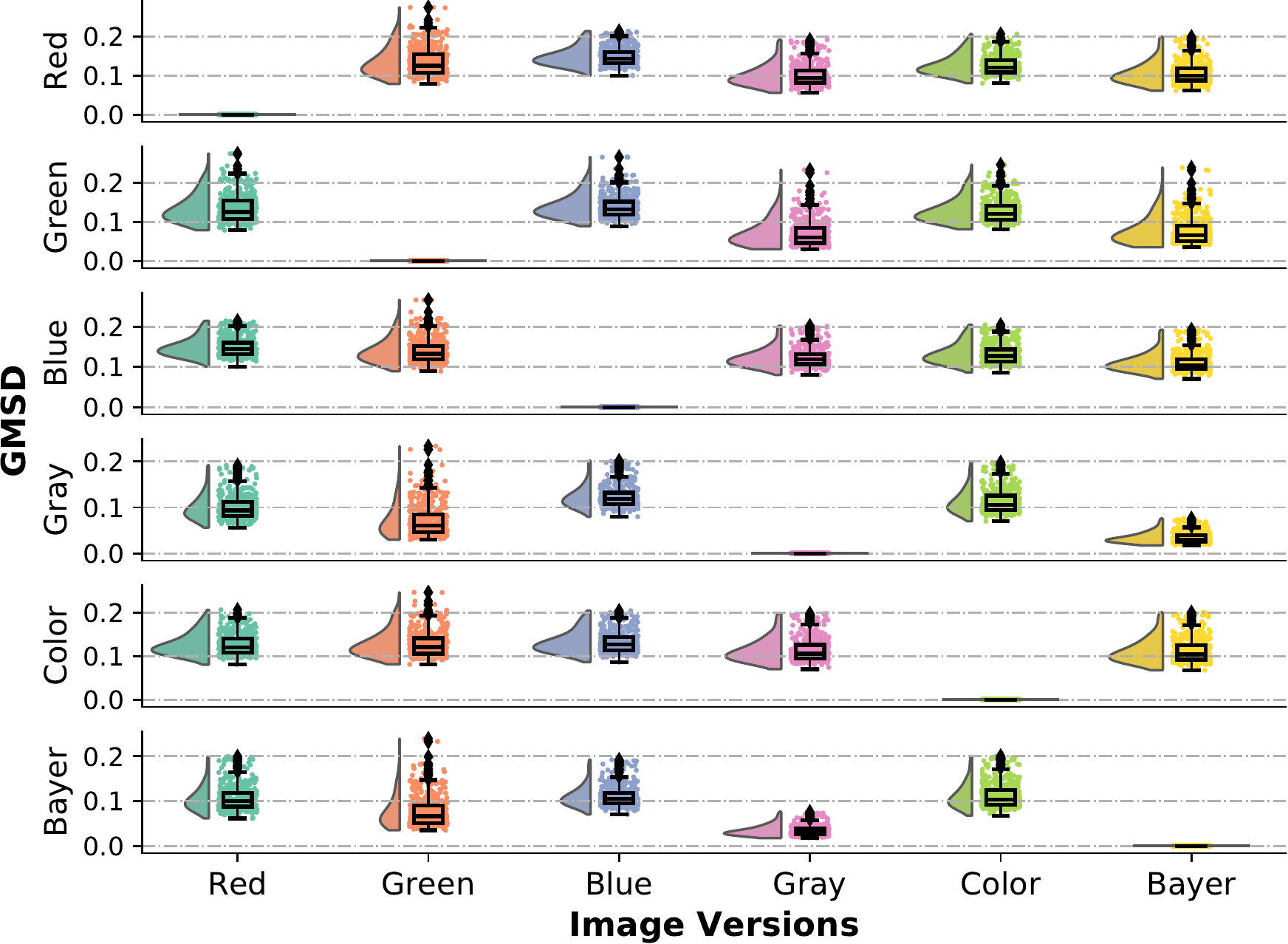}\\
  \caption{GMSD (between different channels and versions of images) distribution of the SHtech Dataset.}
  \label{gmsd}
  \label{fig:violin_SHtech}
\end{figure}

\begin{figure*}[t]
\centering
\subfigure[]{
\label{fig:GMSD:smooth} 
\includegraphics[width=2.3in]{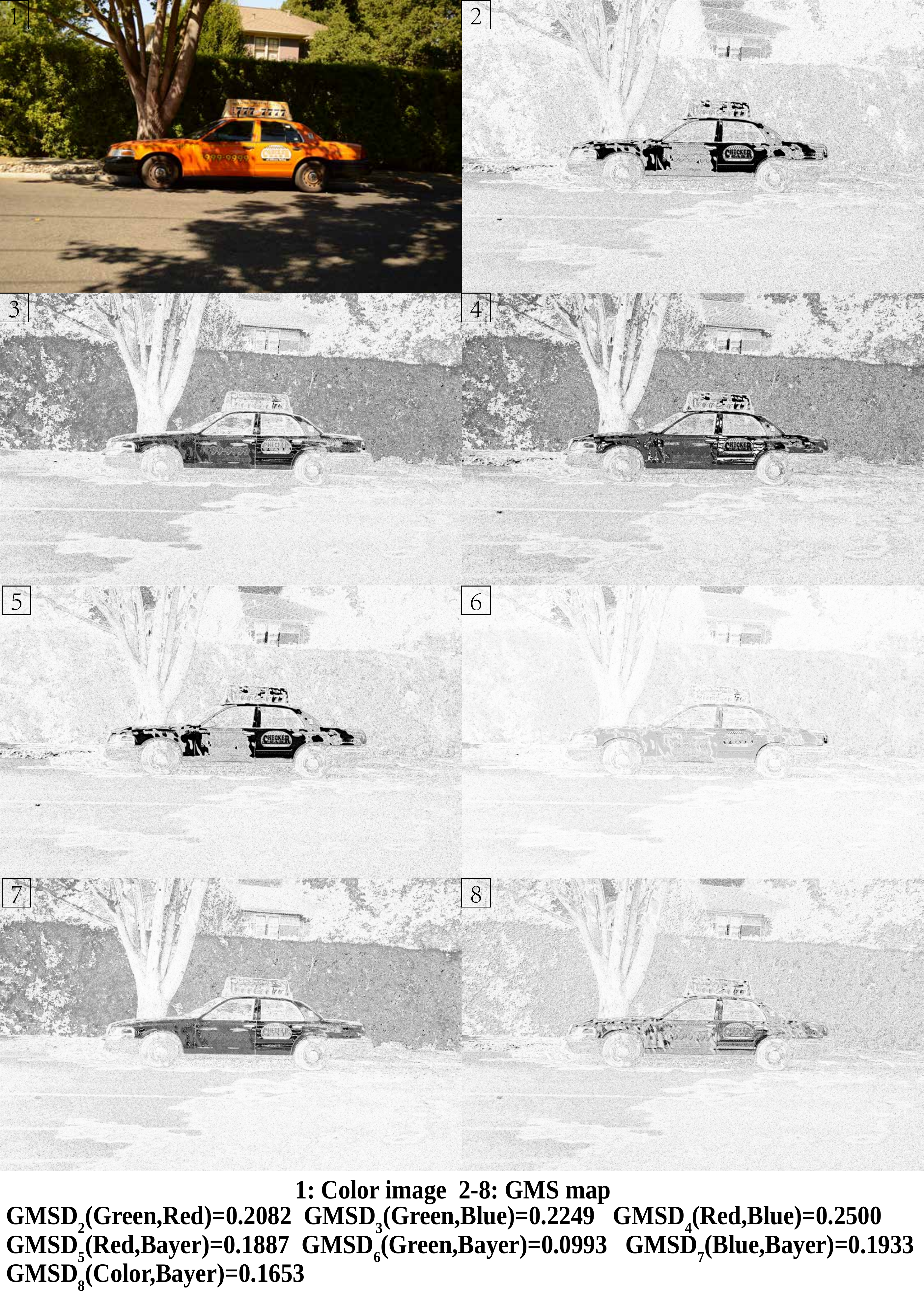}}
\hspace{-1mm}
\subfigure[]{
\label{fig:GMSD:edge} 
\includegraphics[width=2.3in]{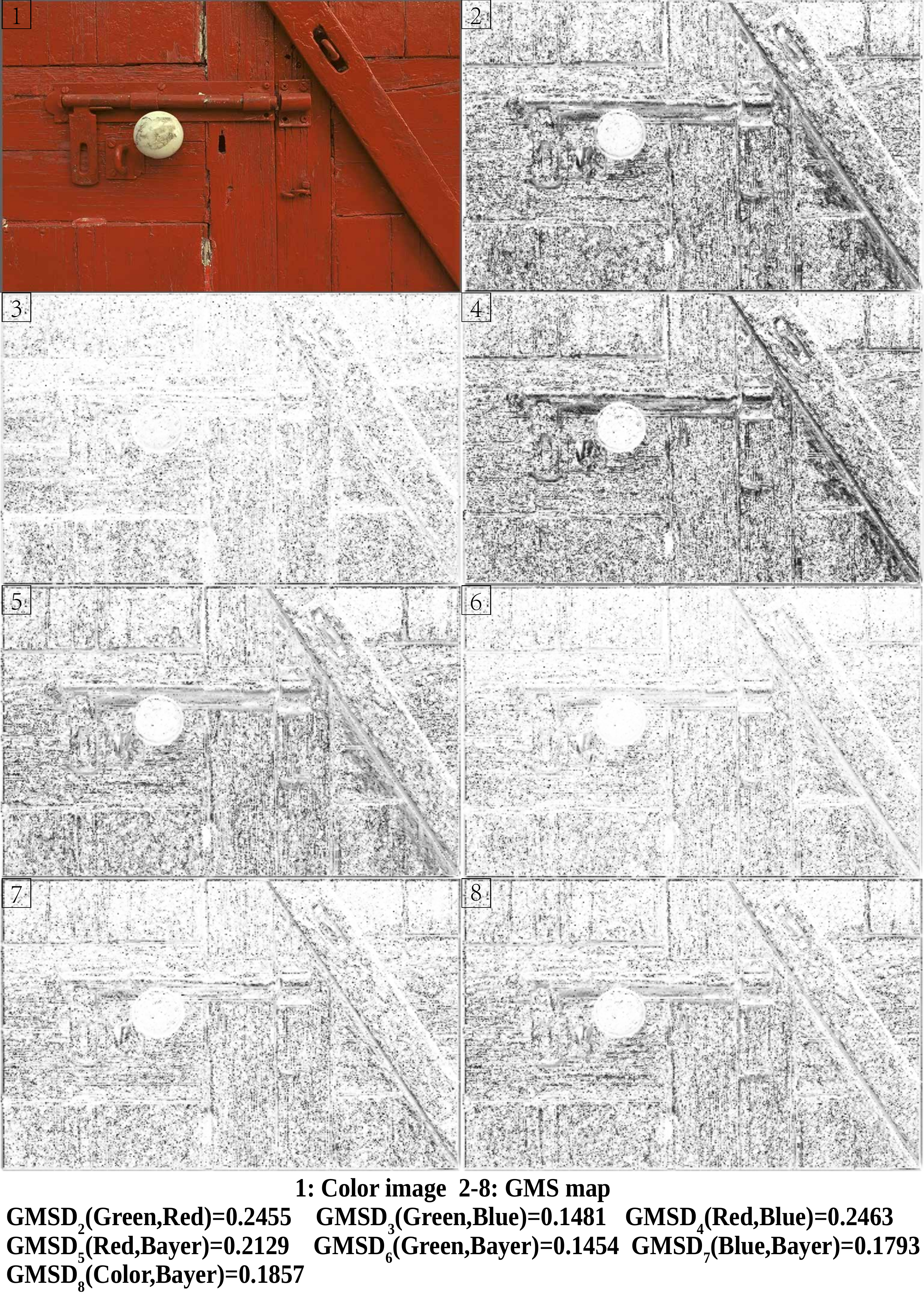}}
\hspace{-1mm}
\subfigure[]{
\label{fig:GMSD:noise} 
\includegraphics[width=2.3in]{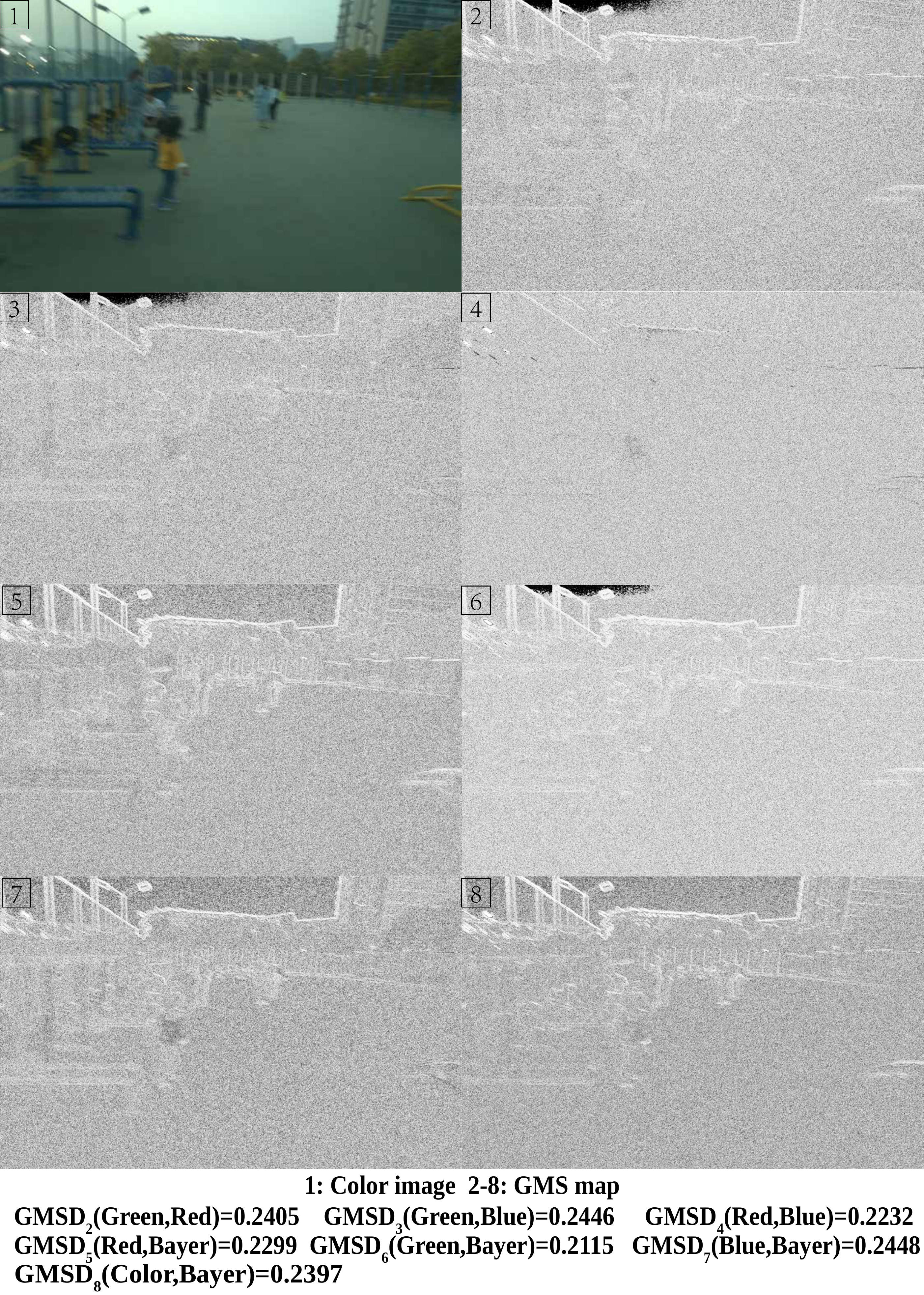}}
\caption{Three situations which cause large gradient difference for different channels. (a) The light source shines directly on a smooth surface. (b) Irregular texture and (c) heavy noise.}
\label{fig:GMSD}
\hspace{-1mm}
\end{figure*}

\begin{figure*}[t]
\centering
\subfigure{
\label{fig:Channels:R1} 
\includegraphics[width=0.9in]{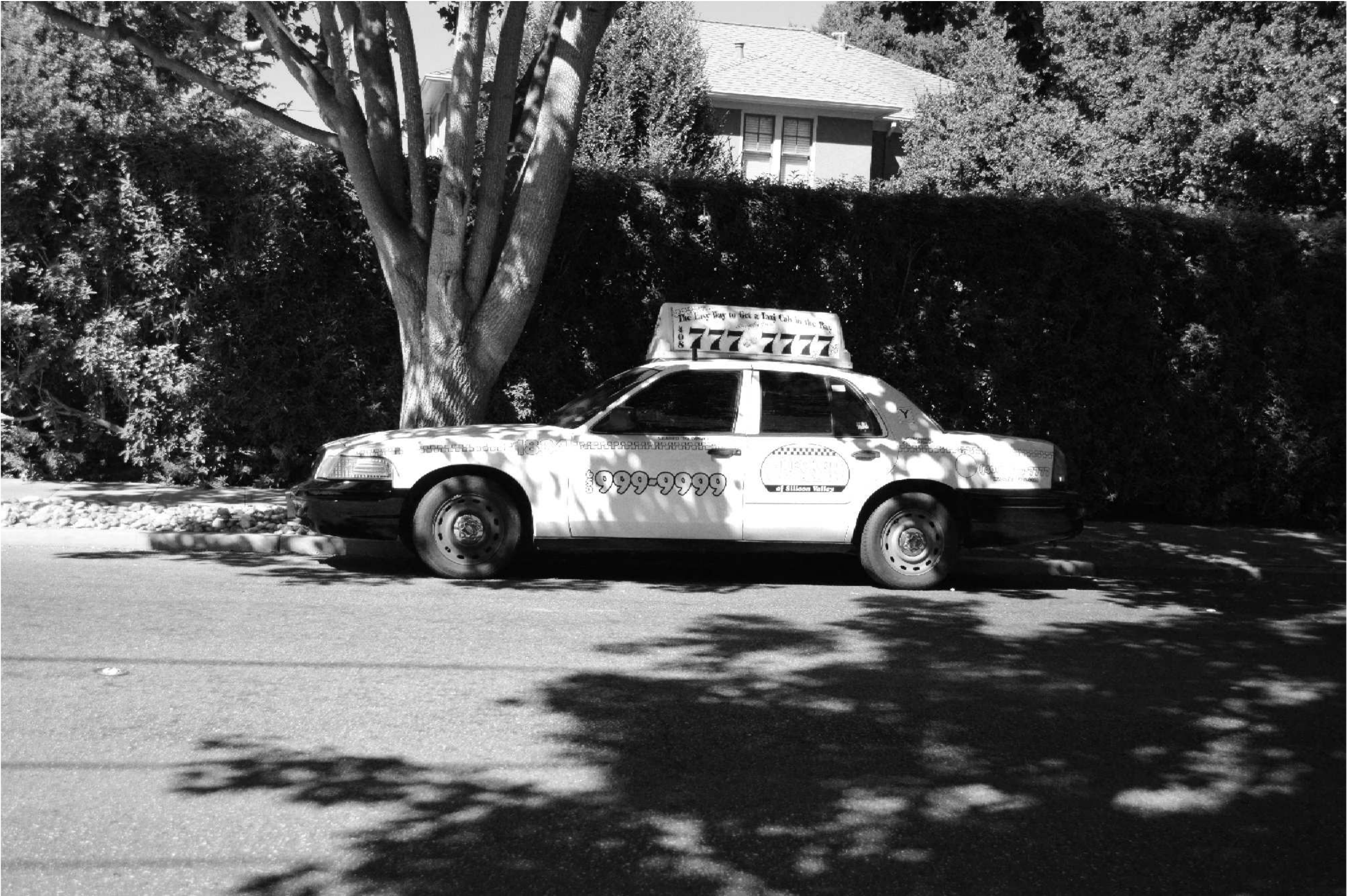}}
\hspace{-4mm}
\subfigure{
\label{fig:Channels:G1} 
\includegraphics[width=0.9in]{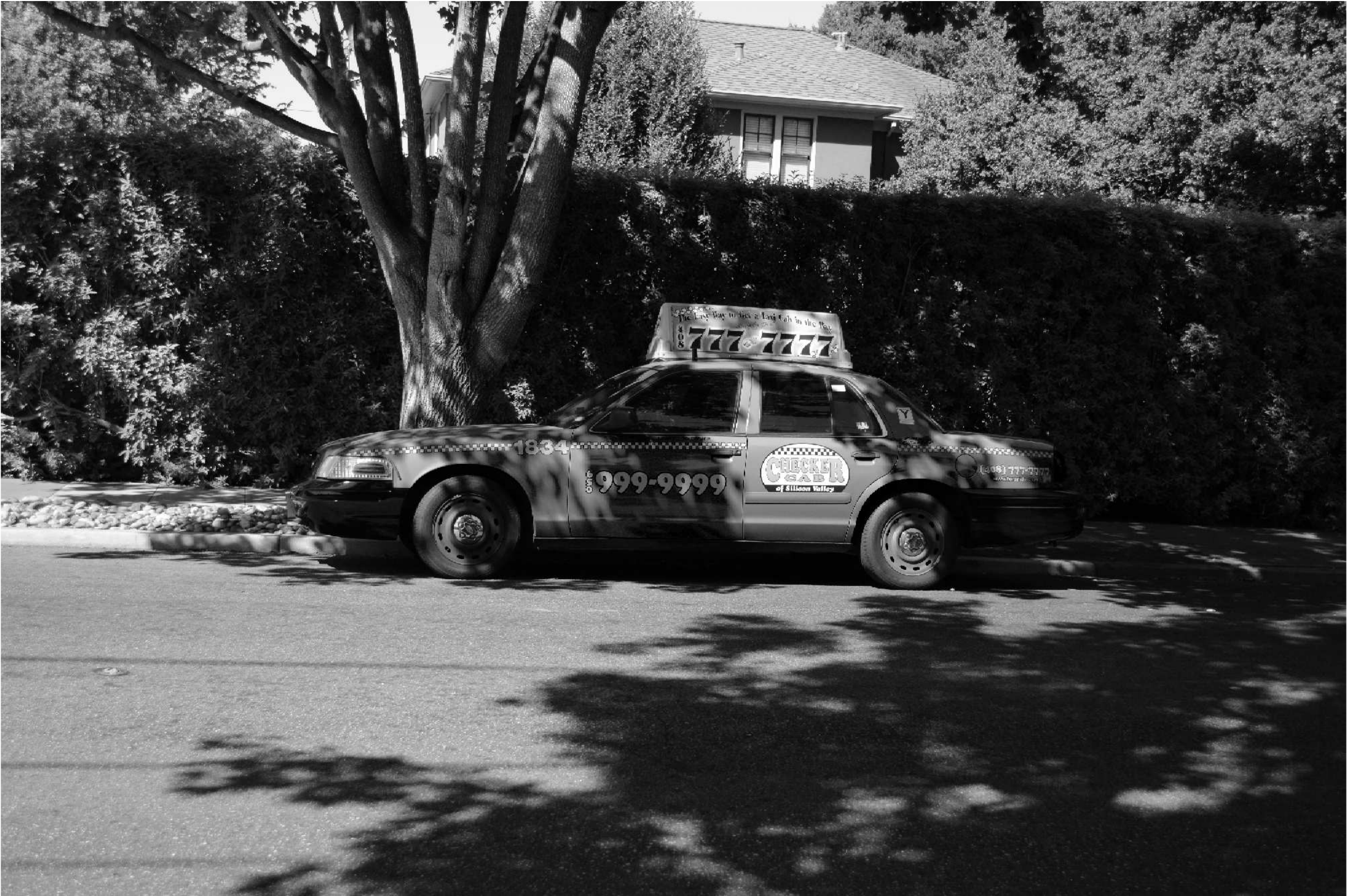}}
\hspace{-4mm}
\subfigure{
\label{fig:Channels:B1} 
\includegraphics[width=0.9in]{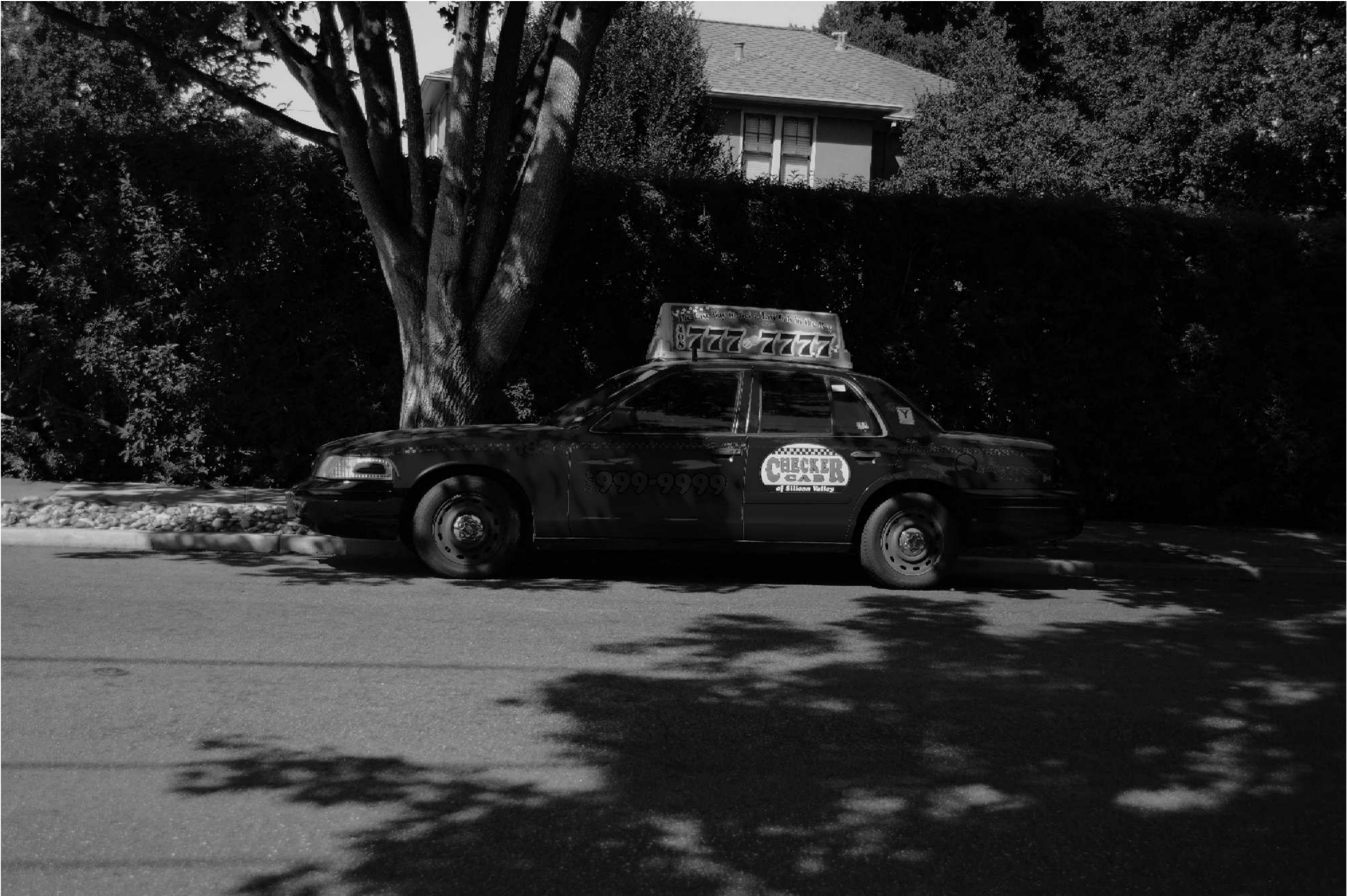}}
\hspace{-4mm}
\subfigure{
\label{fig:Channels:Bayer1} 
\includegraphics[width=0.9in]{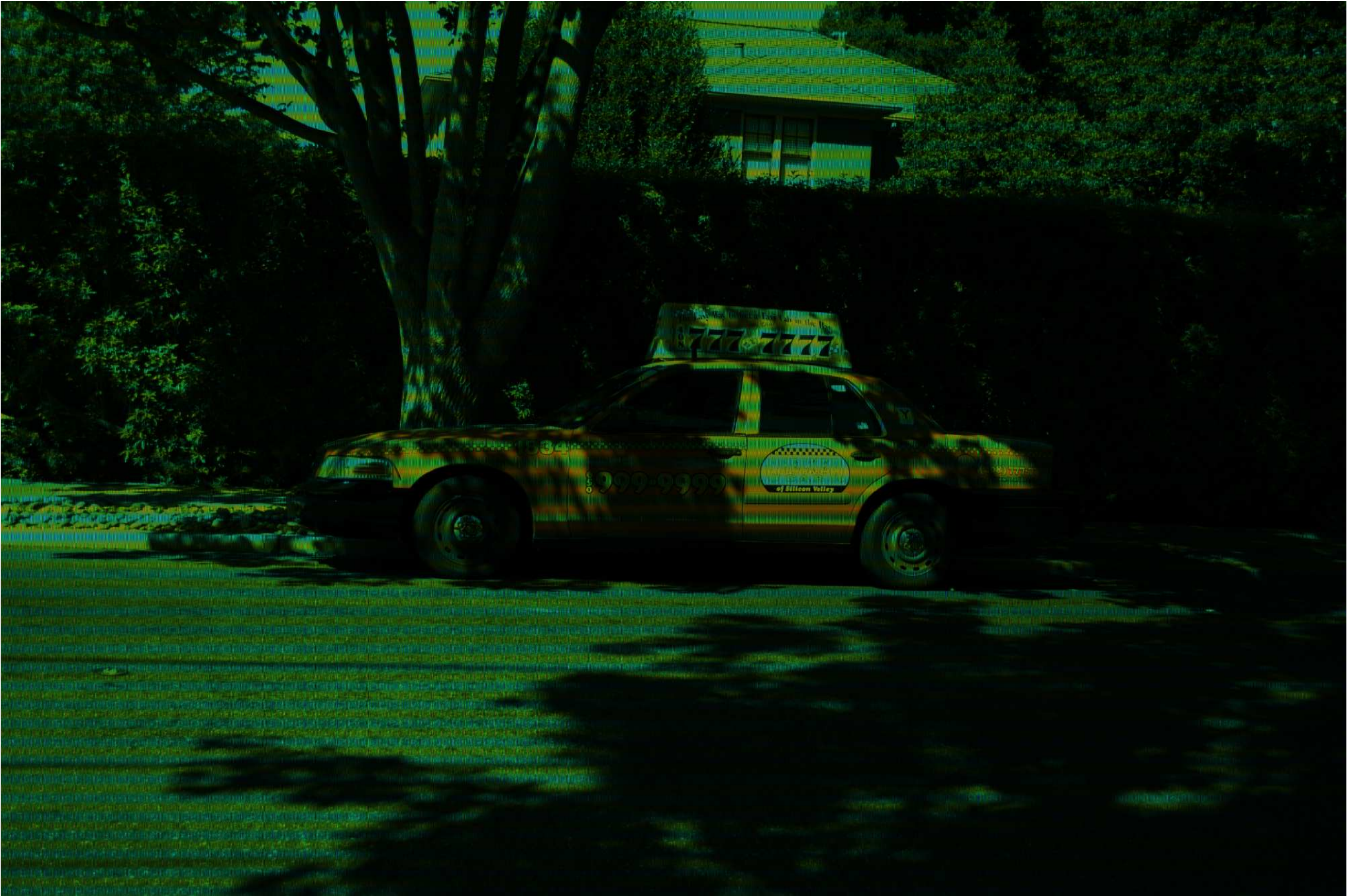}}
\hspace{-4mm}
\subfigure{
\label{fig:Channels:G_R1} 
\includegraphics[width=0.9in]{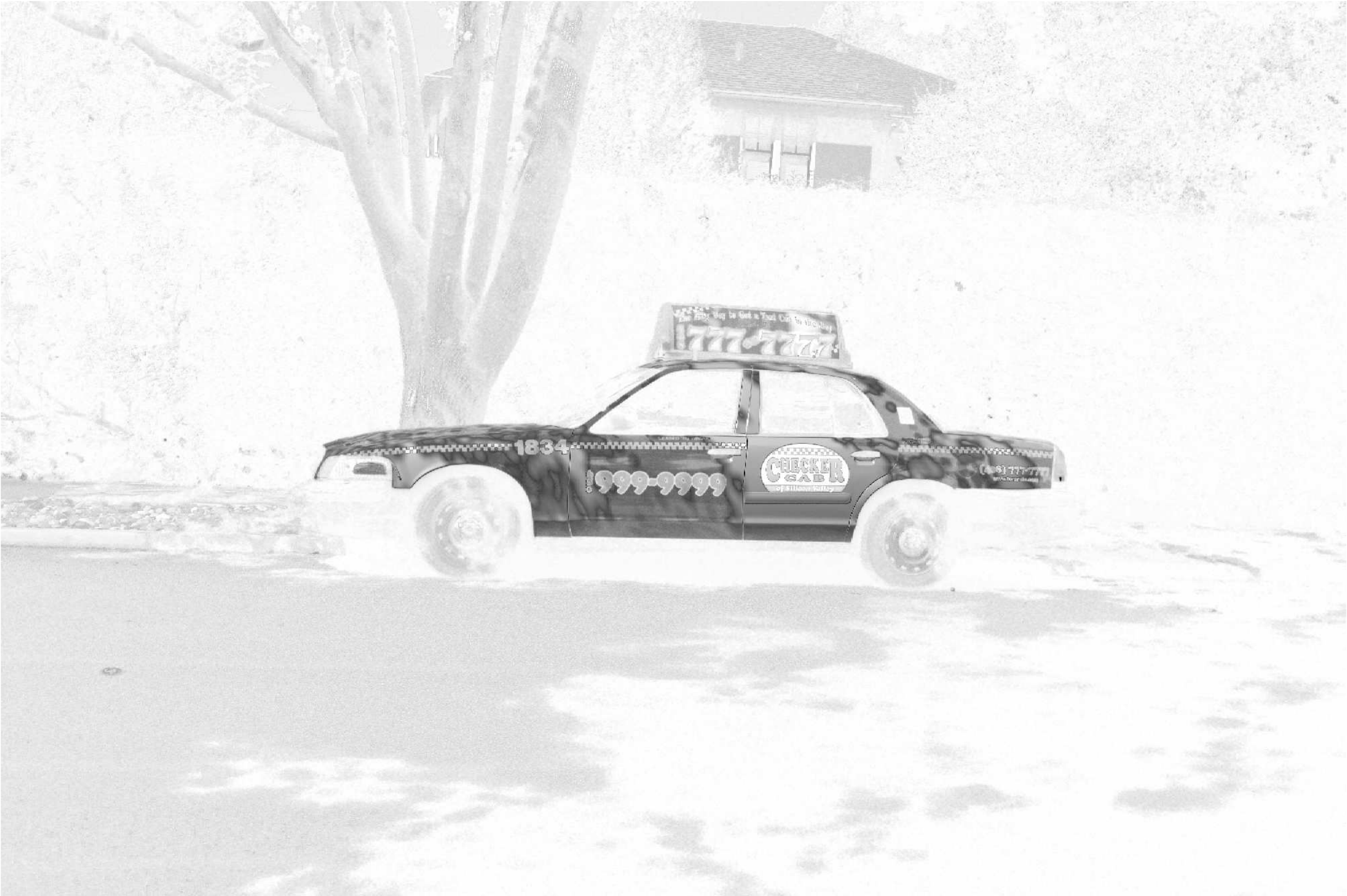}}
\hspace{-4mm}
\subfigure{
\label{fig:Channels:G_B1} 
\includegraphics[width=0.9in]{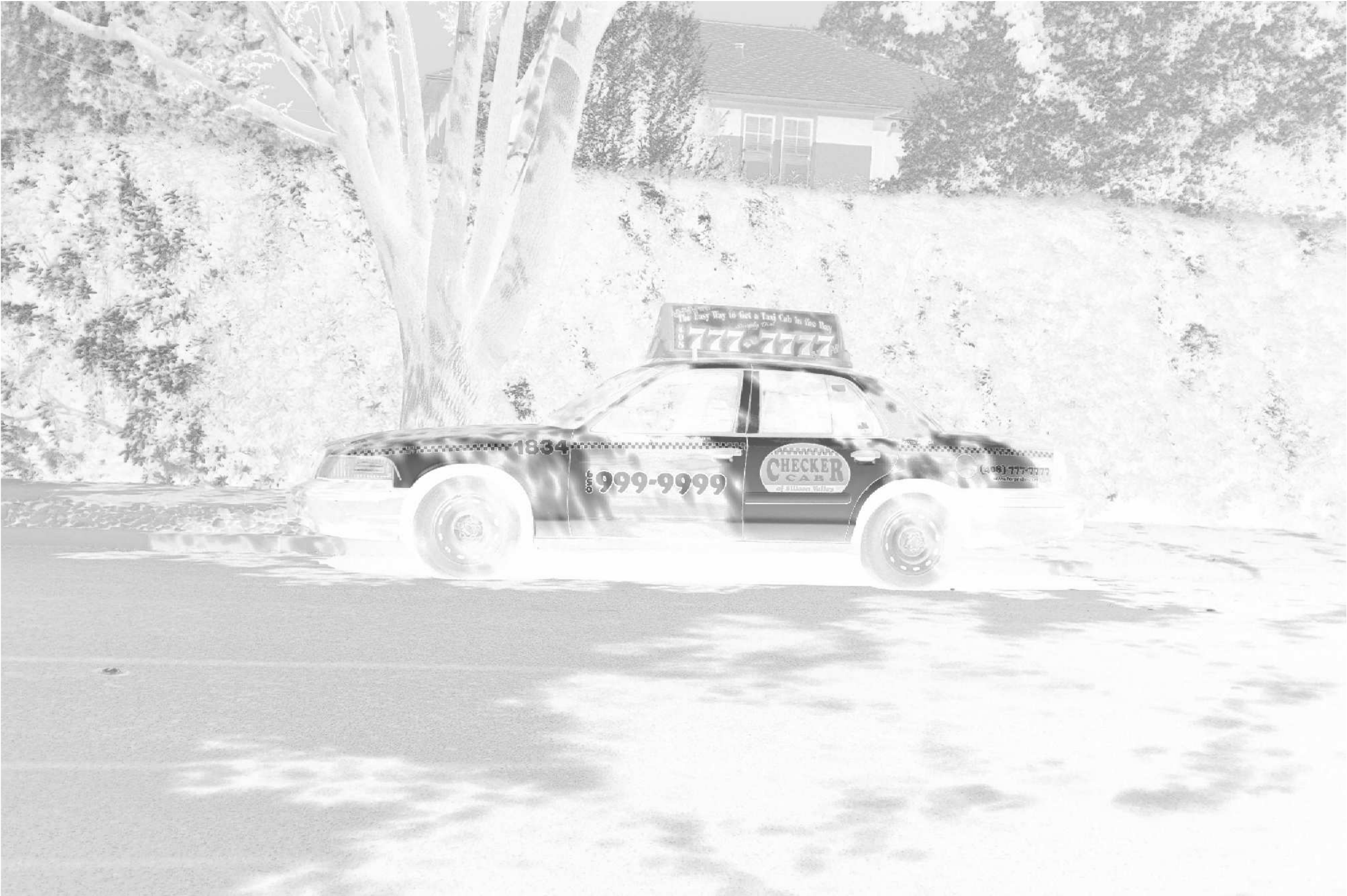}}
\hspace{-4mm}
\subfigure{
\label{fig:Channels:G_R_m1} 
\includegraphics[width=0.9in]{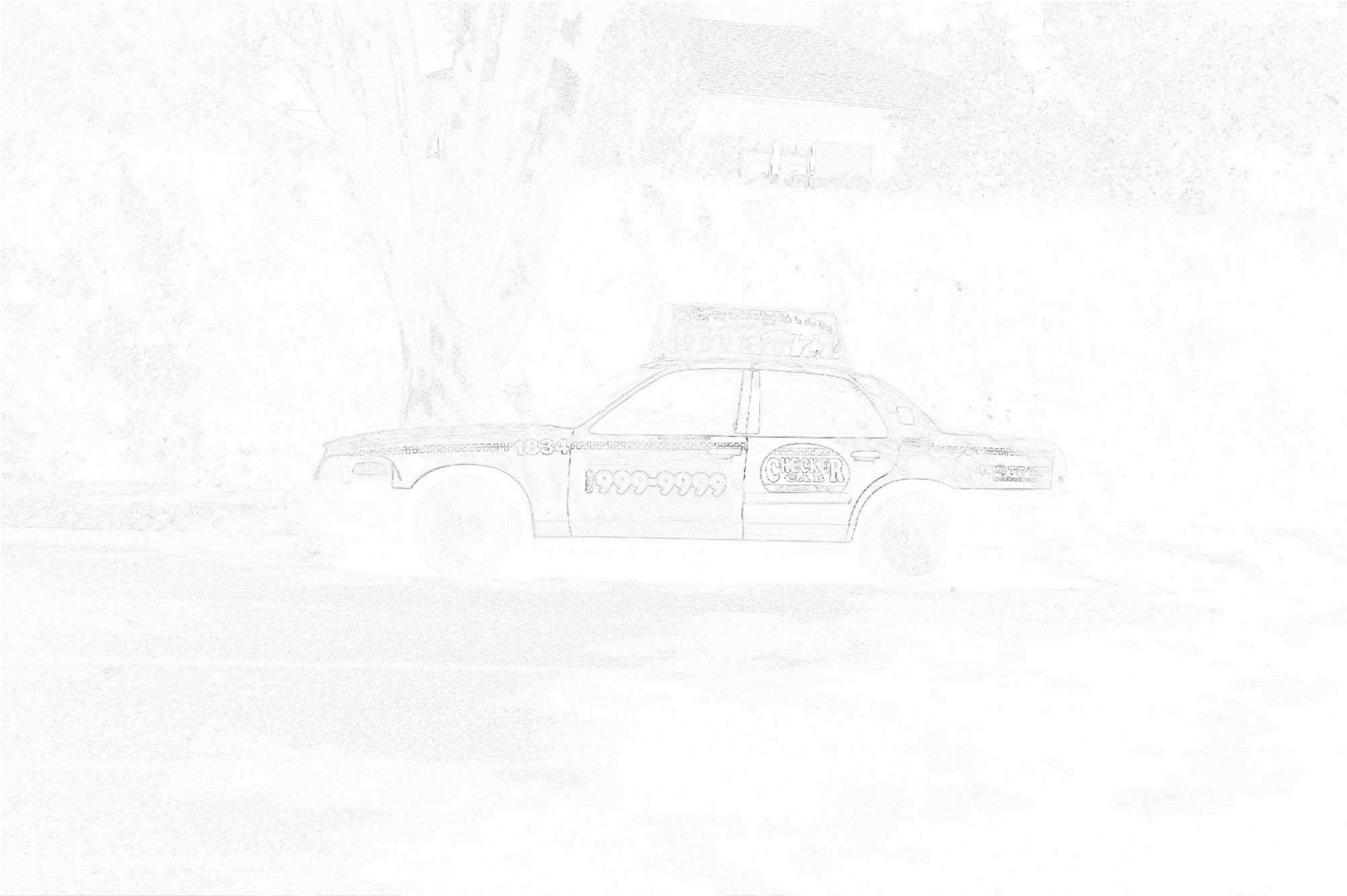}}
\hspace{-4mm}
\subfigure{
\label{fig:Channels:G_B_m1} 
\includegraphics[width=0.9in]{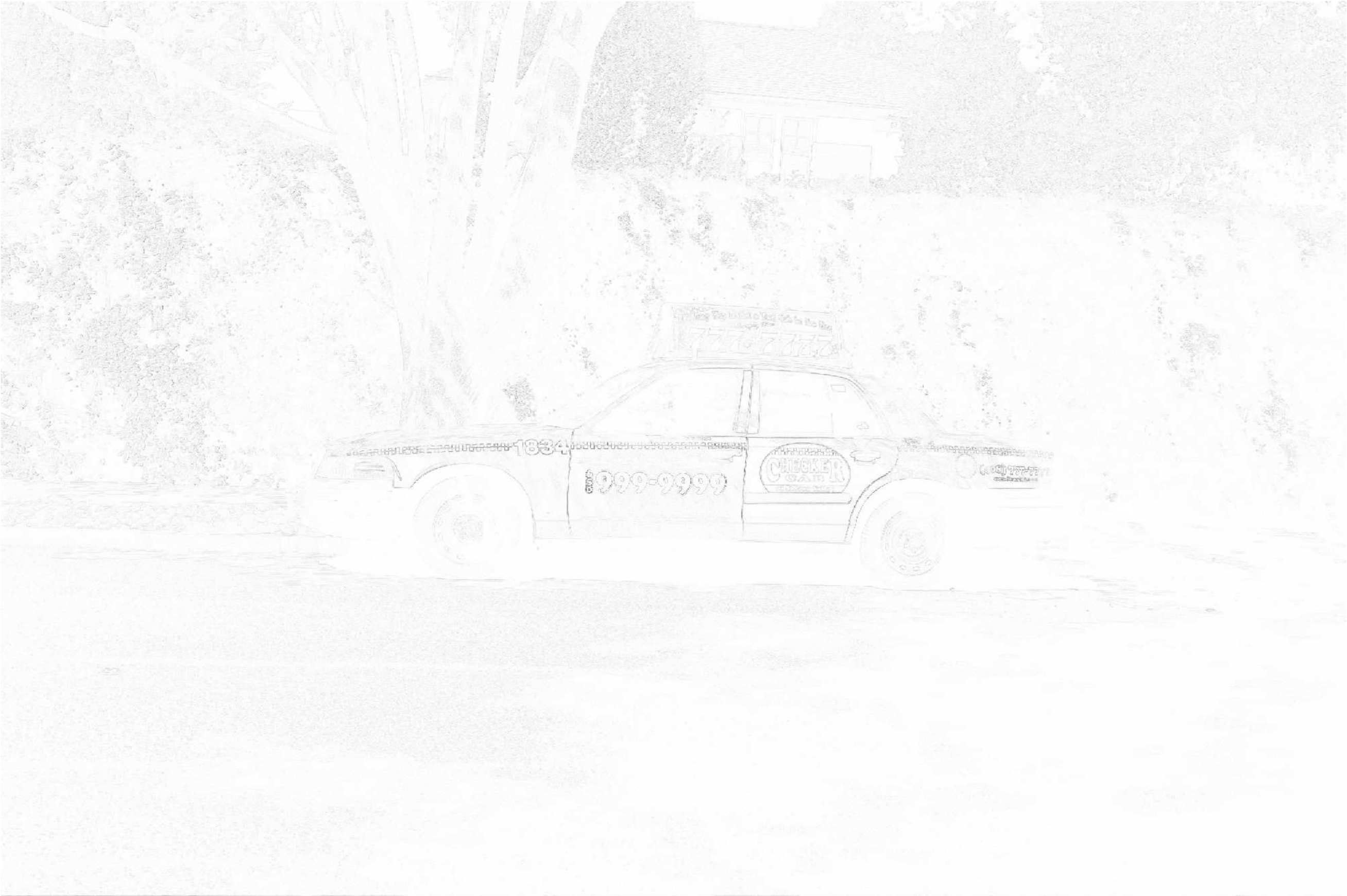}}\\
\vspace{-1mm}
\subfigure{
\label{fig:Channels:R2} 
\includegraphics[width=0.9in]{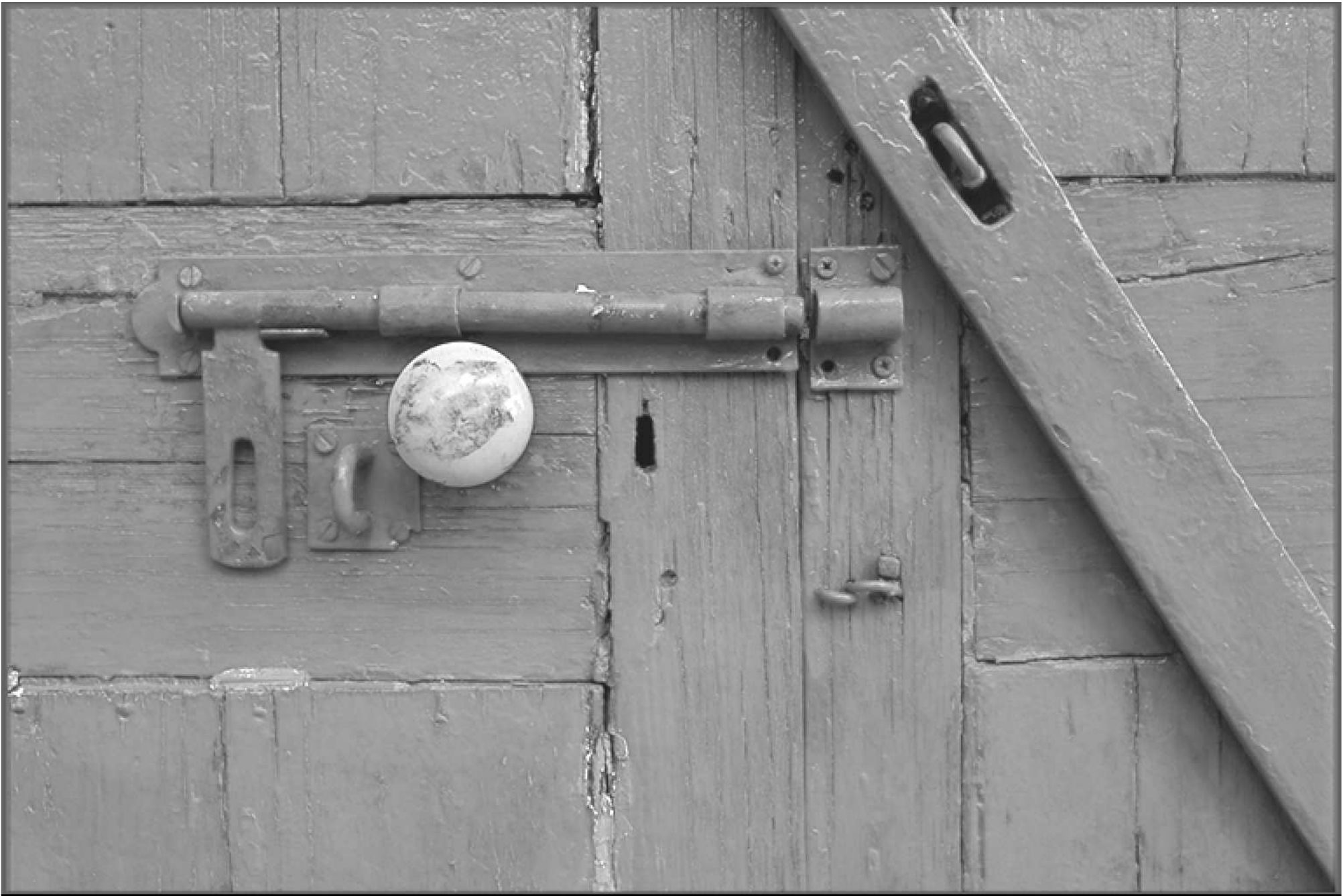}}
\hspace{-4mm}
\subfigure{
\label{fig:Channels:G2} 
\includegraphics[width=0.9in]{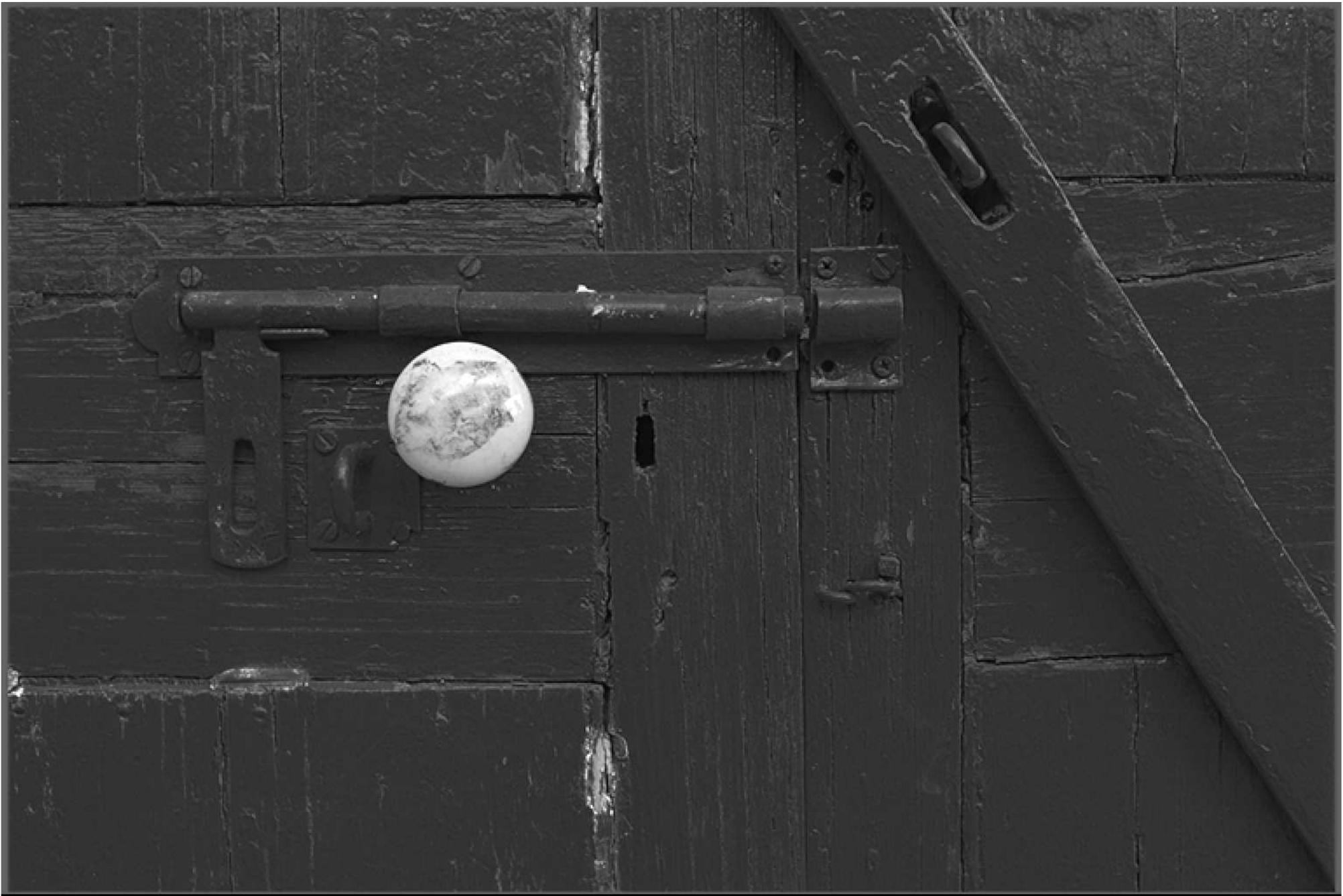}}
\hspace{-4mm}
\subfigure{
\label{fig:Channels:B2} 
\includegraphics[width=0.9in]{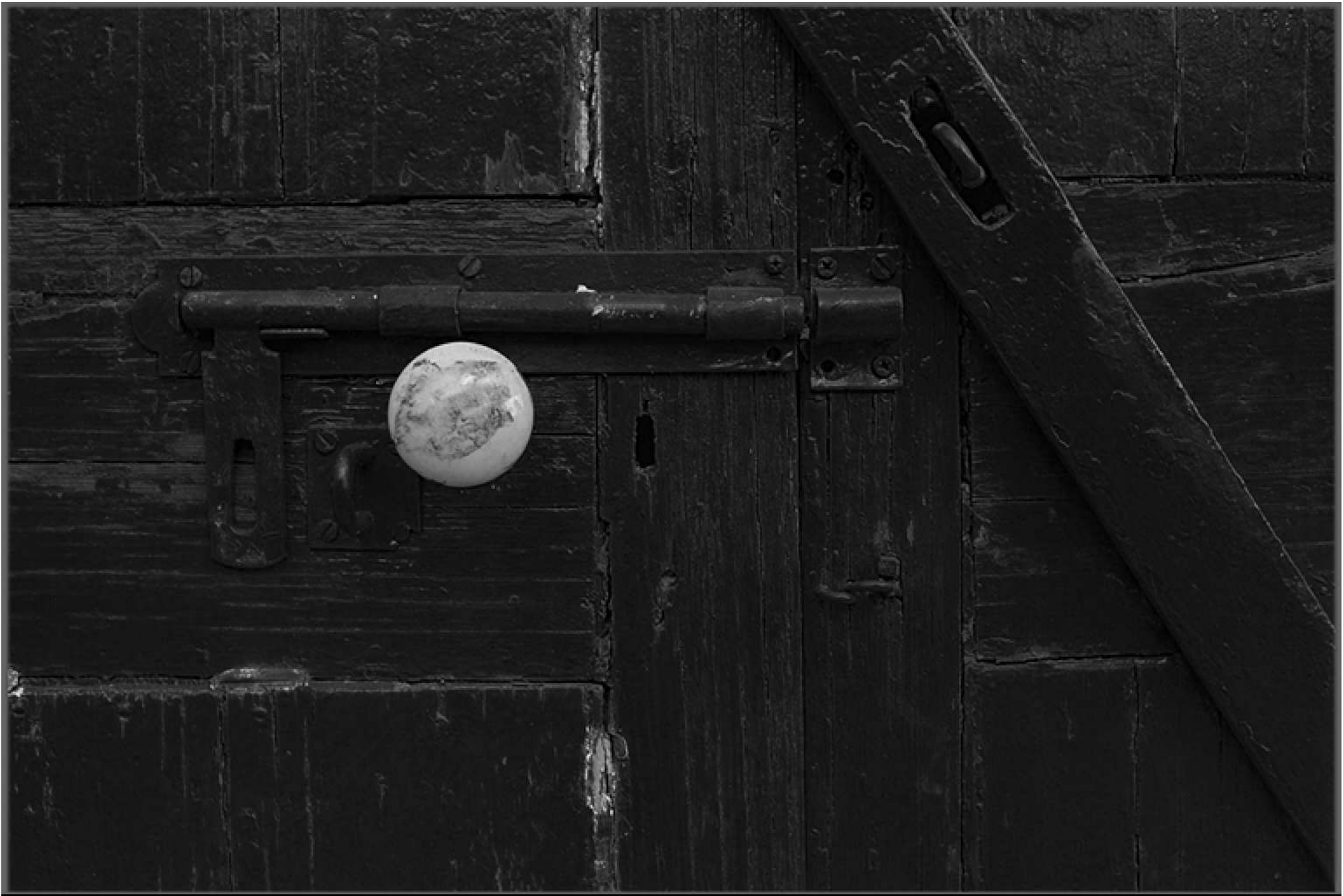}}
\hspace{-4mm}
\subfigure{
\label{fig:Channels:Bayer2} 
\includegraphics[width=0.9in]{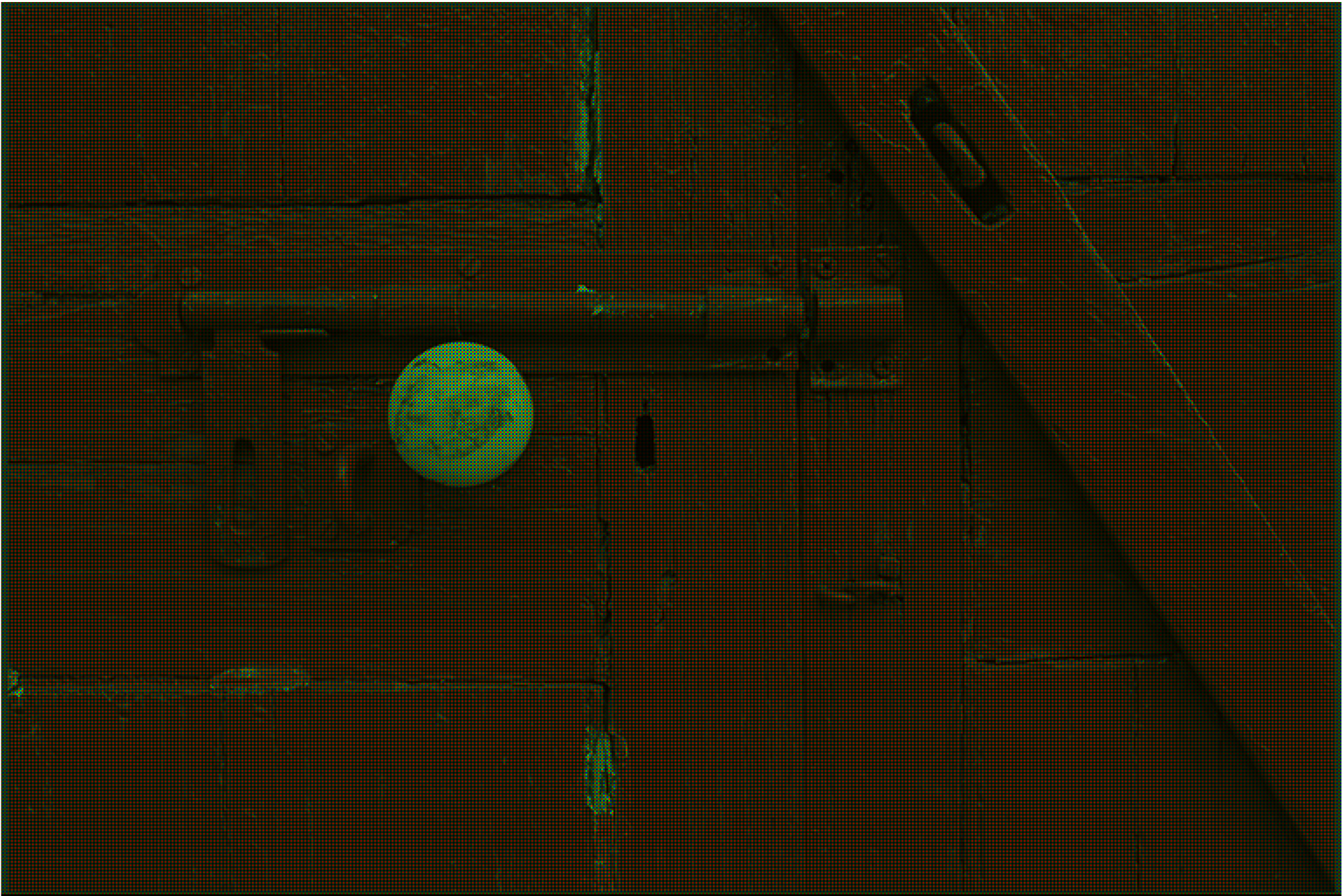}}
\hspace{-4mm}
\subfigure{
\label{fig:Channels:G_R2} 
\includegraphics[width=0.9in]{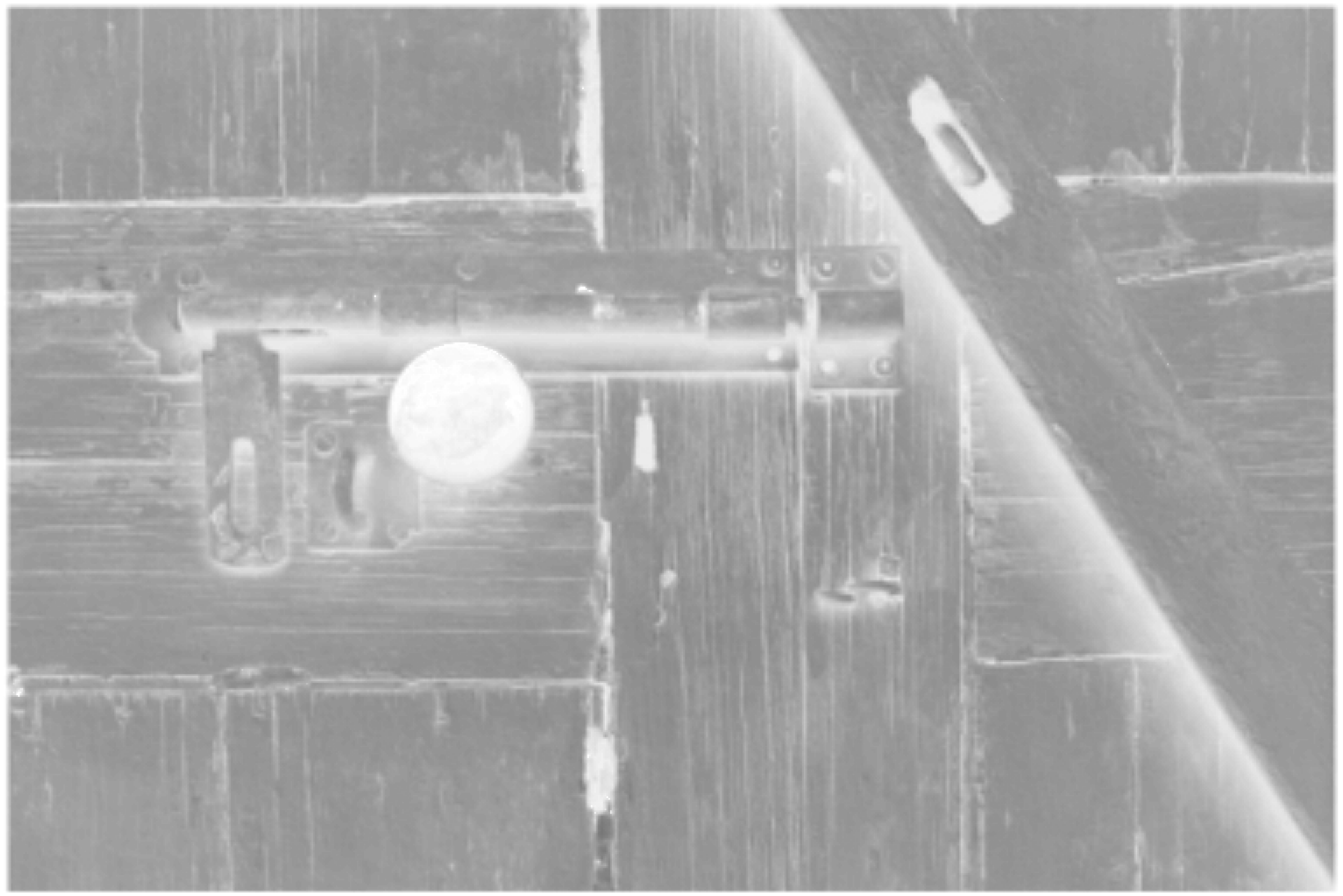}}
\hspace{-4mm}
\subfigure{
\label{fig:Channels:G_B2} 
\includegraphics[width=0.9in]{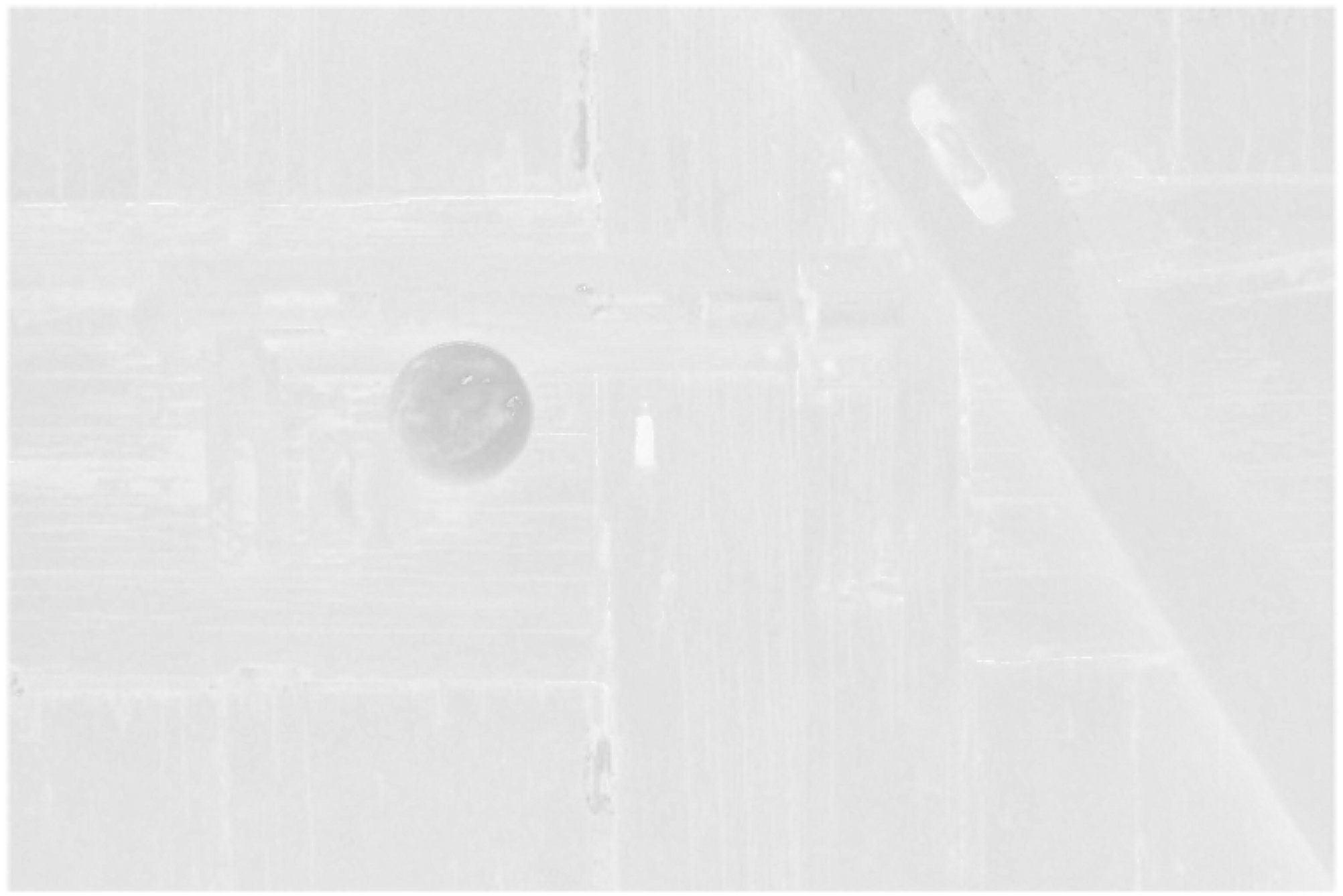}}
\hspace{-4mm}
\subfigure{
\label{fig:Channels:G_R_m2} 
\includegraphics[width=0.9in]{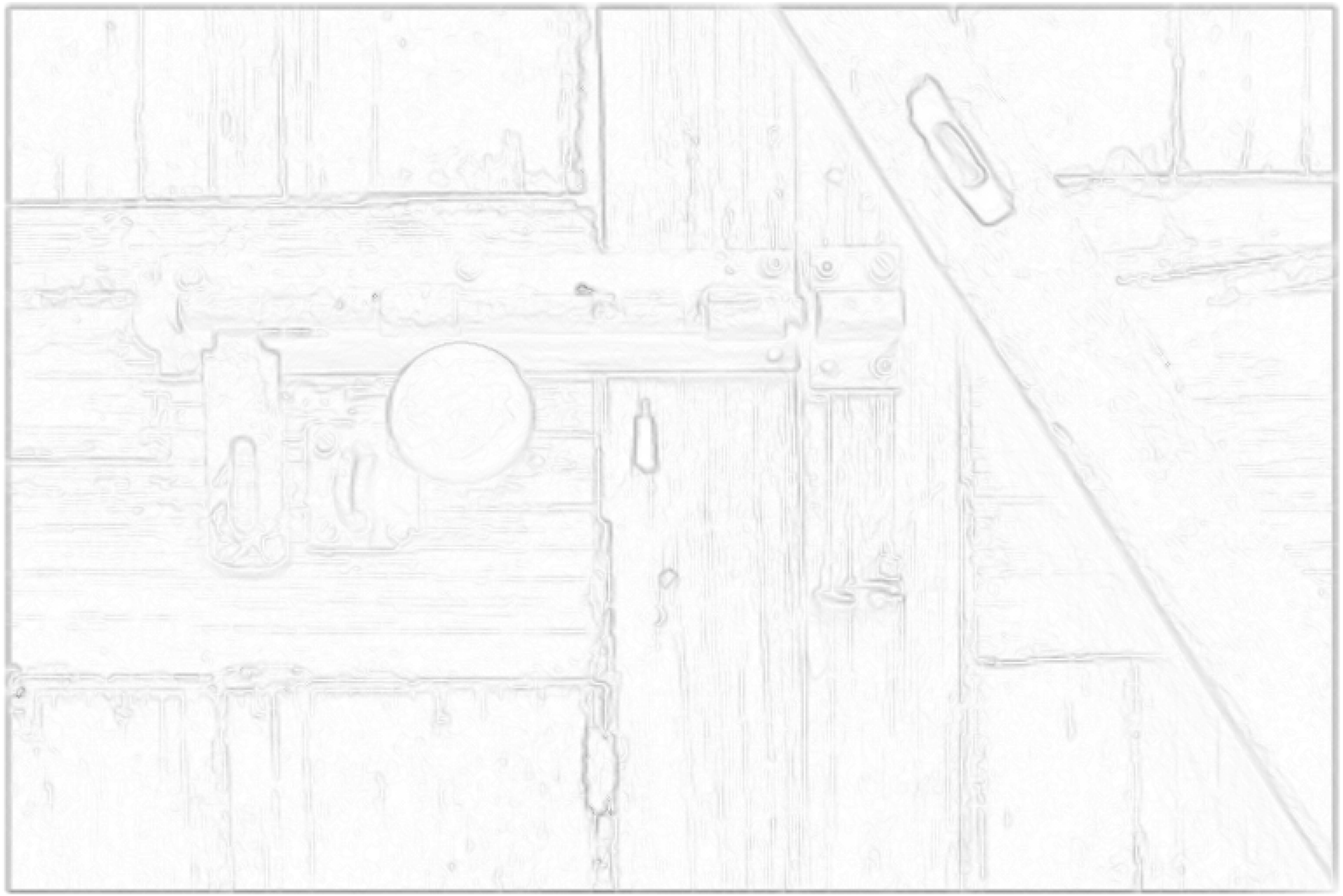}}
\hspace{-4mm}
\subfigure{
\label{fig:Channels:G_B_m2} 
\includegraphics[width=0.9in]{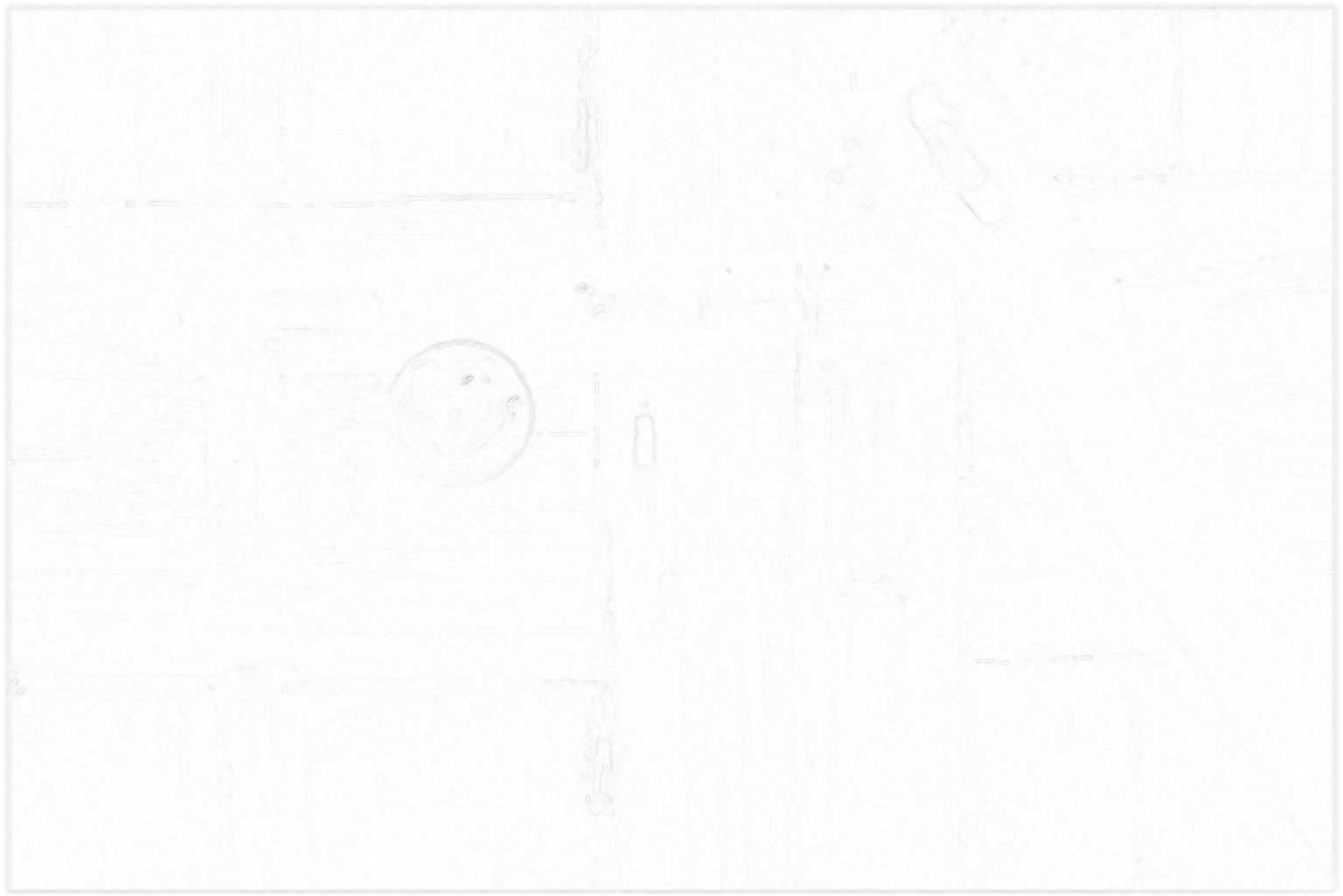}}\\
\setcounter{enumi}{1}
\vspace{-1mm}
\addtocounter{subfigure}{-16}
\subfigure[]{
\label{fig:Channels:R3} 
\includegraphics[width=0.9in]{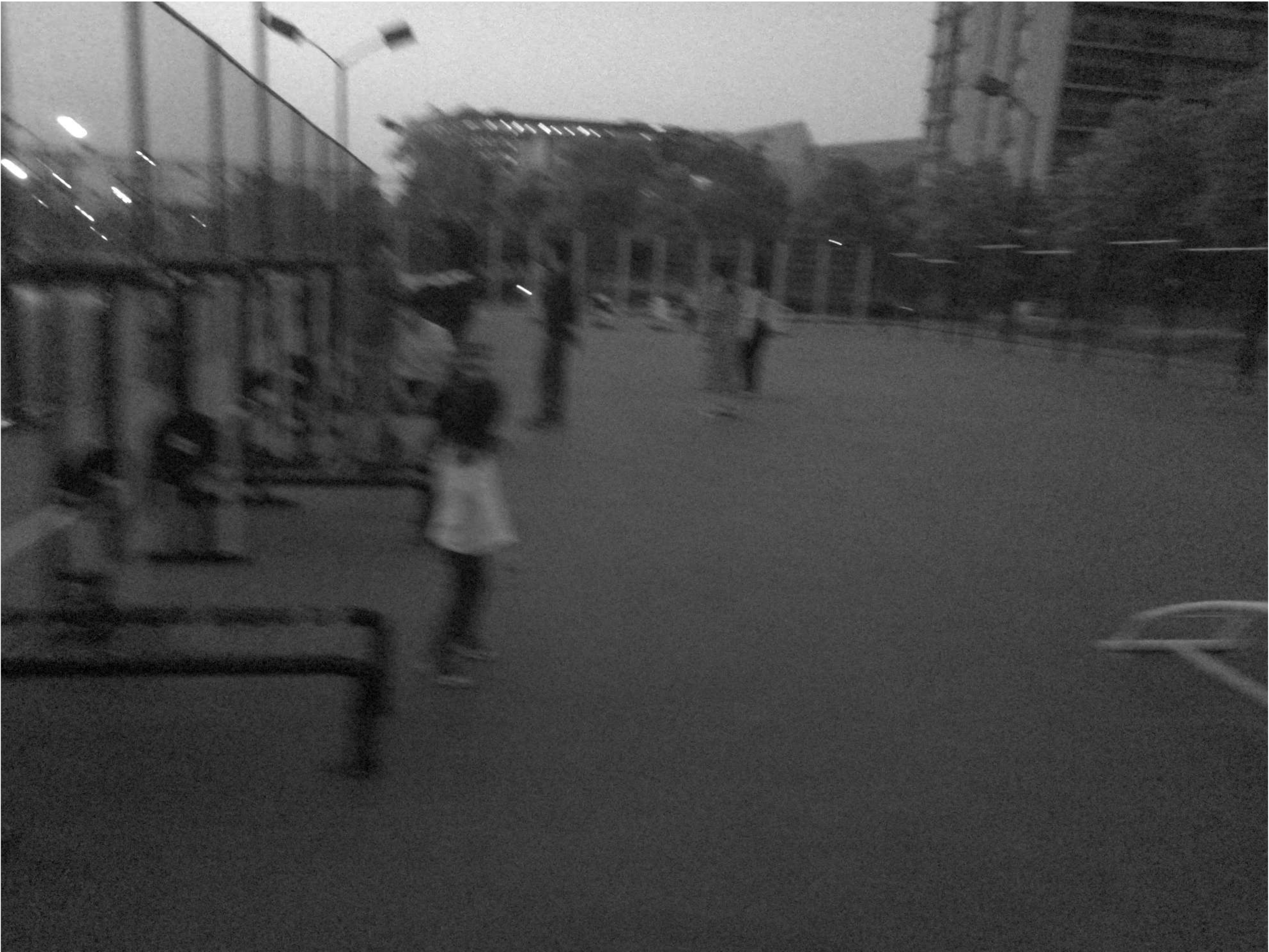}}
\hspace{-4mm}
\subfigure[]{
\label{fig:Channels:G3} 
\includegraphics[width=0.9in]{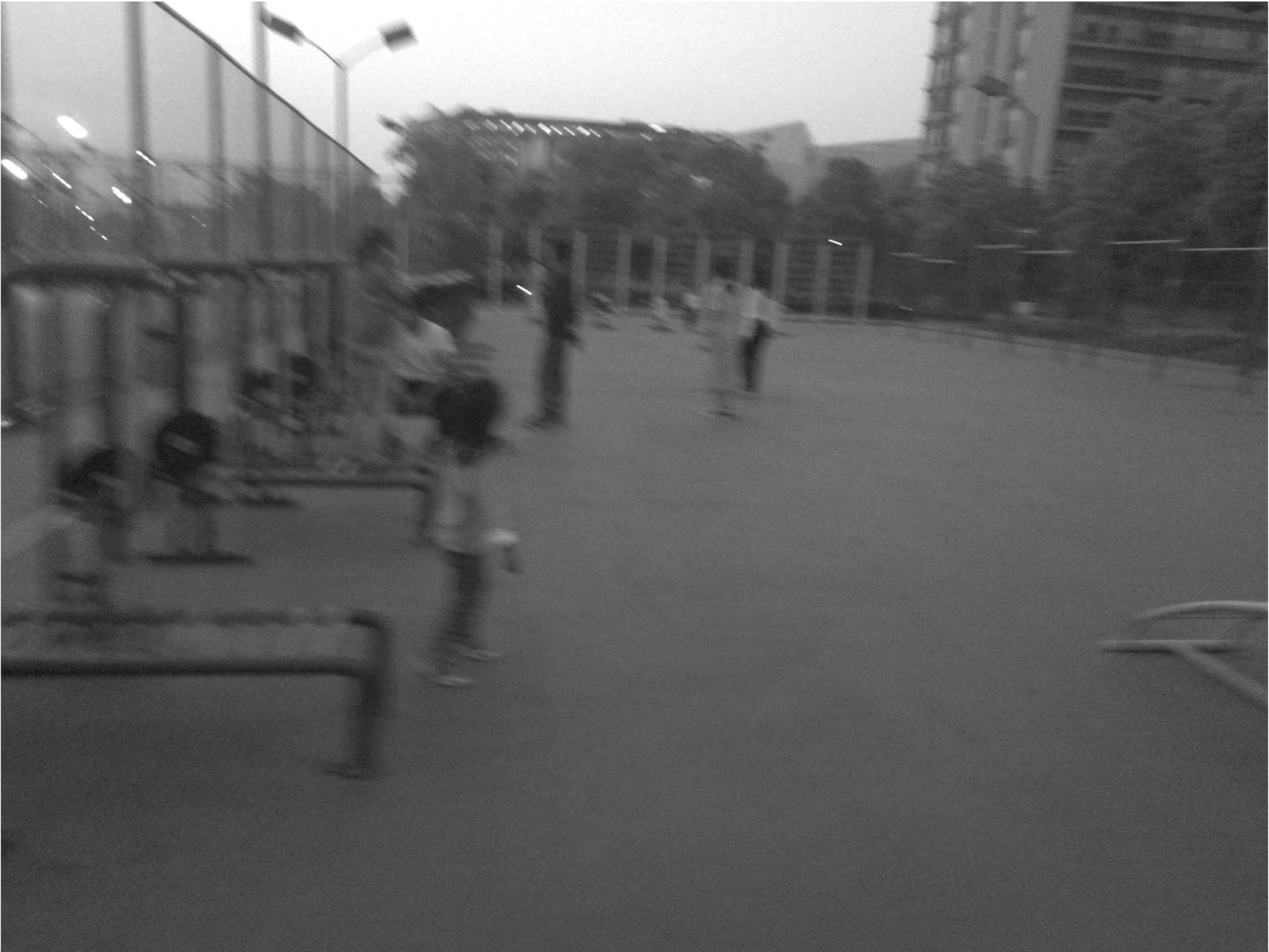}}
\hspace{-4mm}
\subfigure[]{
\label{fig:Channels:B3} 
\includegraphics[width=0.9in]{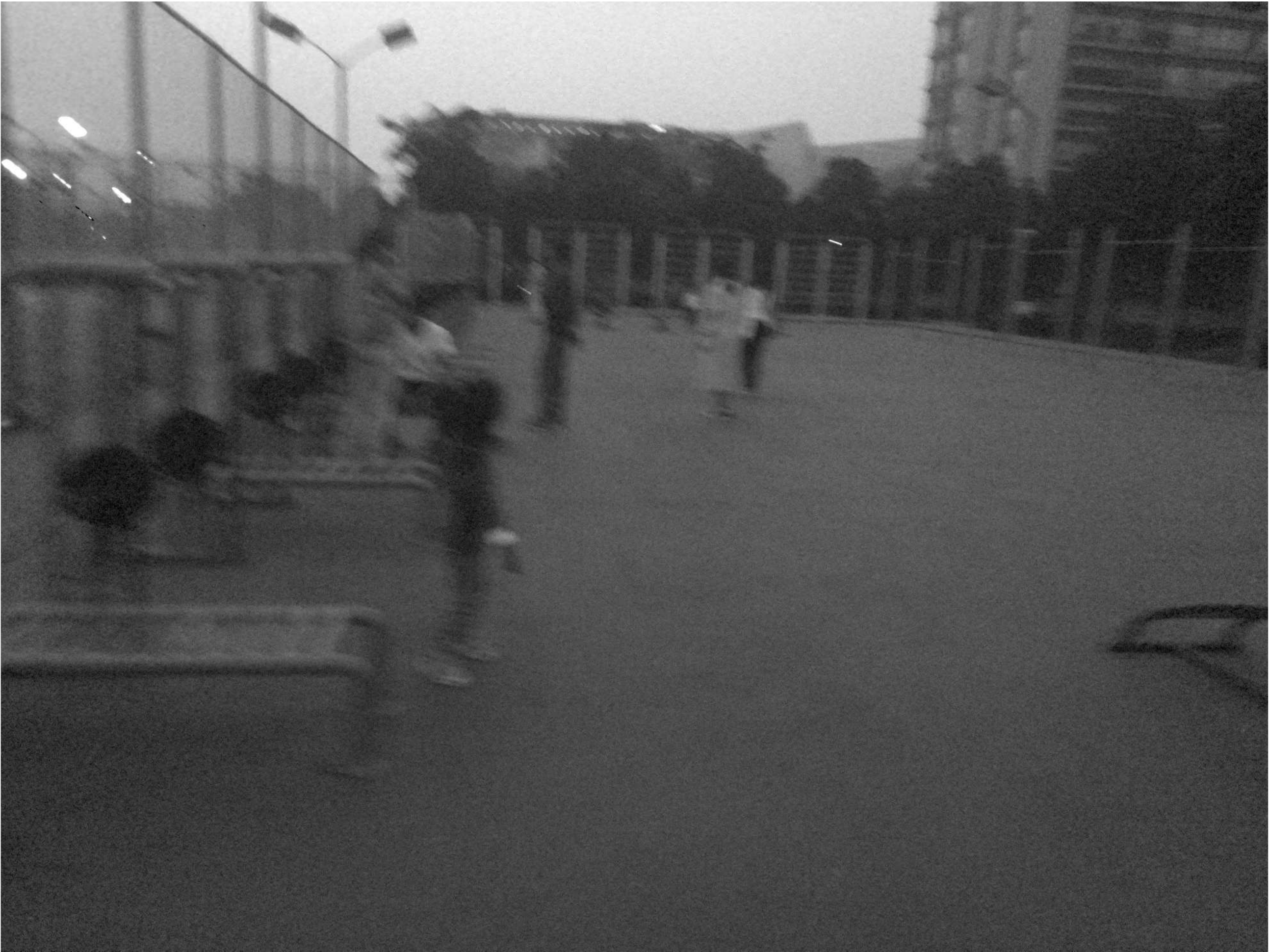}}
\hspace{-4mm}
\subfigure[]{
\label{fig:Channels:Bayer3} 
\includegraphics[width=0.9in]{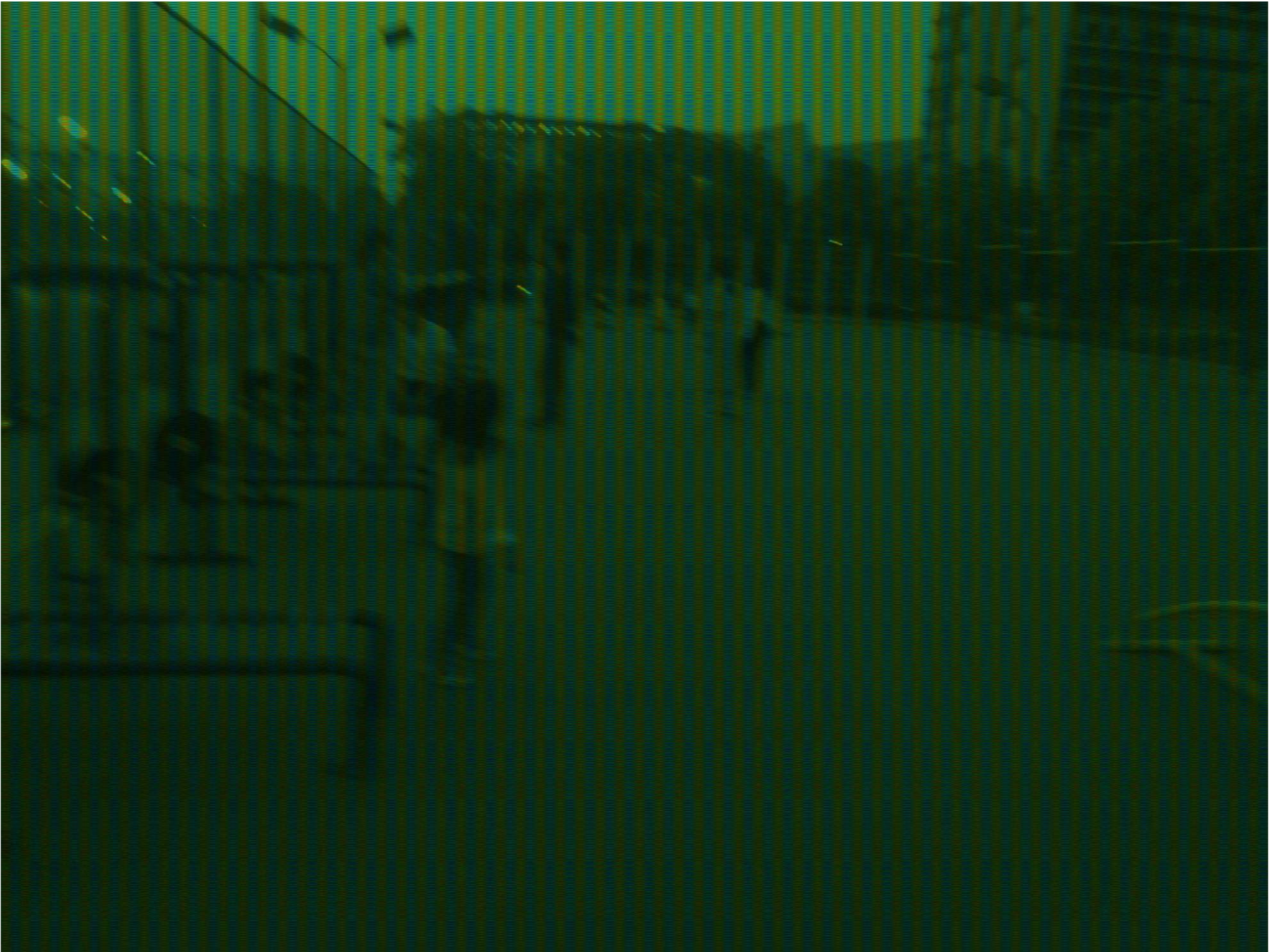}}
\hspace{-4mm}
\subfigure[]{
\label{fig:Channels:G_R3} 
\includegraphics[width=0.9in]{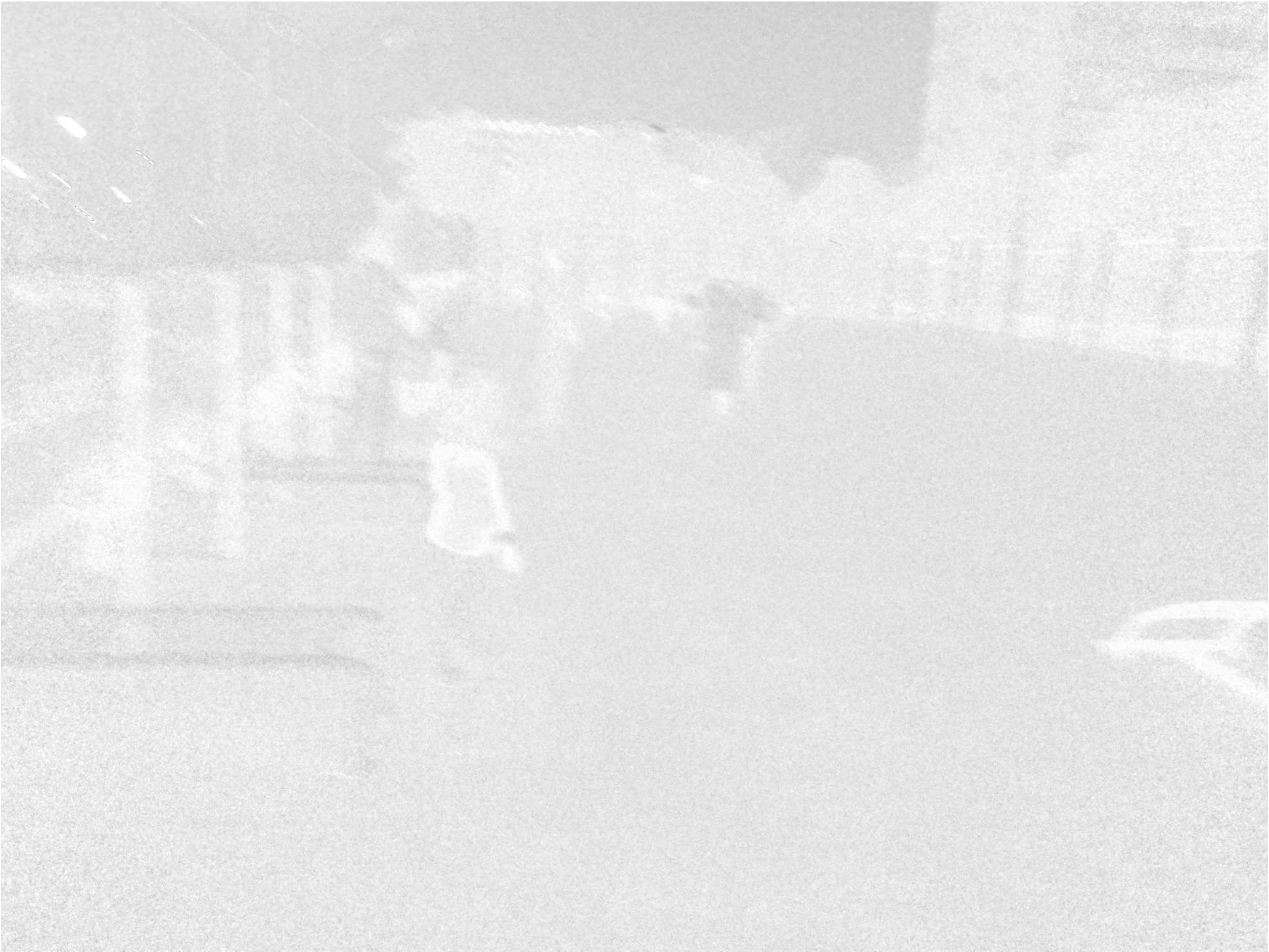}}
\hspace{-4mm}
\subfigure[]{
\label{fig:Channels:G_B3} 
\includegraphics[width=0.9in]{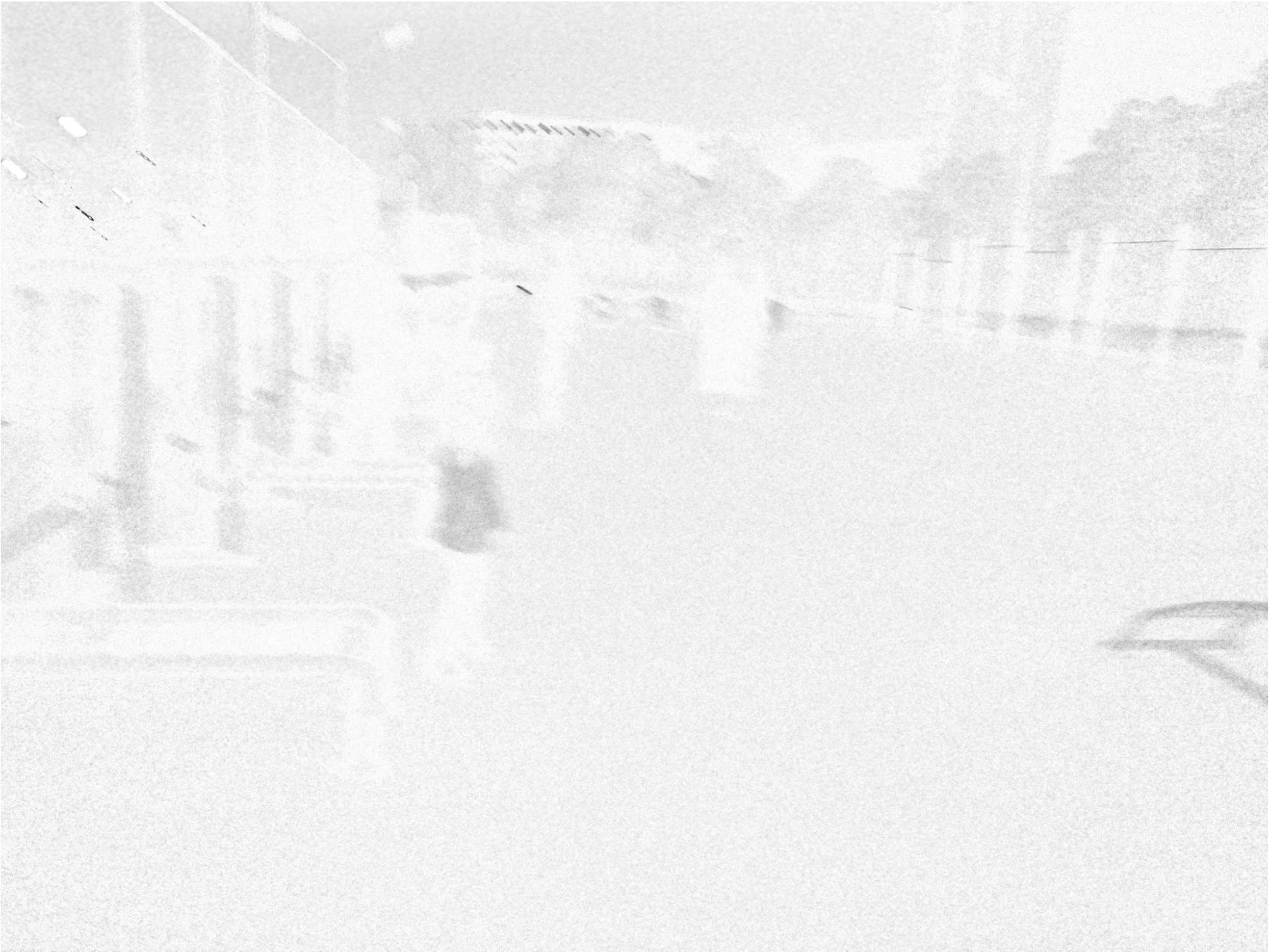}}
\hspace{-4mm}
\subfigure[]{
\label{fig:Channels:G_R_m3} 
\includegraphics[width=0.9in]{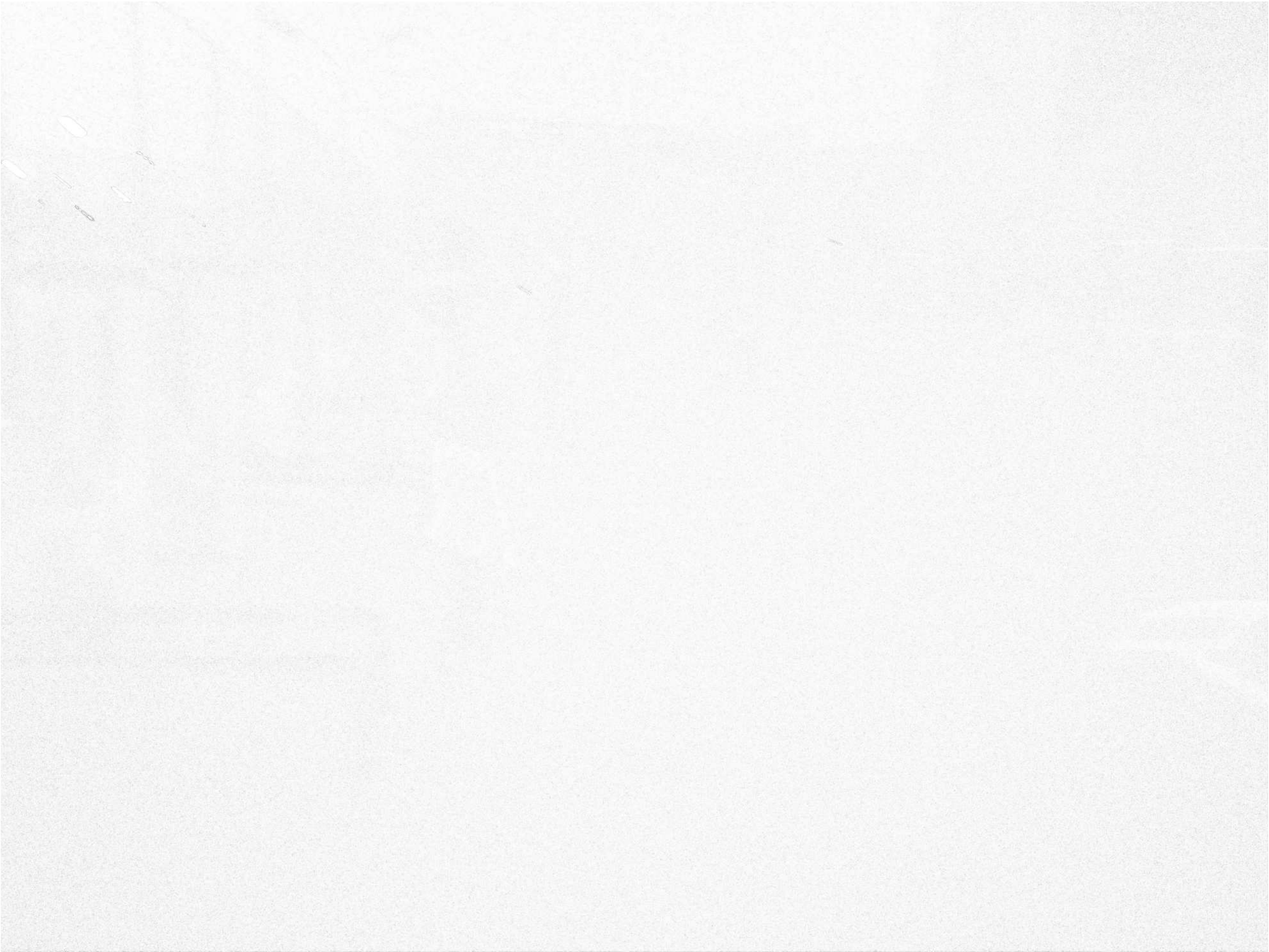}}
\hspace{-4mm}
\subfigure[]{
\label{fig:Channels:G_B_m3} 
\includegraphics[width=0.9in]{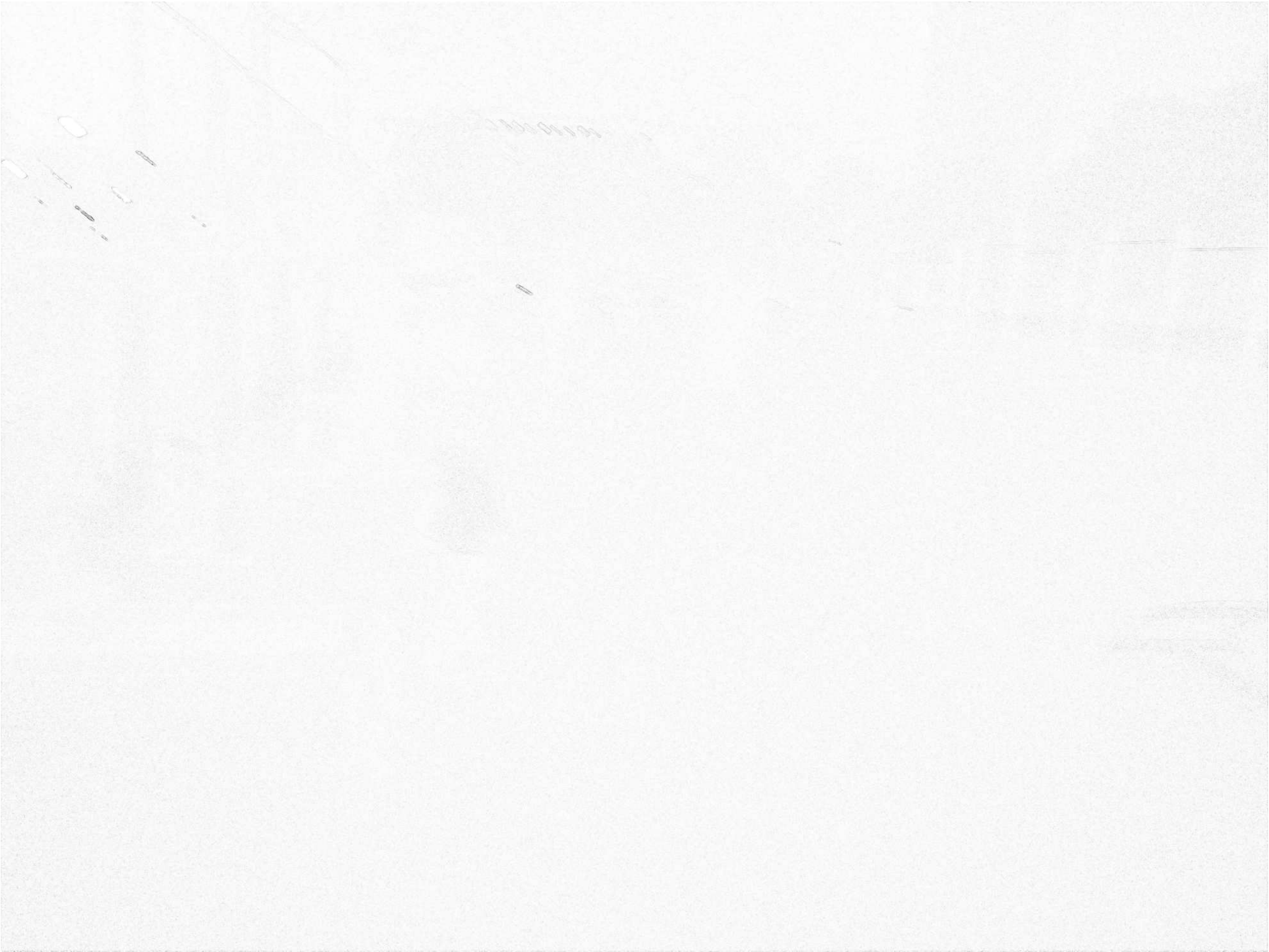}}
\caption{The Corresponding (a) R channel, (b) G channel, (c) B channel, (d) Bayer pattern images of Fig. \ref{fig:GMSD}. (e) G channel $-$ R channel. (f) G channel $-$ B channel. (g) Gradient map of (e). (h) gradient map of (f).}
\label{fig:Channels}
\end{figure*}

\subsubsection{SIFT Descriptor}
\label{section:sec4.2.5}
For the proposed Bayer pattern image-based SIFT feature extraction, extremums are searched among a $5\times5$ neighborhood instead of $3\times3$. To validate the scale and rotation invariant property of the generated SIFT features, key points are detected from the transformed images, i.e., the resized, rotated and blurred images. These key points are matched with the ones detected from the untransformed images. The repeatability criteria introduced in \cite{RN12} are used to evaluate the performance of SIFT descriptors in finding matching points. Given a pair of images, repeatability is defined by
\begin{equation}
\label{equ:repeatability}
P=\frac{M}{min\left ( n_{1},n_{2} \right )},
\end{equation}
where $n_{1}$ and $n_{2}$ are the number of descriptors detected on the images, $T$ is the transform between the original image $I$ and its transformed version $I_{tran}$ \cite{RN2}, $M$ is the number of correct matches. Pixel coordinates $ \left ({x_{1},y_{1}} \right ) $ and $ T^{-1}\left \{ (x_2,y_2) \right \} $ is considered matched within a $t$-neighborhood if
\begin{equation}
\label{equ:distance}
d(\left ( x_{1},y_{1} \right ),T^{-1}\left \{ (x_2,y_2) \right \})< t.
\end{equation}
Here, $d(\cdot)$ is the Euclidean distance between $\left ( x_{1},y_{1} \right )$ and $T^{-1}\left \{ (x_2,y_2) \right \}$. For a given pixel coordinates $(x_{1},y_{1})$, the $\theta$-rotated pixel coordinate $(x_{2},y_{2})$ can be computed as
\begin{equation}
\label{equ:rotate transform}
(x_{2},y_{2},1)=(x_{1},y_{1},1)H
\end{equation}
Here, $H=\begin{pmatrix}
cos\theta & sin\theta  & 0 \\
 -sin\theta& cos\theta  &0 \\
 0& 0  &1
 \end{pmatrix}$
is the homography matrix, corresponding to transform $T$ in \eqref{equ:distance}. Moreover, for two $N\times3$ matrices $A$ and $B$, which consist of  $N (N\geqslant3)$ pairs of matched points $(x_{1}^{j},y_{1}^{j},1)$ and $(x_{2}^{j},y_{2}^{j},1), j=1,2,3...N$, the homography \setlength{\lineskip}{0.1em}{matrix $\widehat{H}$ can be estimated as }
\begin{equation}
\label{equ:estimateH}
\widehat{H}=A^{\dagger} B,
\end{equation}
where $A^{\dagger}$ is the pseudo-inverse of $A$. 

\subsection {Experimental Results}
\label{section:sec4.3}
\subsubsection{Comparison of Gradient Maps}
\label{section:sec4.3.1}
In this experiment, the gradient maps generated from the original color images and the corresponding Bayer versions are compared. Note that for the INRIA dataset, gamma compression is applied to the converted Bayer pattern images to adjust the contrast, while this is not needed for the other two datasets.

The comparison results of different versions of gradient maps are presented in Table \ref{tab:gradient evaluation}. Generally speaking, all the three evaluation criteria reveal similar trends that different versions of gradient maps are close to each other. As shown in Table \ref{tab:gradient evaluation}, for the Kodak dataset, the gradients generated from color as well as Bayer pattern images are almost identical, while for the other two datasets, the similarities are slightly lower. This is because for the Kodak dataset, both the color and Bayer pattern images can be regarded as ``true'' (a true color image dataset with Bayer version generated by resampling), while for the SHTech dataset and the PASCALRAW dataset, images are interpolated from Bayer pattern images using demosaicing algorithm. It is well known that extra errors will be introduced no matter how sophisticated the demosaicing algorithms is. This can be observed from the comparison results of the true color images and images generated using different demosaicing algorithms shown in Table \ref{tab:demosaic methods}.

Moreover, for the INRIA dataset, both the color and Bayer pattern images are ``estimated'' since the color version is interpolated and the Bayer version is reversely converted from the color version. Errors are injected in both forward and reverse ISP pipeline. Therefore, for the evaluation of the proposed Bayer pattern image-based gradient extraction pipeline, the Kodak dataset is more reliable than the other two.

This can be illustrated using Fig. \ref{fig:HOG}, where three versions of HOG descriptors, i.e., HOG from the original image, HOG from the Bayer pattern image without gamma compression and HOG from the Bayer pattern image with gamma compression, are presented. Note that as we mentioned, for the reversed INRIA dataset, proper gamma compression is necessary because the reversed images are at a low bit width, which is a side effect of forward + reverse ISP for Bayer image generation. As shown in Fig. \ref{fig:HOG:singleGammaHOG}, the descriptors cannot find enough features in low contrast Bayer pattern image without gamma compression (in Fig. \ref{fig:HOG:singlehuman}). But after adjusting the contrast by gamma compression, the HOG feature extracted from Fig. \ref{fig:HOG:singleGamma} becomes more stable, and close to the one extracted from the original color image in Fig. \ref{fig:HOG:originalHOG}.

Fig. \ref{fig:violin_SHtech} presents the distribution of per-channel GMSD comparison results of the SHTech dataset, where color means the  maximum gradient among the three channels is selected when computing the gradient maps.  It can be found that the distributions of GMSD between any pairing of the three channels are close. The gradient maps computed from gray images and Bayer images are closer to that computed from the green channel. This is because the green channel contributes a larger proportion in both gray images (60\%) and Bayer images (50\%).

By analyzing the outliers in Fig. \ref{fig:violin_SHtech}, it is found that there are three situations that lead to notable gradient difference, which may harm the gradients generated from Bayer images. These situations are illustrated in Fig. \ref{fig:GMSD}.  Note that GMS is designed to range from 0 to 1, where 1 means no error. Thus, the brighter in GMS map, the higher the similarity.

The first situation is when the light source shines directly on a smooth surface (e.g. smooth wall, metal, etc.), especially for bright colored smooth objects. This situation violates the assumptions of Lambertian non-flat surface patches model because the reflection, in this case, is closer to specular reflection than diffuse reflection. A flat surface cannot be treat as a Lambertian surface, because the brightness of an object is different when seen from different view point. Highlight areas caused by specular reflection make the illuminance no longer slow-varying. Fig. \ref{fig:GMSD:smooth} illustrates this phenomenon. When sunlight hits the car directly, the GMS map shows a big difference among the bodywork (the dark areas in GMS map), leading to a big $C^k(x,y)$ in the car body but a small $C^k(x,y)$ in the background. This is illustrated in Fig. \ref{fig:Channels:G_R3} and Fig. \ref{fig:Channels:G_B3} (top). These areas result in edges as shown in Fig.\ref{fig:Channels:G_R_m3} and Fig.\ref{fig:Channels:G_B_m3} (top), corresponding to the non-zero $\delta^{k}(x+1,y,x-1,y)$ term in \eqref{equ:delta_gradient}. The second situation is when there are irregular textures as shown in Fig. \ref{fig:GMSD:edge}. In this situation,  the $\delta^{k}(x+1,y,x-1,y)$ term in \eqref{equ:delta_gradient} can no longer be ignored, as shown in Fig. \ref{fig:Channels:G_R_m3} and Fig. \ref{fig:Channels:G_B_m3} (middle). However, these kind of violations appear mostly inside objects such that the influence on the edges is relatively small. For example, the edge of the door handle. The last situation is when there exists heavy noise as shown in Fig. \ref{fig:GMSD:noise}. This situation is usually caused by low light condition and motion blur. It can be found from the GMS map that in this case,  the gradient difference is evenly distributed throughout the image.

It can be found from the examples in Fig. \ref{fig:GMSD} that the GMSD values of Bayer pattern images with other images, especially with green channel images, are smaller than other combinations. The failures that caused by a bright spot or saturated colors may not occur between all channels because they may lead to a small $\delta^{R}(x+1,y,x-1,y)$ but a big $\delta^{B}(x+1,y,x-1,y)$ (Fig. \ref{fig:Channels:G_R_m1} and \ref{fig:Channels:G_B_m1} top) or vice versa (Fig. \ref{fig:Channels:G_R_m1} and \ref{fig:Channels:G_B_m1} middle).

\begin{table}[t]
\centering
\caption{Comparison Results of Bayer Image Based and Color Image Based Blur and Resize}
\label{tab:operation}
\setlength\arrayrulewidth{0.6pt}
\renewcommand{\arraystretch}{1.2}
\setlength{\tabcolsep}{4.2pt}{
\begin{tabular}{|c|c|c|c|c|c|}
\hline
\multirow{3}{*}{\textbf{Operation}}                                                 & \multicolumn{2}{c|}{\multirow{2}{*}{\textbf{Parameter}}}                                                                                                      & \multicolumn{3}{c|}{\multirow{2}{*}{\textbf{Average}}} \\
                                                                           & \multicolumn{2}{c|}{}                                                                                                                                & \multicolumn{3}{c|}{}                         \\ \cline{2-6}
                                                                           & Bayer                                                                              & Color                                                           & MSSIM         & MSE           & PSNR          \\ \hline
\multirow{4}{*}{\makecell*[c]{Gaussian blur}}  & \multirow{4}{*}{\makecell[c]{3$\times$3 kernel}}                                                        & 3$\times$3 kernel                                                      & 0.952         & 46.010        & 32.334        \\ \cline{3-6}
                                                                           &                                                                                    & 5$\times$5 kernel                                                      & 0.979         & 18.040        & 36.293        \\ \cline{3-6}
                                                                           &                                                                                    & 7$\times$7 kernel                                                      & 0.988         & 9.807         & 38.962        \\ \cline{3-6}
                                                                           &                                                                                    & 9$\times$9 kernel                                                      & 0.985         & 11.762        & 38.094        \\ \hline
\multirow{2}{*}{Resize}                                                    & \multicolumn{2}{c|}{Scale=0.5}                                                                                                                       & 0.938         & 70.453        & 30.232        \\ \cline{2-6}
                                                                           & \multicolumn{2}{c|}{Scale=2}                                                                                                                         & 0.912         & 93.584        & 30.031        \\ \hline
\multirow{4}{*}{\begin{tabular}[c]{@{}l@{}}Blur \& Resize\end{tabular}} & \multirow{2}{*}{\begin{tabular}[c]{@{}l@{}}3x3 kernel,\\ scale=0.5\end{tabular}} & \multirow{2}{*}{\begin{tabular}[c]{@{}l@{}}7x7 kernel,\\ scale=0.5\end{tabular}} & \multirow{2}{*}{0.977} & \multirow{2}{*}{21.444} & \multirow{2}{*}{35.499} \\
                                                                           &                                                                                  &                                                                                  &                        &                         &                         \\ \cline{2-6}
                                                                           & \multirow{2}{*}{\begin{tabular}[c]{@{}l@{}}3x3 kernel,\\ scale=2\end{tabular}}   & \multirow{2}{*}{\begin{tabular}[c]{@{}l@{}}7x7 kernel,\\ scale=2\end{tabular}}   & \multirow{2}{*}{0.976} & \multirow{2}{*}{15.667} & \multirow{2}{*}{36.691} \\
                                                                           &                                                                                  &                                                                                  &                        &                         &                         \\ \hline
\end{tabular}}
\end{table}

\begin{figure}[t]
  \centering
  \subfigure[]{
  \label{fig:violin_MSSIM} 
  \includegraphics[width=3.3in]{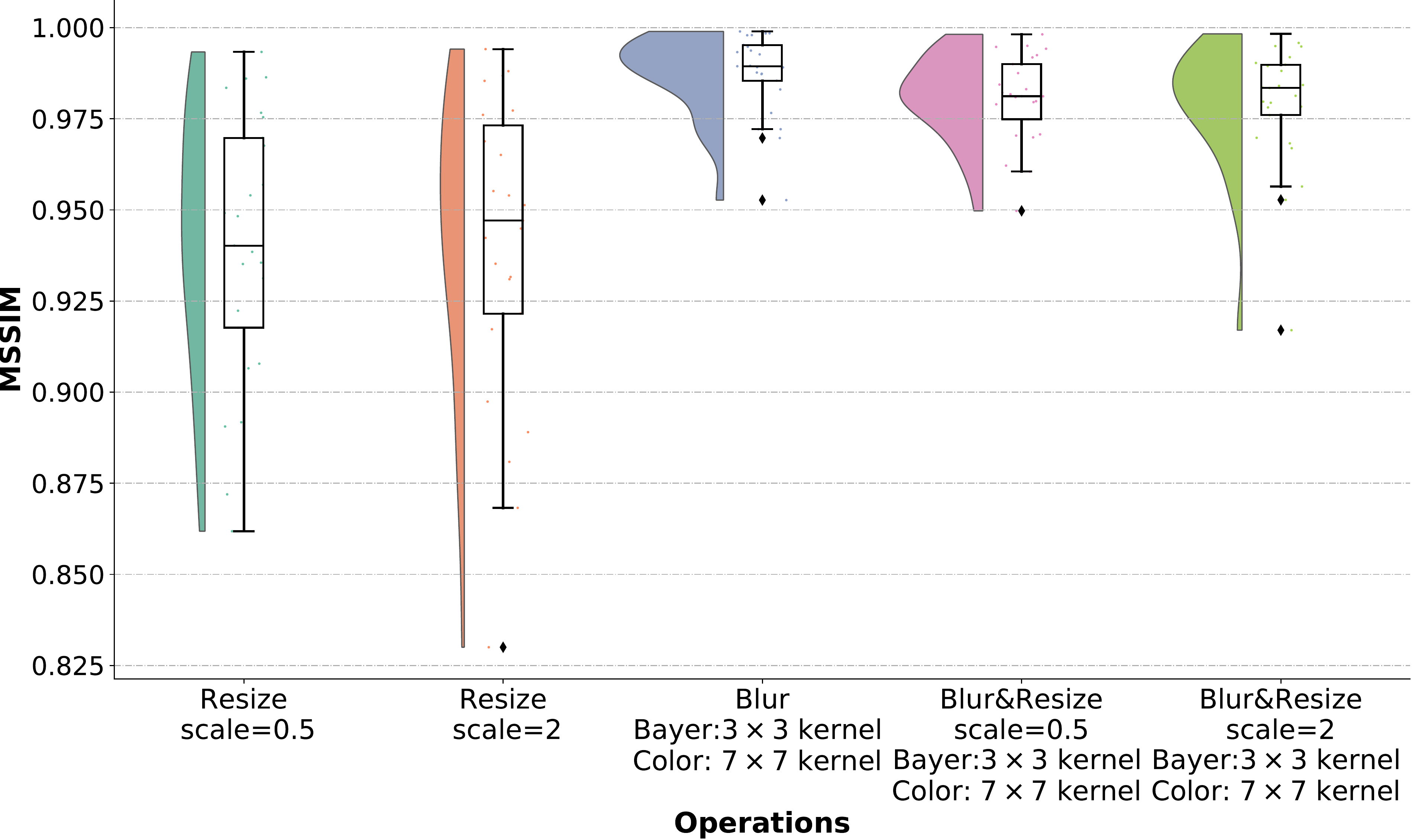}}
  \subfigure[]{
  \label{fig:MSSIM} 
  \includegraphics[width=3in]{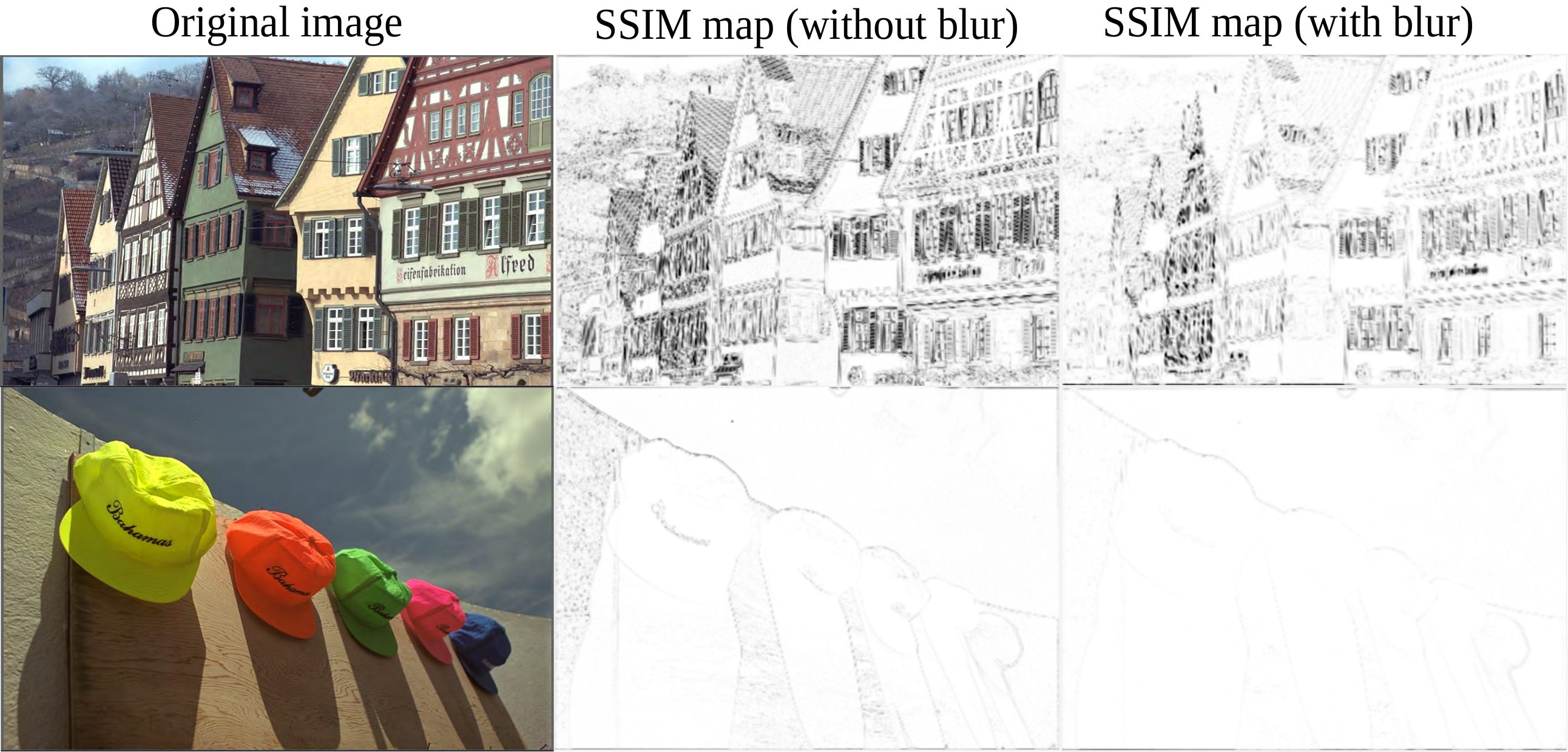}}
  \caption{(a) MSSIM distribution of the corresponding operations in Table \ref{tab:operation} and (b) original images and SSIM maps after scaling operation. The top-left in (b) is an image with rich details and the bottom-left one is an image with less textures. The SSIM maps are generated from the Bayer image scaled by super-pixel structure directly and resampled (as Bayer) scaled color image. The blur parameters are $3\times 3$ kernel and $\sigma=0.8$ for Bayer images and $7\times 7$ kernel and $\sigma=1.4$ for color images.}
  \label{fig:violin_Kodia}
\end{figure}

\begin{table}[t]
\centering
\caption{Evaluation Results of Pipeline 1 and Pipeline 2}
\label{tab:profiling}
\setlength\arrayrulewidth{0.6pt}
\renewcommand{\arraystretch}{1.2}
\setlength{\tabcolsep}{1.6pt}{
\begin{tabular}{|c|c|c|c|c|c|}
\hline
\multicolumn{2}{|c|}{\multirow{2}{*}{}}                                                                                     & \multicolumn{2}{c|}{Original}                        & \multicolumn{2}{c|}{Normalized}                      \\ \cline{3-6}
\multicolumn{2}{|c|}{}                                                                                                      & Time(ms)               & Memory(kb)               & Time                   & Mermory                \\ \hline
\multicolumn{2}{|c|}{Pipeline 1}                                                                                            & 0.87                   & 1148936                  & 1                      & 1                     \\ \hline
\multirow{6}{*}{Pipeline2} & Nearest Neighbor                                                                               & 0.91                   & 1698144                  & 1.05                   & 1.48                  \\ \cline{2-6}
                           & Linear Interpolation                                                                           & 89.45                  & 1673512                  & 102.88                 & 1.46                  \\ \cline{2-6}
                           & Cubic Interpolation                                                                            & 95.62                  & 2212948                  & 109.98                 & 1.93                  \\ \cline{2-6}
                           & \multirow{2}{*}{\begin{tabular}[c]{@{}c@{}}Adaptive Color \\ Plane Interpolation\end{tabular}} & \multirow{2}{*}{65.33} & \multirow{2}{*}{2699868} & \multirow{2}{*}{75.14} & \multirow{2}{*}{2.35} \\
                           &                                                                                                &                        &                          &                        &                       \\ \cline{2-6}
                           & Hybrid Interpolation                                                                           & 57.66                  & 3079720                  & 66.32                  & 2.68                  \\ \hline
\end{tabular}}
\end{table}

\begin{figure}[t]
\centering
\subfigure[]{
\label{fig:kernel:garyimage} 
\includegraphics[width=1.5in]{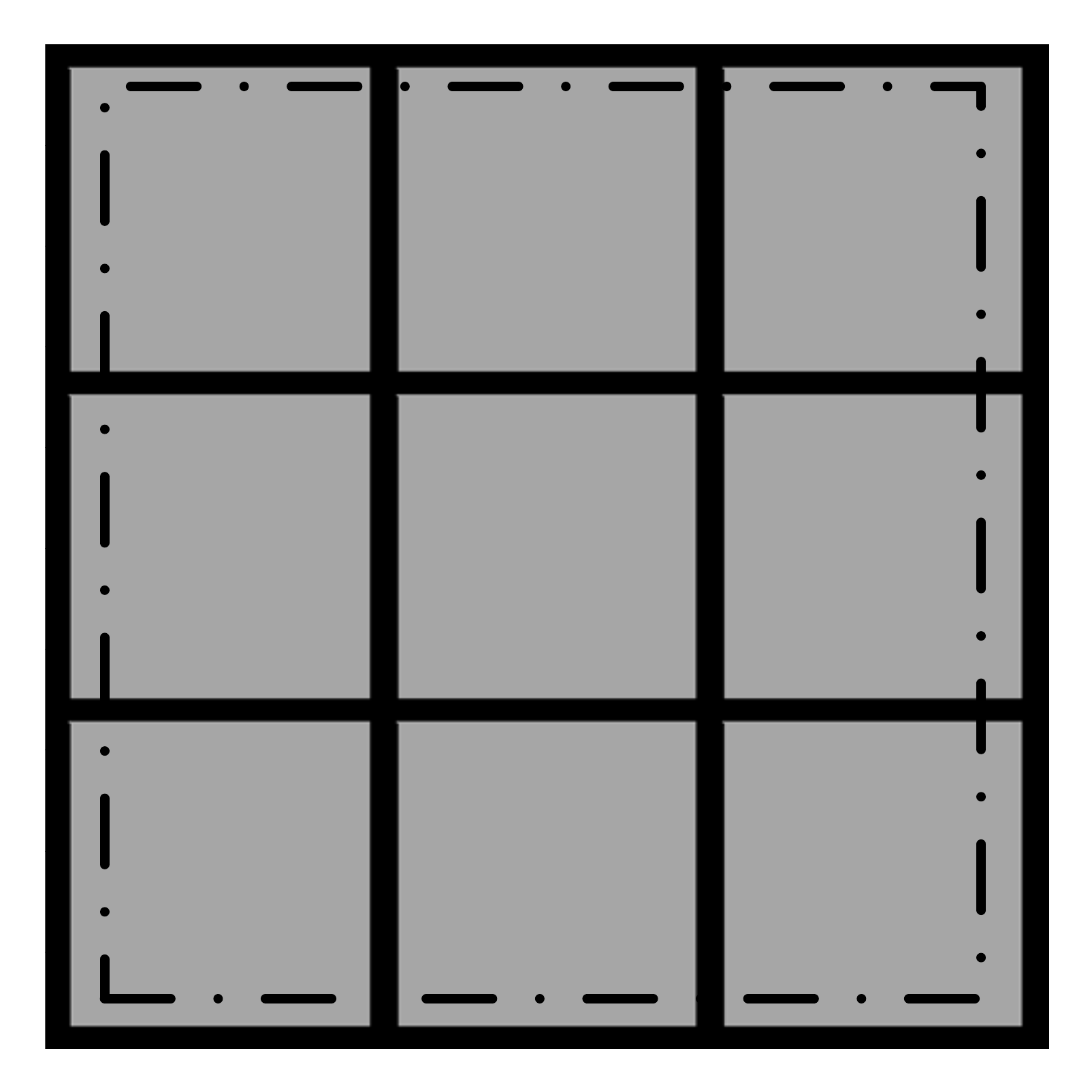}}
\hspace{6mm}
\subfigure[]{
\label{fig:kernel:superpixel} 
\includegraphics[width=1.5in]{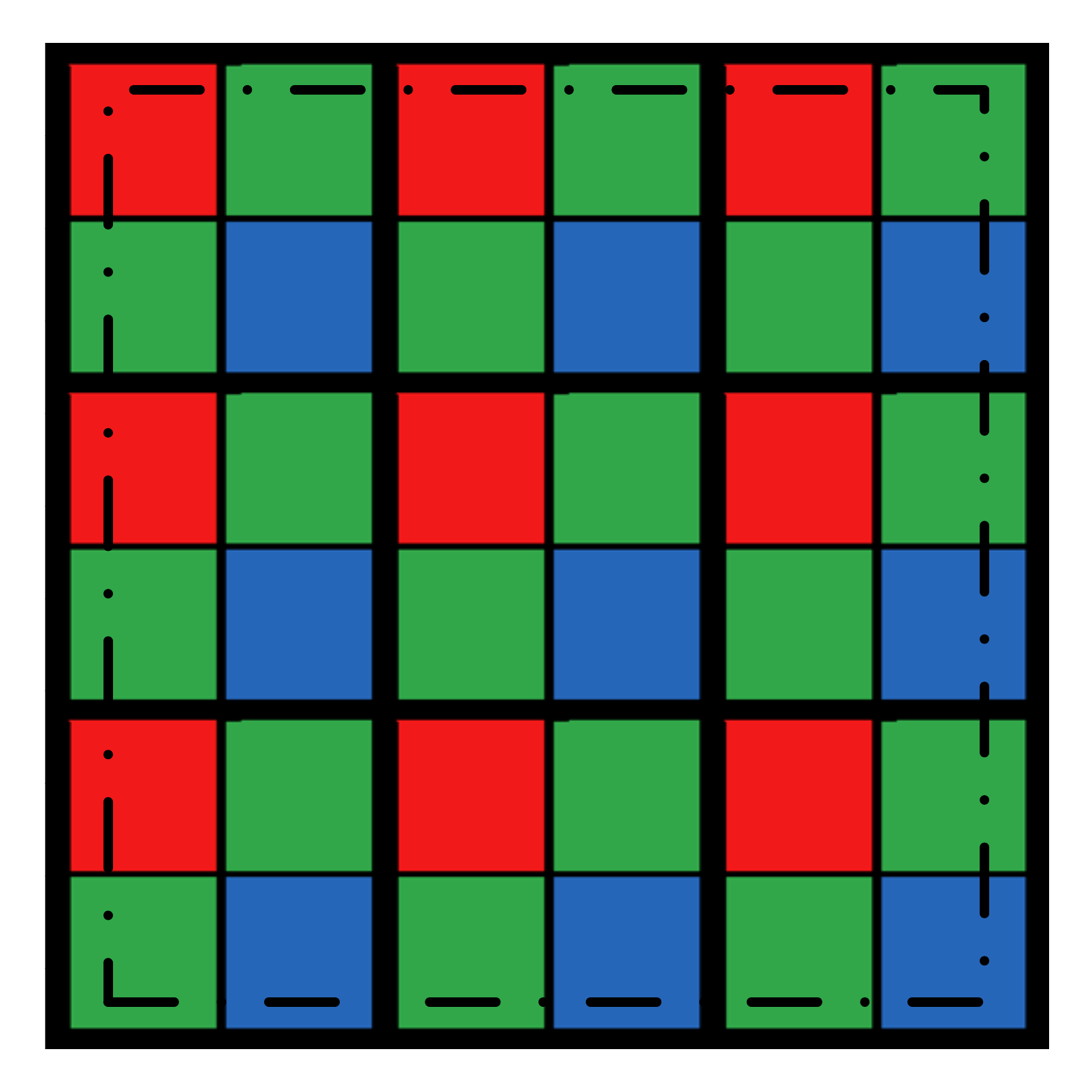}
}\\
\caption{The coverage of a $3 \times 3$ kernel on a (a) gray image and (b) the corresponding super-pixel Bayer pattern image.}
\label{fig:kernel}
\end{figure}

\subsubsection{Blur and Resize}
\label{section:sec4.3.2}

The purpose of this experiment is to show that multiscale model construction (mainly resize and scale operation) can also be performed on Bayer images by super-pixel based resize and scale operations. The operations can either be performed in RGB domain (three-channel) or Bayer domain (single-channel). Since demosaicing affects performance (Table \ref{tab:demosaic methods}), we performed these comparisons in Bayer domain. The resize operation here refers to the change of width and height of a digital image into a specified size, e.g, scale$=$0.5 means to reduce the height and width of a image to half. For Bayer pattern images, blur and resize are directly applied on the super-pixel structure, while for color images, the original images are blurred and resized followed by the generation of Bayer pattern images through resampling, i.e., a blur $+$ resize $+$ resampling pipeline is used to generate Bayer pattern images from color images. The Kodak dataset is used in this experiment.

Presented in Table \ref{tab:operation} are the comparison results of blur and resize. For a certain $a \times a$ kernel, $\sigma$ in Eq. \eqref{equ:DOG} can be determined by the specific application or using the following equation\cite{RN37}
\begin{equation}
\label{equ:sigma}
\sigma =0.3\times \left ( \left ( a-1 \right )\times 0.5-1 \right )+0.8.
\end{equation}
According to the experiment, blur and resize on Bayer pattern images using super-pixel structure generates similar results with that on color images. It can be observed from Table \ref{tab:operation} that the $7\times7$ kernel for color images approach to the $3\times3$ kernel for super-pixel Bayer pattern images. This is because a super-pixel is a collection of pixels in a Bayer pattern which may expand the smooth area. As illustrated in Fig. \ref{fig:kernel}, a $3\times3$ kernel on super-pixel Bayer pattern images covers a $6\times6$  pixel location in the original Bayer pattern image. Therefore, we expand the kernel used in gray images accordingly. Since the length of kernels needs to be odd, we have tried different kernel sizes in our experiments and presented the results in Table V. According to the results in Table \ref{tab:operation}, the Bayer pattern image-based blur and resize generates similar results to the color image-based operations. Fig. \ref{fig:violin_MSSIM} presents the MSSIM distribution of the operations in Table \ref{tab:operation}. 
It can be found that the resize operation makes the distribution more dispersed, while performing blur (low-pass filtering) before resize may alleviate the quality loss caused by the scaling operation. Outliers often appear in images with rich details, e.g., the top left image in Fig. \ref{fig:MSSIM}. The bottom-left image gives an example with less texture. As the SSIM maps show, images with rich details have larger difference among edges and performing blur before resize can improve it.

Moreover, to evaluate the memory access and computation time of the proposed method,  the following two different pipeline configurations are compared.
\begin{itemize}
\item Pipeline 1. Starting from Bayer images, perform blur + resize using the super-pixel method without demosaicing, then compute gradient magnitude images.
\item Pipeline 2. Starting from Bayer images, demosaic it to color image, then perform blur + resize to each channel, generate gray images and compute gradient magnitude image.
\end{itemize}

 The comparison results of pipeline 1 and 2 are presented in Table \ref{tab:profiling}. Five different demosaicing algorithms are used in the evaluation of pipeline 2. All these pipelines are profiled using MATLAB R2020a, on a Windows 10 PC with i7-7700 CPU and 16G memory. This experiment is performed on resampled Kodak dataset. It can be found from Table \ref{tab:demosaic methods} and Table \ref{tab:profiling} that complex interpolation algorithms lead high image quality, but also increase time and memory consumption. By skipping the complex operations, both time and memory can be saved.

\begin{figure}[t]
\centering
\subfigure[]{
\label{fig:key and match:graykey} 
\includegraphics[width=1.4in]{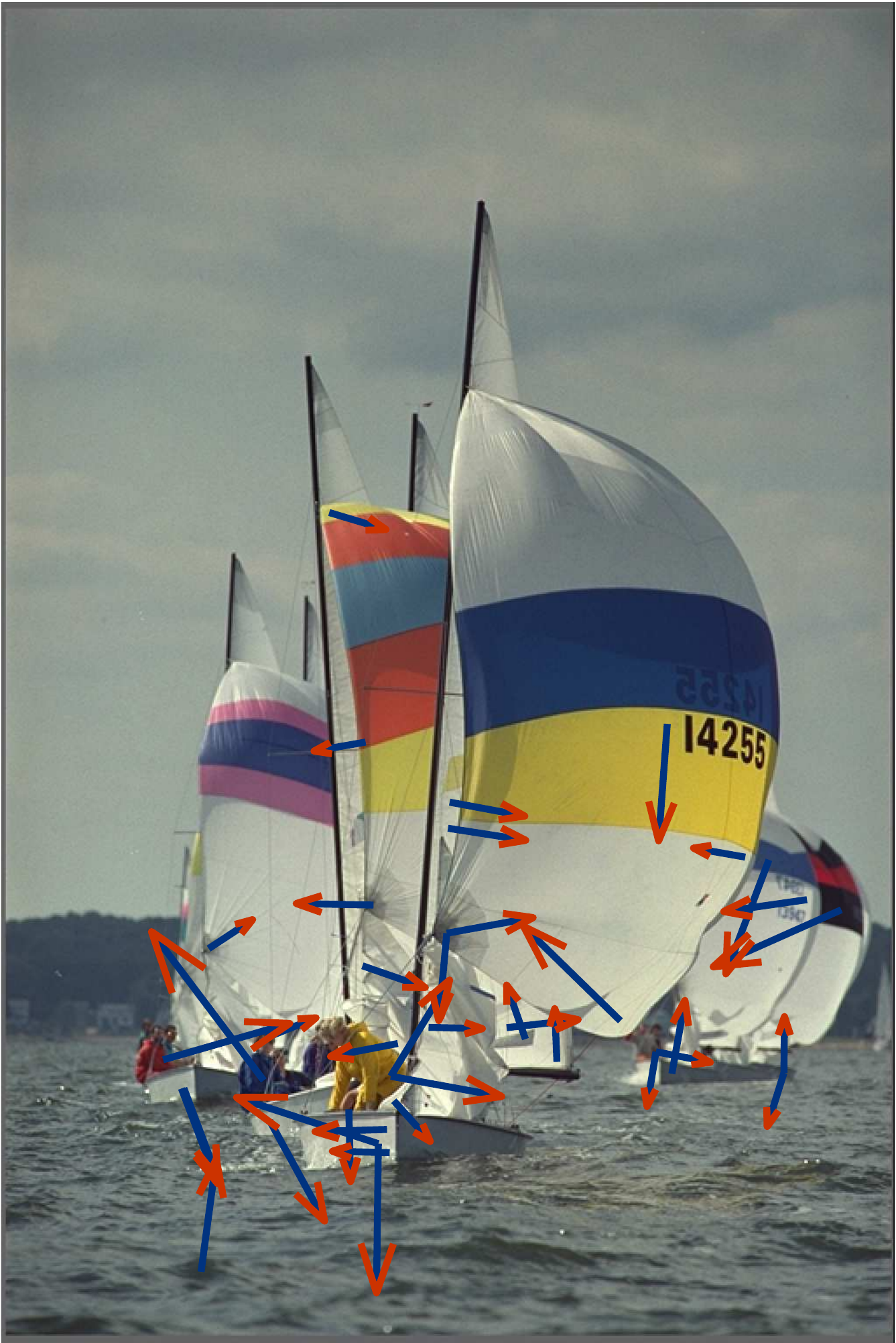}}
\hspace{-1mm}
\subfigure[]{
\label{fig:key and match:bayerkey} 
\includegraphics[width=1.4in]{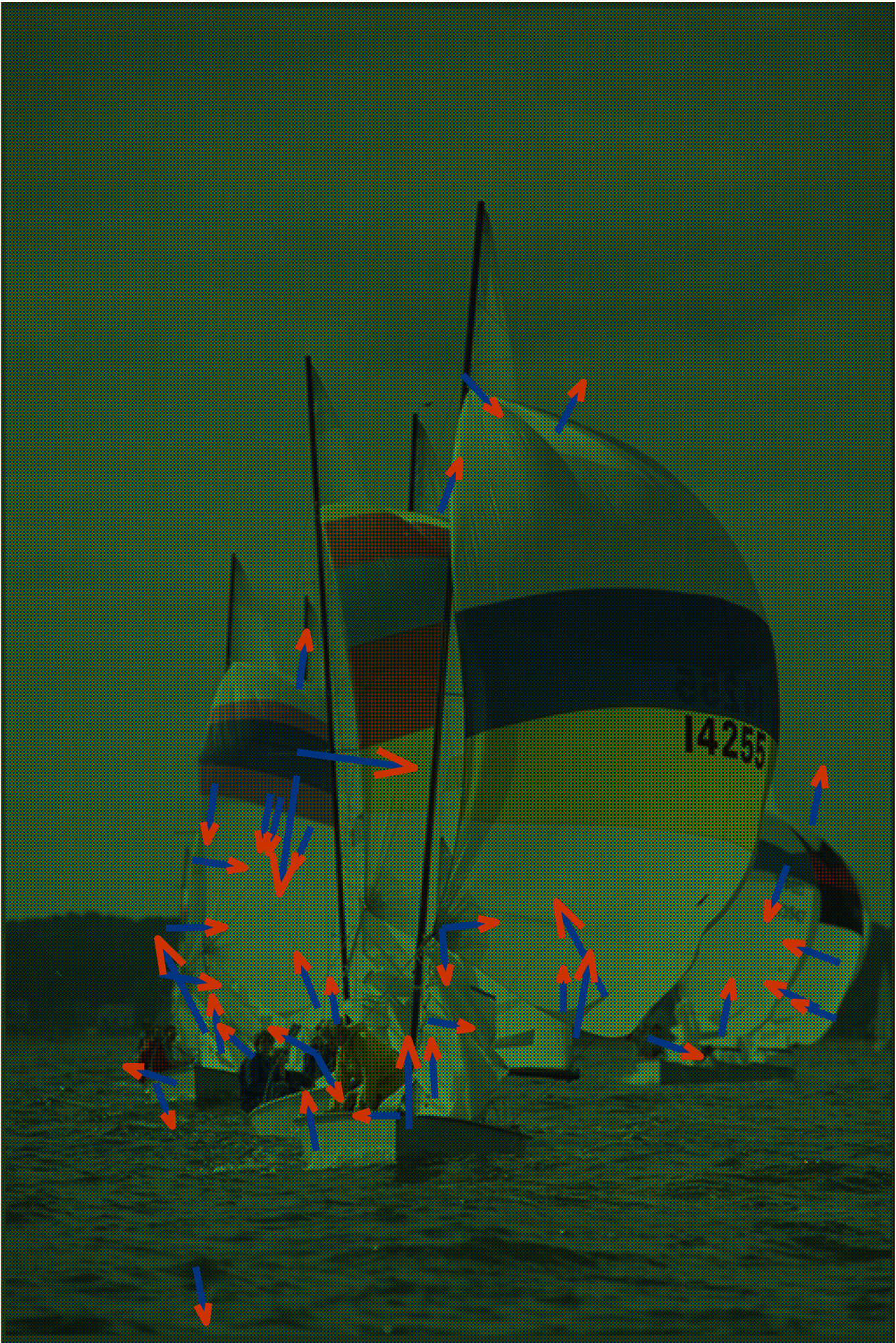}}\\
\subfigure[]{
\label{fig:key and match:graykey20} 
\includegraphics[width=1.4in]{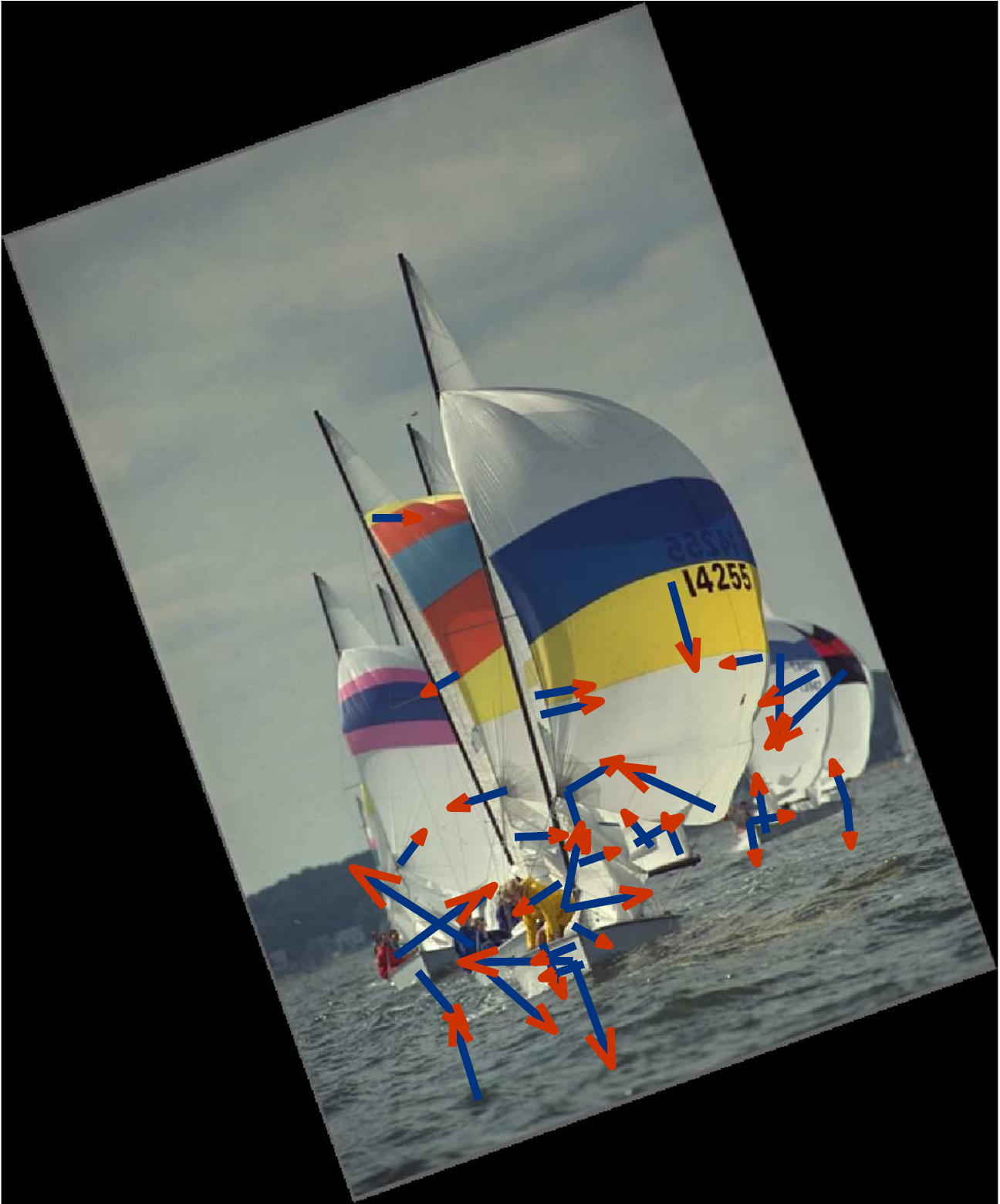}}
\hspace{-1mm}
\subfigure[]{
\label{fig:key and match:bayerkey20} 
\includegraphics[width=1.4in]{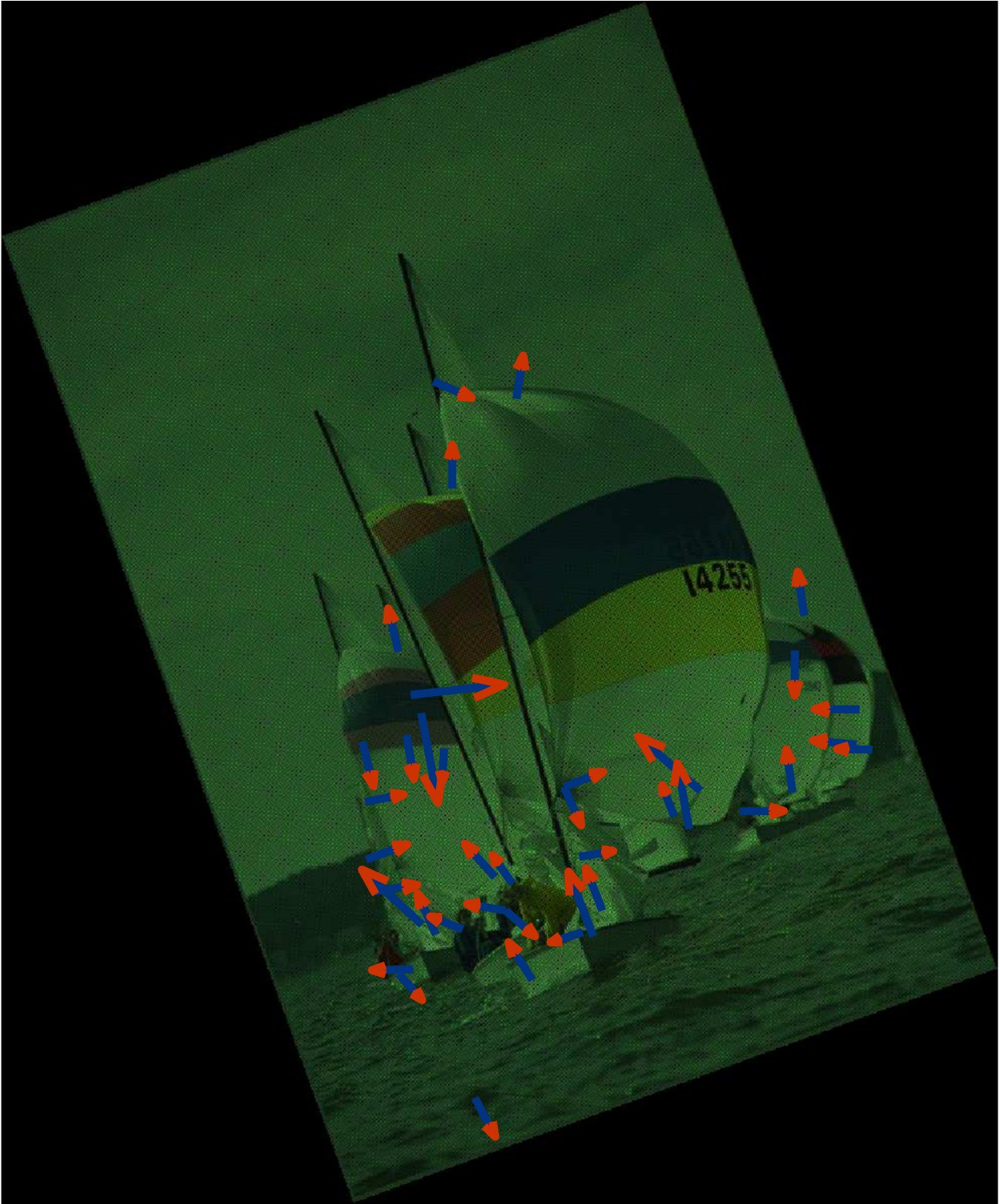}}\\
\subfigure[]{
\label{fig:key and match:graykeyR20} 
\includegraphics[width=1.4in]{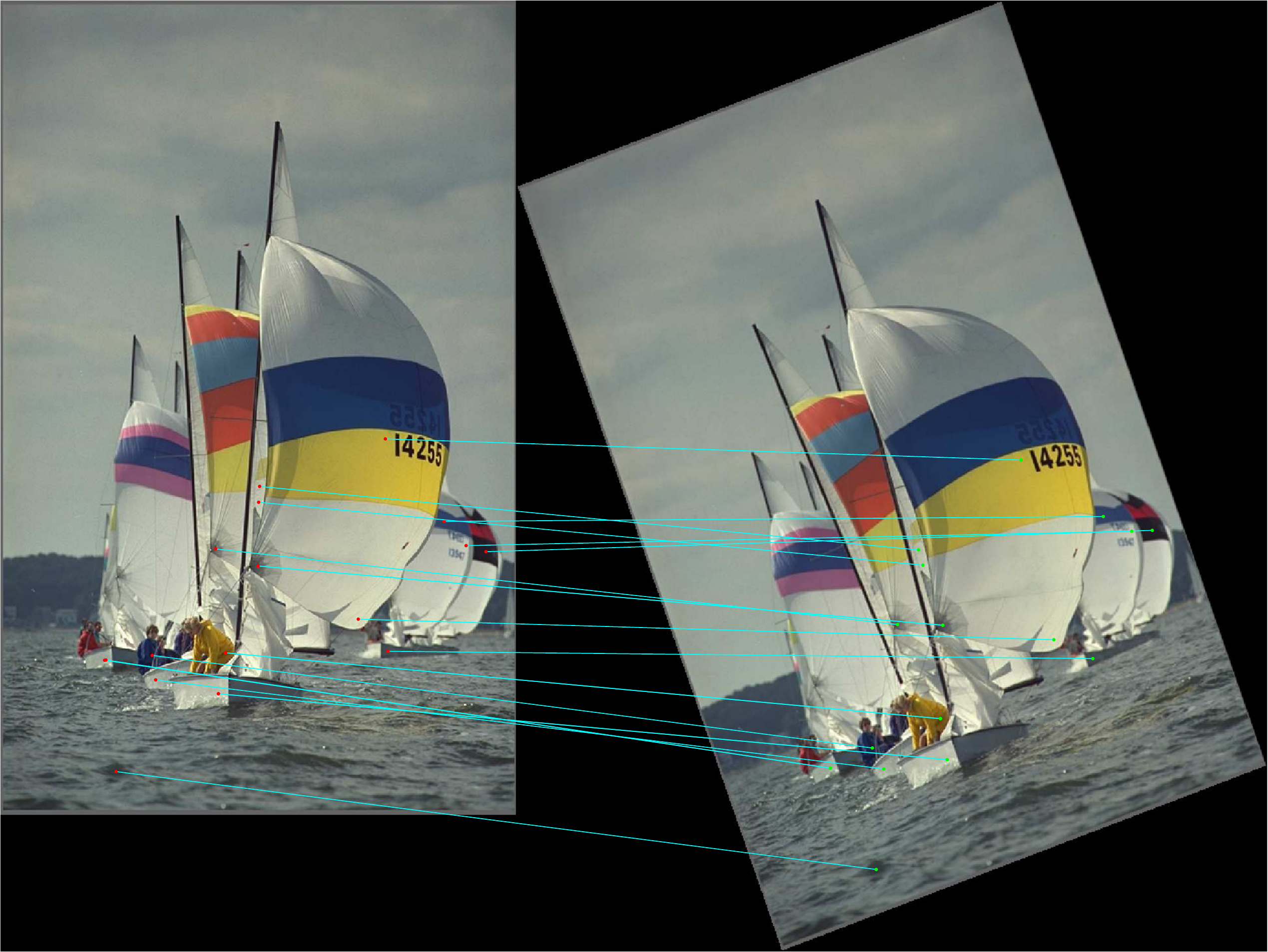}}
\hspace{-1mm}
\subfigure[]{
\label{fig:key and match:bayerkeyR20} 
\includegraphics[width=1.4in]{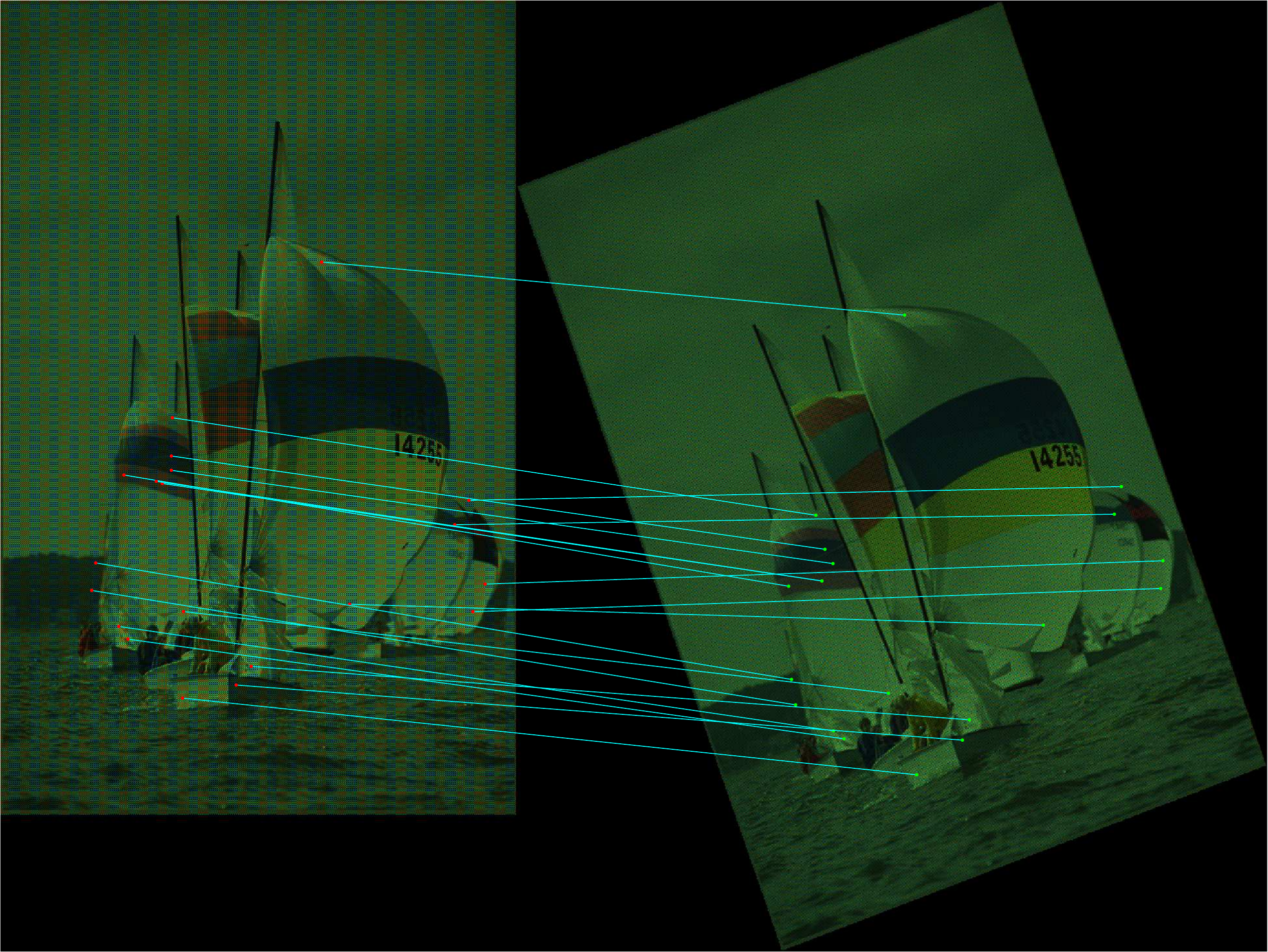}}\\
\centering
\subfigure[]{
\label{fig:key and match:graymatch} 
\includegraphics[width=1.4in]{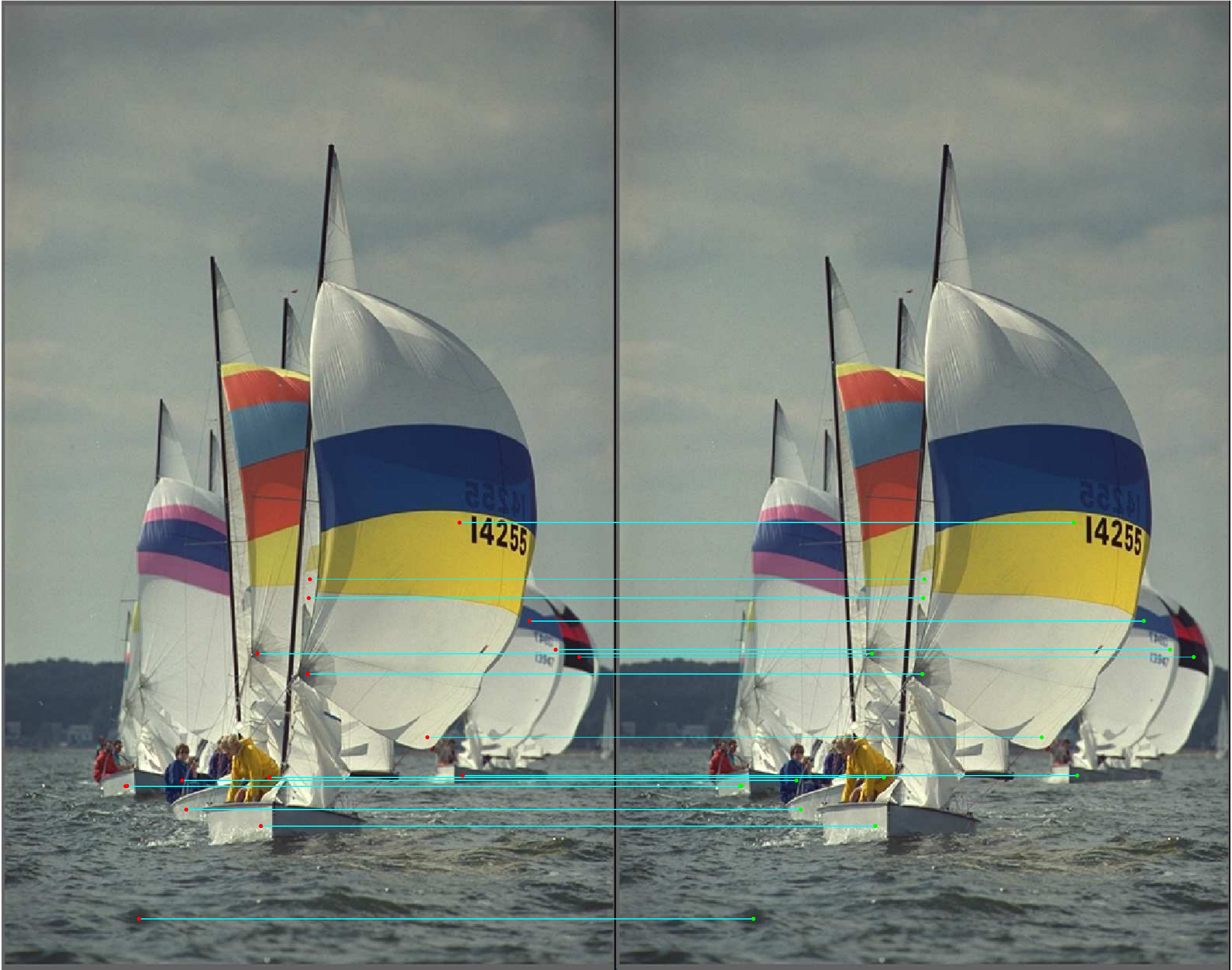}}
\hspace{-1mm}
\subfigure[]{
\label{fig:key and match:bayermatch} 
\includegraphics[width=1.4in]{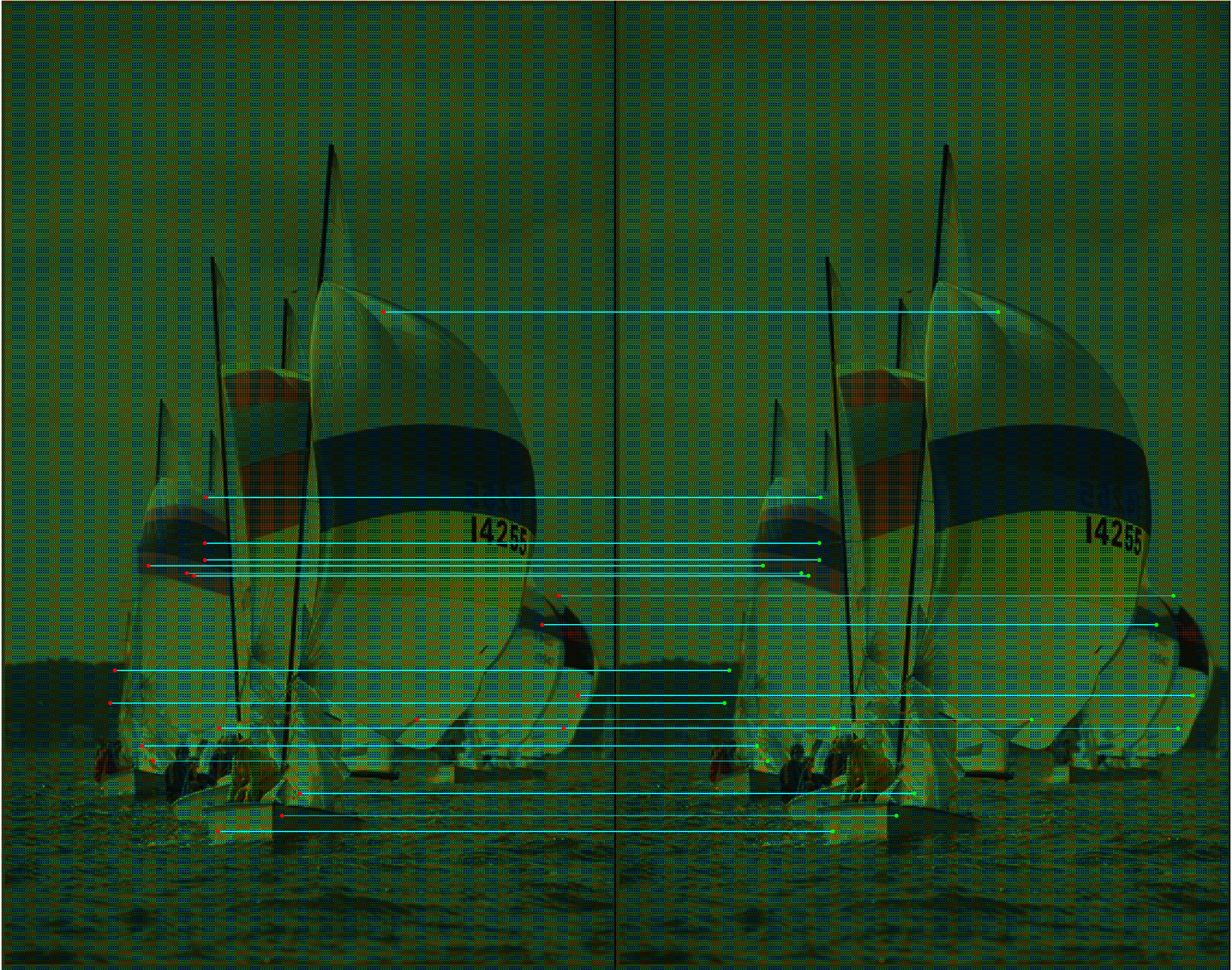}}
\caption{(a)-(d): Part of the SIFT descriptors in the original Kodim09 image, its Bayer version and corresponding 20-degree-rotate version. (e)-(f): Twenty matches in (c) and (d). (g)-(h): Projecting the matches in (e) and (f) back to the location in (a) and (b) by homography matrix $H$.}
\label{fig:key and match}
\vspace{-3mm}
\end{figure}

\begin{figure}
  \centering
  \subfigure[]{
  \label{fig:SIFT Rst1}
  \includegraphics[width=3.5in]{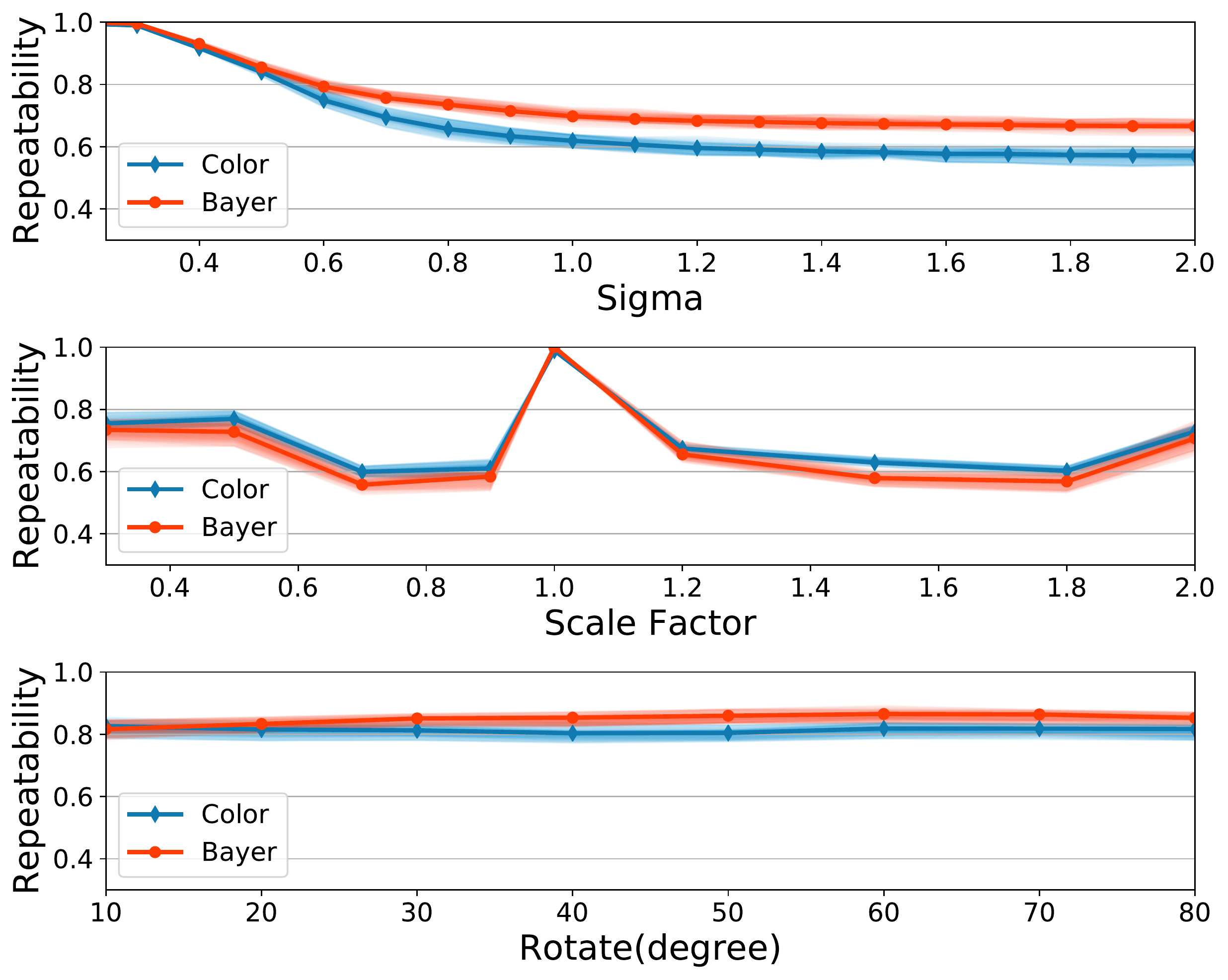}}\\
  \subfigure[]{
  \label{fig:SIFT Rst2}
  \includegraphics[width=3.5in]{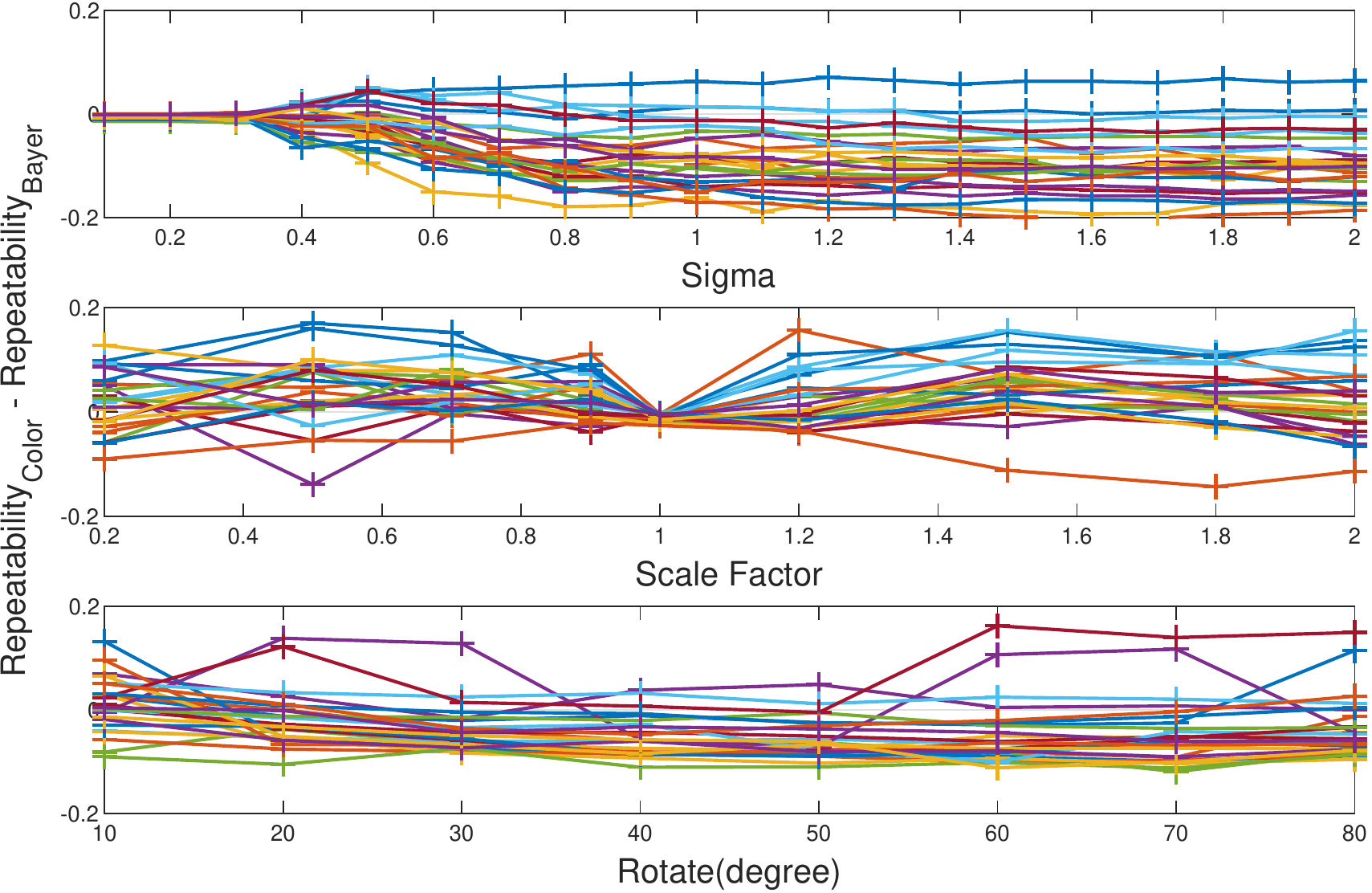}}
  \caption{(a) Average repeatability of SIFT descriptor after blur, scale change and rotate on Kodak dataset. The shaded area indicates the 25-75\% quantile band. (b) $\rm{Repeatability_{Color}}- \rm{Repeatability_{Bayer}}$ for each image.}
  \label{fig:SIFT Rst}
\end{figure}

\begin{figure}[t]
\centering
\subfigure[]{
\label{fig:Pedestrain detection Rst:InriaRst} 
\includegraphics[width=2.5in]{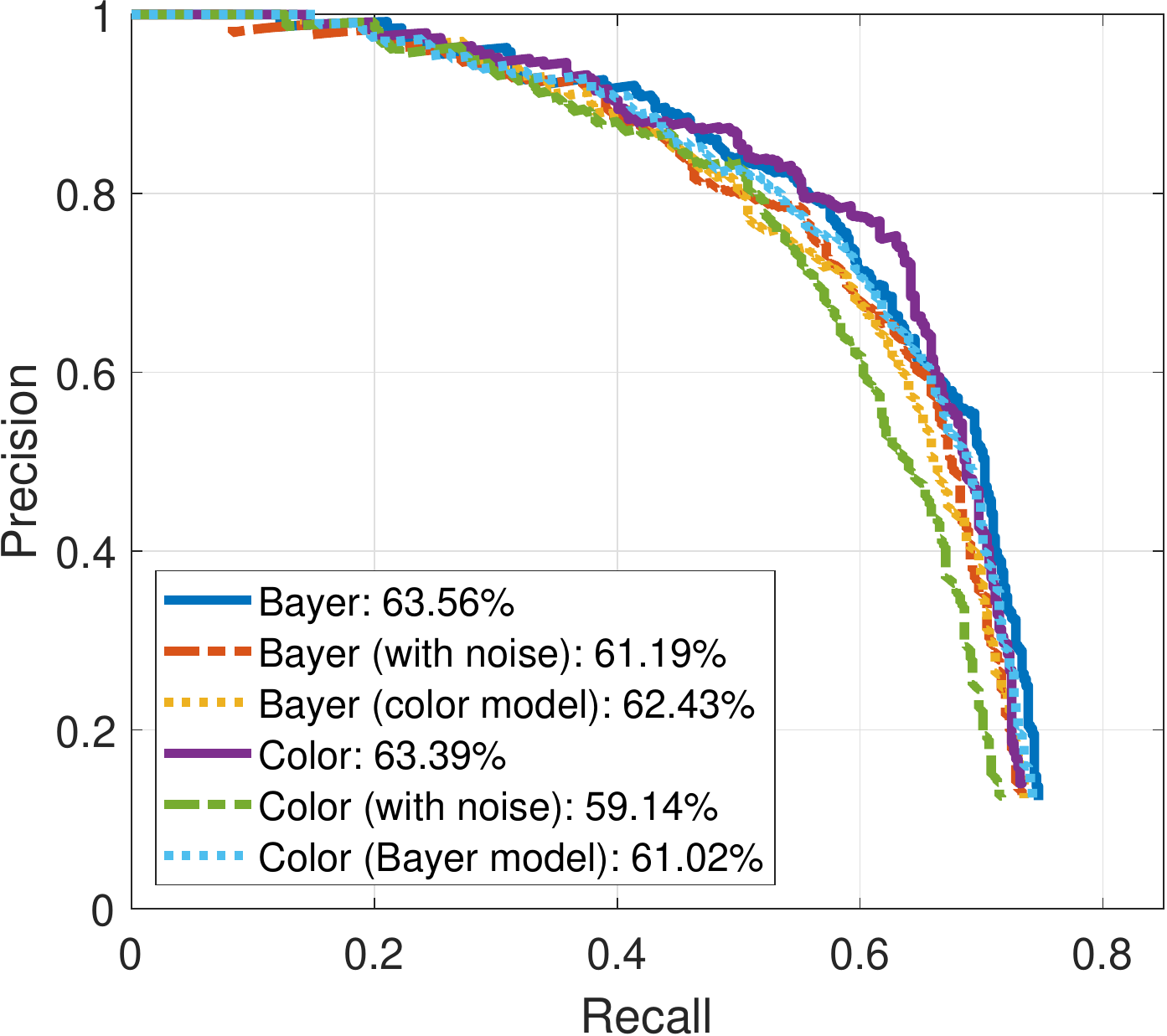}}\\
\vspace{-3mm}
\subfigure[]{
\label{fig:Pedestrain detection Rst:SHtechRst} 
\includegraphics[width=2.5in]{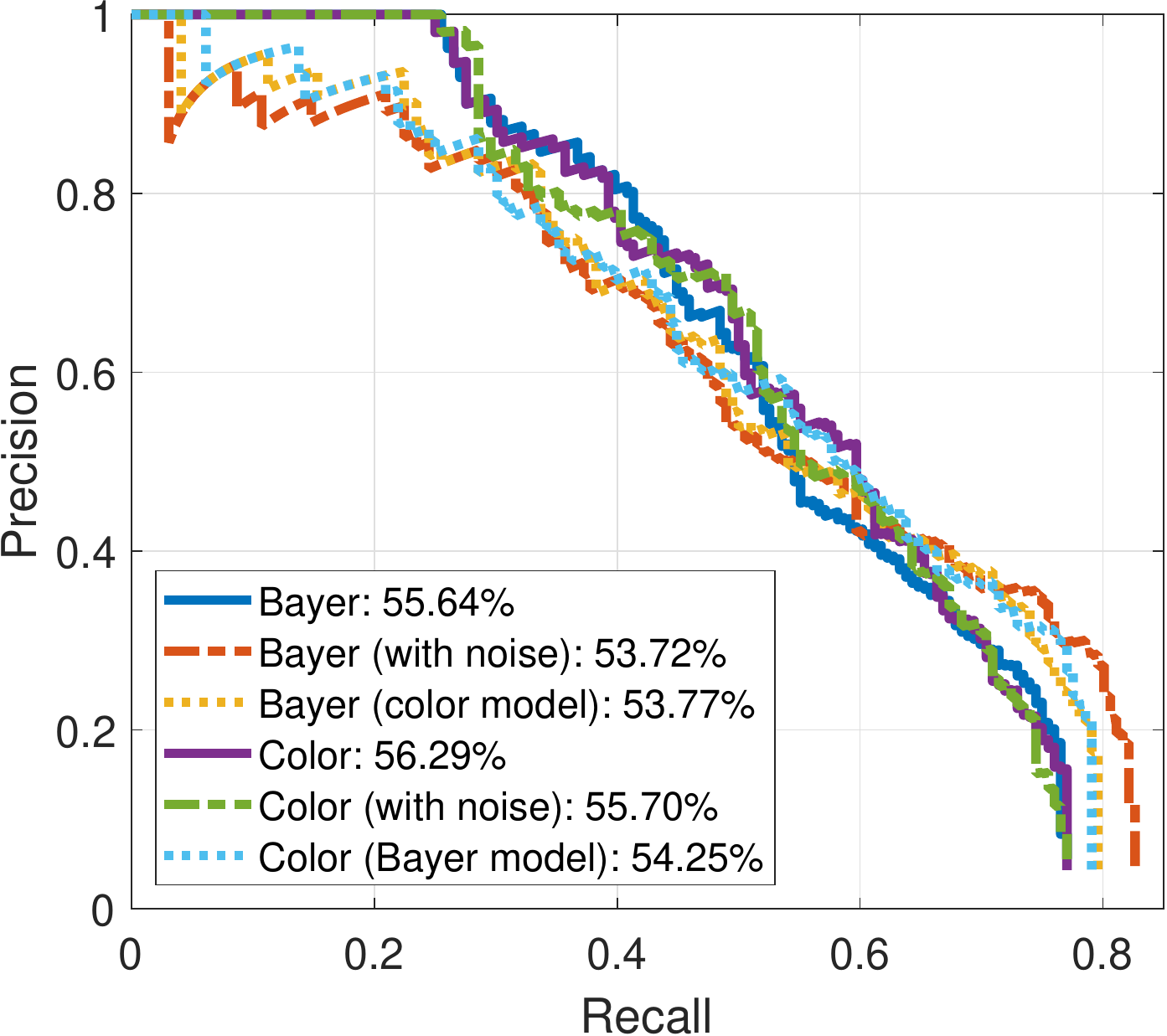}}\\
\vspace{-3mm}
\subfigure[]{
\label{fig:Pedestrain detection Rst:StanfordRst} 
\includegraphics[width=2.5in]{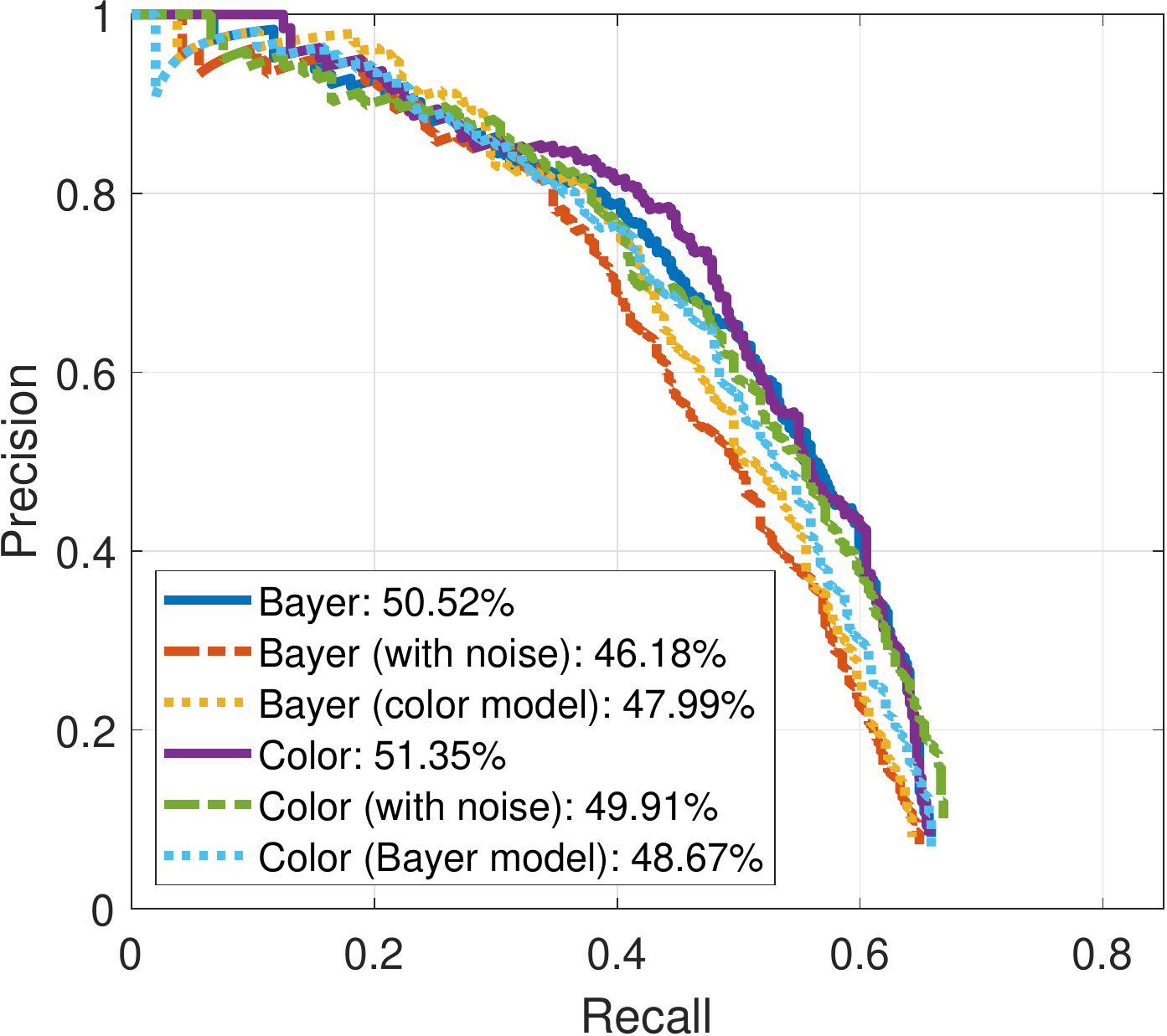}}
\caption{Evaluation of pedestrian detection performance based on different versions of gradients using (a) the INRIA dataset, (b) the SHTech dataset and (c) the PASCALRAW dataset. Here Bayer (color model) means the model is trained on gradient from color images and tested on gradients from Bayer images while Color (Bayer model) means the opposite.}
\label{fig:Pedestrain detection Rst}
\vspace{-4mm}
\end{figure}

\begin{figure}[t]
\centering
\subfigure[]{
\label{fig:noise_rst:original} 
\includegraphics[width=1.7in]{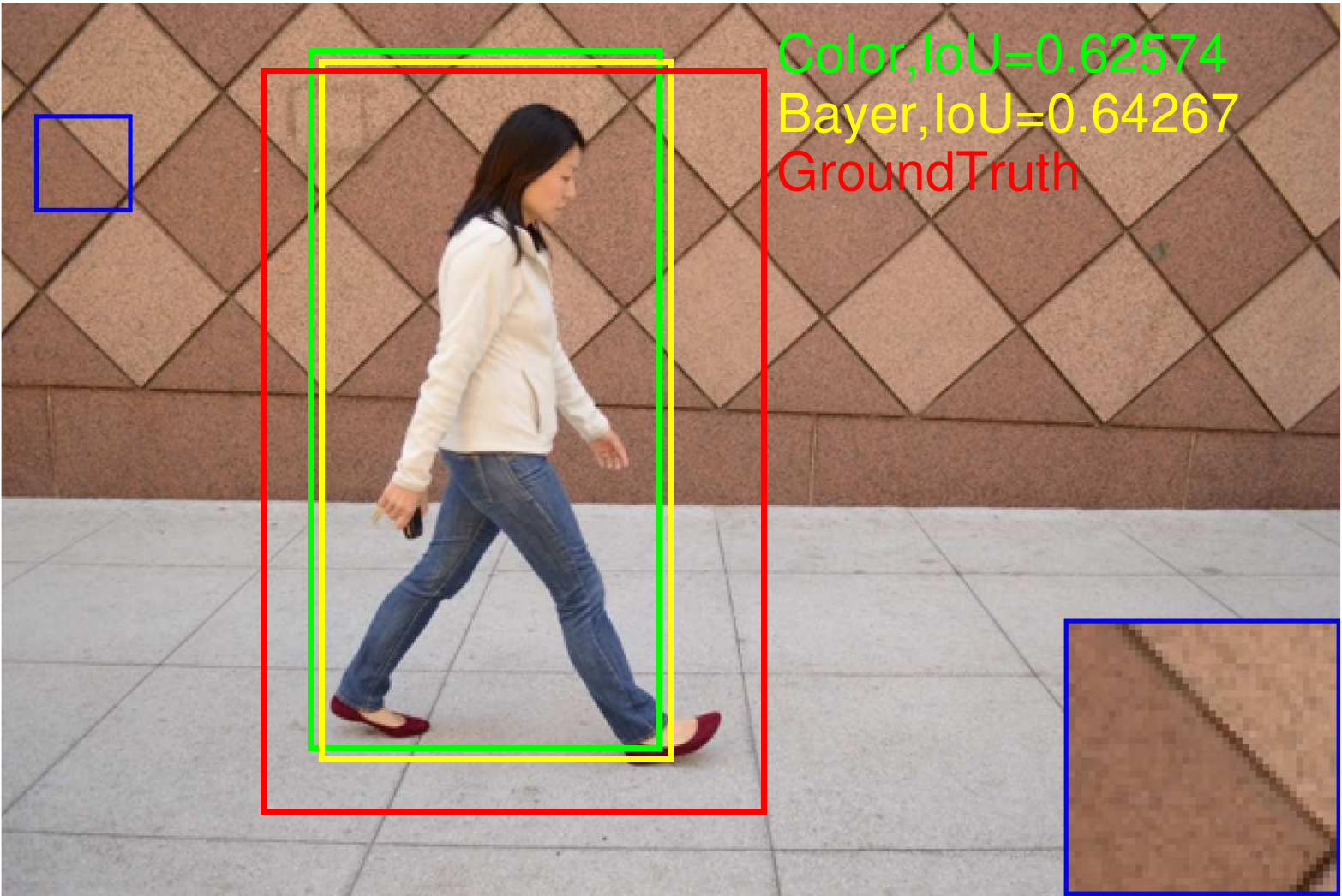}}
\hspace{-3mm}
\subfigure[]{
\label{fig:noise_rst:slight} 
\includegraphics[width=1.7in]{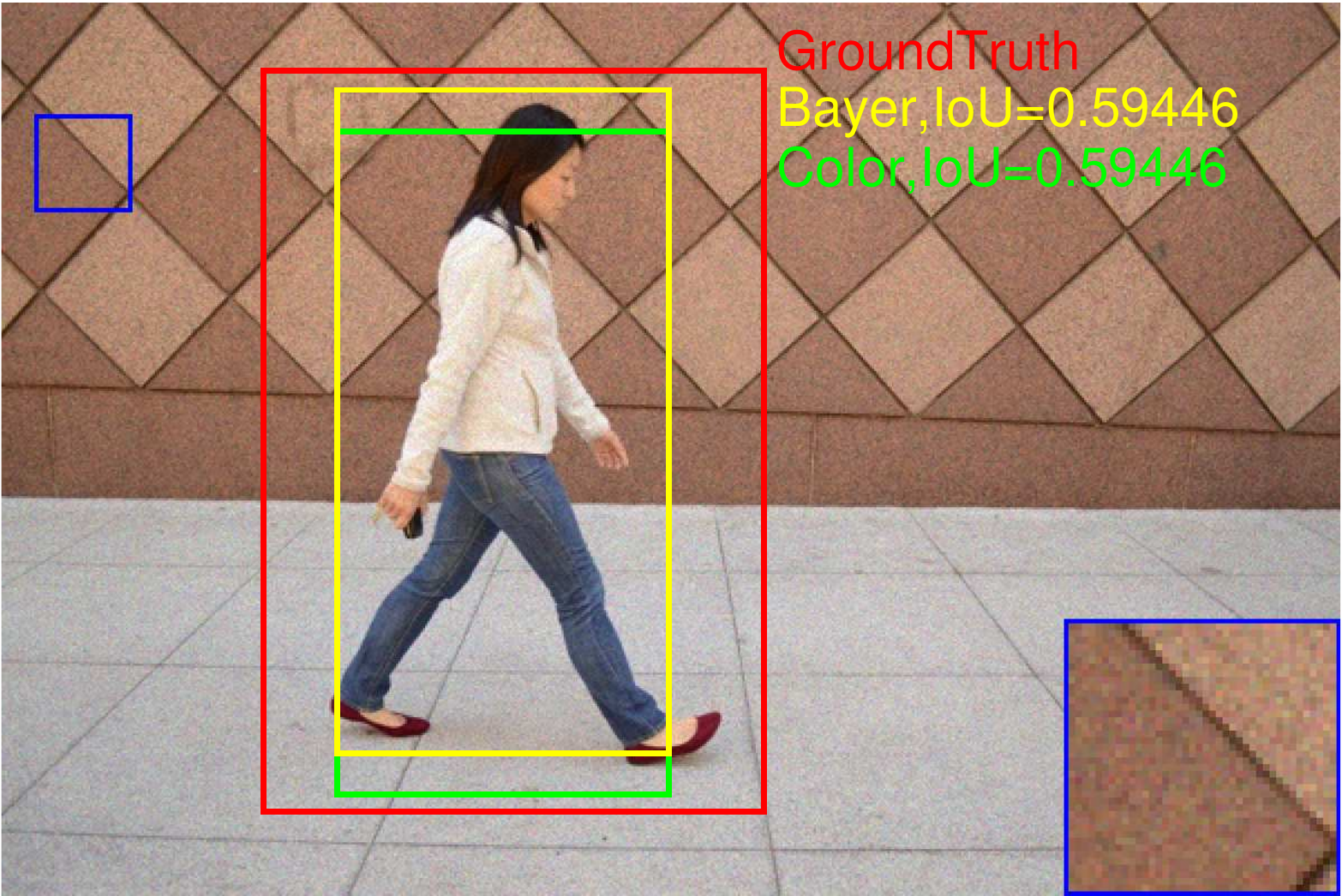}}
\hspace{-3mm}
\subfigure[]{
\label{fig:noise_rst:mid} 
\includegraphics[width=1.7in]{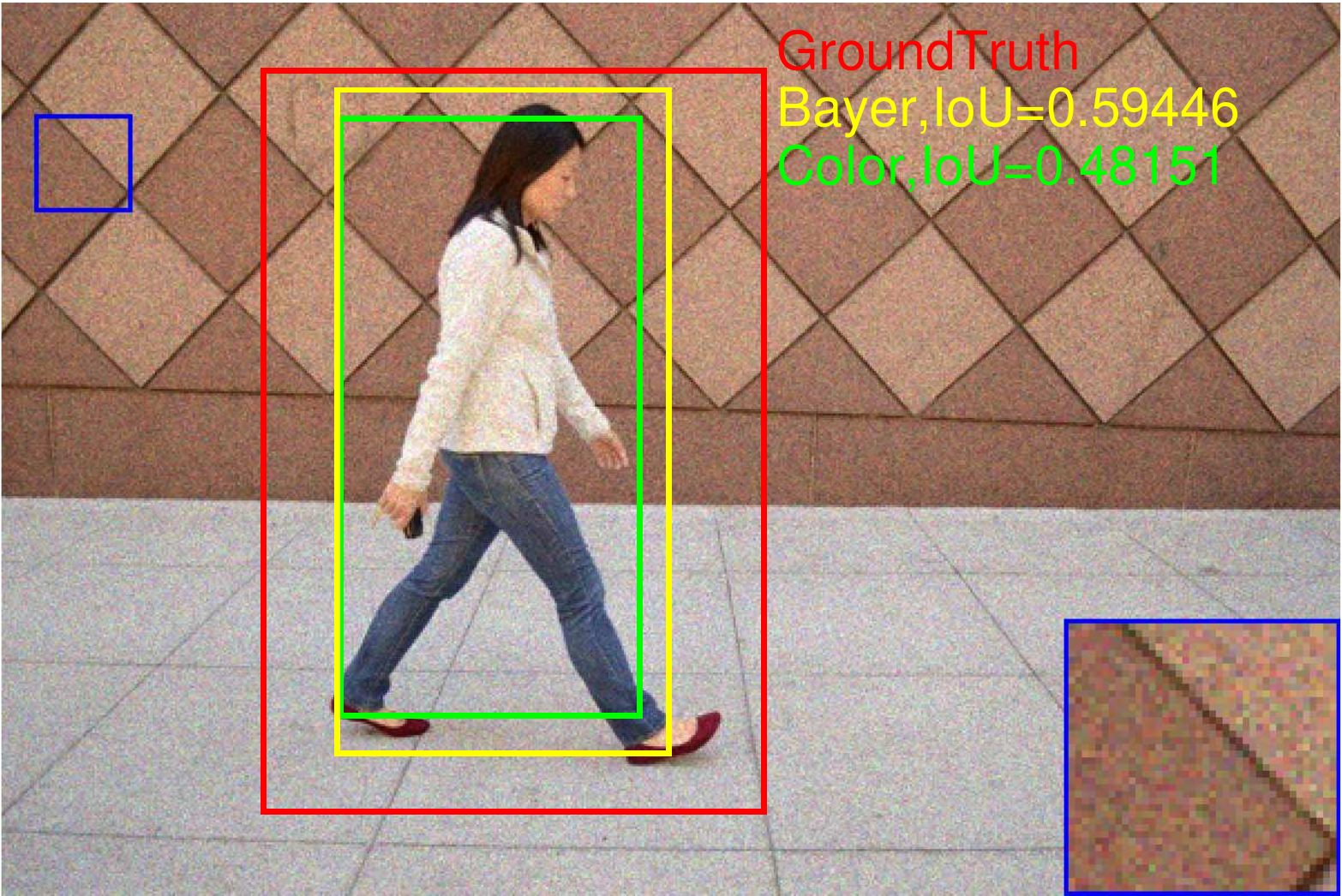}}
\hspace{-3mm}
\subfigure[]{
\label{fig:noise_rst:heavy} 
\includegraphics[width=1.7in]{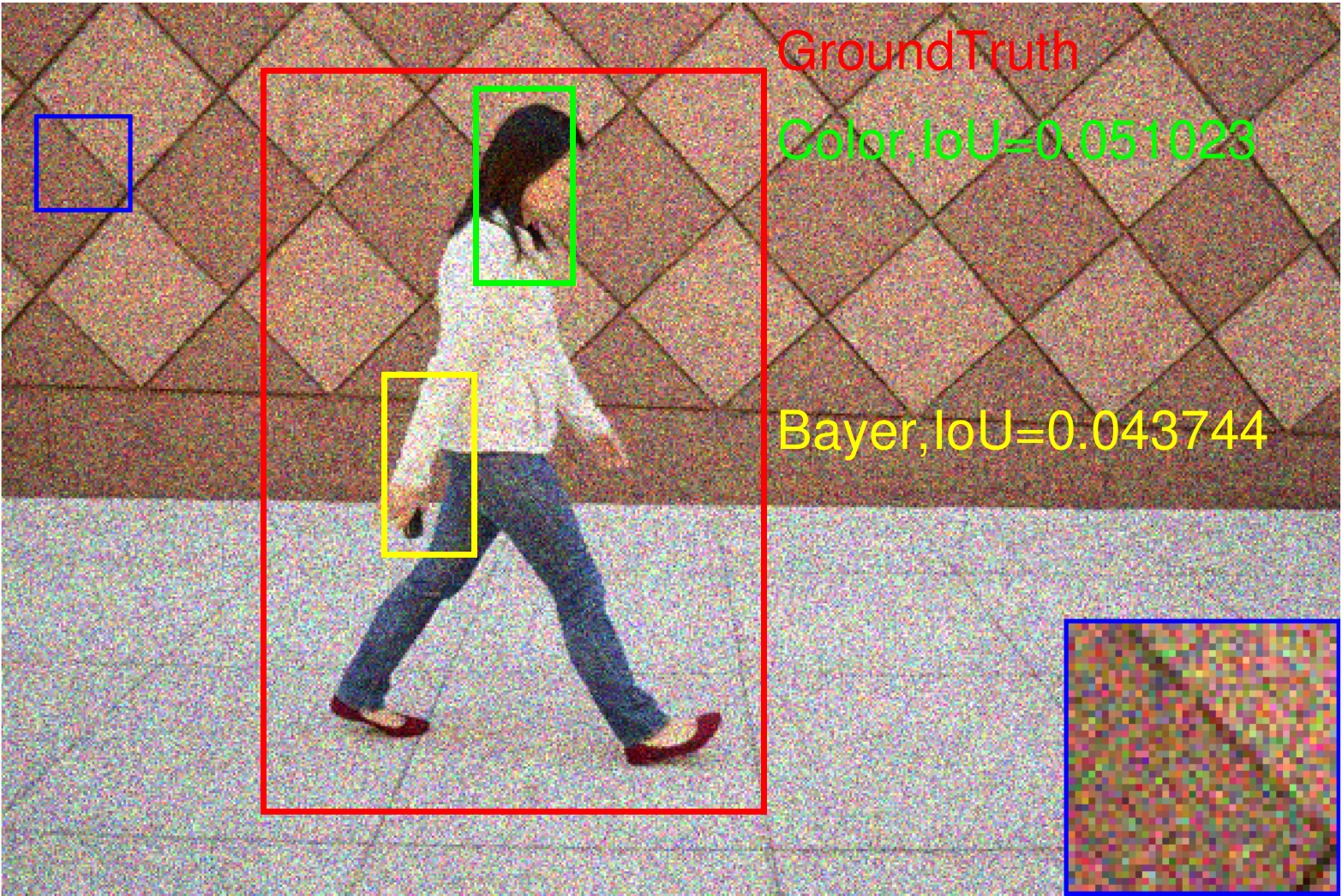}}
\caption{The pedestrian detection results on (a) the original image, (b) noisy image with parameters of $a=9.63\times10^{-4}, b=3.43\times10^{-5}$, (c) $a=4.80\times10^{-3}, b=2.00\times10^{-4}$, (d) $a=3.59\times10^{-2}, b=3.40\times10^{-3}$.} 
\label{fig:noise rst}
\vspace{-3mm}
\end{figure}

\subsubsection{Key points matching}

Fig. \ref{fig:key and match} illustrates the key point matching performance based on SIFT feature using the original color version of  Kodim09 image and its resampled Bayer version. The arrows in Fig. \ref{fig:key and match:graykey}-\ref{fig:key and match:bayerkey20} illustrate the scale and orientation of 20 SIFT descriptors, the cyan lines in Fig \ref{fig:key and match:graykeyR20}-\ref{fig:key and match:bayermatch} indicate the matched point pairs. As we can see, matched points can be identified in both color and Bayer pattern image pairs, meaning that the SIFT features extracted from the Bayer pattern images are robust regardless of the rotate operation. Note that the SIFT descriptors extracted from Bayer pattern image look different from that extracted from gray image. This is because we generate DoG pyramids of Bayer image based on super-pixel structure and extrema are searched among a $5\times5$ neighborhood instead of $3\times3$. It is the difference in DoG pyramids and search area that mainly lead to different descriptors between gray images and Bayer images.

To evaluate the scale and rotation invariance of the SIFT descriptor, the original images and the Bayer pattern images are transformed into different versions by blurring, scaling and rotating. The repeatability among each image is evaluated using the criteria mentioned in Section \ref{section:sec4.2}. To maintain the `mosaic' structure, rotation on Bayer pattern images are performed by extracting the pixels with the same color from the Bayer images to form four sub-images, i.e., $R,G_{1},G_{2}$ and $B$, rotating them separately and reorganizing them back into Bayer pattern images. Note that this process is just for generating experimental samples. Three scales are used in our experiments ($s=3$ in Equation \eqref{equ:DOG}). Euclidean distance is used as the distance measurement between a pair of matching pixels and threshold $t$ in \eqref{equ:distance} is set to 3. Estimation of $H$ from Bayer and Gray images is highly accurate and the difference between both approaches is negligible (see supplemental material).

Fig. \ref{fig:SIFT Rst1} depicts the average repeatability scores for both color and Bayer version. As it can be observed, the  curves in Fig. \ref{fig:SIFT Rst1} are very close to each other, with the Bayer version performs slightly better in the blur and rotation experiment while slightly worse in the scale experiment.
Fig. \ref{fig:SIFT Rst2} illustrates the difference of repeatability for each image. The repeatability on Bayer pattern images is generally better than that on color images for the Blur operation. This may because the stages in the ISP pipeline will introduce some extra `blur' effects. For scale and rotate operation, outliers often appear in images with rich textures, e.g., the top left image in Fig. \ref{fig:MSSIM}, where failure cases are more likely to appear.

\subsubsection{Pedestrian detection}
The HOG+SVM model is used as benchmark framework to evaluate the performance of the proposed Bayer pattern image-based gradients in object detection algorithms. Fig. \ref{fig:Pedestrain detection Rst} shows the pedestrian detection results on INRIA, SHTech and PASCALRAW datasets. As we can see, the performances of detection rate versus false positive per image are very close for different versions of images.

As shown in Fig. \ref{fig:Pedestrain detection Rst:InriaRst}, HOG+SVM achieves 63.56\% average precision (AP) on Bayer version of the INRIA dataset, compared to 63.39\% on the original INRIA dataset. The results are similar on the SHTech dataset, while the average precision in SHTech dataset is worse than that in INRIA dataset for both Bayer and color version. This is due to the difference in the number and posture of the dataset samples. In PASCALRAW datasets, the detection rate for Bayer version is also close to its color version counterpart.
Therefore, the gradients extracted directly from the Bayer pattern images are robust enough to be used in pedestrian detection algorithm, while the performance can be maintained.

We also conduct transfer experiments, i.e., train a classifier on gradients from color images and evaluated on gradients from Bayer images, and vice versa. From the results presented in Fig. \ref{fig:Pedestrain detection Rst}, it is found that there are small decreases in performance when a detector is not trained on the same version. But the decreases are very small, which means the gradients generated from color images are very close to that generated from Bayer images.

The pedestrian detection performance under the influence of noise is also presented in Fig. \ref{fig:Pedestrain detection Rst}. Note that in this experiment, the models are not retrained, i.e., the models trained using the noise-free images are used for pedestrian detection in noisy images. It can be found that the detection performance decreases slightly on all three datasets for both Bayer and color versions. The detection results on one of the images with different noise level are shown in Fig. \ref{fig:noise_rst:original}-(d). It is found that with the increase of noise level, the bounding boxes tend to be smaller. As shown in Fig. \ref{fig:noise_rst:heavy}, where the severest noise parameters are applied, the model seems not working for both Bayer and color versions.

\section{Discussion}
\label{section:sec5}
The objective of computer vision is to obtain high-level understanding from images and videos. Traditional vision algorithms take fully rendered color images as inputs. However, in scenarios where color is not required, such as the gradient-based algorithms discussed in this paper, demosaicing is redundant. It not only costs computing time, but also wastes three times the storage space to get almost the same results.

It has been shown in \cite{RN23} that in a conventional computer vision system consisting of an image sensor, an image signal processor and a vision processor (to run the computer vision algorithms), the image signal processor consumes a significant amount computation resources, processing time and power.
For example, a well-designed HOG processor consumes only 45.3 mW to process 1080P videos at 60 frames per second (FPS) \cite{RN64}, while a typical image signal processor dissipates around 250 mW to process videos with the same resolution and frame rate \cite{RN63}.
Therefore, from the system perspective, if we can skip the ISP pipeline (or most of the ISP steps), the computational complexity and power consumption of the computer vision system can be reduced significantly. Even in some features where color information is necessary, such as integral channel features (ICF) \cite{BMVC.23.91:abbreviated} or color descriptors in SIFT family \cite{RN11}, the location of demosaicing in the ISP pipeline need to be reconsidered. This is because as long as the mosaic structure is maintained, color information can be recovered whenever it is needed, through demosaicing for example. Moreover, though this paper shows that gradients extracted from Bayer pattern images are close to that from color images, the optimality of color image-based gradients extraction deserves a careful reconsideration. According to our understanding, the ISP pipeline and computer vision algorithms need to be co-designed for better performance.

This paper presents a method and corresponding analysis to extract gradient-based features from raw Bayer pattern images. But there are some limitations. The applicability of the proposed method is influenced by the relationship between gradient operators and CFA patterns. To make the proposed approach applicable, it is crucial to ensure that the gradient calculation is performed on pixels from the same color channel, i.e., subtract or add operations are performed on the same color channel, and the coefficients of the subtract or add terms in the gradient operator are equal such that the gradients compute from R/B channel can be approximated to G channel. Moreover, although the method hold in flat areas and some non-smooth texture areas when computing gradients, there are failure cases which not satisfy the model's assumption.

\section{Conclusion}
\label{section:sec6}
In this paper, the impact of demosaicing on gradient extraction is studied and a gradient-based feature extraction pipeline based on raw Bayer pattern images is proposed. It is shown both theoretically and experimentally that the Bayer pattern images are applicable to the central difference gradient-based algorithms with negligible performance degradation. The color difference constancy assumption, which is widely used in various demosaicing algorithms, is applied in the proposed Bayer pattern image-based gradient extraction pipeline. Experimental results show that the gradients extracted from Bayer pattern images are robust enough to be used in HOG-based pedestrian detection algorithms and SIFT-based matching algorithms. Therefore, if gradient is the only information needed in a vision algorithm, the ISP pipeline (or most of the ISP steps) can be eliminated to reduced the computational complexity as well as power consumption of the systems.

\bibliographystyle{ieeetr}
\bibliography{Gradient-based_Feature_Extraction_From_Raw_Bayer_Pattern_Images}

\newpage
\vspace{-8mm}
\begin{IEEEbiography}[{\includegraphics[width=0.83in,clip,keepaspectratio]{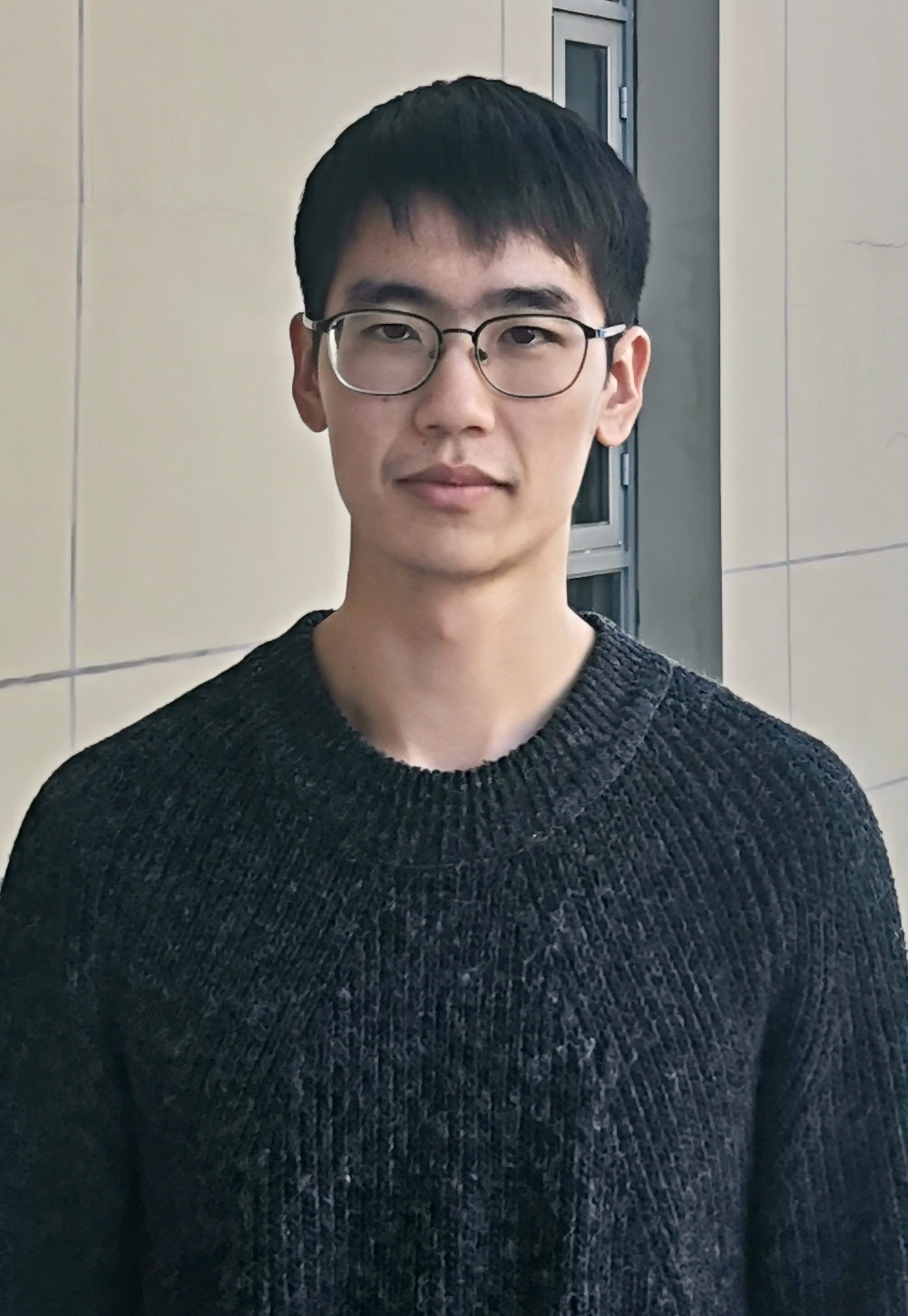}}]{Wei Zhou}
received the B.Eng. degree in instrument science and engineering from Southeast University, Nanjing, China, in 2018. He is currently pursuing his master's degree in electronic science and technology at ShanghaiTech University, Shanghai, China. His research interests include digital image processing and computer vision.
\end{IEEEbiography}

\begin{IEEEbiography}[{\includegraphics[width=1in,height=1.25in,clip,keepaspectratio]{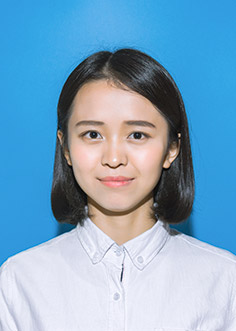}}]{Ling Zhang}  received the B.Eng. degree in electrical engineering from Xidian University, Xi'an, China, in 2018. She is currently working toward the master's degree at ShanghaiTech University, Shanghai, China. From Sep. 2018, she is with the VLSI Signal Processing Lab at School of Information Science and Technology, ShanghaiTech University. Her research interests include computer vision accelerator design, specially pedestrian detection circuits and systems.
\end{IEEEbiography}

\begin{IEEEbiography}[{\includegraphics[width=1in,height=1.25in,clip,keepaspectratio]{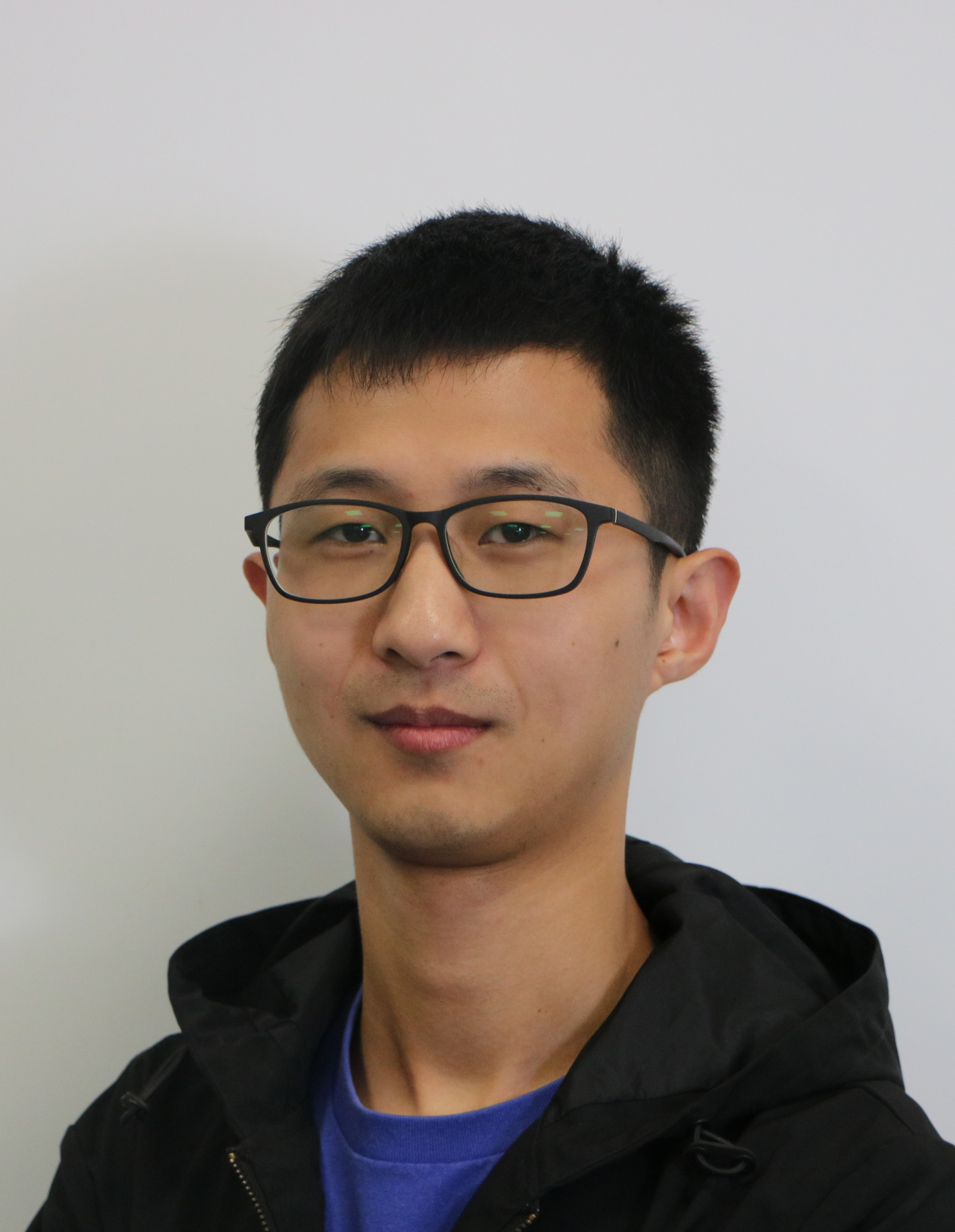}}]{Shengyu Gao}received the B.Eng. degree in electrical engineering from Wuhan University of Technology, China, in 2018. He is currently working toward the master's degree at ShanghaiTech University, Shanghai, China. His research interests include stereo vision-related topics.
\end{IEEEbiography}

\begin{IEEEbiography}[{\includegraphics[width=0.83in]{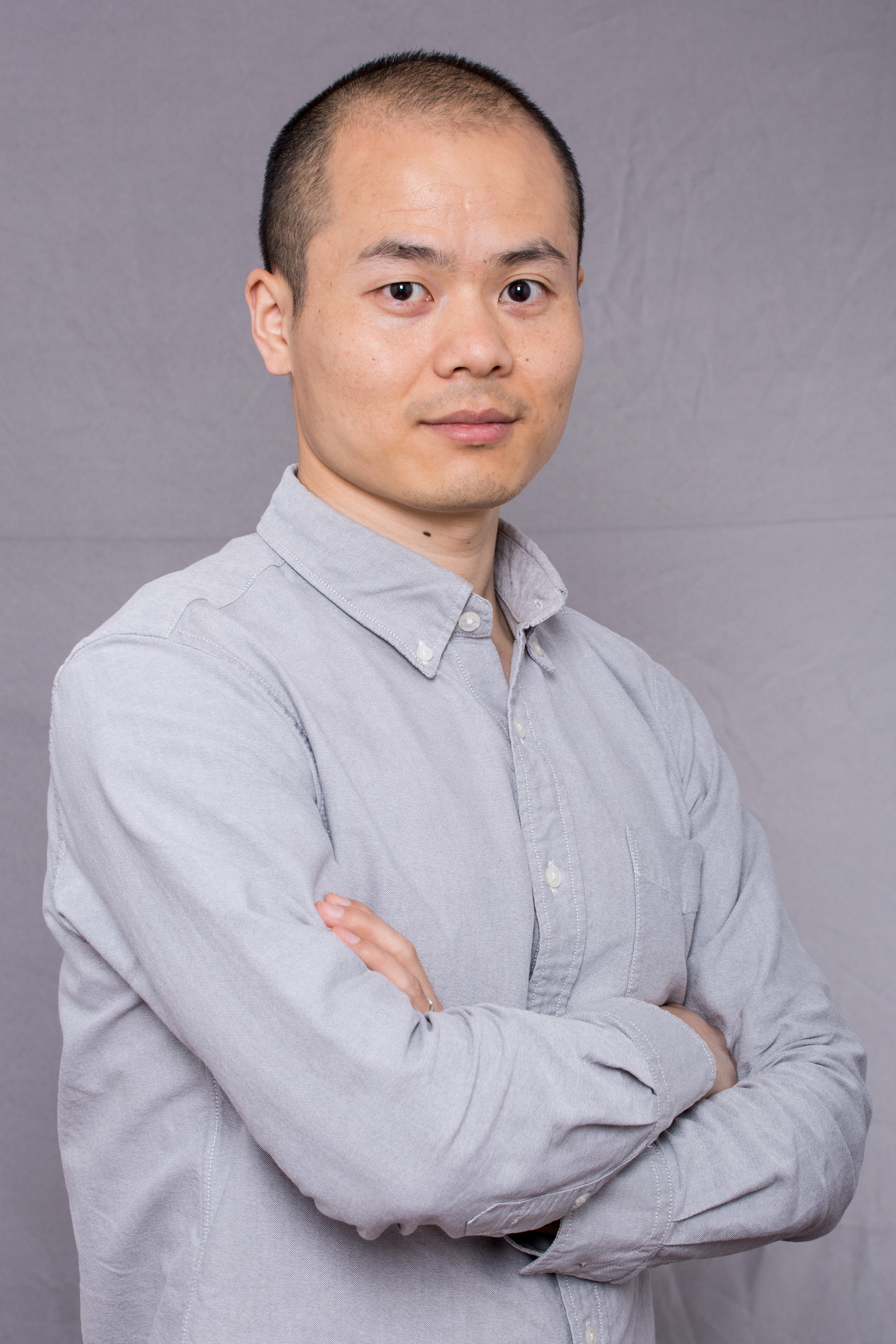}}]{Xin Lou (M’17)}
received the B.Eng. degree in electronic information technology and instruments from Zhejiang University, Hangzhou, China, in 2010, the M.Sc. degree in electrical engineering from the Royal Institute of Technology, Sweden, in 2012, and the Ph.D. degree in electrical and electronic engineering from Nanyang Technological University, Singapore, in 2016. From 2016 to 2017, he was a Research Scientist with Nanyang Technological University, Singapore. Since 2017, he has been with the School of Information Science and Technology, ShanghaiTech University, where he is currently an Assistant Professor. His research interests include VLSI digital signal processing and smart vision circuits and systems.
\end{IEEEbiography}

\end{document}